\documentclass[12pt]{article}
\usepackage{amsmath,amssymb,amsfonts}
\usepackage{fullpage}
\usepackage{enumitem}
\usepackage{caption}
\usepackage{shuffle}
\usepackage{subcaption}
\usepackage{graphicx}
\usepackage{slashed}
\usepackage{booktabs}
\usepackage{mciteplus}

\usepackage[bbgreekl]{mathbbol}
\DeclareSymbolFontAlphabet{\mathbbm}{bbold}
\DeclareSymbolFontAlphabet{\mathbb}{AMSb}

\usepackage{xcolor}
\usepackage{bm}
\usepackage{fancyhdr, dsfont}
\usepackage{float}
\usepackage{url}
\usepackage{setspace}
\usepackage[nosort]{cite}
\usepackage{mathtools}
\allowdisplaybreaks

\usepackage{color}

\usepackage{hyperref}
\definecolor{darkred}{rgb}{0.8,0.1,0.1}
\hypersetup{colorlinks=true, linkcolor=darkred, urlcolor=darkred,citecolor=blue, linktoc=page}

\usepackage{tikz}
\usetikzlibrary{calc,angles} 
\usetikzlibrary{patterns} 
\usetikzlibrary{decorations.pathreplacing} 
\usetikzlibrary{decorations.markings} 
\usetikzlibrary{decorations.pathmorphing} 
\usetikzlibrary{positioning}
\usetikzlibrary{arrows.meta}

\def\beq{\begin{equation}}
\def\eeq{\end{equation}}

\newtheorem{lemma}{Lemma}

\def\nn{\nonumber}

\let\Re\relax
\let\Im\relax
\DeclareMathOperator{\Re}{Re}
\DeclareMathOperator{\Im}{Im}

\newcommand{\nwc}{\newcommand}
\nwc{\ba}  {\begin{array}}
\nwc{\ea}  {\end{array}}
\nwc{\bdm} {\begin{displaymath}}
\nwc{\edm} {\end{displaymath}}
\nwc{\bda} {\bdm\ba{lcl}} 
\nwc{\eda} {\ea\edm}
\nwc{\bc}  {\begin{center}}
\nwc{\ec}  {\end{center}}
\nwc{\ds}  {\displaystyle}
\nwc{\bmat}{\left(\ba}
\nwc{\emat}{\ea\right)}
\nwc{\nnn} {\nonumber \vspace{.2cm} \\ }
\nwc{\ra}  {\rightarrow}
\nwc{\lra} {\longrightarrow}
\nwc{\p} {\partial}
\nwc{\rcr} {\nabla_{\rm alt}}
\nwc{\barrcr} {\overline{\nabla_{\rm alt}}}
\nwc{\ep} {\epsilon}

\def\ad{\mathrm{ad}}
\def\dd{\text{d}}
\def\mot{\mathfrak{m}}
\def\BF{{\rm BF}}

\newcommand{\MMV}[2]{\mathfrak{m}\! \left[\begin{smallmatrix}#1\\#2\end{smallmatrix}\right]}

\newcommand{\ee}[3]{{\cal E}\! \left[\begin{smallmatrix}#1\\#2\end{smallmatrix};#3\right]}

\newcommand{\eeno}[2]{{\cal E}\! \left[\begin{smallmatrix}#1\\#2\end{smallmatrix}\right]}

\newcommand{\ccsv}[2]{c^{\rm sv}\! \left[\begin{smallmatrix}#1\\#2\end{smallmatrix}\right]}
\newcommand{\bsv}[2]{\beta^{\rm sv}\! \left[\begin{smallmatrix}#1\\#2\end{smallmatrix}\right]}
\newcommand{\beqv}[2]{\beta^{\rm eqv}\! \left[\begin{smallmatrix}#1\\#2\end{smallmatrix}\right]}
\newcommand{\beqvtau}[3]{\beta^{\rm eqv}\! \left[\begin{smallmatrix}#1\\#2\end{smallmatrix};#3\right]}
\newcommand{\eeqv}[2]{{\cal E}^{\rm eqv}\! \left[\begin{smallmatrix}#1\\#2\end{smallmatrix}\right]}
\newcommand{\eeqvtau}[3]{{\cal E}^{\rm eqv}\! \left[\begin{smallmatrix}#1\\#2\end{smallmatrix};#3\right]}
\newcommand{\ompm}[3]{\omega_{\pm}\! \left[\begin{smallmatrix}#1\\#2\end{smallmatrix};#3\right]}
\newcommand{\omplus}[3]{\omega_{+}\! \left[\begin{smallmatrix}#1\\#2\end{smallmatrix};#3\right]}
\newcommand{\omminus}[3]{\omega_{-}\! \left[\begin{smallmatrix}#1\\#2\end{smallmatrix};#3\right]}
\newcommand{\nuker}[3]{\nu\! \left[\begin{smallmatrix}#1\\#2\end{smallmatrix};#3\right]}

\newcommand{\bplus}[3]{\beta_+ \! \left[\begin{smallmatrix}#1\\#2\end{smallmatrix};#3\right]}
\newcommand{\bplusno}[2]{\beta_+ \! \left[\begin{smallmatrix}#1\\#2\end{smallmatrix}\right]}
\newcommand{\bminus}[3]{\beta_- \! \left[\begin{smallmatrix}#1\\#2\end{smallmatrix};#3\right]}
\newcommand{\bminusno}[2]{\beta_-\! \left[\begin{smallmatrix}#1\\#2\end{smallmatrix}\right]}
\newcommand{\bpm}[3]{\beta_{\pm} \! \left[\begin{smallmatrix}#1\\#2\end{smallmatrix};#3\right]}
\newcommand{\bpmno}[2]{\beta_{\pm} \! \left[\begin{smallmatrix}#1\\#2\end{smallmatrix}\right]}
\newcommand{\bmpno}[2]{\beta_{\mp} \! \left[\begin{smallmatrix}#1\\#2\end{smallmatrix}\right]}
\newcommand{\chicst}[4]{\chi^{#1}_{#2}\! \left[\begin{smallmatrix}#3\\#4\end{smallmatrix}\right]}
\newcommand{\xicst}[4]{\xi^{#1}_{#2}\! \left[\begin{smallmatrix}#3\\#4\end{smallmatrix}\right]}
\newcommand{\etacst}[4]{\eta^{#1}_{#2}\! \left[\begin{smallmatrix}#3\\#4\end{smallmatrix}\right]}
\newcommand{\wword}[2]{\word\! \left[\begin{smallmatrix}#1\\#2\end{smallmatrix}\right]}
\newcommand{\tce}[2]{\tilde{\cal E}\! \left[\begin{smallmatrix}#1\\#2\end{smallmatrix}\right]}

\newcommand{\GG}{ {\rm G} }
\newcommand{\EE}{ {\rm E} }

\newcommand{\Ip}{\mathbb{I}_+}

\newcommand{\Imint}{{\widetilde{\mathbb{I}}}_-}

\newcommand{\Jeqv}{\mathbb{J}^{\rm eqv}}
\newcommand{\Ieqv}{\mathbb{I}^{\rm eqv}}

\newcommand{\Unew}{{\rm U}_{{\rm SL}_2}}
\newcommand{\Msv}{\mathbb{M}^{{\rm sv}}}
\newcommand{\weyl}{\mathrm{w}}
\newcommand{\CSV}{\mathbb C^{\rm sv}(\eee_k)}
\newcommand{\CSVep}{\mathbb C^{\rm sv}(\ep_k)}

\newcommand{\word}{\mathrm{W}}

\newcommand{\zetaT}{z}
\newcommand{\sigmaT}{\sigma}
\newcommand{\zetaE}{\hat{z}}
\newcommand{\sigmaE}{\hat{\sigma}}

\newcommand{\Pexp}{{\rm P\text{-}exp}}
\newcommand{\Ptexp}{{\rm \widetilde{P}\text{-}exp}}

\newcommand{\pern}{\varpi}

\newcommand{\Fp}[3]{\mathrm{F}_{#1,#2}^{+(#3)}}
\newcommand{\Fm}[3]{\mathrm{F}_{#1,#2}^{-(#3)}}

\newcommand{\SLtwoZ}{\mathrm{SL}(2,\mathbb{Z})}
\newcommand{\UHP}{\mathbb{H}}
\newcommand{\BB}{\mathrm{B}}

\newcommand{\bbsvtau}[3]{b^{\rm sv}\! \left[\begin{smallmatrix}#1\\#2\end{smallmatrix};#3\right]}
\newcommand{\eee}{\mathrm{e}}
\newcommand{\hhh}{\mathrm{h}}
\newcommand{\mf}[1]{\mathfrak{#1}}

\newcommand{\degree}{degree}

\usepackage{stackengine}

\definecolor{cadmiumgreen}{RGB}{60,142,23}
\definecolor{kleinblue}{RGB}{0,46,167}
\definecolor{venetianred}{RGB}{192,6,21}
\definecolor{forestgreen}{RGB}{34,139,34}
\definecolor{electricviolet}{rgb}{0.56, 0.0, 1.0}

\usepackage{titlesec}
\titleformat*{\section}{\large \bfseries}

\begin{document}

 {\flushright  
 UUITP-09/24\\}

\begin{center}

\vspace{-1mm}

{\bf {\Large \sc 
Non-holomorphic modular forms\\[2mm] from zeta generators }}

\vspace{5mm}
\normalsize
{\large  Daniele Dorigoni$^1$,
Mehregan Doroudiani$^2$,
Joshua Drewitt$^3$, \\ \vskip 0.45 em
Martijn Hidding$^{4,5}$, 
Axel Kleinschmidt$^{2,6}$,
Oliver Schlotterer$^{4}$, \\ \vskip 0.45 em
Leila Schneps$^7$
and Bram Verbeek$^{4}$}

\vspace{5mm}
${}^1${\it Centre for Particle Theory \& Department of Mathematical Sciences\\
Durham University, Lower Mountjoy, Stockton Road, Durham DH1 3LE, UK}
\vskip 0.5 em
${}^2${\it Max-Planck-Institut f\"{u}r Gravitationsphysik (Albert-Einstein-Institut)\\
Am M\"{u}hlenberg 1, 14476 Potsdam, Germany}
\vskip 0.5 em
${}^3${\it School of Mathematics, University of Bristol, Queens Road, Bristol, BS8 1QU, UK}
\vskip 0.5 em
${}^4${\it Department of Physics and Astronomy, Uppsala University, 75108 Uppsala, Sweden}
\vskip 0.5 em
${}^5${\it Institute for Theoretical Physics, ETH Zurich, 8093 Z\"urich, Switzerland}
\vskip 0.5 em
${}^6${\it International Solvay Institutes, 
ULB-Campus Plaine CP231, 1050 Brussels, Belgium}
\vskip 0.5 em
${}^7${\it CNRS, Sorbonne Universit\'e, Campus Pierre et Marie Curie\\
4 place Jussieu, 75005 Paris, France}

\vspace{8mm}

\hrule

\vspace{5mm}

\begin{tabular}{p{150mm}}

We study non-holomorphic modular forms built from iterated integrals of holomorphic modular forms for $\SLtwoZ$ known as equivariant iterated Eisenstein integrals.
A special subclass of them furnishes an equivalent description of the modular graph forms appearing in the low-energy expansion of string amplitudes at genus one.
Notably the Fourier expansion of modular graph forms contains single-valued multiple zeta values. We deduce the appearance of products and higher-depth instances of multiple zeta values in equivariant iterated Eisenstein integrals, and ultimately modular graph forms, from the appearance of simpler odd Riemann zeta values. This analysis relies on so-called zeta generators which act on certain non-commutative variables in the generating series of the iterated integrals.
From an extension of these non-commutative variables we incorporate iterated integrals involving holomorphic cusp forms into our setup and use them to construct the modular completion of triple Eisenstein integrals. Our work represents a fully explicit realisation of the modular graph forms within Brown's framework of equivariant iterated Eisenstein integrals and reveals structural analogies between single-valued period functions appearing in genus zero and one string amplitudes.

\end{tabular}

\hrule

\end{center}

\thispagestyle{empty}

\newpage
\setcounter{page}{1}

\numberwithin{equation}{section}

\setcounter{tocdepth}{2}
\tableofcontents

\section{Introduction}
\label{sec:int}

The low-energy expansion of closed-string genus-one amplitudes introduces non-holomorphic modular forms for $\SLtwoZ$ \cite{Green:1999pv, Green:2008uj, DHoker:2015gmr, Gerken:2018jrq} known as {\it modular graph forms} (MGFs) \cite{DHoker:2015wxz, DHoker:2016mwo}. In the physics literature, the study of MGFs informed S-duality properties of low-energy interactions of type IIB superstrings \cite{Green:1999pv, Green:2008uj, DHoker:2015gmr, DHoker:2019blr, Dorigoni:2022iem} and manifested striking connections between open- and closed-string amplitudes \cite{DHoker:2015wxz, Broedel:2018izr, Gerken:2018jrq, Gerken:2020xfv}.

Due to the intriguing behaviour of MGFs both in their expansion at the cusp of the moduli space \cite{Green:2008uj, DHoker:2015gmr, DHoker:2015wxz} and in their differential structures \cite{DHoker:2015gmr, DHoker:2016mwo}, these objects have attracted considerable interest in the mathematics literature \cite{Brown:mmv,Zerbini:2015rss, Brown:2017qwo, Brown:2017qwo2, Zagier:2019eus,Drewitt:2021}.
In particular, the mathematical properties of MGFs stimulated Brown's construction of {\it equivariant iterated Eisenstein integrals}  \cite{Brown:mmv, Brown:2017qwo, Brown:2017qwo2} whose explicit form and valuable implications for MGFs will be the main subject of this work.

The connection between MGFs and equivariant iterated Eisenstein integrals is indirect.
Brown has shown~\cite{Brown:2017qwo, Brown:2017qwo2} how to define non-holomorphic modular forms as combinations of non-modular iterated integrals of holomorphic Eisenstein series and their complex conjugates. 
The main challenge in constructing non-holomorphic modular forms from iterated Eisenstein integrals is to spell out the required admixtures of {\it multiple zeta values} (MZVs) known from the expansion of MGFs around the cusp $\tau \rightarrow i\infty$ \cite{Zerbini:2015rss, DHoker:2015wxz, DHoker:2017zhq, DHoker:2019xef, Zagier:2019eus, Vanhove:2020qtt}. For those iterated Eisenstein integrals compatible with the differential structure of MGFs, the existence of modular completions
via \textit{single-valued} MZVs\footnote{Single-valued MZVs arise as special values of single-valued multiple polylogarithms  where all branch-cuts have been removed, see \cite{Schnetz:2013hqa, Brown:2013gia} for the introduction of single-valued MZVs and \cite{Schlotterer:2012ny, Stieberger:2013wea, Stieberger:2014hba, Schlotterer:2018abc, Vanhove:2018mzv, Brown:2019wna} for their significance in relating closed-string tree amplitudes to those of open strings.} was proven in \cite{Brown:2017qwo, Brown:2017qwo2}. In~\cite{Dorigoni:2022npe}, involving many of the present authors, this prescription was aligned with iterated-integral representations of MGFs that are based on certain generating-series methods~\cite{Gerken:2019cxz, Gerken:2020yii} originating from string theory.

In this work, we shall provide a unified explicit description of the occurrence of MZVs in the modular completions 
of iterated Eisenstein integrals
to their equivariant versions and their connection to MGFs.
One of the key themes of our analysis will be that for equivariant iterated Eisenstein integrals the  appearance of products and higher-depth instances of MZVs can be reconstructed from that of Riemann zeta values by means of certain algebraic structures.
This phenomenon is familiar from the low-energy expansion of string amplitudes at genus zero \cite{Schlotterer:2012ny} which relies on a particular description of (motivic) MZVs \cite{Goncharov:2005sla, Brown:2011mot,Brown:2011ik} reviewed below. 
In the present genus-one context, the information about the Riemann zeta values in equivariant iterated Eisenstein integrals is encoded via certain Lie-algebra generators acting on the generating series. These so-called {\it zeta generators} are determined from genus-zero structures in a companion paper \cite{Dorigoni:2023part1}.

MGFs are related only to specific linear combinations of equivariant iterated Eisenstein integrals, namely those that follow from the so-called Tsunogai relations~\cite{Tsunogai,Pollack}, see~\cite{Brown:2017qwo2, Gerken:2020yii, Dorigoni:2021ngn, Dorigoni:2022npe}. 
In fact, MGFs are expressible in terms of iterated integrals involving only Eisenstein series (and their complex conjugates) and in addition the periods arising in this construction are conjecturally restricted solely to single-valued MZVs. 

By contrast, the modular completion of iterated Eisenstein integrals to \textit{generic} equivariant iterated Eisenstein integrals (i.e.\ including cases beyond MGFs) necessitates iterated integrals of holomorphic cusp forms~\cite{Brown:mmv, Brown:2017qwo}. Referring to the number of iterated integrals as \textit{modular depth}, it was shown in~\cite{Brown:mmv, Brown:2017qwo, Dorigoni:2022npe} that  (critical and non-critical) L-values associated with holomorphic cusp forms appear at modular depth two. In the present work, we shall extend this analysis to modular depth three and also encounter ``hybrid'' double integrals mixing holomorphic Eisenstein series with cusp forms as part of the modular completion. 

As another feature that sets in at modular depth three, we find new periods beyond MZVs and L-values that are closely related to these hybrid double integrals and their associated so-called multiple modular values 
\cite{Brown:mmv, Brown2019}.\footnote{The equivariant iterated Eisenstein integrals of modular depth three involving hybrid double integrals discussed in this work suggest that our results on MZVs apply also beyond MGFs.}

The construction of equivariant iterated Eisenstein integrals and the results of this work are most conveniently presented at the level of generating series. The absence of holomorphic cusp forms in MGFs \cite{DHoker:2016mwo, Gerken:2019cxz, Gerken:2020yii} is implemented through the relations among the non-commutative bookkeeping variables $\ep_k$ in the generating series of equivariant iterated Eisenstein integrals introduced in \cite{Brown:2017qwo2}.\footnote{The generating series of MGFs furnished by closed-string genus-one integrals \cite{Gerken:2019cxz, Gerken:2020yii} conjecturally realise matrix representations of the non-commutative $\ep_k$ such that holomorphic cusp forms cancel automatically.} The underlying derivations $\ep_k$ with even $k\geq 0$  have a long history, encode deep connections between (motivic) fundamental groups, modular forms and mixed elliptic motives \cite{DeligneTBP, Ihara:1990, IharaTakao:1993, Tsunogai, Gonchtalk, GKZ:2006, Schneps:2006, Pollack, Brown:Anatomy, brown_2017, BaumardSchneps:2015, hain_matsumoto_2020} and satisfy the Tsunogai relations mentioned above. In our discussion of generic equivariant iterated Eisenstein integrals beyond MGFs, we shall pass to a variant of the $\ep_k $ that we call $\eee_k$ and that obey no relations other than $({\rm ad}_{\eee_0})^j\eee_k= 0$ if $j\geq k{-}1$. 

The non-commutative bookkeeping variables $\eee_k$ without Tsunogai relations will be seen to govern the modular completions via holomorphic cusp forms. Specialising $\eee_k  \rightarrow \ep_k$ within the generating series of equivariant iterated Eisenstein integrals then reduces it to the generating series of  MGFs where all combinations of multiple modular values conspire to MZVs.

The three main results of the present work can be briefly summarised as:
\begin{itemize}
\item making all sources of MZVs in Brown's construction of equivariant iterated Eisenstein integrals fully explicit;
\item inferring the appearance of arbitrary MZVs in equivariant iterated Eisenstein integrals from that of simple Riemann zeta values;
\item exemplifying the modular completions of iterated Eisenstein integrals to equivariant modular-depth-three combinations featuring double integrals involving holomorphic cusp forms and new periods beyond MZVs or L-values.
\end{itemize}

\subsection{Summary of results}
\label{sec:intro2}

The goal of this work is to construct non-holomorphic modular forms from iterated integrals of
holomorphic modular forms for $\SLtwoZ$, their complex conjugates and real constants such as MZVs.
More specifically, the integration kernels are given by the one-forms $ \tau^j {\rm G}_k(\tau) \,\dd \tau$
and $ \tau^j \Delta_k(\tau)\,\dd \tau$ with holomorphic Eisenstein series,
\beq
{\rm G}_k(\tau)  \coloneqq \sum_{(m,n)\in \mathbb{Z}^2\setminus \{(0,0)\}} \frac{1}{(m\tau {+} n)^{k}} \, , \ \ \ \ k\geq 4\, ,
\label{extintr.01}
\eeq
and Hecke-normalised holomorphic cusp forms $\Delta_k(\tau)$ of modular weight $k\geq 12$.\footnote{Here and later in this work, we abuse notation in that Hecke-normalised cusp forms $\Delta_k$ are not uniquely specified by their weight $k$. We write $\Delta_k$ for an arbitrary  weight-$k$ representative to avoid cluttering  notation, and in all cases appearing in the examples of this paper, the space of cusp forms is one-dimensional.} The integer exponents of $\tau^j$
in the integration kernels are taken to range over  $0\leq j \leq k{-}2$ to ensure that
the one-forms are closed under modular $\SLtwoZ$ transformations. 
The latter act as usual on the variable $\tau$ on the upper half-plane by
\begin{align}
    \gamma\cdot\tau \coloneqq \frac{a\tau+b}{c\tau+d}
    \quad
    \text{for}
    \quad
    \gamma= \begin{pmatrix} a&b \\ c&d \end{pmatrix} \in {\rm SL}(2,\mathbb Z)
\end{align}
and with holomorphic weight $k$ on holomorphic modular forms, i.e., 
\begin{equation}
{\rm G}_k( \gamma\cdot \tau ) = (c\tau{+}d)^k {\rm G}_k(\tau) \, , \ \ \ \  
 \Delta_k( \gamma\cdot \tau ) = (c\tau{+}d)^k \Delta_k(\tau) \, .
\label{extintr.02}
\end{equation}

However, 
it is easy to see that already the simplest integrals $\int_\tau^{i\infty} \dd \tau_1 \, \tau_1^{j_1}
{\rm G}_{k_1}(\tau_1)$ transform inhomogeneously under ${\rm SL}(2,\mathbb Z)$.
After a modular transformation $\tau\to \gamma\cdot\tau$, the change of integration variable $\tau_1\to \gamma \cdot\tau_1$ needed to transform the lower integration limit from
$\gamma\cdot\tau$ back to $\tau$, does not preserve the cusp $i\infty$ in the
upper integration limit, i.e.~in general $\gamma^{-1}\cdot i\infty \neq i \infty$. More specifically, the generators $T: \tau\mapsto \tau{+}1$ and $S: \tau\mapsto -1/\tau$ of
${\rm SL}(2,\mathbb Z)$ map iterated Eisenstein integrals
\beq
\int_\tau^{i\infty} \tau_\ell^{j_\ell}
{\rm G}_{k_\ell}(\tau_\ell)\, \dd \tau_\ell \int_{\tau_\ell}^{i\infty}  \ldots
\int_{\tau_3}^{i\infty}  \tau_2^{j_2}
{\rm G}_{k_2}(\tau_2)\,\dd \tau_2  \int_{\tau_2}^{i\infty}\tau_1^{j_1}
{\rm G}_{k_1}(\tau_1)\, \dd \tau_1 \,,
\label{extintr.03}
\eeq
to combinations of similar iterated integrals over the same path $\int_{\tau}^{i \infty}$ 
and inhomogeneous terms over paths $\int_{i \infty{-}1}^{i \infty}$ in case of $T$
and $\int_0^{i \infty}$ in case of $S$, for instance
\beq
S\, : \ \int_\tau^{i\infty}  \tau_1^{j} {\rm G}_{k}(\tau_1)\,\dd \tau_1 
\rightarrow (-1)^{j} \bigg\{
\int_\tau^{i\infty}  \tau_1^{k-j-2} {\rm G}_{k}(\tau_1)\,\dd \tau_1 
- \int_0^{i\infty}\tau_1^{k-j-2} {\rm G}_{k}(\tau_1)\, \dd \tau_1 
\bigg\} \, .
\label{extintr.aa}
\eeq 
The same kinds of inhomogeneous terms arise for modular transformations of the iterated
integrals (\ref{extintr.03}) with cusp forms $\Delta_{k_i}$ in the place of some of the ${\rm G}_{k_i}$.
For all of these integrals, the endpoint
divergence at $i\infty$ is regularised through the tangential-base-point method \cite{Brown:mmv} which in
the simplest cases amounts to $\int_{\tau}^{i \infty} \tau_1^j\,\dd \tau_1  = - \frac{1}{j+1}\tau^{j+1}$ for $j\geq 0$.

The inhomogeneous terms in the modular transformation of the holomorphic integrals (\ref{extintr.03})
pose a major challenge for combining them to modular forms. 
Adding the complex conjugates suffices to cancel the inhomogeneous terms over $\int_{i \infty{-}1}^{i \infty}$ from the modular $T$-transformation. However, the period integrals over $\int_0^{i \infty}$ from the modular $S$-transformation such as the last term in (\ref{extintr.aa})
turn out to be harder to cancel by further additions. These period integrals  again take the form of (\ref{extintr.03}) with $\tau \rightarrow 0$ (also in case of ${\rm G}_{k_i} \rightarrow \Delta_{k_i}$) and are known as {\it multiple modular values}~\cite{Brown:mmv}.

The existence of non-holomorphic modular forms of this type was proven in Brown's work on equivariant iterated Eisenstein integrals \cite{Brown:mmv, Brown:2017qwo, Brown:2017qwo2} that augment (\ref{extintr.03}) by complex conjugates of the same class of integrals and suitably chosen real coefficients --- MZVs, L-values of holomorphic cusp forms and more general periods. 

\subsubsection{Main results on equivariant iterated Eisenstein integrals}
\label{sec:intro2.1}

The MGFs in the low-energy expansion of genus-one closed-string amplitudes only involve
the equivariant versions of a subset of the general iterated Eisenstein integrals presented in (\ref{extintr.03}).
Brown's work \cite{Brown:mmv, Brown:2017qwo, Brown:2017qwo2} organises the holomorphic
integrals into generating series which are most compactly represented as path-ordered exponentials
\begin{align}
\mathbb I_{+}(\ep_k;\tau) &= \Pexp \bigg(\int _\tau^{i \infty} 
\sum_{k=4}^{\infty} \sum_{j=0}^{k-2} \frac{(k{-}1)}{j!} \, (2\pi i)^{1+j-k}\,  \tau_1^j {\rm G}_{k}(\tau_1) \, \dd \tau_1\,\ep_k^{(j)}
\bigg) \, ,
\label{extintr.04} \\
\widetilde{\mathbb I}_{-}(\ep_k;\tau) &= \Ptexp \bigg({-}\int_{\bar \tau}^{-i \infty} 
\sum_{k=4}^{\infty} \sum_{j=0}^{k-2} \frac{(k{-}1)}{j!} \, (2\pi i)^{1+j-k}\,  \bar \tau_1^j \overline{ {\rm G}_{k}(\tau_1) } \,\dd \bar \tau_1\, \ep_k^{(j)}
\bigg)\, , \notag
\end{align}
without any holomorphic cusp forms $\Delta_k$, see the discussion around (\ref{eq:Pexp}) for our ordering conventions and the meaning of the tilde for the second series.
The non-commutative variables $ \ep_k^{(j)}= ({\rm ad}_{\ep_0})^j\ep_k$ with $k\geq 4$ and $0\leq j \leq k{-}2$, and where ${\rm ad}_{\ep_0}(X)\coloneqq[\ep_0,X]$, in (\ref{extintr.04})
are built from the Tsunogai derivations $\ep_k$ mentioned above.
They obey a wealth of relations such as
\beq
[\ep_4 , \ep_{10}] - 3 [\ep_6 , \ep_{8}] =0\,,
\label{extintr.05}
\eeq
along with corollaries from acting on such identities with $ {\rm ad}_{\ep_0}$ (noting that $\ep_k^{(k-1)}=0$) or more general 
${\rm ad}_{\ep_\ell}$.
Given that the iterated Eisenstein integral (\ref{extintr.03}) enters the generating
series (\ref{extintr.04}) with coefficient $\ep_{k_1}^{(j_1)} \ep_{k_2}^{(j_2)} \ldots \ep_{k_\ell}^{(j_\ell)}$, the
relation (\ref{extintr.05}) eliminates from $\mathbb I_{+}(\ep_k;\tau)$ one linear combination of the double Eisenstein integrals
$\int_{\tau}^{i\infty}   {\rm G}_{k_2}(\tau_2) \,\dd \tau_2  \int_{\tau_2}^{i\infty} 
{\rm G}_{k_1}(\tau_1)\, \dd \tau_1$ at $k_1{+}k_2=14$.
These kinds of dropouts are essential to project the multiple modular values in the modular $S$-transformations
of (\ref{extintr.04}) to the world of MZVs, e.g.\ eliminating two more involved periods $\Lambda(\Delta_{12},12)$ and $c(\Delta_{12},12)$ \cite{Brown2019} present in the
individual $\int_{0}^{i\infty} \tau_2^{j_2} {\rm G}_{k_2}(\tau_2)\, \dd \tau_2  \int_{\tau_2}^{i\infty} \tau_1^{j_1}
{\rm G}_{k_1}(\tau_1)\,\dd \tau_1 $ at $k_1{+}k_2=14$.

Relations such as (\ref{extintr.05}) among Tsunogai's derivations still leave $\mathbb Q[2\pi i]$ combinations of 
MZVs as inhomogeneous terms in the modular transformations of the generating series (\ref{extintr.04}). 
As a first step, one can combine holomorphic iterated Eisenstein integrals and their complex conjugates through the composition $\widetilde{\mathbb I}_{-}(\ep_k;\tau) \mathbb I_{+}(\ep_k;\tau) $ which transforms homogeneously
under $T$. However, the cancellation of
multiple modular values under modular $S$-transformation requires admixtures of single-valued MZVs. This
is achieved in Brown's equivariant generating series \cite{Brown:2017qwo2}
\beq
\Ieqv(\ep_k;\tau)=\hat \psi^{\rm sv}\big( \,\widetilde{\mathbb I}_{-}(\ep_k;\tau)\big) 
\mathbb C^{\rm sv}(\ep_k)
\mathbb I_{+}(\ep_k;\tau)\,,
\label{extintr.06}
\eeq
by adding two ingredients:
\begin{itemize}
\item[(i)] a constant series $\mathbb C^{\rm sv}(\ep_k)$ in $\ep_{k_i}^{(j_i)}$ with $\mathbb Q$-linear combinations of single-valued MZVs as coefficients, and,
\item[(ii)] a change of alphabet $\hat \psi^{\rm sv}: \, \ep_k \rightarrow \ep_k+\ldots$ 
replacing each derivation by similar infinite series in single-valued MZVs and $\ep_{k_i}^{(j_i)}$.
\end{itemize}
While the explicit form of $\hat \psi^{\rm sv}$ was proposed in \cite{Dorigoni:2022npe} for MGFs, a main 
result of the present work is a similar all-order description of the series 
$\mathbb C^{\rm sv}(\ep_k)$ in (\ref{extintr.06}).

The key object to make both $\hat \psi^{\rm sv}$ and $\mathbb C^{\rm sv}(\ep_k)$ explicit
is the following group-like series in single-valued MZVs,
\begin{align}
\mathbb M^{\rm sv}(z_i) \coloneqq& \, \sum_{\ell=0}^{\infty} \sum_{\substack{i_1,i_2,\ldots, i_\ell \\{\in 2\mathbb N+1}}} z_{i_1} z_{i_2}\ldots z_{i_\ell}\, \rho^{-1} \big( {\rm sv}(f_{i_1} f_{i_2} \ldots f_{i_\ell}) \big) 
\label{extintr.07} \\
=& \, 1 +2 \! \! \! \sum_{i_1 \in 2\mathbb N+1} \! \! \! z_{i_1} \zeta_{i_1}
+ 2 \! \! \!  \!  \! \! \sum_{i_1,i_2 \in 2\mathbb N+1} \! \! \! \! \! \! z_{i_1} z_{i_2}  \zeta_{i_1} \zeta_{i_2} 
+ \! \! \! \! \! \!  \! \!  \sum_{i_1,i_2,i_3 \in 2\mathbb N+1} \! \! \! \! \! \! \! \!
z_{i_1} z_{i_2} z_{i_3}  \rho^{-1}\big( {\rm sv}(f_{i_1} f_{i_2}f_{i_3} ) \big) 
+\ldots  \, ,
\notag
\end{align}
where the isomorphism $\rho$, which maps (motivic) MZVs into the
$f$-alphabet \cite{Brown:2011ik, Brown:2011mot}, and the single-valued map sv, which acts on $f_{i_1} f_{i_2} \ldots f_{i_\ell}$ \cite{Schnetz:2013hqa, Brown:2013gia},
are reviewed in appendix \ref{app:A}. The realisation of $\hat \psi^{\rm sv}$ in \cite{Dorigoni:2022npe} involves
derivations $\{z_w, \ w \in 2\mathbb N{+}1\}$ associated with odd zeta values
similar to Tsunogai's~$\ep_k$:
\begin{align}
\hat \psi^{\rm sv}( \ep_k^{(j)}) &=
\mathbb M^{\rm sv}(z_i)^{-1} \ep_k^{(j)} \mathbb M^{\rm sv}(z_i)   \label{extintr.08} \\
& =
\sum_{\ell=0}^{\infty} \sum_{\substack{i_1,i_2,\ldots, i_\ell \\{\in 2\mathbb N+1}}} 
[ [ \ldots [[ \ep_k^{(j)}, z_{i_1} ], z_{i_2} ] ,\ldots ] , z_{i_\ell} ]
 \rho^{-1} \big( {\rm sv}(f_{i_1} f_{i_2} \ldots f_{i_\ell}) \big) \, .
\notag
\end{align}
Given that all brackets $[z_w, \ep_k^{(j)}]$ are expressible via nested commutators of 
$\ep_{k_i}^{(j_i)}$ \cite{Pollack, hain_matsumoto_2020} (see \cite{Dorigoni:2023part1}
and appendix \ref{app:zep3} for
recent higher-order computations), each term in
the series expansion of $\hat \psi^{\rm sv}( \ep_k^{(j)})$ and therefore
$\hat \psi^{\rm sv}\big( \,\widetilde{\mathbb I}_{-}(\ep_k;\tau)\big)$ boils
down to words in $ \ep_k^{(j)}$. 

In order to ensure that the series (\ref{extintr.07}) is well-defined to all orders, canonical choices of the $f$-alphabet and dual canonical choices of the $z_w$ (as well as similar generators $\sigma_w$, $\hat z_w$ and $\hat \sigma_w$ below) are described in \cite{Dorigoni:2023part1}. The choices in the reference for instance eliminate the ambiguity of redefining $z_{11}$ by rational multiples of $[z_3,[z_3,z_5]]$.

As a key result of the present work, the leftover ingredient $\mathbb C^{\rm sv}(\ep_k)$ 
in (\ref{extintr.06}) is identified as a product of group-like series (\ref{extintr.07})
in single-valued MZVs,
\beq
\mathbb C^{\rm sv}(\ep_k) =  \mathbb M^{\rm sv}(z_i)^{-1}  \mathbb M^{\rm sv}(\sigma_i)\, .
\label{extintr.09}
\eeq
Besides the derivations $z_w$ in the expression (\ref{extintr.08}) for $\hat \psi^{\rm sv}$,
our new representation (\ref{extintr.09}) of $\mathbb C^{\rm sv}(\ep_k)$ involves {\it zeta generators}
$\{\sigma_w, \ w \in 2\mathbb N{+}1\}$. As summarised in section \ref{sec:rev.zeta} and
detailed in \cite{Dorigoni:2023part1}, the zeta generators $\sigma_w$ in (\ref{extintr.09})
combine infinite series in nested brackets of $\ep_k^{(j)}$ with
the above $z_w$
\beq
\sigma_w = z_w - \frac{1}{(w{-}1)!}\ep_{w+1}^{(w-1)} + \ \textrm{series in nested brackets of} \ \ep_{k_i}^{(j_i)}\, .
\label{extintr.10}
\eeq
The infinite tower of these nested brackets is computable to any desired order
from the methods of \cite{Dorigoni:2023part1} and follows from the interplay of 
configuration-space integrals at genus zero and genus one. Each $\sigma_w$
already involves infinitely many brackets of two $\ep_{k_i}^{(j_i)}$ which can be 
extracted from the closed-form expression (\ref{zetgen.08}). In principle, the expansion (\ref{extintr.10}) of zeta generators leaves ambiguities to absorb certain types of nested $\ep_{k_i}^{(j_i)}$-brackets into redefinitions of $z_w$. A canonical choice for $z_w$ that fixes these ambiguities is presented in section \ref{sec:4zeta.6} and at the end of section \ref{sec:ngeozvarpi}.

Thanks to the particular combination of $\mathbb M^{\rm sv}$ in (\ref{extintr.09}), all the $z_w$ which 
by themselves go beyond the algebra of $\ep_k^{(j)}$ conspire to nested brackets
$[ [ \ldots [[ \ep_k^{(j)}, z_{i_1} ], z_{i_2} ] ,\ldots ] , z_{i_\ell} ]$. Since the latter boil down
to Lie-algebra valued words in $\ep_k^{(j)}$, the same is true for the series
$\mathbb C^{\rm sv}(\ep_k)$ in (\ref{extintr.09}) as expected \cite{Brown:2017qwo2}.

After combining (\ref{extintr.06}) with (\ref{extintr.08}) and (\ref{extintr.09}), 
our end result for the generating series of equivariant iterated Eisenstein integrals 
can be compactly written as
\beq
\Ieqv(\ep_k;\tau)=  \Msv(\zetaT_i)^{-1} \,\Imint(\ep_k; \tau)\, \mathbb \Msv(\sigma_i) \,\Ip(\ep_k;\tau) \, ,
\label{extintr.11}
\eeq
making the sources of single-valued MZVs in Brown's construction \cite{Brown:2017qwo2} fully explicit.
By the above expressions for the series on the right-hand side, the 
coefficients $\eeqvtau{j_1 &\ldots &j_\ell }{k_1 &\ldots &k_\ell}{\tau} $
 of $\ep_{k_1}^{(j_1)} \ldots \ep_{k_\ell}^{(j_\ell)}$ in (\ref{extintr.11}) are combinations
of holomorphic iterated Eisenstein integrals (\ref{extintr.03}) and their complex conjugates with
$\mathbb Q$-linear combinations of single-valued MZVs as coefficients.

The series
$\Msv(\zetaT_i)^{-1}$ and $\mathbb \Msv(\sigma_i) $ in single-valued MZVs ensure that
the $\eeqvtau{j_1 &\ldots &j_\ell }{k_1 &\ldots &k_\ell}{\tau} $ transform homogeneously under the generators $T$ and $S$ of the modular group,
as seen in (\ref{eeqv.06}). More precisely, these modular transformations match those
of the (product of) integration kernels $\tau_i^{j_i} {\rm G}_{k_i}\,\dd \tau_i $
paired with words in $\ep_{k_i}^{(j_i)}$ via (\ref{extintr.04}). That is why the
$\eeqvtau{j_1 &\ldots &j_\ell }{k_1 &\ldots &k_\ell}{\tau} $ are referred to as
{\it equivariant} iterated Eisenstein integrals. The sum $k_1{+}\ldots{+}k_\ell$ will be called their {\it degree}.

However, by the wealth of relations among the $\ep_{k_i}^{(j_i)}$ starting with (\ref{extintr.05}),
not all the $\eeqvtau{j_1 &\ldots &j_\ell }{k_1 &\ldots &k_\ell}{\tau} $ with $k_i \in 2\mathbb N{+}2$
and $0 \leq j_i \leq k_i{-}2$ appear independently in (\ref{extintr.11}). Instead, the specific relation
(\ref{extintr.05}) implies that only three linear combinations of the four independent $\eeqvtau{0 &0 }{4 &10}{\tau}$, $\eeqvtau{0 &0 }{10 &4}{\tau}$, $\eeqvtau{0 &0 }{6 &8}{\tau}$, $\eeqvtau{0 &0 }{8 &6}{\tau} $ are accessible from 
(\ref{extintr.11}), and similar dropouts occur at all higher degrees $k_1{+}\ldots{+}k_\ell\geq16$.
A separate line of results of this paper, to be summarised in section \ref{sec:intro2.3}, is to fix this shortcoming
of the $\ep_{k}^{(j)}$-valued series (\ref{extintr.11}) and explicitly 
determine the individual $\eeqvtau{j_1 &\ldots &j_\ell }{k_1 &\ldots &k_\ell}{\tau} $ in a wide
range of the $k_i$.

\subsubsection{Connections with MGFs and genus zero}
\label{sec:intro2.2}

The modular properties of equivariant iterated Eisenstein integrals do not yet
line up with the transformation law of modular forms, e.g.\ $\eeqvtau{j}{k}{\tau} $
is not $T$ invariant for $j>0$ (see~\eqref{eeqv.06}). Still, one can systematically generate
linear combinations
\begin{align}
\beqv{\ldots &j &\ldots}{\ldots &k &\ldots} &=  \sum_{p=0}^{k-2-j} \sum_{\ell=0}^{j+p}
\binom{k{-}j{-}2}{p} \binom{j{+}p}{\ell} \frac{({-}2\pi i \bar \tau)^\ell}{(4 \pi \Im \tau)^p} \,
\eeqv{\ldots &j-\ell+p &\ldots}{\ldots &k &\ldots}\, ,
\label{extintr.13}
\end{align}
(with a similar double sum over $p_i$ and $\ell_i$ for each column $\begin{smallmatrix} j_i \\ k_i \end{smallmatrix}$)\footnote{
For the case of two columns~\eqref{extintr.13} reads explicitly
\begin{align*}
\beqv{j_1 &j_2}{k_1 &k_2} &=  \sum_{p_1=0}^{k_1-2-j_1} \sum_{\ell_1=0}^{j_1+p_1} \sum_{p_2=0}^{k_2-2-j_2} \sum_{\ell_2=0}^{j_2+p_2}
\binom{k_1{-}j_1{-}2}{p_1}\binom{k_2{-}j_2{-}2}{p_2} \binom{j_1{+}p_1}{\ell_1}\binom{j_2{+}p_2}{\ell_2} \\
&\quad \times\frac{({-}2\pi i \bar \tau)^{\ell_1+\ell_2}}{(4 \pi \Im \tau)^{p_1+p_2}} \,
\eeqv{j_1-\ell_1+p_1 &j_2-\ell_2+p_2}{k_1 &k_2}\, .
\end{align*}} 
which furnish modular forms of weight $(0,\sum_{i=1}^{\ell} (k_{i}{-}j_i{-}2))$,
\beq
\beqvtau{j_1 &j_2 &\ldots &j_\ell}{k_1 &k_2 &\ldots &k_\ell}{\tfrac{a\tau+b}{c\tau+d}} =
\bigg( \prod_{i=1}^\ell (c\bar \tau{+}d)^{k_i-2j_i-2} \bigg)\beqvtau{j_1 &j_2 &\ldots &j_\ell}{k_1 &k_2 &\ldots &k_\ell}{\tau} \, .
\label{extintr.14}
\eeq
An implementation of the linear combinations (\ref{extintr.13}) at the level of generating series via simple
SL$_2(\mathbb{C})$ transformations is described in section \ref{sec:4.1}. Alternatively, the linear combinations (\ref{extintr.13}) arise from the organisation of equivariant
iterated Eisenstein integrals via polynomials in commutative bookkeeping variables $X_i, Y_i$
by rewriting powers $X^a Y^b$ with $a,b \in \mathbb N_0$ in terms of $(X{-}\tau Y)$ and 
$(X{-}\bar \tau Y)$ \cite{Brown:2017qwo}, see section \ref{sec:8}.

As detailed in \cite{Dorigoni:2022npe}, MGFs are linear combinations of the non-holomorphic modular forms
$\beta^{\rm eqv}$ in (\ref{extintr.13}), conjecturally with $\mathbb Q$-linear combinations of single-valued MZVs
as coefficients. This follows from the differential equations of the generating series
of closed-string genus-one integrals studied in \cite{Gerken:2019cxz, Gerken:2020yii} which contain all MGFs in their
low-energy expansion and involve conjectural matrix representations of the above
derivations $\ep_k$.

The series $\mathbb M^{\rm sv}(z_i)$ and $\mathbb M^{\rm sv}(\sigma_i)$ defined 
by (\ref{extintr.07}) entering the generating series (\ref{extintr.11})
introduce single-valued MZVs into ${\cal E}^{\rm eqv}$ and thereby into $\beta^{\rm eqv}$. This offers a systematic way of 
generating the (conjecturally single-valued) MZVs in the expansion of MGFs at the cusp $\tau \rightarrow i \infty$ 
\cite{Zerbini:2015rss, DHoker:2015wxz}\footnote{See for instance \cite{Green:2008uj, DHoker:2015gmr} for earlier results on the
expansion of MGFs around the cusp and \cite{DHoker:2019xef, Zagier:2019eus, Gerken:2020yii} for closed-form results on the MGFs in two-point genus-one integrals.} which provided the key motivation for their further study in the mathematics literature \cite{Brown:2017qwo, Brown:2017qwo2, Zagier:2019eus, Drewitt:2021}. As a particular virtue of the group-like series $\mathbb M^{\rm sv}$, the coefficients $z_w,\sigma_w$ of odd Riemann
zeta values $\zeta_w$ determine those of all other MZVs composed of two or more letters in the $f$-alphabet
via compositions of zeta generators. This interlocks products $2\zeta_{2m+1}\zeta_{2n+1}=\rho^{-1}( {\rm sv} (f_{2m+1}f_{2n+1} ))$ or (conjecturally) indecomposable higher-depth MZVs (starting with $\zeta^{\rm sv}_{3,3,5}$ involving up to three letters $f_i$) to the appearance of $\zeta_w$ in equivariant iterated Eisenstein integrals.

The pattern in the expansion (\ref{extintr.07}) of $\mathbb M^{\rm sv}$
where MZVs $\rho^{-1}(f_{i_1}f_{i_2}\ldots)$ are accompanied by
compositions $M_{i_1}M_{i_2}\ldots$ of certain operators $M_i$ 
was observed earlier on in the low-energy expansion of 
string tree-level amplitudes  \cite{Schlotterer:2012ny, Mafra:2022wml}. 
In these references, for the case of the configuration-space integrals at genus zero,
the $M_{i \in 2\mathbb N{+}1}$ are matrices whose products determine 
the coefficients of all MZVs at arbitrary depth
from those of Riemann zeta values. Hence, our result (\ref{extintr.11}) that imports all MZVs 
in equivariant iterated Eisenstein integrals from series $\mathbb M^{\rm sv}$ generalises 
these elegant structures of genus-zero integrals to genus one.

Another echo of genus-zero structures in the genus-one results of this work
can be found in the generating series of single-valued iterated Eisenstein integrals \cite{Brown:2017qwo2}
\beq
\mathbb I^{\rm sv}(\ep_k;\tau)=  \Msv(\sigma_i)^{-1} \,\Imint(\ep_k; \tau)\, \mathbb \Msv(\sigma_i) \,\Ip(\ep_k;\tau) \,,
\label{extintr.15}
\eeq
with $ \Msv(\sigma_i)^{-1}$ instead of $ \Msv(z_i)^{-1}$ as the leftmost factor on the right-hand side.
In contrast to their equivariant counterparts in (\ref{extintr.11}), the coefficients of
$\ep_{k_1}^{(j_1)} \ep_{k_2}^{(j_2)} \ldots \ep_{k_\ell}^{(j_\ell)}$ in $\mathbb I^{\rm sv}(\ep_k;\tau)$ 
transform inhomogeneously under both $S$ and $T$. 
At the same time, the single-valued iterated Eisenstein integrals in (\ref{extintr.15}) are
canonically defined and do not depend on specific choices in the definition of 
$\mathbb I^{\rm eqv}(\ep_k;\tau)$
(see \cite{Brown:2017qwo2} and section \ref{sec:4zeta.6} for the ambiguities
in the definition of $\mathbb I^{\rm eqv}(\ep_k;\tau)$
and \cite{Dorigoni:2023part1} for a preferable way of fixing them). 
Moreover, the conjugation of the antiholomorphic 
generating series in (\ref{extintr.15}) by $\mathbb \Msv(\sigma_i)$ is in one-to-one
correspondence to a similar conjugation formula at genus zero \cite{Frost:2023stm}
\beq
 \mathbb G^{\rm sv}_{\{0,1\}}(x_i;z) =
\Msv(M_i)^{-1} 
 \overline{ \widetilde{\mathbb G}_{\{0,1\}}(x_i;z) }
\Msv(M_i)
  \mathbb G_{\{0,1\}}(x_i;z)\,,
\label{extintr.16}
\eeq 
for the generating series $ \mathbb G^{\rm sv}_{\{0,1\}}(x_i;z)$ of single-valued polylogarithms in one variable \cite{svpolylog}.
The series $\mathbb G_{\{0,1\}}(x_i;z)$ and $ \overline{ \widetilde{\mathbb G}_{\{0,1\}}(x_i;z) }$ on the right-hand side are path-ordered exponentials of the Knizhnik--Zamolodchikov connection $(\frac{x_0}{z}+ \frac{x_1}{z{-}1})\dd z $ similar to (\ref{extintr.04}). The operators $M_i$ in the series $\Msv$ of (\ref{extintr.16}) now refer to zeta generators at genus zero\footnote{In slight abuse of notation, we employ the same symbol $M_i$ for the zeta generators
in (\ref{extintr.16}) and their matrix representations relevant to the aforementioned string tree-level amplitudes.} which normalise the braid operators $x_0$ and $x_1$ \cite{Ihara:stable, Furusho2000TheMZ}. Similar to the reasoning below (\ref{extintr.08}), $\Msv(M_i)^{-1}  \overline{ \widetilde{\mathbb G}_{\{0,1\}}(x_i;z) }
\Msv(M_i)$ is expressible in terms of $x_0,x_1$ after iteratively
inserting commutators such as $[x_0, M_w]=0$ or $[x_1,M_3] = [[[x_0,x_1],x_0{+}x_1],x_1]$,
also see the discussion of section \ref{sec:4zeta.8}.

Hence, our explicit description of the series of single-valued MZVs in the equivariant generating series
(\ref{extintr.11}) manifests an important analogy between genus zero and genus one: The construction
of single-valued iterated Eisenstein integrals (\ref{extintr.15}) at genus one closely follows that of single-valued 
genus-zero polylogarithms (\ref{extintr.16}) in one variable. In both cases, the zeta generators in the
conjugations of the antiholomorphic series need to be adapted to the variables $x_i$ or
$\epsilon_k^{(j)}$ relevant to genus zero or one, respectively.

\subsubsection{Main results beyond MGFs}
\label{sec:intro2.3}

Another body of results in this work concerns the generalisations of MGFs to
non-holomor\-phic modular forms involving iterated integrals of (anti-)holomorphic cusp forms.
Our starting point is to uplift the Tsunogai derivations $\ep_{k}^{(j)}$ in the path-ordered exponential
(\ref{extintr.04}) to a free algebra of symbols $\eee_k^{(j)} =({\rm ad}_{\eee_0})^j \eee_k$ with $k\geq 4$ and $0\leq j \leq k{-}2$,
\begin{align}
\mathbb I_{+}(\eee_k;\tau) &= \Pexp \bigg(\int _\tau^{i \infty} \! 
\sum_{k=4}^{\infty} \sum_{j=0}^{k-2} \frac{(k{-}1)}{j!} \, (2\pi i)^{1+j}\,  \tau_1^j \bigg[  \frac{ {\rm G}_{k}(\tau_1)}{(2\pi i)^k} \, \eee_k^{(j)} + \! \sum_{\Delta \in {\cal S}_k} \! \Delta(\tau_1)  \eee_{\Delta^+}^{(j)} \bigg]\dd \tau_1
\bigg)\, ,
\label{extintr.00} \\
\widetilde{\mathbb I}_{-}(\eee_k;\tau) &= \Ptexp \bigg({-}\int_{\bar \tau}^{-i \infty}  \! 
\sum_{k=4}^{\infty} \sum_{j=0}^{k-2} \frac{(k{-}1)}{j!} \, (2\pi i)^{1+j}\,  \bar \tau_1^j \bigg[  \frac{ \overline{ {\rm G}_{k}(\tau_1) }  }{(2\pi i)^k}\, \eee_k^{(j)} + \!  \sum_{\Delta \in {\cal S}_k}   \! \overline{ \Delta(\tau_1) }  \eee_{\Delta^-}^{(j)} \bigg]\dd \bar \tau_1
\bigg)\, .
\notag
\end{align}
The admixtures of holomorphic cusp forms are needed to find an equivariant completion
of the additional iterated Eisenstein integrals in $\mathbb I_{+}(\eee_k;\tau)$
that are absent from $\mathbb I_{+}(\ep_k;\tau)$ due to Tsunogai's relations. The vanishing combinations
of $\ep_k^{(j)}$ such as (\ref{extintr.05}) translate into non-zero combinations
of the free-algebra generators such as $P_{14}^2 \sim [\eee_4 , \eee_{10}] - 3 [\eee_6, \eee_8] $.
More general combinations $P_{w}^d$ of this type described in section \ref{sec:tsalg}
appear as coefficients of various periods beyond MZVs in the modular $S$ transformations 
of the iterated Eisenstein integrals in $\mathbb I_{+}(\eee_k;\tau)$. The
role of the cusp forms in (\ref{extintr.00}) is to compensate many of these
new periods which necessitates relations between the letters $\eee_{\Delta^{\pm}}$ and
the combinations $P_w^d$ of $\eee_k^{(j)}$, for instance
\beq
\eee_{\Delta_{12}^{\pm}} = \mp  \frac{\Lambda(\Delta_{12}, 12) }{\Lambda(\Delta_{12}, 10)} \,
 P_{14}^2
+ \frac{\Lambda(\Delta_{12}, 13)}{\Lambda(\Delta_{12}, 11)} \,  P_{16}^3
+\ldots\, ,
\label{extintr.18}
\eeq
with an infinite tower of $P^d_w$ and their brackets with $\eee_k^{(j)}$ in the ellipsis.
As detailed in section \ref{sec:noedelta}, terms of higher degree in the expansion of $\eee_{\Delta^{\pm}}$
involve new periods, beyond ratios of L-values $\Lambda(\Delta_{12}, t) $ defined by (\ref{defLfct}).
Most importantly, the goal of constructing an equivariant series from $\mathbb I_{+}(\eee_k;\tau)$ and $\widetilde{\mathbb I}_{-}(\eee_k;\tau) $ in (\ref{extintr.00}) completely 
reduces the letters $\eee_{\Delta^{\pm}}$ associated with holomorphic cusp forms to infinite series in 
the Eisenstein letters $\eee_k^{(j)}$. 

The most ambitious claim of this work concerns the equivariant completion of the
series (\ref{extintr.00}) of iterated integrals of (anti-)holomorphic modular forms. 
The key idea is to uplift each ingredient in the series (\ref{extintr.06}) of 
Eisenstein integrals with generators $\ep_k^{(j)}$ (subject to relations such as (\ref{extintr.05}))
to the free algebra of $\eee_k^{(j)}$ via\footnote{The first part of this expression also realises an uplift of $\hat\psi^{\rm sv}$ to generating series in the $\eee_k^{(j)}$; equation~\eqref{extintr.06} only applies to expressions series in $\ep_k^{(j)}.$}
\beq
\Ieqv(\eee_k;\tau) =
\Msv( \zetaE_i,\zetaE_{\pern})^{-1} \, \Imint(\eee_k;\tau) \,
\mathbb \Msv(  \sigmaE_i, \sigmaE_{\pern}) \, \Ip(\eee_k;\tau)\, .
\label{extintr.19}
\eeq
Apart from the augmentation (\ref{extintr.00}) of the (anti-)holomorphic generating series 
$\mathbb I_{\pm}$ by cusp forms, also the series $\mathbb M^{\rm sv}$ in (\ref{extintr.07})
combining single-valued MZVs with zeta generators requires an uplift 
in passing from $\ep_k^{(j)}$ to $\eee_k^{(j)}$. The series
\beq
\Msv(  \sigmaE_i, \sigmaE_{\pern}) = \Msv(  \sigmaE_i) +\ldots \, , \ \ \ \
\Msv( \zetaE_i,\zetaE_{\pern}) = \Msv( \zetaE_i)+\ldots
\label{extintr.99}
\eeq
in (\ref{extintr.19}) first of all contain the single-valued MZVs of (\ref{extintr.07}),
where the associated zeta generators $\sigma_i \rightarrow \sigmaE_i$ and
$z_i \rightarrow \zetaE_i$ are adapted to 
$\ep_k^{(j)}\rightarrow \eee_k^{(j)}$ as described in section \ref{sec:rev.up}.
The extra terms in the ellipsis of (\ref{extintr.99}) comprise new periods beyond MZVs that 
are collectively denoted by $\varpi$, see section \ref{sec:sivarpi} for the
simplest examples.

Similar to the letters $\eee_{\Delta^{\pm}}$ multiplying cusp forms in (\ref{extintr.00}),
the generators $\sigmaE_{\pern}= \zetaE_{\pern}+\ldots$ associated with the new periods (see \cite{Brown2019} for the association of generators with general primitive periods)
are needed to ensure that the right-hand side of (\ref{extintr.19}) is expressible via words in $\eee_k^{(i)}$.
We expect that both the terms $\sigmaE_{\pern}{-} \zetaE_{\pern}$, usually referred to as the \textit{geometric part} of $\sigmaE_{\pern}$, and the
brackets $[ \zetaE_{\pern}, \eee_k^{(j)}]$ boil down to nested brackets in $\eee_k^{(j)}$.
Moreover, both of $\sigmaE_{\pern}{-} \zetaE_{\pern}$ and $[ \zetaE_{\pern}, \eee_k^{(j)}]$
have to vanish in the specialisation $\eee_k^{(j)} \rightarrow \ep_k^{(j)}$ in order to recover
the generating series (\ref{extintr.11}) relevant to MGFs when adapting
(\ref{extintr.19}) to Tsunogai's derivation algebra. Hence, similar to the expansion
of $\eee_{\Delta^{\pm}}$ exemplified in (\ref{extintr.18}), both of $\sigmaE_{\pern}{-} \zetaE_{\pern}$ 
and $[ \zetaE_{\pern}, \eee_k^{(j)}]$
are naturally expressed in terms of the combinations $P_w^d$ that vanish upon $\eee_k^{(j)} \rightarrow \ep_k^{(j)}$,
see for instance (\ref{svieis.14}) and (\ref{svieis.15}).

In practice, the expansions of $\eee_{\Delta^{\pm}}$ as well as 
$\Msv(  \sigmaE_i, \sigmaE_{\pern}) $ and $[ \zetaE_{\pern}, \eee_k^{(j)}]$ in terms
of $\eee_k^{(i)}$ is determined by imposing modularity of the series (\ref{extintr.19})
at each modular depth and degree:
Following the discussion below (\ref{extintr.11}), the coefficients $\eeqvtau{j_1 &\ldots &j_\ell }{k_1 &\ldots &k_\ell}{\tau} $
of $\eee_{k_1}^{(j_1)} \ldots \eee_{k_\ell}^{(j_\ell)}$ in (\ref{extintr.19}) are now individually accessible by the absence of bracket relations among the $\eee_{k}^{(j)}$. By adding suitable brackets of $\eee_{k}^{(j)}$ 
to the expressions for $\eee_{\Delta^{\pm}}$, 
$\Msv(  \sigmaE_i, \sigmaE_{\pern}) $ and $[ \zetaE_{\pern}, \eee_k^{(j)}]$,
these coefficients ${\cal E}^{\rm eqv}$ are made to transform equivariantly under SL$(2,\mathbb Z)$.
Equivalently, their linear combinations $\beta^{\rm eqv}$
defined by (\ref{extintr.13}) are imposed to be modular forms. 

The conditions for the modular properties (\ref{extintr.14})
of $\beqvtau{j_1 &j_2 &\ldots &j_\ell}{k_1 &k_2 &\ldots &k_\ell}{\tau}$ are investigated separately for each choice $(k_1,\ldots,k_\ell)$. At modular depth $\ell =2$, the known results for 
$\beqvtau{j_1 &j_2}{k_1 &k_2}{\tau}$ \cite{Dorigoni:2022npe} are translated into contributions
$[\eee_{k_1}^{(j_1)} , \eee_{k_2}^{(j_2)}]$ to $\eee_{\Delta^{\pm}}$ with ratios of L-values $\Lambda(\Delta,t)$ as coefficients. At modular depth $\ell=3$, modularity of $\beqvtau{j_1 &j_2 &j_3}{k_1 &k_2 &k_3}{\tau}$
up to and including degree $k_1{+}k_2{+}k_3=20$ is used to infer contributions
$[\eee_{k_1}^{(j_1)} , [\eee_{k_2}^{(j_2)} , \eee_{k_3}^{(j_3)}]]$ to both 
$\eee_{\Delta^{\pm}}$ and $\Msv(  \sigmaE_i, \sigmaE_{\pern})$, identifying
new periods beyond MZVs and L-values in both cases. The contributions $\zetaE_{\pern}$ are usually referred to as the \textit{arithmetic part}
of the new generators $ \sigmaE_{\pern}$ and do not yet enter the 
$\beqvtau{j_1 &j_2 &j_3}{k_1 &k_2 &k_3}{\tau}$ at any degree $k_1{+}k_2{+}k_3$. 
Instead, their leading contributions are easily anticipated from $T$-invariance 
at modular depth $\ell=4$, see sections \ref{sec:ngeozvarpi} and \ref{sec:sequiv}.

By assembling the modular forms $\beta^{\rm eqv}$ from the coefficients in the generating series
(\ref{extintr.19}), we obtain explicit examples of the fact that the known modular 
completions $\beqvtau{j_1 &j_2}{k_1 &k_2}{\tau}$ of double Eisenstein integrals
\cite{Dorigoni:2022npe} solely involve:\footnote{In the following we do not list separately the fact that the coefficients contain rational numbers and factors of $2\pi i \bar\tau$ and $\pi\Im\tau$, see for example \eqref{extintr.13}.} 
\begin{itemize}
\item rational combinations of 
odd zeta values (possibly multiplied by Eisenstein integrals of modular depth one) 
and products $\zeta_{2a+1}\zeta_{2b+1}$;
\item cusp form integrals over $ \tau_1^j \Delta_k(\tau_1)$ or $\bar \tau_1^j 
\overline{\Delta_k(\tau_1)}$ along with rational multiples of
$\frac{ \Lambda(\Delta_k,t_1)}{ \Lambda(\Delta_k,t_2)}$, where the L-values in the numerator are non-critical
with $t_1\in \mathbb{N}$ and $t_1\geq k$ and those in the denominator are critical with $t_2 \in \{k{-}1, k{-}2 \}$.\footnote{The restriction to these integer values in the critical strip for $t_2$ is possible due to Manin's theorem~\cite{Manin:1973}.}
\end{itemize}
More importantly, we pinpoint a variety of new ingredients in the 
modular completions $\beqvtau{j_1 &j_2 &j_3}{k_1 &k_2 &k_3}{\tau}$ of triple Eisenstein integrals:
\begin{itemize}
\item indecomposable single-valued MZVs $\rho^{-1}({\rm sv}(f_{2a+1}f_{2b+1}f_{2c+1}))$ beyond
depth one in the
constant terms (i.e.\ without any accompanying $\tau_i$ integrals) at degree $k_1{+}k_{2}{+}k_3\geq 14$;
\item new periods $\varpi$ associated with the new generators $\hat \sigma_\varpi$ discussed below (\ref{extintr.19}) in the constant terms at degree $k_1{+}k_{2}{+}k_3\geq 18$;
\item double integrals involving an Eisenstein series and a holomorphic cusp form
with $\mathbb Q[ \frac{ \Lambda(\Delta_k,t_1)}{ \Lambda(\Delta_k,t_2)}]$ coefficients
at degree $k_1{+}k_{2}{+}k_3\geq 18$ (where again $\mathbb N \ni t_1 \geq k$ and $t_2 \in \{k{-}1, k{-}2 \}$);
\item additional new periods along with the integrals of a single cusp form
at $k_1{+}k_{2}{+}k_3\geq 18$.
\end{itemize}
Up to and including the coefficients of $\eee_{k_1}^{(j_1)}\eee_{k_2}^{(j_2)} \eee_{k_3}^{(j_3)}$ 
at degree $k_1{+}k_2{+}k_3=20$, our results for the generating series (\ref{extintr.19}) of 
equivariant iterated Eisenstein integrals are confirmed both analytically and numerically.
At higher degree or modular depth, equivariance is contingent on the relation
(\ref{beyond1708}) interlocking multiple modular values, the $\ep_k^{(j)} \rightarrow \eee_k^{(j)}$ 
uplift of zeta generators as well as
the extension of $\mathbb M^{\rm sv}$ by new periods and generators $\varpi$ 
and $\hat \sigma_\varpi$. In section
\ref{sec:gen}, we prove the existence of modular completions of
iterated Eisenstein integrals over $ \tau_i^{j_i} {\rm G}_{k_i}\,\dd \tau_i$ 
of arbitrary degree and modular depth. 

Nevertheless, it will be important to determine the explicit form of $\beta^{\rm eqv}$ of modular
depth $\geq 4$ in future work in order to gather information on the structure of the letters $\eee_{\Delta^{\pm}}$ and the series
$\Msv(  \sigmaE_i, \sigmaE_{\pern})$: First, it will be interesting to understand the systematics of the new periods encountered at different modular depths and degrees. Second, by analogy with the group-like series (\ref{extintr.07}) of
MZVs and zeta generators, one may expect contributions of the form
$\hat \sigma_i \hat \sigma_\varpi$ or $\hat \sigma_\varpi \hat \sigma_i $ 
to $\Msv(  \sigmaE_i, \sigmaE_{\pern})$. It will be interesting to investigate these
terms $\hat \sigma_i \hat \sigma_\varpi$ or $\hat \sigma_\varpi \hat \sigma_i $
from the modularity conditions for $\beqvtau{j_1 &j_2 &j_3 &j_4}{k_1 &k_2 &k_3 &k_4}{\tau}$
and to relate their coefficients to $f$-alphabet descriptions of $\zeta_i$ and $\varpi$. Third, it remains to prove our conjecture that the MZV sector of $\Msv(  \sigmaE_i, \sigmaE_{\pern})$ in (\ref{extintr.99}) follows the group-like series (\ref{extintr.07}) which is so far checked up to including modular depth three and degree 20.

\subsection{Outline}
\label{sec:intro3}

The remainder of this work will elaborate on the ideas and results summarised in the
previous section, supported by a multitude of examples. In section \ref{sec:rev},
we start by reviewing the key ingredients of our later construction of non-holomorphic modular
forms and by fixing conventions on iterated integrals as well as the non-commuting variables in
their generating series. Our new results on generating series of equivariant iterated Eisenstein
integrals are presented in section \ref{sec:4}, with a discussion of SL$_2$ frames for the non-commutative
variables in section \ref{sec:4.1}, an all-order account on the series in
MGFs and Tsunogai derivations $\ep_k^{(j)}$ in section~\ref{sec:4zeta} and their extension
to include holomorphic cusp in section~\ref{sec:4.2} (focusing on modular depth three).
In the same way as equivariant iterated Eisenstein integrals of modular depth two 
\cite{Brown:mmv, Brown:2017qwo} were made explicit \cite{Dorigoni:2022npe} on the basis of Laplace 
eigenfunctions ${\rm F}^{\pm(s)}_{m,k}$ \cite{Dorigoni:2021ngn},
their counterparts ${\rm F}^{\pm(s)}_{m,k,\ell}$ at modular depth three are briefly discussed
in section \ref{sec:7}. Section~\ref{sec:6} provides a conditional proof for the modular properties 
of the equivariant generating series~(\ref{extintr.19}) where the key assumption is inductively proven
in section \ref{sec:8}. A variety of appendices contain background information on MZVs, explicit results
on zeta generators and proofs of important lemmas.

The arXiv submission of this work is accompanied by several ancillary files which include: \begin{itemize}
    \item a {\tt Mathematica} implementation (with examples) of the $\mathfrak{sl}_2$ projectors $t^d_{p,q}$ and $s^d_{p,q}$, introduced in section \ref{sec:rev.2.1};
    \item all Pollack combinations $P^w_d$ of degree $w\leq 20$ and $d\leq 5$, discussed in section \ref{sec:tsalg};
    \item uplifted zeta generators $\hat{\sigma}_w$ up to degree $20$ and arithmetic relations up modular depth $\leq 3$ and degree $\leq 21$, as discussed in section \ref{sec:rev.up};
    \item expressions for all $\beta^{\rm{eqv}}$ at  modular depth $\leq 3$ and degree $\leq 20$, and relevant new periods $\varpi$, $\Lambda$ introduced in section \ref{sec:4.2} typically to $300$ digits of precision;
    \item the coefficients $c^{\rm sv}$ of the  generating series \eqref{defcsv} in single-valued MZVs and new periods for modular depth $\leq 3$ and degree $\leq 20$;
    \item all modular invariant solutions ${\rm F}_{m,k,\ell}^{\pm(s)}$ with degree $2m+2k+2\ell\leq20$ of inhomogeneous Laplace eigenvalue equations expressed in terms of $\beta^{\rm{eqv}}$, as discussed in section \ref{sec:7};
    \item explicit isomorphism between single-valued MZVs and the $f$-alphabet following the method of \cite{Dorigoni:2023part1} (also see appendix \ref{app:A}) as well as the dictionary between MZVs and their single-valued versions -- both up to weight $17$.
\end{itemize}

\section{Review and basics}
\label{sec:rev}

In this section, we review the basic constructions appearing in this paper. A key ingredient are iterated integrals of holomorphic or anti-holomorphic modular forms of $\SLtwoZ$ which we introduce in section~\ref{sec:rev.1}. A recap of modular graph forms and more general non-holomorphic modular forms can be found in section~\ref{sec:rev.3}. The Tsunogai derivation algebra is briefly reviewed in section~\ref{sec:rev.2} while in  section~\ref{sec:rev.zeta} we present derivations related to zeta values or {\it zeta generators} in short. Both types of derivations enter the generating series of real-analytic modular forms that we review in section~\ref{sec:rev.6}.

\subsection{Iterated integrals of holomorphic modular forms}
\label{sec:rev.1}

Iterated integrals of modular forms have become a driving force in numerous recent developments in particle physics~\cite{Bourjaily:2022bwx, Abreu:2022mfk} and string perturbation theory~\cite{Gerken:review, Berkovits:2022ivl, DHoker:2022dxx}. Our conventions here follow most closely the ones used in~\cite{Gerken:2020yii, Dorigoni:2022npe}. 

\subsubsection{Definitions}

We define holomorphic Eisenstein series $\GG_k$ for $k\geq 4 $ an even integer by
\begin{align}
\label{eq:GGk}
\GG_k(\tau) \coloneqq \sum_{(m,n)\in \mathbb{Z}^2\setminus \{(0,0)\}} \frac{1}{(m\tau {+} n)^{k}} = 2\zeta_k + \frac{2(2\pi i)^k}{(k{-}1)!}\sum_{n=1}^{\infty} \sigma_{k-1}(n) q^n
\end{align}
with $\tau$ on the upper half plane $\UHP\coloneqq \{\tau\in\mathbb{C} \,|\, \Im(\tau)>0\}$ and $q\coloneqq e^{2\pi i \tau}$. The Fourier coefficients in (\ref{eq:GGk}) are divisor sums $\sigma_{k-1}(n) \coloneqq \sum_{d|n} d^{k-1}$. A relation between the zeta values $2\zeta_k$ and the Bernoulli numbers $\BB_k$ for even $k$ that we will make frequent use of is
\begin{align}
2\zeta_k = - \BF_k (2\pi i)^k\,,\quad \BF_k \coloneqq \frac{\BB_k}{k!}\,,
\end{align}
introducing a factorially rescaled version $\BF_k$ of Bernoulli numbers $\BB_k$, which will appear quite often in the rest of the paper.

As already stated in \eqref{extintr.02}, under a modular transformation $\gamma\in\SLtwoZ$ we have
\begin{align}
\GG_k(\gamma \cdot\tau) = (c\tau{+}d)^{k} \GG_k(\tau)\,,
\label{eq:Sl2G}
\end{align}
which means that $\GG_k$ is a modular form of (holomorphic/anti-holomorphic) weight $(k,0)$. More generally, a non-holomorphic modular form of weight $(w,\bar{w})$ will transform similarly with factors $(c\tau{+}d)^{w} (c\bar\tau{+}d)^{\bar{w}}$ and we call $w$ and $\bar{w}$ its holomorphic and anti-holomorphic modular weights, respectively. We recall that the modular group $\SLtwoZ$ is generated by the two transformations $T: \tau\mapsto \tau{+}1$ and $S: \tau\mapsto -1/\tau$.


We will be interested in iterated integrals of holomorphic modular forms, starting
from the integration kernels
\begin{align}
\nuker{j}{k}{\tau} \coloneqq (2\pi i)^{1+j-k} \tau^j {\rm G}_{k}(\tau) \,\dd \tau\,,
\quad\quad
\overline{\nuker{j}{k}{\tau}}\coloneqq    (-1)^{j+1} (2\pi i)^{1+j-k} \bar\tau^j \overline{{\rm G}_{k}(\tau)} \,\dd \bar\tau\,,
\label{nuker}
\end{align}
for which it is very easy using \eqref{eq:Sl2G} to compute the two $\SLtwoZ$ transformations
\begin{align}
\nuker{j}{k}{\tau{+}1} &\label{eq:nukerT}= \sum_{p=0}^j \binom{ j}{p} (2\pi i )^{j-p} \nuker{p}{k}{\tau}\,,\\
\nuker{j}{k}{-\tfrac{1}{\tau}} &\label{eq:nukerS} = (-1)^j (2\pi i)^{2+2j-k}  \nuker{k-2-j}{k}{\tau}\,.
\end{align}
From these kernels we construct homotopy invariant iterated integrals, given by the recursive definition
\begin{align}
\ee{j_1 &j_2 &\ldots &j_\ell}{k_1 &k_2 &\ldots &k_\ell}{\tau} &\coloneqq \int_\tau^{i \infty} \nuker{j_\ell}{k_\ell}{\tau_\ell} \cdots \int_{\tau_3}^{i\infty} \nuker{j_2}{k_2}{\tau_2}\int_{\tau_2}^{i\infty} \nuker{j_1}{k_1}{\tau_1}\nn\\*
&\phantom{:}=(2\pi i)^{1+j_\ell-k_\ell}
 \int^{i\infty}_{\tau}\tau_\ell^{j_\ell} {\rm G}_{k_\ell}(\tau_\ell) \,
 \ee{j_1 &\ldots &j_{\ell-1}}{k_1  &\ldots &k_{\ell-1}}{\tau_\ell}\, \dd \tau_\ell\,  ,
 \label{MMVsec.07}
\end{align}
with the convention that $\mathcal{E}[\emptyset; \tau]=1$. Here, we restrict the integers $j_i$ to the range $0\leq j_i\leq k_i{-}2$ for all of $i=1,2,\ldots,\ell$. Note that in our convention the right-most columns correspond to the outer-most integrations of the iterated integral. 

In the rest of this paper we will introduce different flavours of iterated integrals. When referring to objects such as $\ee{j_1 &j_2 &\ldots &j_\ell}{k_1 &k_2 &\ldots &k_\ell}{\tau}$, we will consistently use the following terminology:
\begin{itemize}
\item The number $\ell$ of iterated integrals, i.e.\ the number of different integration kernels present, will be referred to as the \textit{modular depth} of the iterated integral;
 \item The grading introduced by the sum $\sum_{i=1}^\ell k_i$ of the modular weights of the modular forms being integrated will be referred to as the \textit{degree} of the iterated integral.
 \end{itemize}
 
 The simplest examples at modular depth one and two are
\begin{align}
\ee{j }{k}{\tau} &=
\int_\tau^{i\infty} \nuker{j}{k}{\tau_1}= (2\pi i)^{1+j-k}
 \int^{i\infty}_{\tau}  \tau_1^{j} {\rm G}_{k}(\tau_1)\,\dd \tau_1\, \,, \label{doubleint}\\
 \ee{j_1&j_2 }{k_1&k_2}{\tau} &=\int_\tau^{i\infty} \nuker{j_2}{k_2}{\tau_2}\int_{\tau_2}^{i\infty} \nuker{j_1}{k_1}{\tau_1} \nn\\
 &=(2\pi i)^{2+j_1+j_2-k_1-k_2}
 \int^{i\infty}_{\tau}  \tau_2^{j_2} {\rm G}_{k_2}(\tau_2)\,\dd \tau_2
 \int^{i\infty}_{\tau_2}  \tau_1^{j_1} {\rm G}_{k_1}(\tau_1)\,\dd \tau_1\,.
 \notag
\end{align}
We note that the integrals as written are not well-defined since the zero mode $2\zeta_k$ of $\GG_k$ introduces endpoint divergences at the cusp $\tau_i \rightarrow i\infty$. These divergences can be dealt with by using the method of tangential-base-point regularisation~\cite{Brown:mmv} that we will employ throughout. Its simplest incarnation $\int_{\tau}^{i\infty} \tau_1^j \dd\tau_1 = -\frac{\tau^{j+1}}{j+1}$ propagates to higher modular depth~$\ell$ by imposing the regularisation to preserve the shuffle relations of (\ref{MMVsec.07}).

The vector space of holomorphic modular forms of weight $k$ can be decomposed into the one-dimensional space spanned by the  holomorphic Eisenstein series $\GG_k$, and the vector space $\mathcal{S}_k$ of holomorphic cusp forms. 
The first holomorphic cusp form arises for $\mathcal{S}_{12}$ and is given by the Ramanujan cusp form $\Delta_{12}(\tau)$.
Even though $\mathcal{S}_{k=24}$ and $\mathcal{S}_{k\geq 28}$ have dimensions $\geq 2$, we shall employ the abusive placeholder notation $\Delta_k$ for an unspecified cusp form of weight $k$ to avoid cluttering. Moreover, the examples in this paper only involve modular weight $k\leq 20$ where no ambiguity arises.

Definitions similar to~\eqref{MMVsec.07} apply when the holomorphic Eisenstein series $\GG_k(\tau)$ is replaced by a holomorphic cusp form $(2\pi i)^k \Delta_{k}(\tau)$ of weight $k$.
Throughout this paper, we assume $\Delta_k\in \mathcal{S}_k$ to be a  normalised Hecke eigenform, so that its Fourier expansion starts as $\Delta_k= q + O(q^2)$, and the factor $(2\pi i)^k$ is included everywhere for uniformity with the Fourier coefficients $\in \mathbb Q \pi^k$ of $\GG_k$.
Our notation
\begin{align}
\nuker{j}{\Delta_k}{\tau}  &\label{eq:nudelta}\coloneqq (2\pi i )^{j+1} \tau^j \Delta_k(\tau)\,\dd\tau\,,\\
\ee{j}{\Delta_k}{\tau} &\coloneqq\int_\tau^{i\infty} \nuker{j}{\Delta_k}{\tau_1}=  (2\pi i)^{1+j}  \int_\tau^{i\infty} \tau_1^j \Delta_k(\tau_1)\,\dd\tau_1\,  ,
\notag
\end{align}
closely follows (\ref{nuker}) and (\ref{MMVsec.07}), replacing the letter $k$ in the lower line by $\Delta_k$ for every instance of a cusp form instead of a holomorphic Eisenstein series. The range of the upper index is still $0\leq j\leq k{-}2$. At higher modular depth, for instance, $\ee{j_1&j_2 }{k_1&\Delta_{k_2}}{\tau}$ denotes a double integral as in (\ref{doubleint}) over
$\nuker{j_1}{k_1}{\tau_1}$ and $\nuker{j_2}{\Delta_{k_2}}{\tau_2}$.

\subsubsection{Modular properties}
\label{sec:modE}

The iterated integral~\eqref{MMVsec.07} does not have clean modular properties under the action of $\SLtwoZ$. Under the $T$-transformation we get for instance at modular depth one
\begin{align}
\ee{j}{k}{\tau{+}1} &= \int_{\tau{+}1}^{i\infty} \nuker{j_1}{k_1}{\tau_1}= \int_\tau^{i\infty -1} \nuker{j_1}{k_1}{\tau_1{+}1} \label{eettrf} \\
&= \int_\tau^{i\infty}\nuker{j_1}{k_1}{\tau_1{+}1}+  \int_{i\infty}^{i\infty-1}\nuker{j_1}{k_1}{\tau_1{+}1}   \nn\\
&= \sum_{p=0}^j (2\pi i)^{j-p} \binom{j}{p} \ee{p}{k}{\tau}
+ (2\pi i)^{1+j-k} \sum_{p=0}^j (-1)^{p+1} \binom{j}{p} \frac{2\zeta_k  }{p{+}1}\,, \notag
\end{align}
where we used the kernels' transformation property \eqref{eq:nukerT}, while the $T$-period integral is regularised using $\int_{i\infty}^{i\infty-1} \tau_1^j \GG_k(\tau_1) \dd \tau_1 = 2\zeta_k (-1)^{j+1}/(j{+}1)$ in agreement with the tangential-base-point scheme. It only has contributions from the zero mode of $\GG_k$. For holomorphic cusp forms the corresponding $T$-period integral vanishes and $\ee{j}{\Delta_k}{\tau{+}1}$ is a linear combination of $\ee{p}{\Delta_k}{\tau}$ with $0\leq p\leq j$.

For higher modular depth, the general behaviour of iterated integrals under compositions  
of paths (in this case $(\tau,i\infty{-}1)$ versus $(\tau,i\infty)$ followed 
by $(i\infty,i\infty{-}1)$) leads to
\begin{align}
\ee{j_1 &\ldots &j_\ell}{k_1 &\ldots &k_\ell}{\tau{+}1}  &= \sum_{r=0}^\ell
 \int_{\tau}^{i\infty} \nuker{j_r}{k_r}{\tau_r{+}1}
 \int_{\tau_{r}}^{i\infty} \ldots  \int_{\tau_{2}}^{i\infty} 
\nuker{j_1}{k_1}{\tau_1{+}1}
\label{Tathigherdepth} \\
&\quad \times
\int_{i\infty}^{i\infty-1} \nuker{j_\ell}{k_\ell}{\tau_\ell{+}1} \int_{\tau_\ell}^{i\infty-1} \ldots
\int_{\tau_{r+2}}^{i\infty-1} \nuker{j_{r+1}}{k_{r+1}}{\tau_{r+1}{+}1}
\, ,
\notag
\end{align}
where the entries of the forms $\nuker{j_{i}}{k_{i}}{\tau_{i}{+}1}$ on the right-hand side 
deconcatenate the word on the left-hand side in all ordering-preserving ways.
The iterated integral over a path $(\tau,i\infty)$ in the first line of 
(\ref{Tathigherdepth}) yields combinations of $\ee{p_1 &\ldots &p_r}{k_1 &\ldots &k_r}{\tau}$ 
with $0\leq p_i\leq j_i$ and $\mathbb Q[2\pi i]$ coefficients similar to the first term 
in (\ref{eettrf}).
The $T$-period integral over a path $(i\infty ,i\infty{-}1)$ in the second line of (\ref{Tathigherdepth}) 
evaluates to a rational multiple of $(2\pi i)^{\ell-r + j_{r+1}+\ldots +j_\ell}$, generalising 
the second term in (\ref{eettrf}), and vanishes in passing to cusp forms 
$k_i\rightarrow \Delta_{k_i}$ for one or more of $i=r{+1},\ldots,\ell$. 

Under the $S$-modular transformation, the analogue of the result (\ref{eettrf}) at modular depth one  is
\begin{align}
\ee{j}{k}{-\tfrac1{\tau}} &=  \int_{-1/\tau}^{i\infty} \nuker{j_1}{k_1}{\tau_1}=  \int_\tau^{0}  \nuker{j_1}{k_1}{-\tfrac{1}{\tau_1}} \label{strfd1} \\
&\nn= (-1)^j (2\pi i)^{2+2j-k}\int_\tau^{i\infty}   \nuker{k-2-j}{k}{\tau}-(-1)^j (2\pi i)^{2+2j-k}\int_0^{i\infty}   \nuker{k-2-j}{k}{\tau} \nn\\
&= (-1)^j  (2\pi i)^{2+2j-k} \ee{k{-}2{-}j}{k}{\tau}  - (-1)^j (2\pi i)^{1+j-k} \MMV{j}{k}\,,
\notag
\end{align}
where we used the kernel transformation property \eqref{eq:nukerS} and have defined the regularised $S$-period integral\footnote{Since Riemann zeta values $\zeta_{n}$ at negative integers $n$ are rational, vanish for negative even integers and are rational multiples of powers of $\pi$ for positive even integers, the expression for $\MMV{j}{k}$ in (\ref{mmvd1}) evaluates to rational multiples of $\pi^k$ if $1\leq j \leq k{-}1$ and involves the odd zeta value $\zeta_{k-1}$ at $j\in \{0,k{-}2\}$, e.g.\ $\MMV{k-2}{k}=\frac{2\pi i \zeta_{k{-}1}}{k{-}1}$. The transcendental weight of $\MMV{j}{k}$ is $k$ for all values of $j$.}
\begin{align}
\MMV{j}{k} \coloneqq \int_0^{i\infty} \tau_1^{j} \GG_k(\tau_1)\,\dd\tau_1  = \left\{\begin{array}{cl}  \displaystyle
- \frac{2\pi i \zeta_{k{-}1}}{k{-}1} &: \ \text{$j=0 \, $,}\\[2mm] \displaystyle
\frac{2 (-1)^{j{+}1} j! (2\pi i)^{k{-}1{-}j}}{(k{-}1)!} \zeta_{j{+}1}\zeta_{j{+}2{-}k} &: \ \text{$0<j\leq k{-}2 \,, $}
\end{array}\right.
\label{mmvd1}
\end{align}
that in this case has contributions from the non-zero modes of the modular form.
The quantity $\MMV{j}{k}$ introduced here is a known period integral~\cite{ZagierF} but can be also thought of as the simplest instance of a \textit{multiple modular value} (MMV)~\cite{Brown:mmv}. For the analogous iterated integrals over a path $(0,i\infty)$ at higher modular depth, numbers more complicated than (multiple) zeta values arise and their appearances is a central theme of our investigation: Given the definition
\beq
\MMV{j_1 &j_2 &\ldots &j_\ell}{k_1 &k_2 &\ldots &k_\ell} = \int_0^{i \infty}  \tau_\ell^{j_\ell} \GG_{k_\ell}(\tau_\ell) \,\dd \tau_\ell\,\int_{\tau_\ell}^{i \infty}  \cdots  \int_{\tau_3}^{i \infty}  \tau_2^{j_2} \GG_{k_2}(\tau_2)\,\dd \tau_2 \int_{\tau_2}^{i \infty}  \tau_1^{j_1} \GG_{k_1}(\tau_1) \,\dd \tau_1 \,,\label{eq:mmvdef}
\eeq
of higher modular depth MMVs, the modular $S$-transformation of generic iterated Eisenstein integrals (\ref{MMVsec.07}) can be derived from the composition-of-paths formula similar to (\ref{Tathigherdepth}) (here applied to $(\tau,0)$ versus $(\tau,i\infty)$ followed 
by $(i\infty,0)$),
\begin{align}
&\ee{j_1 &\ldots & j_\ell}{k_1 &\ldots & k_\ell}{{-}\tfrac{1}{\tau}} = \sum_{r=0}^{\ell} (2 \pi i)^{r+j_1+\ldots+j_r-k_1-\ldots-k_r}  \MMV{j_1 &\ldots &j_r}{k_1 &\ldots &k_r} \label{eestrf} \\
&\quad \times (2\pi i)^{2(\ell - r)+ 2 j_{r+1}+\ldots+2 j_\ell-k_{r+1}-\ldots-k_\ell}(-1)^{j_{r+1}+\ldots+j_\ell} 
\ee{k_{r+1}-j_{r+1}-2 &\ldots & k_\ell-j_\ell-2}{k_{r+1} &\ldots & k_\ell}{ \tau} \, ,
\notag
\end{align}
for instance
\begin{align}
\ee{j_1 & j_2}{k_1 & k_2}{{-}\tfrac{1}{\tau}} &=
(2\pi i)^{4+2j_1+2j_2-k_1-k_2} (-1)^{j_1+j_2} \ee{k_1-j_1-2 & k_2-j_2-2}{k_1 & k_2}{\tau}   \label{mt.27} \\
&\quad
+(2\pi i)^{1+j_1-k_1} \MMV{j_1}{k_1}
(2\pi i)^{2+2j_2-k_2} (-1)^{j_2} \ee{ k_2-j_2-2}{k_2}{\tau}  
\notag \\
&\quad +   (2\pi i)^{2+j_1+j_2-k_1-k_2} \MMV{j_1 &j_2}{k_1 &k_2} \, .
\notag
\end{align}
Iterated integrals involving holomorphic cusp forms $\Delta_k \in \mathcal{S}_k$ as in (\ref{eq:nudelta}) give rise to generalisations of MMVs in their modular $S$-transformations. Our conventions
\begin{align}
\MMV{j_1}{\Delta_{k_1}} &=(2\pi i)^{k_1} \int_0^{i\infty}  \tau_1^{j_1} \Delta_{k_1}(\tau_1)\, \dd \tau_1\,,
\label{MMVsec.17b} \\
\MMV{j_1 &j_2}{k_1 &\Delta_{k_2}} &= (2\pi i)^{k_2}  \int_0^{i\infty}  \tau_2^{j_2} \Delta_{k_2}(\tau_2)\,\dd \tau_2
\int_{\tau_2}^{i\infty} \tau_1^{j_1} {\rm G}_{k_1}(\tau_1)\,\dd \tau_1\, ,
\notag
\end{align}
ensure that (\ref{eestrf}) readily generalises to situations with some of the labels $k_i$ replaced by $\Delta_{k_i}$. In both cases, the substitution rule for the integration kernels reads ${\rm G}_{k}\rightarrow (2\pi i)^k \Delta_k$, leading to Fourier coefficients in $\mathbb K \pi^k$ with $\mathbb K$ the number-field extension of $\mathbb Q$ defined by the Fourier coefficients $\{a_k(n), n\in \mathbb{N}\}$ of the Hecke normalised cusp form $\Delta_{k}(\tau) = \sum_{n=1}^\infty a_k(n) q^n\in \mathcal{S}_{k}$.\footnote{The relation between $\mathbb{K}$ and the period polynomial of the cusp form is discussed in~\cite{Manin:1973}. Note that the number field $\mathbb{K}$ differs from $\mathbb{Q}$ when ${\rm dim} \,\mathcal{S}_{k}\neq 1$.  For example at $k=24$ we have ${\rm dim}\,\mathcal{S}_{24}=2$ and the associated number-field extension of $\mathbb{Q}$ is given by $\mathbb{K}= \mathbb{Q}(\sqrt{144169})$.}

Note that at modular depth one, the MMVs associated with a cusp form $\Delta_k$ simply correspond to critical completed L-values, i.e.
\begin{equation}
\MMV{j}{\Delta_{k}} = (2\pi )^k i^{k+j+1} \Lambda(\Delta_{k},j{+}1)\,,
\end{equation}
where given a holomorphic cusp form $\Delta_{k}(\tau) = \sum_{n=1}^\infty a_k(n) q^n$, we defined its completed L-function via meromorphic analytic continuation in the variable $t$ of the Dirichlet series
\begin{equation}
\Lambda(\Delta_k, t) \coloneqq \frac{\Gamma(t)} {(2\pi)^{t}} \sum_{n=1}^\infty \frac{a_k(n)}{n^t}\,.
\label{defLfct}
\end{equation}
Our aim is to build combinations of iterated integrals that have good modular properties. This will be achieved by considering very specific combinations of the iterated integrals above and their complex conjugates. 

\subsubsection{Alternative basis of integration kernels}
\label{sec:altker}

The construction of modular forms from iterated Eisenstein integrals is facilitated
by reorganising the integration kernels $\sim \tau_1^j \GG_k(\tau_1)$ in (\ref{nuker}) to
\cite{Dorigoni:2022npe}
\begin{align}
\omplus{j}{k}{\tau,\tau_1} &\coloneqq   \bigg(\frac{\tau{-}\tau_1}{4y}\bigg)^{k-2-j}
 (\bar \tau{-}\tau_1)^j {\rm G}_k(\tau_1)\,\frac{  \dd \tau_1 }{2\pi i }\, ,
 \label{cuspat3.01} \\
 \omminus{j}{k}{\tau,\tau_1} &\coloneqq {-}   \bigg(\frac{\tau{-}\bar \tau_1}{4y}\bigg)^{k-2-j}
 (\bar \tau{-}\bar \tau_1)^j \overline{{\rm G}_k(\tau_1)}\,\frac{  \dd \bar \tau_1 }{2\pi i }\, ,
 \notag
\end{align}
where we restrict the integer $j$ to the range $0\leq j\leq k{-}2$ and introduced
the shorthand
\begin{align}
y \coloneqq \pi \Im \tau = \frac{\pi}{2i} \,(\tau{-}\bar\tau)\,.
\end{align}
By their holomorphic (antiholomorphic) dependence on the integration variable $\tau_1$, the 
integration kernels $\omega_{+}$ and  $\omega_{-}$ in (\ref{cuspat3.01}) give
rise to homotopy invariant iterated integrals \cite{Dorigoni:2022npe}
\begin{align}
 \bplus{j_1 &j_2& \ldots &j_\ell}{k_1 &k_2 &\ldots &k_\ell}{\tau} &\coloneqq \int_\tau^{i\infty} 
 \omplus{j_\ell}{k_\ell}{\tau,\tau_\ell}  
  \ldots  \int_{\tau_3}^{i\infty}\omplus{j_2}{k_2}{\tau,\tau_2}\int_{\tau_2}^{i\infty} \omplus{j_1}{k_1}{\tau,\tau_1}\,,
\label{cuspat3.03}
\\
\bminus{j_1 &j_2 &\ldots &j_\ell}{k_1 &k_2 &\ldots &k_\ell}{\tau} &\coloneqq \int  _{\bar \tau}^{-i\infty} 
 \omminus{j_\ell}{k_\ell}{\tau,\tau_\ell}  
  \ldots \int_{{\bar{\tau}}_3}^{-i\infty} \omminus{j_2}{k_2}{\tau,\tau_2}\int_{{\bar{\tau}}_2}^{-i\infty}\omminus{j_1}{k_1}{\tau,\tau_1}\, ,
  \notag
\end{align}
despite their non-holomorphic dependence on $\tau$. By isolating the powers of the integration variables $\tau_i$
in (\ref{cuspat3.01}) via binomial expansion, one can straightforwardly relate the iterated integrals $\beta_{\pm}$ to the earlier iterated Eisenstein integrals ${\cal E}$ in
(\ref{MMVsec.07}) and their complex conjugates, with simple rational functions of $\tau$ and $\bar \tau$ as coefficients.

In the previous basis of integration kernels $\sim \tau_1^j \GG_k(\tau_1)$, modular transformations
mixed different values of $j$, see (\ref{eq:nukerT}) and (\ref{eq:nukerS}). The kernels in (\ref{cuspat3.01}) by contrast are engineered to transform as modular forms
of purely antiholomorphic modular weight $k{-}2{-}2j$,
\begin{align}
\ompm{j}{k}{\tau{+}1,\tau_1{+}1} &= \ompm{j}{k}{\tau,\tau_1} \, , \ \ \ \ \ \
\ompm{j}{k}{{-}\tfrac{1}{\tau},{-}\tfrac{1}{\tau_1}} = \bar \tau^{k-2-2j}\ompm{j}{k}{ \tau, \tau_1}
\label{modompm}
\,.
\end{align}
As a consequence, the terms of highest modular depth
 in the modular transformation of the iterated integrals (\ref{cuspat3.03}) are those of modular forms, with lower-modular-depth corrections
from $T$- and $S$-period integrals over $\int_{i\infty}^{i\infty-1}$ and $\int_{i\infty}^{0}$.
At modular depth one, for instance, the integrals $\beta_{\pm}$ on the right-hand side of  
\begin{align}
\bpm{j}{k}{\tau{+}1} &= \bpm{j}{k}{\tau} 
\mp \frac{2\zeta_k}{2\pi i} \sum_{p_1=0}^{k{-}2{-}j} \sum_{p_2=0}^j \binom{k{-}2{-}j}{p_1}\binom{j}{p_2}\frac{\tau^{k{-}2{-}j{-}p_1} \bar\tau^{j{-}p_2}}{(p_1{+}p_2{+}1)(4y)^{k{-}2{-}j}} \,,\label{eq:bminusT}\\
\bpm{j}{k}{-\tfrac{1}{\tau}} &= \bar\tau^{k{-}2{-}2j}  \bpm{j}{k}{\tau} 
{-} \frac{ (\tau \bar\tau)^{k{-}2{-}j} }{2\pi i(4y)^{k{-}2{-}j}} 
\sum_{p_1=0}^{k{-}2{-}j} \!\sum_{p_2=0}^j \binom{k{-}2{-}j}{p_1}\binom{j}{p_2} 
\frac{(\mp 1)^{p_1+p_2} }{\tau^{p_1} \bar \tau^{p_2}}
 \MMV{p_1{+}p_2}{k},
\notag
\end{align}  
no longer share the mixing of different $j$ as seen in (\ref{strfd1}) and (\ref{eettrf}).
The $T$-periods in (\ref{eq:bminusT}) are easily seen to cancel from the
sum $\bplus{j}{k}{\tau} + \bminus{j}{k}{\tau} $, see section
\ref{sec:6} for a discussion of higher modular depth analogues in terms of generating series. The $S$-periods in (\ref{eq:bminusT}) are combinations of the MMVs in (\ref{mmvd1}) where $ \MMV{j}{k}$ at odd values of $j$ cancel from $\bplus{j}{k}{\tau} + \bminus{j}{k}{\tau} $.
As a consequence of (\ref{modompm}), the leading-modular-depth contributions to $\bpm{j_1 &\ldots & j_\ell}{k_1&\ldots &k_\ell}{\frac{a \tau{+}b}{c\tau{+}d}}$ line up with modular forms of weight $(0, \sum_{i=1}^\ell (k_i{-}2{-}2j_i))$. 

In close analogy with (\ref{eq:nudelta}), we also introduce a variant of
the forms $\omega_{\pm}$ and iterated integrals $\beta_{\pm}$ with
holomorphic cusp forms $(2\pi i)^k\Delta_k$ in place of the Eisenstein series $\GG_k$,
for instance 
\begin{align}
\omplus{j}{\Delta_k}{\tau,\tau_1} &\coloneqq   \bigg(\frac{\tau{-}\tau_1}{4y}\bigg)^{k-2-j}
 (\bar \tau{-}\tau_1)^j (2\pi i)^k \Delta_k(\tau_1)\,\frac{  \dd \tau_1 }{2\pi i }\, ,
 \label{cuspat3.02} \\
 \omminus{j}{\Delta_k}{\tau,\tau_1} &\coloneqq{-}  \bigg(\frac{\tau{-}\bar \tau_1}{4y}\bigg)^{k-2-j}
 (\bar \tau{-}\bar \tau_1)^j (2\pi i)^k \overline{\Delta_k(\tau_1)}\, \frac{  \dd \bar \tau_1 }{2\pi i }\, ,
 \notag
\end{align}
as well as $\bplus{j}{\Delta_k}{\tau} = \int_\tau^{i\infty} 
 \omplus{j}{\Delta_k}{\tau,\tau_1} $ and $\bminus{j}{\Delta_k}{\tau} = \int  _{\bar \tau}^{-i\infty}
 \omminus{j}{\Delta_k}{\tau,\tau_1}$ with obvious generalisations to higher modular depth and arbitrary combinations of ${\rm G}_k$ and $\Delta_k$ kernels.

\subsubsection{Differential properties}

{}From their definition, the iterated integrals~\eqref{MMVsec.07} and their cusp-form generalisations satisfy simple differential equations with respect to $\tau$-derivatives:
\begin{align}
\partial_\tau \ee{j_1&\ldots&j_{\ell-1}&j_\ell}{k_1&\ldots&k_{\ell-1}&k_\ell}{\tau} &= - (2\pi i)^{1+j_\ell-k_\ell}\,  \tau^{j_\ell} \GG_{k_\ell}(\tau) \ee{j_1&\ldots&j_{\ell-1}}{k_1&\ldots&k_{\ell-1}}{\tau} \,,\nn\\
\partial_\tau \ee{j_1&\ldots&j_{\ell-1}&j_\ell}{k_1&\ldots&k_{\ell-1}&\Delta_{k_\ell}}{\tau} &= - (2\pi i)^{1+j_\ell}\,  \tau^{j_\ell} \Delta_{k_\ell}(\tau) \ee{j_1&\ldots&j_{\ell-1}}{k_1&\ldots&k_{\ell-1}}{\tau} \,,\nn\\
\partial_{\bar\tau} \ee{j_1&\ldots&j_{\ell-1}&j_\ell}{k_1&\ldots&k_{\ell-1}&k_\ell}{\tau} &= 0\,.
\label{diffcalE}
\end{align}
The differential equations of the iterated integrals $\beta_{\pm}$ in (\ref{cuspat3.03})
are modified by the $\tau$-dependence of the modular kernels $\omega_{\pm}$ in (\ref{cuspat3.01}) 
\begin{align}
2\pi i(\tau{-}\bar \tau)^2 \partial_\tau  \bplus{j_1 &j_2& \ldots &j_\ell}{k_1 &k_2 &\ldots &k_\ell}{\tau} &= \sum_{i=1}^\ell (k_i{-}j_i{-}2) \bplus{j_1 & \ldots &j_i+1 &\ldots &j_\ell}{k_1 &\ldots &k_i &\ldots &k_\ell}{\tau}\label{dtaubeta} \\
&\quad  - \delta_{j_\ell,k_\ell-2} (\tau{-}\bar \tau)^{k_\ell} \GG_{k_\ell}(\tau) 
 \bplus{j_1 &j_2& \ldots &j_{\ell-1}}{k_1 &k_2 &\ldots &k_{\ell-1}}{\tau} \, ,
 \notag \\
2\pi i(\tau{-}\bar \tau)^2 \partial_\tau  \bminus{j_1 &j_2& \ldots &j_\ell}{k_1 &k_2 &\ldots &k_\ell}{\tau}
&= \sum_{i=1}^\ell (k_i{-}j_i{-}2) \bminus{j_1 & \ldots &j_i+1 &\ldots &j_\ell}{k_1 &\ldots &k_i &\ldots &k_\ell}{\tau} \,,\notag
\end{align} 
where the analogous differential equations for $\begin{smallmatrix} j_\ell \\ k_\ell \end{smallmatrix}\rightarrow \begin{smallmatrix} j_\ell \\ \Delta_{k_\ell} \end{smallmatrix}$ are again obtained by replacing
${\rm G}_{k_\ell} \rightarrow (2\pi i)^{k_{\ell}}\Delta_{k_\ell}$ on the right-hand side.
The derivatives with respect to $\bar\tau$ are given by
\begin{align}
2\pi i(\tau{-}\bar \tau)^2 \partial_{\bar\tau}  \bplus{j_1 &j_2& \ldots &j_\ell}{k_1 &k_2 &\ldots &k_\ell}{\tau} 
&= 
-(4y) (k{-}2{-}2j) \bplus{j_1 &j_2& \ldots &j_\ell}{k_1 &k_2 &\ldots &k_\ell}{\tau}
\label{dtaubarbeta}\\*
&\quad 
- (4y)^2
\sum_{i=1}^\ell j_i  \,\bplus{j_1 & \ldots &j_i-1 &\ldots &j_\ell}{k_1 &\ldots &k_i &\ldots &k_\ell}{\tau}\,, \notag \\
2\pi i(\tau{-}\bar \tau)^2 \partial_{\bar\tau}  \bminus{j_1 &j_2& \ldots &j_\ell}{k_1 &k_2 &\ldots &k_\ell}{\tau}
&=
-(4y) (k{-}2{-}2j) \bminus{j_1 &j_2& \ldots &j_\ell}{k_1 &k_2 &\ldots &k_\ell}{\tau}\nn\\*
&\quad 
- (4y)^2
\sum_{i=1}^\ell j_i  \,\bminus{j_1 & \ldots &j_i-1 &\ldots &j_\ell}{k_1 &\ldots &k_i &\ldots &k_\ell}{\tau}
\notag\\*
&\quad  - \delta_{j_\ell,0} \frac{(4y)^{2}}{(2\pi i)^{k_\ell}}  \overline{\GG_{k_\ell}(\tau)} 
 \bminus{j_1 &j_2& \ldots &j_{\ell-1}}{k_1 &k_2 &\ldots &k_{\ell-1}}{\tau} \, . \notag
\end{align} 
The first terms on the right-hand sides are due to the non-vanishing anti-holomorphic modular weight of $\beta_\pm$
and can be accounted for by promoting $ \partial_{\bar\tau} $ to a Maa\ss{} derivative \cite{Maass}.

We will later on construct real-analytic
combinations of $\mathcal{E}[\cdots]$ and $\overline{\mathcal{E}[\cdots]}$
as well as $\beta_+[\cdots]$ and $\beta_-[\cdots]$ with good modular
properties that preserve the differential equations 
in $\partial_\tau$ at the cost of more complicated $\bar\tau$ derivatives. 


\subsection{Modular graph forms and beyond}
\label{sec:rev.3}

Modular graph forms (MGFs) \cite{DHoker:2015wxz, DHoker:2016mwo} are non-holomorphic modular forms that arise from the low-energy expansion of configuration-space integrals in one-loop closed-string amplitudes, see \cite{Gerken:review, Berkovits:2022ivl, Dorigoni:2022iem, DHoker:2022dxx} for overview references. More precisely, MGFs are defined through a graphical organisation of the underlying conformal-field-theory correlators, assigning an edge for each (generalised) scalar propagator connecting pairs of punctures on the genus-one world-sheet. 

\begin{figure}[t!]
\centering
\begin{tikzpicture}
  \draw[-,draw=black,thick] (-2,0) circle (1cm);
  \filldraw[black] ({-2+1*cos(0)},{1*sin(0)}) circle (2pt);
  \filldraw[black] ({-2+1*cos(40)},{1*sin(40)}) circle (2pt);
  \filldraw[black] ({-2+1*cos(80)},{1*sin(80)}) circle (2pt);  
  \filldraw[black] ({-2+1*cos(-40)},{1*sin(-40)}) circle (2pt);  
 \draw[black] ({-2+1*cos(20)},{1*sin(20)}) node[anchor=west] {\footnotesize (2,0)};
 \draw[black] ({-2+1*cos(60)-.1},{1*sin(60)+0.2}) node[anchor=west] {\footnotesize (2,0)};
 \draw[black] ({-2+1*cos(-20)},{1*sin(-20)}) node[anchor=west] {\footnotesize (2,0)};
  \draw[black, thick, dashed] ({-2+1.1*cos(90)},{1.1*sin(90)})  arc (90:320:1.1);
  \draw[black] (-2,-2) node[anchor=south] {(a): ${\rm G}_{2k}$};
\end{tikzpicture}
\hspace{10mm}
\begin{tikzpicture}
  \draw[-,draw=black,thick] (-2,0) circle (1cm);
  \filldraw[black] ({-2+1*cos(0)},{1*sin(0)}) circle (2pt);
  \filldraw[black] ({-2+1*cos(40)},{1*sin(40)}) circle (2pt);
  \filldraw[black] ({-2+1*cos(80)},{1*sin(80)}) circle (2pt);  
  \filldraw[black] ({-2+1*cos(-40)},{1*sin(-40)}) circle (2pt);  
 \draw[black] ({-2+1*cos(20)},{1*sin(20)}) node[anchor=west] {\footnotesize (1,1)};
 \draw[black] ({-2+1*cos(60)-.1},{1*sin(60)+0.2}) node[anchor=west] {\footnotesize (1,1)};
 \draw[black] ({-2+1*cos(-20)},{1*sin(-20)}) node[anchor=west] {\footnotesize (1,1)};
\draw[black, thick, dashed] ({-2+1.1*cos(90)},{1.1*sin(90)})  arc (90:320:1.1);
  \draw[black] (-2,-2) node[anchor=south] {(b):  ${\rm E}_{k}$};
\end{tikzpicture}
\hspace{10mm}
\begin{tikzpicture}
  \filldraw[black] (0,0) circle (2pt);
  \filldraw[black] (2,0) circle (2pt);
  \filldraw[black] (1,1) circle (2pt);
  \draw[black,thick] (0,0) -- (2,0);
  \draw[-,draw=black,thick] (1,0) circle (1cm);
  \draw [black] (0.1,0.7) node[anchor=south] {\footnotesize (1,1)};
  \draw [black] (1.9,0.7) node[anchor=south] {\footnotesize (1,1)};
  \draw [black] (1,-0.1) node[anchor=south] {\footnotesize (1,1)};
  \draw [black] (1,-1.1) node[anchor=south] {\footnotesize (1,1)};  
  \draw[black] (1,-2) node[anchor=south] {(c):  $C_{2,1,1}$};
\end{tikzpicture}
\caption{\label{fig:mgfs}\textit{Some modular graphs with associated MGFs indicated. The one-loop graphs in panels (a) and (b) have $k$ links each while the two-loop graph in panel (c) has four links.
}}
\end{figure}
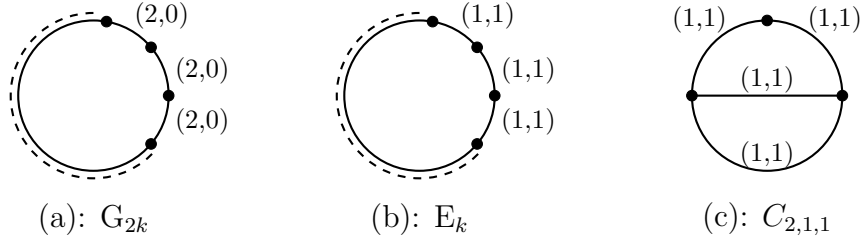

Momenta on the compact world-sheet torus with complex modulus $\tau \in \mathbb H$ are 
quantised as $p=m\tau{+}n$ with $m,n\in\mathbb{Z}$. The Feynman integrals corresponding to a given graph are then discretised to sums over such momenta, where the conformal-field-theory building blocks of MGFs exclude the value $p=0$ as an automatic infrared regulator. For this reason, a rather generic expression for MGFs associated with a dihedral graph $\Gamma$ with $r$ edges is given (when it converges) by
\begin{align}
C_\Gamma(\tau) = 
    N_\Gamma \sum_{p_1\neq 0}\cdots \sum_{p_r\neq 0} \frac{\delta_\Gamma(\{p_i\})}{p_1^{a_1} \bar{p}_1^{b_1}\cdots p_r^{a_r} \bar{p}_r^{b_r}}\,,
    \label{cplusconv}
\end{align}
see \cite{DHoker:2016mwo, Gerken:2018zcy} for discussions of trihedral MGFs and \cite{Basu:2015ayg,Kleinschmidt:2017ege,Gerken:review} for more general graph topologies.
We have allowed for different exponents $a_i,b_i \in \mathbb Z$ of the momenta $p_i=m_i\tau{+}n_i$ and their complex conjugates $\bar p_i=m_i\bar \tau{+}n_i$. The numerator $\delta_\Gamma(\{p_i\})$ gathers the momentum-conserving Kronecker deltas of the graph arising at each vertex and thereby encodes its adjacency relations. The convention-dependent normalisation factor $N_\Gamma$ usually comprises integer powers of $\pi$ and $\Im \tau$ where the choices in numerous physics references assign modular weight $\big(0,\sum_{i=1}^r (b_i{-}a_i)\big)$ to the MGF in (\ref{cplusconv}). In graphical representations as in figure~\ref{fig:mgfs}, the exponents of the momentum $p_i$ translate into labels $(a_i,b_i)$ of the associated edge.

On the one hand, lattice-sum representations as in (\ref{cplusconv}) expose that MGFs are modular forms of $\SLtwoZ$. On the other hand, the wealth of algebraic and differential relations among MGFs \cite{DHoker:2015gmr, DHoker:2015sve, DHoker:2016mwo, Basu:2016kli, DHoker:2016quv, Gerken:2018zcy} are typically not evident from lattice sums. 

\subsubsection{Illustrative examples}

The holomorphic Eisenstein series $\GG_k(\tau)$ introduced in~\eqref{eq:GGk} is related to a modular graph form associated with the one-loop graph in the left panel (a) of figure~\ref{fig:mgfs} via 
\begin{align}
    \GG_{2k}(\tau) = \sum_{p_1\neq 0} \cdots \sum_{p_k\neq 0} \frac{\delta(p_1{-}p_2)\cdots \delta(p_{k-1}{-}p_k)\delta(p_k{-}p_1)}{p_1^2 \cdots p_k^2} = \sum_{p\neq 0} \frac{1}{p^{2k}}\,.
\end{align}
(Here, the factor $N_\Gamma$ was chosen to match the definition of the holomorphic Eisenstein series~\eqref{eq:GGk}.)
Similarly, we can view the non-holomorphic Eisenstein series as a modular graph form associated with the same one-loop graph but different edge labels $(1,1)$,
\begin{align}
\label{eq:EEk}
    \EE_k (\tau)= \frac{(\Im\tau)^k}{\pi^k} \sum_{p_1\neq 0} \cdots \sum_{p_k\neq 0} \frac{\delta(p_1{-}p_2)\cdots \delta(p_{k-1}{-}p_k)\delta(p_k{-}p_1)}{|p_1|^2 \cdots |p_k|^2} = \frac{(\Im\tau)^k}{\pi^k} \sum_{p\neq 0} \frac{1}{|p|^{2k}}\,,
\end{align}
see the middle panel (b) of figure \ref{fig:mgfs}. As a final example, we consider the two-loop graph in the right panel (c) of figure \ref{fig:mgfs}, where simplifications of the momentum-conserving delta functions in the associated modular graph form yield the modular invariant
\begin{align}
  C_{2,1,1}(\tau)=   \frac{(\Im\tau)^4}{\pi^4} \sum_{\substack{p_1,p_2\neq0\\p_1+p_2\neq 0}} \frac{1}{|p_1|^2 |p_2|^2  |p_1{+}p_2|^4}\,.
  \label{2loopmgf}
\end{align}
The study of closely related two-loop MGFs denoted in general by $C_{a,b,c}(\tau)$ with more general
lattice summands $|p_1|^{2b} |p_2|^{2c} |p_1{+}p_2|^{2a}$ considerably advanced the
low-energy expansion of the one-loop four-graviton 
amplitude of type-II superstrings~\cite{Green:1999pv, Green:2008uj, DHoker:2015gmr, DHoker:2019blr}, also
see \cite{Richards:2008jg, Green:2013bza, Basu:2016mmk} for the analogous expansions at five points.

\subsubsection{Differential equations}

MGFs satisfy a rich network of algebraic relations that make it hard to find a canonical basis in terms of graphs and lattice sums.\footnote{See \cite{Gerken:2020aju} for bases of MGFs with exponents $\sum_i(a_i{+}b_i)\leq 12$ and a {\tt Mathematica} package with a variety of MGF manipulations and integration routines for one-loop closed-string amplitudes.} In order to understand the space of MGFs, it is instructive to study their differential equations with respect to $\tau$ and $\bar\tau$. There exists standard differential operators, called Maa\ss{} operators, that act on the non-holomorphic modular forms of weight $(A,B)$~\cite{Maass}. Since the definition includes an explicit dependence on $A$ and $B$, we will here use adapted versions~\cite{DHoker:2016mwo}
\begin{align}
\nabla \coloneqq 2i (\Im \tau)^2 \partial_\tau \,,\hspace{10mm}
\overline{\nabla} \coloneqq -2i (\Im \tau)^2 \partial_{\bar\tau} 
\label{mops}
\end{align}
that act on forms of weight $(0,w)$ and $(w,0)$, respectively. In particular,
these differential operators shift the modular weights according to 
$\nabla: \, (0,w)\rightarrow(0,w{-}2)$ or $\overline{\nabla}: \, (w,0) \rightarrow (w{-}2,0)$,
which preserves the vanishing of the holomorphic or anti-holomorphic weight, respectively.
In this way, $\nabla$ and $\overline{ \nabla }$ can be applied several times. 

As an example, we can act with $\nabla$ on the non-holomorphic Eisenstein series $\EE_k$ in (\ref{eq:EEk}) that is modular invariant and thus has weight $(0,0)$. One can check that
\begin{align}
\label{eq:EGk}
(\pi \nabla)^k \EE_k = \frac{(2k{-}1)!}{(k{-}1)!}  (\Im \tau)^{2k} \GG_{2k}  \, ,
\end{align}
where both sides have modular weight $(0,-2k)$ since $\Im\tau$ has weight $(-1,-1)$ and $\GG_{2k}$ has weight $(2k,0)$. On MGFs associated with higher-loop graphs, repeated action of $\nabla$ brings out holomorphic Eisenstein series multiplying simpler MGFs as for instance seen in
\beq
\label{eq:C211Gk}
(\pi \nabla)^3 \big( C_{2,1,1} -\tfrac{9}{10} \EE_4 \big) = - 6 (\Im \tau)^4 \GG_4 \pi \nabla \EE_2 \, .
\eeq
 
Differential equations of the type~\eqref{eq:EGk} and (\ref{eq:C211Gk}) are at the heart of the sieve algorithm for analysing MGFs~\cite{DHoker:2016mwo} and can be extended to generating functions of closed-string integrals to compactly address infinite families of MGFs through their low-energy expansion \cite{Gerken:2019cxz}.

\subsubsection{Iterated-integral representations}
\label{sec:223}

One consequence suggested by~\eqref{eq:EGk} is that the non-holomorphic Eisenstein series $\EE_k$ should be expressible as a $\tau$-integral over $\GG_{2k}$. Based on the integrals introduced in~\eqref{MMVsec.07} and their differential equations (\ref{diffcalE}), one can indeed verify that for instance \cite[Eq.\ (4.25)]{Gerken:2020yii}  \begin{align}
\pi \Im \tau \,\EE_2(\tau) = 12 \pi^2 \tau \bar\tau  \Re \ee{0}{4}{\tau}-   
 6  \pi ( \tau{+}\bar \tau)  \Im \ee{1} {4}{\tau} 
 - 3  \Re \ee{2}{4}{\tau}
+  \zeta_3  \, ,
\label{exE2int}
\end{align}
in line with earlier representations of $\EE_k$ as iterated Eisenstein integrals \cite{Ganglzagier, DHoker:2015wxz, Brown:2017qwo, Broedel:2018izr}. However, (\ref{exE2int}) and similar iterated-integral representations of more general MGFs substantially simplify once we change our basis of integration kernels towards (\ref{cuspat3.01}).
The alternative organisation (\ref{cuspat3.03}) of iterated Eisenstein integrals brings
the expression (\ref{exE2int}) for the non-holomorphic Eisenstein series $\EE_2$ into the
more illuminating form 
\begin{align}
 \EE_2(\tau) &= - 6 \big(  \bplus{1}{4}{\tau} +  \bminus{1}{4}{\tau} \big) + \frac{ \zeta_3 }{y} 
\label{E2new}
\\
&= - \frac{3}{2\pi y} \Im \int_\tau^{i\infty} \dd \tau_1 \, (\tau{-}\tau_1) (\bar \tau{-}\tau_1) 
\GG_4(\tau_1)+ \frac{ \zeta_3 }{y} \, .
\notag
\end{align}
For the iterated-integral representation of the two-loop MGF (\ref{2loopmgf}),
passing from the basis of ${\cal E}$ and $\overline{{\cal E}}$
to the $\beta_{\pm}$ considerably streamlines the expressions and results in
\begin{align} 
C_{2,1,1}(\tau) &= -18 \big( \bplus{ 2 &0 }{4 &4}{\tau}
 + \bplus{ 0 }{4}{\tau} \bminus{ 2  }{4 }{\tau} +  \bminus{ 0 &2 }{4 &4}{\tau}  \big)  
 \label{intc211}  \\
 &\quad -126   \big(  \bplus{3}{8}{\tau} +  \bminus{3}{8}{\tau} \big)
 +  12 \zeta_3 \bigg(  \bplus{ 0 }{4 }{\tau} + \frac{\bminus{ 2  }{4 }{\tau}}{(4y)^2} \bigg) 
  \notag \\
  &\quad  + \frac{ \pi^2 \zeta_3 \tau \bar \tau }{60 y} 
 + \frac{ 5 \zeta_5}{12 y}- \frac{ \zeta_3^2}{ 4 y^2} 
 + \frac{ 9 \zeta_7 }{16 y^3}\, , \notag
\end{align}
identifying this two-loop MGF as an iterated Eisenstein integral at modular depth two.

As these two examples illustrate, MGFs are closely related to iterated Eisenstein integrals
and most conveniently written in their basis of $\beta_{\pm}$ instead of the integrals $\mathcal{E}$.

At this point, we caution the reader that the terms of modular depth zero are not the same as the Laurent polynomial (LP) of  a modular graph form. The latter is given by the purely $y$-dependent terms in the Fourier zero mode of an MGF and for the case of $C_{2,1,1}$ is~\cite{DHoker:2015gmr}
\begin{align}
    C_{2,1,1} \, \big|_{\text{LP}} = 
    \frac{2}{14175}y^4  + \frac{\zeta_3}{45} y  + \frac{5\zeta_5}{12y}  - \frac{\zeta_3^2}{4y^2}+ \frac{9\zeta_7}{16 y^3}\,.
    \label{lpc211}
\end{align}
The terms of modular depth zero in the last line of~\eqref{intc211} contain only some of the terms of the Laurent polynomial since the integrals $\beta_\pm$ have non-trivial contributions at the cusp. In particular, the first term $\sim \tau \bar \tau \zeta_3/y$ in the last line of~\eqref{intc211} conspires with the leading terms $\bplusno{ 0 }{4 }=  \frac{i \pi^3 \tau^3}{4320 y^2} +O(q)$ and  $\bminusno{ 2 }{4 }= -\frac{i  \pi^3\bar \tau^3 }{270}  +O(\bar q)$ at the cusp to build up the $T$-invariant contribution $\frac{\zeta_3}{45}y$ to the Laurent polynomial in (\ref{lpc211}).
In general, the notion of modular depth introduced below (\ref{MMVsec.07}) does not give the Laurent polynomial when projecting to modular depth zero.

\subsubsection{Modular combinations of iterated integrals}

By the modular $T$- and $S$-transformations of $\beta_{\pm}$ as in (\ref{eq:bminusT}), 
iterated-integral representations of MGFs no longer manifest that they transform as modular
forms. For instance, modular invariance of (\ref{exE2int}), (\ref{E2new}) and (\ref{intc211}) is tied
to the interplay of iterated Eisenstein integrals of different modular depths with multiple zeta values (MZVs) reviewed in appendix \ref{app:A}.

Non-holomorphic Eisenstein series and their derivatives under the Maa\ss\ operators in (\ref{mops}) realise the real-analytic modular completions of modular-depth-one integrals \cite{Brown:2017qwo, Gerken:2020yii}
\begin{align}
{\rm E}_k &= - \frac{(2k{-}1)!}{(k{-}1)!^2} \beqv{k-1}{2k}\,,
\label{eq:Cab}  \\
\sum_{p \neq 0} \frac{1}{p^a \bar p^b} &= - \frac{ \pi^b (2 i)^{b-a} (a{+}b{-}1)! }{(\Im \tau)^a (a{-}1)!(b{-}1)!} \beqv{a-1}{a+b} 
 \,, \notag
\end{align}
with $a,b \in \mathbb{N}$ and where the $T$- and $S$-cocycles in (\ref{eq:bminusT}) cancel between the different
contributions to
\begin{align}
\label{eq:beqv1}
\beqv{j}{k} &\coloneqq   \bplusno{j}{k} +  \bminusno{j}{k} 
- \frac{2 \zeta_{k{-}1}}{(k {-} 1) (4 y)^{k{-}2{-}j}} \,.
\end{align}
Similarly, (\ref{intc211}) identifies a modular-depth-two example of modular iterated integrals
\beq
C_{2,1,1}=
-18   \beqv{ 2 &0 }{4 &4} 
-126    \beqv{3}{8}\,,
\label{c221again}
\eeq
with more intricate cancellations between the $T$- and $S$-cocycles of
\begin{align}
\beqv{ 2 &0 }{4 &4} &= \bplusno{ 2 &0 }{4 &4}  + \bplusno{ 0 }{4} \bminusno{ 2  }{4 }  +  \bminusno{ 0 &2 }{4 &4} \label{beqvd2} \\
&\quad - \frac{2 \zeta_3}{3} \bigg(  \bplusno{ 0 }{4 } + \frac{ \bminusno{ 2  }{4 } }{(4y)^2} \bigg)
-\frac{ \pi^2 \tau \bar \tau \zeta_3}{1080 y} - \frac{ 5 \zeta_5}{216 y} + \frac{ \zeta_3^2 }{72y^2}\, .
\notag
\end{align}
The main goal of this work is to explicitly construct the modular completions
of arbitrary iterated Eisenstein integrals $\beta_+$,
\beq
\beqv{j_1 &j_2 &\ldots &j_\ell}{k_1 &k_2 &\ldots &k_\ell}
= \sum_{r=0}^\ell \bplusno{j_{r+1} &j_{r+2} &\ldots &j_\ell}{k_{r+1} &k_{r+2} &\ldots &k_\ell}
\bminusno{j_r &j_{r-1}&\ldots &j_{1}}{k_r &k_{r-1}&\ldots &k_{1}}+ \ldots\, ,
\label{beqvdell}
\eeq
where the ellipsis comprises iterated integrals of lower modular depth $\leq \ell{-}1$ and
rational functions of $\tau,\bar \tau$ with constant coefficients including MZVs
such as to attain the modular equivariant transformation
\beq
\beqvtau{j_1 &j_2 &\ldots &j_\ell}{k_1 &k_2 &\ldots &k_\ell}{\tfrac{a\tau+b}{c\tau+d}} =
\bigg( \prod_{i=1}^\ell (c\bar \tau{+}d)^{k_i-2j_i-2} \bigg)\beqvtau{j_1 &j_2 &\ldots &j_\ell}{k_1 &k_2 &\ldots &k_\ell}{\tau} \, .
\label{modbeqv}
\eeq
While the modular-depth $\ell$ terms in (\ref{beqvdell}) are by themselves $T$ invariant,
the modular-depth two example (\ref{beqvd2}) illustrates that the cancellation of the $T$-cocycle of
$\beta^{\rm eqv}$ generically relies on the interplay of different modular depths. Modular $S$ 
transformation of $\beta^{\rm eqv}$ in turn is tied to even more intricate conspiracies between
iterated integrals and MMVs (\ref{eq:mmvdef}) of different modular depths.
The superscript of $\beta^{\rm eqv}$ alludes to Brown's equivariant iterated Eisenstein
integrals \cite{Brown:2017qwo2} where several modular iterated integrals (\ref{beqvdell})
are combined with bookkeeping variable to be reviewed below that also transform 
under $\SLtwoZ$. Following Brown's approach, we will construct the modular 
completion in (\ref{beqvdell}) via generating series and pinpoint the structure and 
explicit form of the series in MZVs of \cite{Brown:2017qwo2}.

In slight abuse of terminology, we will sometimes attribute modular depth $\ell$ to the modular forms $\beqv{j_1 &j_2 &\ldots &j_\ell}{k_1 &k_2 &\ldots &k_\ell}$ even though some of the contributing products of holomorphic and antiholomorphic iterated Eisenstein integrals have combined modular depth $<\ell$, see for instance (\ref{eq:beqv1}) and (\ref{beqvd2}). By the linear-independence result for holomorphic iterated Eisenstein integrals in \cite{Nilsnewarticle}, there are no {\it linear} relations among $\beta^{\rm eqv}$ with different entries $j_i,k_i$, though the generating-series construction of \cite{Brown:2017qwo2} and this work imply standard (non-linear) shuffle relations.

Note that the approach of \cite{Gerken:2020yii} to the low-energy expansion of closed-string genus-one integrals organises MGFs in terms of real-analytic and $T$-invariant combinations $\bsv{j_1 &j_2 &\ldots &j_\ell}{k_1 &k_2 &\ldots &k_\ell}$ of $\beta_{\pm}$ that share the leading-modular-depth terms on the right-hand side of (\ref{beqvdell}). However, the $\beta^{\rm sv}$ are lacking most of the $S$-invariant completion by terms of lower modular depth in the ellipsis of (\ref{beqvdell}), e.g.\ the zeta values in (\ref{eq:beqv1}) at modular depth one are absent from $\bsv{j}{k}$. The systematics of
MZV contributions to $\beta^{\rm sv}$ at modular depth two and three was described in \cite{Dorigoni:2022npe}, and we will review the translation to $\beta^{\rm eqv}$ in (\ref{dictsveqv}) below.

\subsubsection{Modular iterated integrals beyond MGFs}
\label{sec:rev.4}

All MGFs can be expressed by definition in terms of (manifestly modular) lattice sums
over torus momenta $p_i$, while from their differential equations they can also be written in terms of iterated Eisenstein integrals. More precisely, the differential equations of MGFs \cite{DHoker:2016mwo} and their generating series \cite{Gerken:2019cxz} imply that the $\beta_{\pm}$ in their integral representations only involve Eisenstein series $\GG_k$ as their integration kernels and no holomorphic cusp forms. However, some of the modular completions of the iterated Eisenstein integrals in (\ref{beqvdell}) necessitate cuspidal integration kernels 
\cite{Brown:mmv, Brown:2017qwo}, say integrals $\bpmno{j &\ldots}{\Delta_k &\ldots}$ 
of modular depth $\ell{-}1$ (see the discussion below (\ref{cuspat3.02})) contributing to $\beqv{j_1 &j_2 &\ldots &j_\ell}{k_1 &k_2 &\ldots &k_\ell}$.
Even if their leading modular depth terms are solely built from Eisenstein kernels, 
modular integrals $\beta^{\rm eqv}$ with cusp-form admixtures cannot be
realised by MGFs.

Nevertheless, modular integrals $\beqv{j_1 &j_2}{k_1 &k_2}$ at modular depth two 
including their cusp-form contributions have been constructed in
\cite{Dorigoni:2021jfr, Dorigoni:2021ngn} using methods that
prominently feature in the MGF literature, namely Laplace equations and
Poincar\'e series. Two-loop MGFs $C_{a,b,c}$ including the example
(\ref{2loopmgf}) obey inhomogeneous Laplace eigenvalue equations such as \cite{DHoker:2015gmr}
\begin{equation}
  (\Delta - 2) C_{2,1,1}
= 9 \EE_4 - \EE_2^2  \,,
\label{FCEex}
\end{equation}
with $\Delta \coloneqq 4 (\Im \tau)^2 \partial_\tau \partial_{\bar \tau}$ the $\rm{SL}_2$ invariant Laplace-Beltrami operator. This inspired the assembly of all
the $\beqv{j_1 &j_2}{k_1 &k_2}$ at modular depth two from
solutions ${\rm F}_{m,k}^{\pm(s)}$ to inhomogeneous Laplace
equations with source terms built from ${\rm E}_m,{\rm E}_k$
and eigenvalues $s(s{-}1)$ with $s,m,k \in \mathbb N$ and
$|k{-}m|{+}1 \leq s \leq k{+}m{-}1$. In case of (\ref{FCEex}), we can eliminate the source term ${\rm E_4}$ by redefining ${\rm F}_{2,2}^{+(2)}= -C_{2,1,1}+\frac{9}{10}\mathrm{E}_4$ such that $(\Delta {-} 2) {\rm F}_{2,2}^{+(2)}
= \EE_2^2$.
The functions $\mathrm{F}_{m,k}^{\pm(s)}$ and their generalisation to modular depth three are discussed in more detail in section~\ref{sec:7}.

Integration
kernels involving holomorphic cusp forms arise for eigenvalues $s(s{-}1)\geq 30$, 
for example in the solution to
the modular Laplace equation \cite{Dorigoni:2021ngn}
\begin{align}
(\Delta -6\cdot 5) {\rm F}_{4,4}^{+(6)} = \EE_4 \EE_4\, .
\end{align}
The iterated-integral representation of ${\rm F}_{4,4}^{+(6)}$ and its
Maa\ss\ derivatives involves modular-depth-one integrals
$\bpmno{j}{\Delta_{12}}$ of the Hecke normalised Ramanujan discriminant cusp form $\Delta_{12}$ and their complex conjugates for various values of $j$. More generally, the
cuspidal contributions to the modular invariant ${\rm F}_{m,k}^{\pm (s)}$ are
packaged into Laplace eigenfunctions \cite{Dorigoni:2021ngn}
\begin{align}
{\rm H}_{\Delta_{2s}}^{\pm}(\tau)  &=  
- \frac{(i\pi)^{2s-1} }{(s{-}1)! y^{s-1}}
 \int_\tau^{i\infty} \dd\tau_1\, (\tau{-}\tau_1)^{s-1} (\bar\tau{-}\tau_1)^{s-1} {\Delta_{2s}}(\tau_1) \, \pm \, \text{c.c.} \notag \\
&= -\frac{1}{2(s{-}1)!} \Big(\bplus{s-1}{\Delta_{2s}}{\tau}\pm \bminus{s-1 }{ \Delta_{2s}}{\tau}  \Big) \, ,
\label{Hdelta}
\end{align}
with $(\Delta-s(s{-}1)){\rm H}_{\Delta_{2s}}^\pm =0$.
The associated $\bpmno{j}{\Delta_{2s}}$ at larger and smaller values of $j$ 
correspond to the $(j{+}1{-}s)^{\rm th}$ $\tau$-derivative
and $(s{-}j{-}1)^{\rm th}$ $\bar \tau$-derivative of ${\rm H}^{\pm}_{\Delta_{2s}}$, respectively:
\begin{align}
\bplus{s-1+m}{\Delta_{2s}}{\tau}  \pm \bminus{s-1+m}{\Delta_{2s}}{\tau} 
&= -2(-4)^m (s{-}1{-}m)! (\pi \nabla)^m  {\rm H}^{\pm}_{\Delta_{2s}}(\tau)\,, \notag \\
\bplus{s-1-m}{\Delta_{2s}}{\tau} \pm \bminus{s-1-m}{\Delta_{2s}}{\tau} 
&= \frac{ -2 (s{-}1{-}m)! (\pi \overline{\nabla})^m  {\rm H}^{\pm}_{\Delta_{2s}}(\tau) }{ (-4)^m y^{2m} }\,.
\label{nablaH}
\end{align}
Interestingly, the coefficients of these functions in ${\rm F}_{m,k}^{\pm (s)}$
involve non-critical values of the completed L-function $\Lambda(\Delta_{2s},t)$ in (\ref{defLfct})
associated with the cusp form $\Delta_{2s}$, where `non-critical' refers to an evaluation 
at $t\geq 2s$.  For instance, for the function $ {\rm F}_{4,4}^{+(6)}$ one has
\begin{align}
 {\rm F}_{4,4}^{+(6)} &= 2450\big( \bplusno{4 &2}{8 &8}+\bplusno{2 }{8 }\bminusno{4 }{8 }+\bminusno{2 &4}{8 &8}\big)
 + 784 \big( \bplusno{5 &1}{8 &8}+\bplusno{1 }{8 }\bminusno{5 }{8 }+\bminusno{1 &5}{8 &8}\big) \notag \\
 &\quad
 + \tfrac{98}{3} \big( \bplusno{6 &0}{8 &8}+\bplusno{0 }{8 }\bminusno{6 }{8 }+\bminusno{0 &6}{8 &8}\big) +\ldots + \frac {7 \Lambda(\Delta_{12},13)}{10365 \Lambda(\Delta_{12},11)} {\rm H}_{\Delta_{12}}^{+} \notag \\
 &= 2450 \beqv{4 &2}{8 &8} +  784 \beqv{5 &1}{8 &8} + \tfrac{98}{3} \beqv{6 &0}{8 &8}  \, ,
 \label{f44eqv}
\end{align}
where the ellipsis denotes iterated Eisenstein integrals of lower modular depth and
coefficients built from $\tau,\bar \tau$ and odd zeta values \cite{Dorigoni:2021jfr, Dorigoni:2021ngn}. 
The analogous iterated-integral representations of general ${\rm F}_{m,k}^{\pm (s)}$
inform the contributions of ${\rm H}_{\Delta_{2s}}^{\pm}$ or equivalently
$\bpmno{j}{\Delta_{2s}}$ to $\beqv{j_1 &j_2}{2m &2k} $, see section 3.4 of \cite{Dorigoni:2022npe}\footnote{The combinations of $\bpmno{j}{\Delta_{2s}}$ in (\ref{Hdelta}) and (\ref{nablaH}) are denoted by $\bsv{j}{\Delta_{2s}^{\pm}}=\bplusno{j}{\Delta_{2s}}\pm \bminusno{j}{\Delta_{2s}}$ in \cite{Dorigoni:2022npe}. In order to avoid unwieldy notation at higher modular depth (e.g.\ double integrals of a cusp form and an Eisenstein series)
we do not use or extend the $\beta^{\rm sv}$-notation for cuspidal contributions in this work.} for further details. An earlier discussion of the modular completion of double Eisenstein integrals via modular-depth-one integrals of holomorphic cusp forms can be found in Brown's work \cite{Brown:mmv, Brown:2017qwo}.

The critical and non-critical L-values $\Lambda(\Delta_{12},t)$ in (\ref{f44eqv}) can be
inferred from the $S$-cocycles arising in the iterated-integral representation of 
$ {\rm F}_{4,4}^{+(6)} $. These L-values exemplify that the MMVs
governing the construction of modular iterated integrals (\ref{beqvdell}) are in general
periods outside the realm of (single-valued) MZVs but are essential ingredients of the 
full theory of iterated integrals of holomorphic modular 
forms~\cite{Brown:mmv, Brown:2017qwo, Brown2019}. Due to the connection of the cusp 
forms and their periods to Tsunogai's derivation algebra to be reviewed in section~\ref{sec:rev.2} 
below, the appearance of new periods in equivariant iterated Eisenstein integrals can be 
anticipated from the commutator relations among these derivations. These structures are
most conveniently incorporated by unifying the iterated integrals of Eisenstein series and
cusp forms in generating series which will be a central theme of this paper.


\subsection{\texorpdfstring{$\mf{sl}_2$ representations and Tsunogai's derivations}{sl(2) representations and Tsunogai's derivations}}
\label{sec:rev.2}

In preparation for the generating series of iterated integrals in section \ref{sec:rev.6} below,
we shall now introduce their non-commutative expansion variables. The constructions
are based on
certain non-commuting letters $\eee_k$ for even $k\geq 4$ as well as 
an $\mf{sl}_2$ algebra generated by
\begin{align}
&\eee_0 \ {\rm (raising} \ {\rm operator)} \, ,  \label{sl2gens}\\
& \eee_0^\vee \ {\rm (lowering} \ {\rm operator)} \, ,  \notag\\
&\hhh=[\eee_0,\eee_0^\vee]\,\ (\rm{Cartan\,operator})\, .
\notag
\end{align}
We can use this $\mf{sl}_2$ algebra to define irreducible $\mf{sl}_2$ representations of dimension $k{-}1$ by letting for $0\leq j\leq k{-}2$
\begin{align}
\eee_k^{(j)} &\coloneqq\ad_{\eee_0}^j \eee_k = \underbrace{ [\eee_0, [\eee_0,\ldots [\eee_0,}_{j}\eee_k]\ldots ]] 
\quad \text{with nilpotency}\quad \eee_k^{(k-1)} = 0 \,.\label{HW}
\end{align}
The $\mf{sl}_2$ algebra acts on these elements by 
\begin{align}
\ad_{\eee_0} \eee_k^{(j)} &= \eee_k^{(j+1)}\,,\label{eq:ade0e0chH}\\
\ad_{\eee_0^\vee} \eee_k^{(j)} &= j(k{-}1{-}j) \eee_k^{(j-1)}\,,\nn\\
\ad_{\hhh} \eee_k^{(j)} &= (2j{+}2{-}k) \eee_k^{(j)}\,, \notag
\end{align}
which imply the following commutator relations for the diagonalised generator $\hhh$
\begin{align}
[\hhh , \eee_0] = 2 \eee_0\,,\quad
[\hhh , \eee_0^\vee] = -2 \eee_0^\vee\,.
\end{align}
Given these relations, it is rather straightforward, and shortly extremely useful, to compute $(\ad_{\eee_0^\vee})^\ell$ on a generic element:

\begin{lemma}\label{lemma1}
If we iterate $\ell$-times the action of $(\ad_{\eee_0^\vee})$ on $ \eee_k^{(j)}$ we obtain
\begin{equation}
(\ad_{\eee_0^\vee})^\ell \eee_k^{(j)} = \frac{j!\, (k{+}\ell{-}2{-}j)!}{(j{-}\ell)!\,(k{-}2{-}j)!}\eee_k^{(j-\ell)}\,,\label{eq:adeochIter}
\end{equation}
which can be easily proven by induction on $\ell$ starting from \eqref{eq:ade0e0chH} and this is a standard result in $\mathfrak{sl}_2$ representation theory.
\end{lemma}

Given the $\mathfrak{sl}_2$ action \eqref{HW} and \eqref{eq:ade0e0chH} on the  $(k{-}1)$-dimensional $\mathfrak{sl}_2$-module spanned by $\eee_k^{(j)}$ with $0\leq j\leq k{-}2$, we will refer to $\eee_k^{(0)}= \eee_k$ as the  \textit{lowest-weight vector} of the module, and to $\eee_k^{(k-2)}$ as its \textit{highest-weight vector}. Note that the highest-/lowest-weight vectors have highest/lowest eigenvalues with respect to the action of $\ad_{\hhh} $, namely $k{-}2$ and $-(k{-}2)$.

We also define the Casimir operator
\begin{align}
\Omega \coloneqq \eee_0 \eee_0^\vee + \frac14 \hhh (\hhh{-}2)
\end{align}
that commutes with the whole $\mathfrak{sl}_2$ algebra and is related to the modular Laplace operator in functional realisations. On the representation generated from the lowest-weight vector $\eee_k$ the eigenvalue is 
\begin{equation}\Omega\, \eee_k^{(j)} = \frac{k}{2} \left(\frac{k}{2}-1\right) \eee_k^{(j)}\,,\qquad \text{for all $0\leq j\leq k{-}2$}\,.
\label{caseig}
\end{equation}

It is also useful to define the standard $\mathfrak{sl}_2$ (Weyl) reflection $\weyl$ that acts as follows on the members of any given $(k{-}1)$-dimensional module:
\begin{align}
\weyl\coloneqq e^{\eee_0^\vee} e^{-\eee_0} e^{\eee_0^\vee} \, , \ \ \ \ 
\weyl\, \eee_k^{(j)} \weyl^{-1} = (-1)^j \frac{j!}{(k{-}2{-}j)!} \eee_k^{(k-2-j)}\, .
\label{eq:Weyl}
\end{align}
This expression for $\weyl\, \eee_k^{(j)} \weyl^{-1}$ can be explicitly checked for example by using matrix representatives or by expanding the exponentials and using the relations \eqref{eq:ade0e0chH}.

\subsubsection{Tensor products and highest-/lowest-weight vectors}
\label{sec:rev.2.1}

We shall next review a few standard facts about tensor products of finite-dimensional $\mf{sl}_2$ multiplets. Let $\eee_k^{(j)}$ denote the basis elements of a $(k{-}1)$-dimensional multiplet (with even $k$) for $0\leq j\leq k{-}2$ and $V(\eee_k)$ the associated irreducible representation. Then we have the well-known decomposition of the tensor product of two representations
\begin{align}
\label{eq:sl2tens}
V(\eee_p) \otimes V(\eee_q) = \bigoplus_{\substack{r=|p-q|+2\\r \in 2\mathbb{Z}}}^{p+q-2} V(\eee_r)\,,
\end{align}
where each irreducible representation $V(\eee_r)$ on the right-hand side
occurs with multiplicity one.

Many of our expressions will be written in terms of generating series with coefficients made out of words of the form $\eee_{k_1}^{(j_1)} \cdots \eee_{k_\ell}^{(j_\ell)}$, where the order of the non-commutative letters $\eee_{k_i}^{(j_i)}$ is important. For both of $\eee_{k_1}^{(j_1)} \eee_{k_2}^{(j_2)}$ and $\eee_{k_2}^{(j_2)} \eee_{k_1}^{(j_1)}$, the $(k_1{-}1)(k_2{-}1)$ words associated with $0\leq j_i \leq k_i{-}2$ fall into the irreducible representations in the tensor product $V(\eee_{k_1}) \otimes V(\eee_{k_2})$, i.e.\ the decomposition (\ref{eq:sl2tens}) applies to both orderings of the non-commutative $\eee_{k_i}^{(j_i)}$.

As will become clear in later sections, most of the salient features of modular iterated integrals enter their generating series with nested commutators $\eee_{k_i}^{(j_i)}$ as coefficients. For instance, the key information about $\beqv{j_1 &j_2}{k_1 &k_2}$ and $\beqv{j_1 &j_2 &j_3}{k_1 &k_2 &k_3}$ at modular depth two and three will be accompanied by brackets $[\eee_{k_1}^{(j_1)}, \eee_{k_2}^{(j_2)}]$ and $[[\eee_{k_1}^{(j_1)}, \eee_{k_2}^{(j_2)}], \eee_{k_3}^{(j_3)}]$, respectively. Accordingly, we will be particularly interested in the tensor-product
decomposition (\ref{eq:sl2tens}) applied to commutators of $\eee_{k_i}^{(j_i)}$.
In this case, the explicit form of the projectors isolating the lowest-weight vectors 
in the irreducible representations of (\ref{eq:sl2tens}) is given~by
\begin{align}
\label{eq:tdpq}
t^d_{p,q} \coloneqq t^d(\eee_p,\eee_q) \coloneqq  \frac{(d{-}2)!}{(p{-}2)!(q{-}2)!}\sum_{i=0}^{d-2} (-1)^i \frac{(p{-}2{-}i)! (q{-}d{+}i)!}{i! (d{-}2{-}i)!} \left[ \eee_p^{(i)} , \eee_q^{(d-2-i)} \right] \,,
\end{align}
where we have introduced $d=\tfrac12( p{+}q{-}r{+}2)$ to match the notation in~\cite{Dorigoni:2021ngn,Pollack} and have fixed an overall normalisation factor. One can check that $[\eee_0^\vee, t^d_{p,q} ] = 0$ and ${\rm ad}_{\eee_0}^{p+q-2d+1}(t^d_{p,q})=0$ such that the vector generates an $(r{-}1)$-dimensional representation under the action of $\mf{sl}_2$ via ${\rm ad}_{\eee_0}^j$ with $0 \leq j \leq r{-}2$, where $r=p{+}q{-}2d{+}2$.

For $p\neq q$, the commutators $[\eee_{p}^{(j_1)}, \eee_{q}^{(j_2)}]$ realise all representations on the right-hand side of (\ref{eq:sl2tens}) with multiplicity one. The integer parameter $d$ 
labelling the irreducible representations in (\ref{eq:tdpq}) then covers the full range 
$2\leq d\leq \min(p,q)$. The largest representation of dimension $p{+}q{-}3$ that arises in the tensor product corresponds to $d=2$. 

For $p=q$, however, the antisymmetrisation in $[\eee_{p}^{(j_1)}, \eee_{p}^{(j_2)}]$ projects out the  irreducible representations $V(\eee_2),V(\eee_6),\ldots,V(\eee_{2p-2})$ of dimensions $\in 4\mathbb N{+}1$, and one is left with $V(\eee_4),$ $V(\eee_8),$ $\ldots,V(\eee_{2p-4})$ of dimensions $\in 4\mathbb N{+}3$ in the antisymmetrised tensor product of two identical representations $V(\eee_p)$. Accordingly, the lowest-weight vectors (\ref{eq:tdpq}) at $p=q$ vanish for even $d$, and their non-zero instances are
$t_{p,p}^{3},t_{p,p}^{5},\ldots,t_{p,p}^{p-1}$.
The letter $\eee_2$ associated with the singlet
of $\mf{sl}_2$ does not enter our generating series (or anywhere else in the construction), though we do encounter one-dimensional representations $V(\eee_2)$ among the tensor products of two or more $\eee_{k\geq 4}^{(j)}$.

We can also project to a highest-weight vector proportional to $\eee_r^{(r-2)}$ in the same module $V(\eee_r)$ using a formula closely related to (\ref{eq:tdpq})
\begin{align}
\label{eq:sdpq}
s^d_{p,q}\coloneqq s^d(\eee_p,\eee_q) \coloneqq 
\frac{(d{-}2)! }{(p{-}2)! (q{-}2)!} \sum_{i=0}^{d-2} (-1)^i  \left[ \eee_p^{(p-2-i)}, \eee_q^{(q-d+i)}\right] \,.
\end{align}
This vector satisfies $[\eee_0, s^d_{p,q}]=0$ as well as ${\rm ad}_{\eee_0^\vee}^{p+q-2d+1}(s^d_{p,q})=0$ and generates the same $(r{-}1)$-dimensional representation of $\mf{sl}_2$ as $t^d_{p,q}$ through the repeated action of ${\rm ad}_{\eee_0^{\vee}}$. This formula is sometimes more convenient than~\eqref{eq:tdpq} since it does not develop singular denominators when taking some of the parameters outside their standard range $2\leq d \leq \min(p,q)$. When taking a $d>\min(p,q)$, one still obtains an element of the tensor product, however, it will not belong to a single irreducible representation $V(\eee_r)$ but will be a linear combination of vectors of different irreducible representations. 

In the ancillary file, numerous main results of this work are presented in terms of the projectors $t^d_{p,q}$ and $s^d_{p,q}$, along with a {\tt Mathematica} implementation of (\ref{eq:tdpq}), (\ref{eq:sdpq}) and their iterations.

\subsubsection{Tsunogai's derivation algebra}
\label{sec:tsalg}

The above $\eee_k^{(j)}$ with $0\leq j \leq k{-}2$ are taken to generate a free 
Lie algebra, i.e.\ to obey no commutation relations other than $\eee_k^{(k-1)}=0$.
In this way, they can be later on used to gather {\it all} iterated Eisenstein integrals 
$\eeno{j_1  &\ldots &j_\ell}{k_1  &\ldots &k_\ell},\, \bpmno{j_1 & \ldots &j_\ell}{k_1  &\ldots &k_\ell}$
with $0 \leq j_i\leq k_i{-}2$ in generating series, without any `dropouts'. However,
the construction of MGFs only involves those iterated Eisenstein integrals
whose modular completion (\ref{beqvdell}) does not require any cusp forms, or equivalently,
which allow for configuration-space representations as iterated integrals over
punctures on a torus \cite{Broedel:2015hia}. The restriction of generating series
to this subclass of iterated Eisenstein integrals is attained by sending the letters $\eee_k$
to certain derivations $\epsilon_k$ dual to holomorphic Eisenstein series. We will refer to the algebra generated by
these $\{\epsilon_k, \ k \in 2\mathbb N_0\}$ as Tsunogai's derivation algebra and it has been
studied from a multitude of perspectives in the mathematics literature \cite{DeligneTBP, Ihara:1990, IharaTakao:1993, Tsunogai, Gonchtalk, GKZ:2006, Schneps:2006, LNT, Pollack, Brown:Anatomy, Hain, brown_2017, BaumardSchneps:2015, hain_matsumoto_2020}. The $\mathfrak{sl}_2$ algebra is simply mapped to $\eee_0\to \ep_0\,,\,\eee^\vee_0\to \ep^\vee_0$, while the derivation $\epsilon_2$ does not enter in our construction.

The derivations $\epsilon_k$ share the $\mf{sl}_2$ multiplet structure (\ref{HW}) and the action of
the lowering operator $\ep^\vee_0$ of the $\eee_k$,
\begin{align}
\epsilon_k^{(j)} = \ad_{\epsilon_0}^j \epsilon_k &= \underbrace{ [\epsilon_0, [\epsilon_0,\ldots [\epsilon_0,}_{j}\epsilon_k]\ldots ]] 
\quad \text{with nilpotency}\quad \epsilon_k^{(k-1)} = 0\,,    \notag\\
(\ad_{\ep_0^\vee})^\ell \ep_k^{(j)} &= \frac{j!\, (k{+}\ell{-}2{-}j)!}{(j{-}\ell)!\,(k{-}2{-}j)!}\ep_k^{(j-\ell)} \, , \label{HWeps}
\end{align}
but satisfy a variety of additional commutator 
relations \cite{LNT, Pollack, Broedel:2015hia} such as 
\begin{align}
0 &= [\ep_4,\ep_{10}] - 3 [\ep_6,\ep_8] \, ,
\label{tsurels} \\
0 &= 80 [\ep_4^{(1)}, \ep_{12} ] + 16 [\ep_{12}^{(1)},\ep_4]
- 250 [\ep_6^{(1)},\ep_{10}] -125 [\ep_{10}^{(1)},\ep_6] + 280 [\ep_8^{(1)},\ep_8] \notag \\
&\quad
- 462[\ep_4, [\ep_4,\ep_8]] - 1725 [\ep_6,[\ep_6,\ep_4]] \, .\notag
\end{align}
One way of generating or checking such relations is based on
the action of the derivations $\ep_k$ on the freely generated
Lie algebra ${\rm Lie}[a,b]$ of the fundamental group of the once-punctured torus, see
(\ref{TSaction}) below. 
The formal signpost variables $\eee_k$ by contrast are not realised as derivations of ${\rm Lie}[a,b]$.

When replacing $\eee_k^{(j)} \rightarrow \ep_k^{(j)}$ in the generating series we will shortly introduce, 
relations like (\ref{tsurels}) lead to dropouts of the accompanying iterated Eisenstein 
integrals and project
to their subclass that enters MGFs.\footnote{More details will be provided in a later section, but for the moment as an example
of such dropouts we have that the relation $[\ep_4,\ep_{10}] - 3 [\ep_6,\ep_8]=0$ among the
non-commutative variables in the generating series $\mathbb J_{+}$ in (\ref{gser.01}) below
implies that not all of the four accompanying integrals $\bplusno{2 &8}{4 &10},\, \bplusno{8 &2}{10 &4},\, \bplusno{4 &6}{6 &8}$ and $\bplusno{6 &4}{8 &6}$ appear independently.
Instead, only three linearly independent combinations of the four integrals remain
in the expansion of $\mathbb J_{+}$ after eliminating one of 
$\ep_4\ep_{10},\, \ep_{10}\ep_{4}, \, \ep_6\ep_{8}$
or $\ep_{8}\ep_6$ by means of the commutation relation (\ref{tsurels}), see~\cite[Eq.~(3.17)]{Gerken:2020yii} for more details.} This can for instance be seen from the conjectural
matrix representations of $\ep_k^{(j)}$ in the differential equations of closed-string
integrals \cite{Gerken:2019cxz, Gerken:2020yii} that generate all MGFs through 
their low-energy expansion. When retaining the variables $\eee_k^{(j)}$ 
without any relation like (\ref{tsurels}), the generating series accommodate 
more iterated integrals or modular forms and can be used to construct 
integral representations such as (\ref{f44eqv}) also for those
modular invariant ${\rm F}^{\pm(s)}_{m,k}$ of section \ref{sec:rev.4} that
receive contributions from holomorphic cusp forms 
\cite{Dorigoni:2021ngn, Dorigoni:2022npe}. One of the key results in
this work is the modular-depth three generalisation of ${\rm F}^{\pm(s)}_{m,k}$
in sections \ref{sec:4.2} and \ref{sec:7}, i.e.\ real-analytic combinations of triple Eisenstein integrals with
double integrals of a cusp form and an Eisenstein series in their modular completions.

The relations (\ref{tsurels}) in the derivation algebra are governed by period
polynomials of holomorphic cusp forms \cite{Pollack}. Accordingly, the
structure of these relations will be later on used to organise the appearance of
iterated integrals involving cusp forms such as (\ref{eq:nudelta}) in equivariant
iterated Eisenstein integrals. For this purpose, we shall promote the 
commutators in (\ref{tsurels}) to non-vanishing 
combinations of free-Lie-algebra generators $\eee_k^{(j)}$, 
\begin{align}
P_{14}^2 &= \frac{1}{49896000}\big([ \eee_4, \eee_{10} ] 
- 3 [ \eee_6, \eee_8 ] \big) \, ,\label{P142}
\\
P_{16} ^3 &= \frac{1}{183883392000}\big( 
80  [ \eee_4^{(1)}, \eee_{12} ]  + 16 [ \eee_{12}^{(1)} ,  \eee_4 ] 
- 250 [ \eee_6^{(1)}, \eee_{10} ] - 125  [ \eee_{10}^{(1)} ,  \eee_6]
+ 280 [ \eee_8^{(1)} , \eee_8]
 \notag \\
&\quad\quad\quad\quad\quad\quad\quad\quad  -462 [ \eee_4, [ \eee_4, \eee_8 ] ]  - 1725 [ \eee_6, [ \eee_6, \eee_4 ] ] \big)\, , \notag
\end{align}
where the normalisation here and below is chosen for later convenience (see
section \ref{sec:hybint}). Relations in the derivation algebra can then be presented as
\beq
P_{w}^d  \, \big|_{ \eee_k^{(j)} \rightarrow \epsilon_k^{(j)} }  = 0\, .
\label{prerel}
\eeq
The non-vanishing combinations $P_w^d$ of $\eee_k^{(j)}$ 
that are mapped to zero upon specialisation $ \eee_k^{(j)} \rightarrow \epsilon_k^{(j)}$ to
Tsunogai's derivations will be referred to as {\it Pollack combinations} in the rest of this work. 
Pollack combinations $P_{w}^d$
are labelled by two integers $w,d$ that account for the expected systematics
of $\ep_k$ relations: the subscript tracks the sum $w=k_1{+}k_2{+}\ldots$ of the 
labels of the $\eee_{k_i}$ in each term and will be referred to as 
{\it \degree} (as it will be directly related to the \degree\ of iterated integrals). 
The superscript $d$ in turn matches the total number of $\eee_k$ 
generators, counting both $\eee_{k \geq 4}$ and $\eee_0$ on equal footing. This is
different from the counting of $\eee_{k_i}^{(j_i)}$ or $\ep_{k_i}^{(j_i)}$ at $k_i\geq 4$
which is not uniform in the expression (\ref{P142}) for $P_{16} ^3$.
In slight abuse of terminology, we shall refer to the number of $\eee_{k_i}^{(j_i)}$ 
and $\ep_{k_i}^{(j_i)}$ at nonzero $k_i$ as {\it modular depth}, i.e.\ $P_{16} ^3$
is a combination of modular-depth-two and modular-depth-three terms.

\subsubsection{$\mf{sl}_2$ structure of $\ep_k$ relations and Pollack combinations}

In order to compactly represent generalisations of 
$P_{14}^2,P_{16}^3$ in (\ref{P142}) to higher \degree, it
is convenient to employ iterations of the
operation $t^{d_1}_{p,q}= t^{d_1}(\eee_p,\eee_q)$ in (\ref{eq:tdpq}).
Given that $t^{d_1}_{p,q}$ is the lowest-weight vector of a
$(p{+}q{-}2d_1{+}1)$-dimensional $\mf{sl}_2$ module, it can 
be treated on the same footing as $\eee_r$ with $r=p{+}q{-}2d_1{+}2$
when entering as an input for another operation $t^{d_2}(\eee_u,\eee_r)$
or $s^{d_2}(\eee_u,\eee_r)$ in (\ref{eq:sdpq}).
We can therefore write for example \begin{align}
t^{d_2}( \eee_u, t^{d_1}_{p,q})= t^{d_2}\big(\eee_u,t^{d_1}(\eee_p,\eee_q)\big) \,,\qquad\qquad  s^{d_2}( \eee_u, t^{d_1}_{p,q})= s^{d_2}\big(\eee_u,t^{d_1}(\eee_p,\eee_q)\big) \,.
\end{align}
In particular, the contributions $\eee_r^{(j)}$ to $t^{d_2}(\eee_u,\eee_r)$ or
$s^{d_2}(\eee_u,\eee_r)$ are then promoted to ${\rm ad}_{\eee_0}^j( t^{d_1}_{p,q} )$ 
which can be simplified by means of the Leibniz rule
\beq
{\rm ad}_{\eee_0}^j [\eee_p^{(j_1)}, \eee_q^{(j_2)} ]
= \sum_{m=0}^j \binom{ j}{m }  [\eee_p^{(j_1+m)}, \eee_q^{(j_2+j-m)} ]
\label{leibniz}
\eeq
while discarding contributions outside the contributing $\mf{sl}_2$ multiplets 
via $\eee_p^{(p-1)} = \eee_q^{(q-1)} = 0$. In terms of these iterated 
$t^d_{p,q}$-operations, we can present the combinations (\ref{P142}) encoding
the simplest $\ep_k$ relations and their generalisations to higher \degree\ $\leq 20$ as follows:

\noindent
$\bullet$ total of $d=2$ letters $\eee_k$:
\begin{align}
P_{14}^2 &=
\frac{4}{5\cdot 11!}\,
\big( t^2_{4,10} - 3 t^2_{6,8} \big)\, ,
 \label{P182}  \\
P_{18}^2 &= 
\frac{12}{35 \cdot 7! \cdot 11!} \,
(2 t_{4,14}^2 - 7 t_{6,12}^2 + 11 t_{8,10}^2)  \, ,
\notag \\
P_{20}^2 &=  
\frac{2}{17!} \,
(-8 t_{4,16}^2 + 25 t_{6,14}^2 - 26 t_{8,12}^2)
\,. \notag
\end{align}
$\bullet$ total of $d=3$ letters $\eee_k$:
\begin{align}
P_{16} ^3 &=
\frac{3}{13820\cdot 11!}\, 
\big(   {-} 160 t_{4,12}^3+ 1000 t_{6,10}^3 - 840 t_{8,8}^3 -462 t^2(\eee_4, t^2_{4,8}) + 1725 t^2(\eee_6, t^2_{4,6}) \big)\, ,
\notag \\
P_{20}^3 &= 
\frac{1}{253190\cdot 7!\cdot 11!}\, \big(
{-} 7560 t_{4,16}^3 + 51450 t_{6,14}^3 - 113190 t_{8,12}^3 + 69300 t_{10,10}^3
 \label{P203}\\*
 &\quad+ 10970 t^2(\eee_4, t^2_{4,12}) - 166675 t^2(\eee_4, t^2_{6,10}) 
 + 500675 t^2(\eee_6, t^2_{6,8}) + 80388 t^2(\eee_8, t^2_{8,4}) \big)\,.
 \notag
 \end{align}
Upon comparison with the earlier expression for  $P_{16} ^3$ in
(\ref{P142}), the commutators $[\eee_p,\eee_q^{(1)}]$
and $[\eee^{(1)}_p,\eee_q]$ have been combined to $t^3_{p,q}$.
 
\noindent
$\bullet$ total of $d=4$ letters $\eee_k$: 
\begin{align}
&P_{18}^4 
=\frac{4}{3455 \cdot 13!} \big( 
36 t^{4}_{4,14} - 691 t^{4}_{6,12} + 2073 t^{4}_{8,10}\big) \notag \\
&\ +\frac{48}{3455 \cdot 11!\cdot 11!}\big( 
 26603500 t^{2} ( \eee_{6}, t^{3}_{6,6} )
-2404395 t^{2}( \eee_{6}, t^{3}_{4,8}) - 63679140 t^{3}( \eee_{6}, t^{2}_{4,8})  
\notag \\
&\quad \ \ - 17133660 t^{2}( \eee_{8}, t^{3}_{4,6})  
+ 86454270 t^{3}( \eee_{8}, t^{2}_{4,6}) 
+2166948 t^{2} ( \eee_{4}, t^{3}_{4,10} ) 
+ 1805790 t^{3} ( \eee_{4}, t^{2}_{4,10} )  \big)\notag \\
&\ + \frac{10771}{39718812672000} [ \eee_4, [ \eee_4, [ \eee_4, \eee_6 ] ] ]   \, ,
\label{Patd4}
\end{align}
where the last commutator can be written as the nested $t^d_{p,q}$-operations
$[ \eee_4, [ \eee_4, [ \eee_4, \eee_6 ] ] ]=t^2(\eee_4,t^2(\eee_4,t^2_{4,6}))$.

\noindent
$\bullet$ total of $d=5$ letters $\eee_k$: 
\begin{align}
&P_{20}^5 = 
 \frac{36}{4837\cdot 13!} \big(  4 t^5_{6,14} 
	- 25 t^5_{8,12} + 21 t^5_{10,10} \big)  
 +\frac{192}{322887025\cdot11!\cdot 13!} \label{Patd5}\\
 &\ \ \notag \times \big(
7106178167028 t^{2} (\eee_4, t^{4}_{4,12}) 
- 4682105275344 t^{3} (\eee_4, t^{3}_{4,12})
- 749415645000 t^{4} (\eee_4, t^{2}_{4,12})  \notag \\
&\quad- 16678946440520 t^{2} (\eee_6, t^{4}_{4,10}) 
+ 30606884011392 t^{3} (\eee_6, t^{3}_{4,10}) 
+ 16726165060230 t^{4} (\eee_6, t^{2}_{4,10})\notag \\
&\quad - 1766591296938 t^{2} (\eee_{10}, t^{4}_{4,6}) 
+ 13691844269480 t^{3} (\eee_{10}, t^{3}_{4,6}) 
+ 15799996899120 t^{4} (\eee_{10}, t^{2}_{4,6})  \notag \\
&\quad - 31509458355375 t^{2} (\eee_6, t^{4}_{6,8}) 
+ 1515484971000 t^{3} (\eee_6, t^{3}_{6,8})
- 11919872175750 t^{4} (\eee_6, t^{2}_{6,8}) \notag \\
&\quad + 15073861990239 t^{2} (\eee_8, t^{4}_{4,8})
 - 43428965937084 t^{3} (\eee_8, t^{3}_{4,8}) 
 - 20009178181944 t^{4} (\eee_8, t^{2}_{4,8})  \big) \notag\\
 &\ \ + \ldots\,.\notag
\end{align} \normalsize
The ellipsis in the last line refers to modular-depth-four terms -- nested brackets
of four $\eee_{k\geq 4}$ -- which can be found in the ancillary files 
and can be reconstructed from the
website \cite{WWWe}. However, in contrast to the presentation
of the $\ep_k$ relations on the website, all of (\ref{P182}) to
(\ref{Patd5}) are organised into linear combinations of
lowest-weight vectors 
\beq
[\eee^\vee_0, P_w^d] = 0 
\label{PwdHW}
\eeq
which give rise to $(w{-}2d{+}1)$-dimensional
$\mf{sl}_2$ multiplets under repeated action of ${\rm ad}_{\eee_0}$.
For instance, shifting the above $P_{20}^5$ by multiples of $({\rm ad}_{\eee_0})^2P_{20}^3$ or
$({\rm ad}_{\eee_0})^3P_{20}^2$ as done on the website 
would preserve the counting associated with the labels $d=5$ and $w=20$
but conflict with the highest-weight property (\ref{PwdHW}).

Nevertheless, (\ref{PwdHW}) still does not fix a unique form of $P_w^d$
at $w\geq 18$. Both the lowest-weight property and the counting encoded
in $d,w$ are preserved by adding $[\eee_4, P_{w-4}^{d-1}]$ to $P_w^d$, or 
more general (iterated) brackets of $\eee_{k\geq 4}$ with $P_{w'}^{d'}$
of suitable lower \degree\ $w'\leq w{-}4$. As we will see in section \ref{sec:noedelta},
this ambiguity in the lowest-weight vectors $P^d_{w \geq 18}$ reflects the
freedom of redefining certain new periods by rational multiples of 
non-critical L-values (\ref{defLfct}). Since each term of the ambiguities 
$[\eee_k, P_{w-k}^{d}]$ has modular depth three or more, these
effects kick in with the modular completions of triple Eisenstein integrals
$ \bplusno{j_1 &j_2&j_3}{k_1 &k_2 &k_3}$ of \degree\ $k_1{+}k_2{+}k_3\geq 18$ 
and do not affect the ${\rm F}^{\pm(s)}_{m,k}$ of modular depth two.

In the same way as the $(k{-}1)$-dimensional multiplets $\eee_k^{(j)}$ will be later on associated
with the integration kernels $\tau^j \GG_k$ with $j=0,1,\ldots,k{-}2$,
the size of the multiplets $({\rm ad}_{\eee_0})^j P_w^d$ singles out
the modular weights of holomorphic cusp forms: Since the $\mf{sl}_2$ module
generated by $P_w^d$ has dimension $(w{-}2d{+}1)$, it is associated
with some $\Delta_{2s}$ at modular weight $2s=w{-}2d{+}2$. By the range
$d\geq 2$, each cusp form $\Delta_{2s}$ is thereby expected to
induce an infinite tower of lowest-weight vectors of \degree\
$2(s{+}d{-}1) = 2(s{+}1), \, 2(s{+}2), \, 2(s{+}3),\, \ldots$ which
yields relations in Tsunogai's derivation algebra under $\eee_k^{(j)} \rightarrow \ep_k^{(j)}$.
The above examples of $P^d_w$ are associated with the simplest holomorphic 
cusp forms $\Delta_{12},\Delta_{16},\Delta_{18}$ \cite{Pollack},
\begin{align}
\Delta_{12} &\leftrightarrow P_{14}^2 , \, P_{16}^3 , \, P_{18}^4, \, P_{20}^5, \, \ldots &&: \ 11\textrm{-dim $\mf{sl}_2$ multiplets} \, ,
\notag  \\
\Delta_{16} &\leftrightarrow P_{18}^2 , \, P_{20}^3 , \, \ldots &&: \ 15\textrm{-dim $\mf{sl}_2$ multiplets}\, ,
\notag  \\
\Delta_{18} &\leftrightarrow P_{20}^2, \, \ldots &&: \ 17\textrm{-dim $\mf{sl}_2$ multiplets}\, ,
\label{deltatoP}
\end{align}
and the dimensions of the multiplets are in one-to-one correspondence to
the $11,15$ and $17$ choices of integration kernels $\tau^j \Delta_{2s}$ with $0\leq j\leq 2s{-}2$ for
$s=6,8$ and $9$ in the generating series we will introduce in section \ref{sec:rev.6}. 
 
Note that, whenever the vector space ${\cal S}_{2s}$ of cusp forms at weight $2s$ has dimension greater than one, the integration kernels $\tau^j \Delta_{2s}$ run over a basis of Hecke-normalised cusp forms, thus leading to different linearly independent instances of~$P_w^d$, one for each basis element. For example, $\rm {dim}\, {\cal S}_{24}=2$ leads to two linearly independent instances of $P_{26}^2, P_{28}^3,\ldots$, with similar results for the higher-dimensional ${\cal S}_{2s}$ at $2s\geq 28$.


\subsection{Zeta generators associated with primitive zeta values}
\label{sec:rev.zeta}

The appearance of primitive zeta values $\zeta_w$ in MGFs for $w\in2\mathbb{N}{+}1$ can be understood from derivations on the free Lie algebra on two elements $a$ and $b$. These derivations will be referred to as \textit{zeta generators}, and the goal of this section is to review their rich interplay with Tsunogai's derivations $\ep_k$. We shall also discuss an uplift of these zeta generators to the free Lie algebra of $\eee_k^{(j)}$ which no longer admits any known 
description in terms of ${\rm Lie}[a,b]$.

\subsubsection{Zeta generators $\sigmaT_{w}$ for Tsunogai's derivation algebra}
\label{sec:rev.5}

Tsunogai's derivations $\ep_k$ with $k\in 2\mathbb N_0$
are realised through their action on free-Lie-algebra generators $a,b$ of the fundamental group
of the once-punctured torus \cite{Tsunogai, KZB, EnriquezEllAss, Hain}:
\begin{align}
\ep_k([a,b]) &= 0  \,, \quad \ep_k(a) = \ad_a^k b
\quad \text{and} \quad
\ep_k(b) = \sum_{j=0}^{k/2} (-1)^j \left[ \ad_a^j b, \ad_a^{2k-1-j}b\right]
\quad \text{for} \quad k \geq 2 \, , \notag
\\
\ep_0(a) &= b \,, \quad \ep_0(b) = 0 \quad \text{and} \quad
\ep_0^\vee(a) = 0 \,, \quad \ep^\vee_0(b) = a
\,.
\label{TSaction}
\end{align}
 Another infinite family $ \sigmaT_{w}$ of derivations acting on $a,b$ is associated with odd Riemann zeta values $\zeta_{w}$ with $w \in 2\mathbb N{+}1$.
The explicit form of the action $ \sigmaT_{w}(a), \sigmaT_{w}(b)$ is determined
in \cite{EnriquezEllAss, Brown:Anatomy, brown_2017, Schneps:2015mzv} and the companion
paper \cite{Dorigoni:2023part1} by
the tight interplay between configuration-space integrals at genus zero and
at the boundary of moduli space at genus one. Upon comparison of 
$ \sigmaT_{w}(a), \sigmaT_{w}(b)$ with the action (\ref{TSaction}) of Tsunogai's 
derivation algebra on $a,b$, the zeta generators can be written as
infinite series of nested brackets of $\epsilon_k$,
up to a tightly constrained so-called `arithmetic' or `non-geometric' contribution $ \zetaT_{w}$, e.g.\ \cite{Dorigoni:2023part1} 
\begin{align}
 \sigmaT_3 &=  \zetaT_3 - \frac{1}{2} \ep_4^{(2)} + \frac{1}{480} [\ep_4,\ep_4^{(1)}]
+ \sum_{k=6}^\infty \BF_k \bigg( [\ep_4^{(1)} , \ep_k ] -\frac{ [ \ep_4 , \ep_k^{(1)} ] }{k{-}2} \bigg) \label{zetgen.01} \\
&\quad + \sum_{m=4}^{\infty} \sum_{r=6}^{\infty}
\frac{(m{-}1)\BF_m \BF_r}{m{+}r{-}2} \big[ \ep_m ,[\ep_4, \ep_r]\big]
 \notag
\end{align}
with $\BF_k= {\rm B}_k/k!$.
The arithmetic parts $\zetaT_{w}$ are $\mf{sl}_2$ singlets and
therefore commute with the raising and lowering operators,
\beq
[  \zetaT_{w} , \ep_0 ] = [  \zetaT_{w} , \ep_0^\vee ] = 0 \, .
\label{zetgen.02}
\eeq
Moreover, they normalise the derivation algebra of the $\ep_k$ in the sense that
$[ \zetaT_{w} , \ep_k ] $ are expressible in terms of nested brackets
of $\ep_{k_i}^{(j_i)}$ in a $(k{-}1)$-dimensional representation of $\mf{sl}_2$, for instance
\beq
[ \zetaT_3,\ep_4 ] = \tfrac{1}{504} \big( [\ep_6^{(2)} , \ep_4] 
- 3 [\ep_6^{(1)} , \ep_4^{(1)}]  +6 [\ep_6 , \ep_4^{(2)}]\big)\, .
\label{zetgen.03}
\eeq
As detailed in the companion paper \cite{Dorigoni:2023part1}, a finite
number of contributions to $ \sigmaT_{w}$ is sufficient
to determine both the infinite tower of $\epsilon_{k_i}^{(j_i)}$-brackets and
the entirety of commutators $[  \zetaT_{w} , \ep_k ] $: Both of them
can be extracted from the contributions of $ \epsilon_{k_i}^{(j_i)} $ to $\sigma_w$ up
to and including key \degree\ $\sum_i k_i = 2w$ by using
that the $ \sigmaT_{w}$ commute with
\beq
 N \coloneqq - \ep_0 + \sum_{k=4}^\infty (k{-}1) \BF_{k} \ep_k \, , \ \ \ \ 
[  N ,  \sigmaT_{w}] = 0 \, .
\label{zetgen.04}
\eeq
This is equivalent to section 7.1 (iii) of Brown's work \cite{Brown:2017qwo2} and
will also be used to pinpoint the modular properties of our generating series
in section \ref{sec:Tequiv}. 

We note that, starting from $\sigmaT_7$, there
are ambiguities in shifting the arithmetic parts $ \zetaT_{w}$ by $\mf{sl}_2$-invariant
nested brackets of $\ep_{k_i}^{(j_i)}$ of total \degree\ $2w$, see
section \ref{sec:4zeta.6} for further details.
A preferred choice of $ \zetaT_{w}$ resolving this ambiguity is
described in section \ref{sec:4zeta.6}, at the end of section \ref{sec:ngeozvarpi} and the companion paper \cite{Dorigoni:2023part1}.


\subsubsection{Zeta generators $ \sigmaE_{w}$ beyond Tsunogai's derivation algebra}
\label{sec:rev.up}

In order to use zeta generators for the construction of equivariant
iterated Eisenstein integrals beyond MGFs, i.e.\ with cusp-form contributions,
we need to uplift  the derivations $\ep_k^{(j)}$ in their representations and 
bracket relations to the free-algebra generators $\eee_k^{(j)}$. 
Since the $\mf{sl}_2$ representation theory of the $\ep_k^{(j)}$ and $\eee_k^{(j)}$
is identical, the only challenge in this uplift stems from the combinations 
$P_w^d$ at \degree\ $w\geq 14$ that vanish in the 
$\ep_k^{(j)}$-incarnation of zeta generators.
Throughout this work, we shall describe guiding principles and (not necessarily canonical)
choices on how to reinstate $P_w^d$ into uplifted versions of zeta generators.
Still, as will be seen in section \ref{sec:sivarpi}, passing from zeta generators for $\ep_k^{(j)}$ 
to their uplifts for $\eee_k^{(j)}$
generically introduces ambiguities involving $P_w^d$ at modular depth $\geq 3$ and \degree\ $\geq 18$.

We will use the notation $\sigmaE_{w},\zetaE_{w}$ for the uplifted zeta
generators and their arithmetic parts which reduce to the
$ \sigmaT_{w},  \zetaT_{w}$ of the previous section
upon the replacement $\eee_k^{(j)} \rightarrow \ep_k^{(j)}$,
\beq
\sigmaT_{w} = \sigmaE_{w} \, \big|_{ \eee_k^{(j)} \rightarrow \ep_k^{(j)} }
\, , \ \ \ \
[ \zetaT_{w}, \ep_k ] = [  \zetaE_{w}, \eee_k] \, \big|_{ \eee_k^{(j)} \rightarrow \ep_k^{(j)} } \, .
\label{zetgen.05}
\eeq
As a first guiding principle, the uplifted zeta generators are still taken to commute 
with the canonical uplift of $N$, similarly denoted by $\hat{N}$, in (\ref{zetgen.04}),
\beq
 \hat{N} = - \eee_0 + \sum_{k=4}^\infty (k{-}1) \BF_{k} \eee_k \, , \ \ \ \ 
[ \hat{N} , \sigmaE_{w}] = 0 \, .
\label{zetgen.06}
\eeq
Second, the closed formula of \cite{hain_matsumoto_2020} for the
modular-depth-two contributions to $[\zetaT_{w},\ep_k]$ is taken to
uplift in its simplest representation, i.e.
\beq
[\zetaE_w,\eee_k] = \frac{ \BF_{w+k-1} }{\BF_k } t^{w+1}(\eee_{w+1},\eee_{w+k-1})+ \ldots\, ,
\label{zetgen.07}
\eeq 
where each term in the ellipsis features modular depth three or higher, see 
appendix \ref{app:zep3} for examples, and
the arithmetic part is once more an $\mf{sl}_2$ singlet,
\beq
[  \zetaE_{w} , \eee_0 ] = [  \zetaE_{w} , \eee_0^\vee ] = 0 \, .
\label{zetgen.09}
\eeq
Third, the closed formula for the complete modular-depth-two contributions to $ \sigmaT_{w}$
determined in \cite{Dorigoni:2023part1}
is taken to uplift in its simplest representation,
\begin{align}
&\sigmaE_w = \zetaE_w - \frac{1}{(w{-}1)!} \eee_{w+1}^{(w-1)}
-\frac{1}{2} \sum_{d=3}^{w-2} \frac{ \BF_{d-1} }{\BF_{w-d+2} }
\sum_{k=d+1}^{w-1} \BF_{k-d+1} \BF_{w-k+1} s^d_{k,w-k+d}
\label{zetgen.08}\\
&\ \
- \sum_{d=5}^w \BF_{d-1} s^d_{d-1,w+1} 
 - \frac{1}{2} \BF_{w+1} s^{w+2}_{w+1,w+1}
 + \! \!  \sum_{k=w+3}^\infty \! \! 
\BF_k \sum_{j=0}^{w-2}  \frac{(-1)^j  \binom{k{-}2}{j}^{-1} }{j! (w{-}2{-}j)! } \, 
[  \eee_{w+1}^{(w-2-j)}  , \eee_k^{(j)} ] + \ldots\, ,
\notag
\end{align}
where the terms at $d=3$ in the first line are known from \cite{brown_2017}
and the ellipsis again gathers all contributions of modular depth three and higher.
Setting $w=3$ in (\ref{zetgen.07}), (\ref{zetgen.08}) and specialising the generators according to (\ref{zetgen.05}) reproduces the earlier examples (\ref{zetgen.01}) and (\ref{zetgen.03}).
As detailed in \cite{Dorigoni:2023part1}, the closed modular-depth-two formula (\ref{zetgen.08}) 
together with the vanishing condition $[ \hat{N} , \sigmaE_{w}]=0$ determine the modular-depth-three contributions
to $\sigmaE_w$ up to highest-weight vectors, see appendix \ref{app:exd3} for examples. Moreover, (\ref{zetgen.08}) and $[ \hat{N} , \sigmaE_{w}]=0$ fix the complete modular-depth-three contributions
to $[\zetaE_w,\eee_k]$ in the ellipsis of (\ref{zetgen.07}).


\subsection{Generating series}
\label{sec:rev.6}

We shall finally review generating-series constructions combining
the non-commuting variables $\ep_k,z_w$ relevant to MGFs with
the iterated Eisenstein integrals of section \ref{sec:rev.1}. The uplifts
of the series in this subsection to $\eee_k,\hat z_w$ and iterated integrals involving
$\Delta_k$ will be the main subject of section \ref{sec:4}. Following
the conventions of \cite{Dorigoni:2022npe}, the combinations $\beta_{\pm}$
of holomorphic iterated Eisenstein integrals in (\ref{cuspat3.03}) entering MGFs are
organised into generating series
\begin{align}
\mathbb J_{\pm} ( \ep_k ;\tau ) &\coloneqq 1 
+ \sum_{k_1=4}^\infty \sum_{j_1=0}^{k_1-2} \frac{(-1)^{j_1} (k_1{-}1) }{(k_1{-}2{-}j_1)!}
\bpmno{j_1 }{ k_1}  \epsilon_{k_1}^{(k_1-j_1-2)} 
\label{gser.01} \\
&\quad+  \sum_{k_1=4}^\infty  \sum_{k_2=4}^\infty \sum_{j_1=0}^{k_1-2}   \sum_{j_2=0}^{k_2-2} \frac{(-1)^{j_1+j_2} (k_1{-}1) (k_2{-}1) }{(k_1{-}2{-}j_1)!(k_2{-}2{-}j_2)!} 
\bpmno{  j_1 &j_2 }{ k_1 &k_2}
 \epsilon_{k_2}^{(k_2-j_2-2)} \epsilon_{k_1}^{(k_1-j_1-2)}  + \ldots
\notag
\end{align}
with iterated integrals of modular depth $\geq 3$ in the ellipsis.
It will be convenient to introduce a shorthand notation $\epsilon[P]$ for the reoccurring
combinations of derivations
\begin{align}
\mathbb J_{\pm} ( \ep_k ;\tau ) &= \sum_{P \in \big\{ \begin{smallmatrix} j \\ k \end{smallmatrix} \big\}^\times} \epsilon[P] \beta_{\pm}[P;\tau] \, , \label{MGFtoBR22}
\\
\epsilon[P]&=\epsilon\! \left[\begin{smallmatrix}  j_1 &j_2 &\ldots &j_\ell \\ k_1 &k_2 &\ldots &k_\ell \end{smallmatrix}\right]\coloneqq
\bigg( \prod_{i=1}^\ell \frac{ (-1)^{j_i}(k_i{-}1) }{(k_i{-}j_i{-}2)!} \bigg)
\epsilon^{(k_\ell-2-j_\ell)}_{k_\ell} \cdots \epsilon^{(k_2-2-j_2)}_{k_2}\epsilon^{(k_1-2-j_1)}_{k_1}\, ,
\notag
\end{align}
where the summation range $P \in \big\{ \begin{smallmatrix} j \\ k \end{smallmatrix} \big\}^\times$
refers to all words in $j_i,k_i$ of length $\ell \in \mathbb N_0$ specified in the second line
such that $k_i\geq 4$ and $0\leq j_i \leq k_i{-}2$. Our convention for the empty word at $\ell=0$
is $\epsilon[\emptyset]=1$.

Brown described a composition of the generating series $\mathbb J_{\pm}$
whose coefficients are modular forms after invoking the relations among the
derivations $\ep_k$ \cite{Brown:2017qwo2}: In the notation 
of~\cite{Dorigoni:2022npe}, Brown's construction of equivariant iterated
Eisenstein integrals takes the form
\begin{align}
\Jeqv (\epsilon_k;\tau) &\coloneqq
\sum_{P \in \big\{ \begin{smallmatrix} j \\ k \end{smallmatrix} \big\}^\times} \epsilon[P] \beta^{\rm eqv}[P;\tau]
= \mathbb J_{+} ( \ep_k ;\tau ) \mathbb B^{\rm sv}(\epsilon_k;\tau)
 \phi^{\rm sv}\big( \widetilde{\mathbb J_{-}} ( \ep_k ;\tau )  \big)\, ,
\label{select.1}
\end{align}
where the tilde in the second line reverses the relative order of the derivations
and the iterated integrals in $\mathbb J_{-} ( \ep_k ;\tau )$ and can be implemented by replacing
$\bminusno{  j_1 &j_2 &\ldots &j_\ell }{ k_1 &k_2 &\ldots &k_\ell}
\rightarrow \bminusno{  j_\ell &\ldots &j_2 &j_1 }{ k_\ell &\ldots &k_2 &k_1}$ 
in the expansions (\ref{gser.01}) and (\ref{MGFtoBR22}).
The coefficients $ \beta^{\rm eqv}[P;\tau]$ of $\epsilon[P]$ are the modular forms with ${\rm SL}(2,\mathbb Z)$ weights
in (\ref{modbeqv}) and whose terms at highest modular depth in (\ref{beqvdell})
are generated by the simpler composition 
$ \mathbb J_{+} ( \ep_k ;\tau ) \widetilde{\mathbb J_{-}} ( \ep_k ;\tau ) $.
However, the modularity of $\beta^{\rm eqv}$ relies on the admixture of MZVs
multiplying integrals of lower modular depth as exemplified in (\ref{eq:beqv1}) and 
(\ref{beqvd2}). Brown's prescription to insert appropriate combinations of MZVs
is based on two ingredients: 
\begin{itemize}
\item[(i)]an $\ep_{k}^{(j)}$-valued series
$\mathbb B^{\rm sv}$ of single-valued MZVs (see appendix \ref{app:A} for the latter notion);
\item[(ii)]
a change of alphabet $\phi^{\rm sv}$ for the derivations
$\epsilon_k$ multiplying the antiholomorphic integrals $\beta_-$ 
in the conventions of~\cite{Dorigoni:2022npe}.
\end{itemize}

\subsubsection{The series $\mathbb B^{\rm sv}$ of single-valued MZVs}
\label{subsec:2.4.1}

The expansion of $\mathbb B^{\rm sv}$ in words in $\ep_{k_i}^{(j_i)}$ 
follows the form of $ \mathbb J_{\pm}, \Jeqv$ in (\ref{MGFtoBR22}), (\ref{select.1}), and its
components $b^{\rm sv}[P;\tau]$ only depend rationally on $\tau,\bar \tau$:
\begin{align}
\mathbb B^{\rm sv}(\epsilon_k;\tau) &\coloneqq
\sum_{P \in \big\{ \begin{smallmatrix} j \\ k \end{smallmatrix} \big\}^\times} \epsilon[P] b^{\rm sv}[P;\tau] \, ,
\label{defbsv} \\
\bbsvtau{\ldots &j &\ldots}{\ldots &k &\ldots}{\tau} &\coloneqq  \sum_{p=0}^{k-2-j} \sum_{\ell=0}^{j+p}
\binom{k{-}j{-}2}{p} \binom{j{+}p}{\ell} \frac{({-}2\pi i \bar \tau)^\ell}{(4y)^p} \,
\ccsv{\ldots &j-\ell+p &\ldots}{\ldots &k &\ldots}\, .
\notag
\end{align}
The double sum over $p,\ell$ is understood to apply separately to
each column $ \begin{smallmatrix} j_i \\ k_i \end{smallmatrix} $ of $b^{\rm sv}[\ldots;\tau]$.
The $\tau$-independent $\ccsv{j_1 &\ldots &j_\ell}{k_1 &\ldots &k_\ell}$ 
are projected to $\mathbb Q$-linear combinations of single-valued MZVs of 
transcendental weight $\ell{+}j_1{+}\ldots{+}j_\ell$ through the relations
of the accompanying $\ep_k$ in (\ref{defbsv}).\footnote{Starting from section
\ref{sec:csv}, the $\ccsv{j_1 &\ldots &j_\ell}{k_1 &\ldots &k_\ell}$ will be dressed
with the free-Lie-algebra generators $\eee_k^{(j)}$ and therefore be individually
accessible. In this setting, the single-valued MZVs of 
transcendental weight $\ell{+}j_1{+}\ldots{+}j_\ell$ in $\ccsv{j_1 &\ldots &j_\ell}{k_1 &\ldots &k_\ell}$ of modular depth $\ell\geq 3$ and degree $\sum_i k_i \geq 18$ will be augmented by
new periods to be discussed in section \ref{sec:sivarpi}.} Their modular-depth-one
instances take the closed form $\ccsv{j}{k} = - \frac{ 2 \zeta_{k-1} }{k{-}1} \delta_{j,k-2}$, 
and they are responsible for the zeta value in the expression (\ref{eq:beqv1}) for $\beqv{j}{k}$.
Numerous examples of $c^{\rm sv}$ at modular depth two and three
can be found in the ancillary files and in section 3.1 of \cite{Dorigoni:2022npe}.

Note that the constants $c^{\rm sv}$ give rise to a convenient dictionary between the building blocks $\beta^{\rm sv}$ for MGFs introduced in \cite{Gerken:2020yii} and the modular forms $\beta^{\rm eqv}$ in (\ref{select.1}) \cite{Dorigoni:2022npe}
\begin{align}
\bsv{j_1 &j_2 &\ldots &j_\ell}{k_1 &k_2 &\ldots &k_\ell} &= \sum_{r=0}^{\ell}(-1)^r d^{\rm sv}\!\left[\begin{smallmatrix} j_r &\ldots &j_2 &j_1 \\ k_r &\ldots &k_2 &k_1 \end{smallmatrix} \right] \beqv{j_{r+1} &\ldots &j_\ell}{k_{r+1} &\ldots &k_\ell} + \ldots \, ,
\label{dictsveqv} \\
d^{\rm sv}\!\left[\begin{smallmatrix} j_1 &j_2 &\ldots  &j_\ell \\ k_1 &k_2 &\ldots &k_\ell \end{smallmatrix} \right] &=
\sum_{p_1=0}^{k_1-2-j_1}\ldots \sum_{p_\ell=0}^{k_\ell-2-j_\ell} \frac{(\smallmatrix k_1-2-j_1 \\ p_1 \endsmallmatrix) (\smallmatrix k_2-2-j_2 \\ p_2 \endsmallmatrix) \ldots (\smallmatrix k_\ell-2-j_\ell \\ p_\ell \endsmallmatrix)}{(4y)^{p_1+p_2+\ldots+p_\ell}} \ccsv{j_1+p_1 &j_2+p_2 &\ldots &j_\ell+p_\ell}{k_1 &k_2 & \ldots &k_\ell} \, .
\notag
\end{align}
The ellipsis in the first line refers to terms with at least one holomorphic cusp forms in the integrand which always drop out from the combinations of $\beta^{\rm sv}$ encountered in MGFs.

\subsubsection{The change of alphabet $\phi^{\rm sv}$}
\label{subsec:2.4.2}

While theorem 7.2 of \cite{Brown:2017qwo2} implicitly
determines $\mathbb B^{\rm sv}$ and $\phi^{\rm sv}$ in terms of multiple modular
values, the following series in single-valued MZVs gives an explicit characterisation
of the change of alphabet in terms of the arithmetic parts $\zetaT_w$ of
zeta generators in section \ref{sec:rev.5}, see section 4 of \cite{Dorigoni:2022npe},
\begin{align}
\phi^{\rm sv}(\mathbb X) &\coloneqq   \mathbb M^{\rm sv}(\zetaT_i) \, \mathbb X \, \mathbb M^{\rm sv}(\zetaT_i)^{-1} \, ,
\label{defphisv}
\end{align}
where $\mathbb X$ refers to an arbitrary series in $\ep_k$  and
\begin{align}
\mathbb M^{\rm sv}(\zetaT_i) \coloneqq& \, \sum_{\ell=0}^{\infty} \sum_{\substack{i_1,i_2,\ldots, i_\ell \\{\in 2\mathbb N+1}}} \zetaT_{i_1} \zetaT_{i_2}\ldots \zetaT_{i_\ell}\, \rho^{-1} \big( {\rm sv}(f_{i_1} f_{i_2} \ldots f_{i_\ell}) \big) 
\label{Msvz} \\
=& \, 1 +2 \! \! \! \sum_{i_1 \in 2\mathbb N+1} \! \! \! \zetaT_{i_1} \zeta_{i_1}
+ 2 \! \! \!  \!  \! \! \sum_{i_1,i_2 \in 2\mathbb N+1} \! \! \! \! \! \! \zetaT_{i_1} \zetaT_{i_2}  \zeta_{i_1} \zeta_{i_2} 
+ \! \! \! \! \! \!  \! \!  \sum_{i_1,i_2,i_3 \in 2\mathbb N+1} \! \! \! \! \! \! \! \!
\zetaT_{i_1} \zetaT_{i_2} \zetaT_{i_3}  \rho^{-1}\big( {\rm sv}(f_{i_1} f_{i_2}f_{i_3} ) \big) 
+\ldots \, .\notag
\end{align}
We stress that formula~\eqref{defphisv} is only applicable to series $\mathbb{X}$ in the generators $\ep_k$ and a central part of our work will be to discuss a more general change of alphabet for series $\mathbb{X}$ in the unconstrained $\eee_k$ in section~\ref{sec:4}.

Expression~\eqref{Msvz} contains both the isomorphism $\rho^{-1}$ from the $f$-alphabet to real MZVs \cite{Brown:2011ik, Brown:2011mot} and the single-valued map on the $f$-alphabet given by \cite{Schnetz:2013hqa, Brown:2013gia}
\begin{align}
  {\rm sv}( f_{i_1} f_{i_2}\cdots f_{i_\ell}) =  \sum_{j=0}^\ell f_{i_j}\cdots f_{i_2} f_{i_1} \shuffle f_{i_{j+1}} \cdots f_{i_\ell} \, , \ \ \ \ \ \
  i_1,\ldots,i_\ell \in 2\mathbb N{+}1\,.
\end{align}
In passing to the second line of (\ref{Msvz}), we have used 
$\rho^{-1}( {\rm sv}(f_{i_1}) ) = 2 \zeta_{i_1} $ and
$\rho^{-1}( {\rm sv}(f_{i_1} f_{i_2}) ) = 2 \zeta_{i_1} \zeta_{i_2}$
for words of length one and two in the $f$-alphabet.
Starting from length three, $f$-alphabet expressions such as
 $\rho^{-1}( {\rm sv}(f_{i_1} f_{i_2}f_{i_3} ) )$ generically comprise
 (conjecturally) indecomposable single-valued MZVs beyond depth one, see appendix \ref{app:A.2} for examples at weight $\leq 15$ and the ancillary file for weight $\leq 17$.
 The expansion (\ref{Msvz}) of $\mathbb M^{\rm sv}(\zetaT_i)$ reorganises
 the conjugation in (\ref{defphisv}) into a series in nested commutators,
\begin{align}
\phi^{\rm sv}(\mathbb X) = \sum_{\ell=0}^{\infty} \sum_{\substack{i_1,i_2,\ldots, i_\ell \\ \in 2\mathbb N+1}} [\zetaT_{i_1} ,[\zetaT_{i_2}, \ldots [\zetaT_{i_\ell},\mathbb X]\ldots ]]\, \rho^{-1} \big( {\rm sv}(f_{i_1} f_{i_2} \ldots f_{i_\ell}) \big) \, .
\label{finalphisv}
\end{align} 
Since $[\zetaT_{w},\ep_k]$ is expressible via nested brackets of $\ep_{k}^{(j)}$,
the coefficients of each ${\rm sv}(f_{i_1}  \ldots f_{i_\ell})$ take values in Tsunogai's derivation
algebra and do not obstruct an extraction of coefficients as in (\ref{select.1}) after modding
out by Pollack-relations. Based on the examples of the uplifted $[ \zetaE_{w},\eee_k]$ in appendix
\ref{app:zep3}, we find the contributions
\begin{align}
\phi^{\rm sv}(\ep_4) &=  \ep_4 + 2 \zeta_3 [\zetaT_3,\ep_4] + 2 \zeta_5 [\zetaT_5,\ep_4]
+ 2 \zeta_3^2 \big[ \zetaT_3, [\zetaT_3, \ep_4] \big] + \ldots
\label{phiex} \\
&= \ep_4 + 2 \zeta_3 \frac{ \BF_6 }{\BF_4} t^4(\ep_4,\ep_6) 
+ 2 \zeta_5 \frac{ \BF_8 }{\BF_4} t^6(\ep_6,\ep_8) 
+ \zeta_5 \frac{ \BF_6 \BF_2^3 }{ \BF_4^2 } t^4(\eee_6,t^3(\eee_4,\eee_4)) 
  \notag \\
&\quad -  \zeta_5 \BF_4 \bigg\{
\frac{9}{5} t^3(\ep_4,t^4(\ep_4,\ep_6))
+\frac{2}{5} t^4(\ep_4,t^3(\ep_4,\ep_6))
\bigg\} 
 +  2 \zeta_3^2 \frac{ \BF_6^2 }{\BF_4^2} t^4( t^4(\ep_4,\ep_6) ,\ep_6 )
 \notag \\
 &\quad  +2 \zeta_3^2 \frac{ \BF_8 }{\BF_4} t^4(\ep_4, t^4(\ep_4,\ep_8))
 - \frac{9 \zeta_3^2}{5} \BF_4 t^4 (\ep_4, t^2(\ep_4,t^3(\ep_4,\ep_4))) + \ldots
\notag
\end{align}
up to and including transcendental weight six of the accompanying MZVs.

\section{Generating series of equivariant iterated Eisenstein integrals}
\label{sec:4}

In this section, we analyse $\SLtwoZ$-equivariant iterated Eisenstein integrals in more detail and arrange them into the generating series presented in~\eqref{select.1}. These generating series are expressed in terms of the non-commutative letters $\eee_k^{(j)}$ introduced in section~\ref{sec:rev.2}. The equivariant integrals can be presented in two different `frames' that we discuss in section~\ref{sec:4.1}. One of our main results is to identify the way periods (such as multiple zeta values or L-values) enter equivariant expressions and this can be traced back to the zeta generators of section~\ref{sec:rev.zeta} as we explain in section~\ref{sec:4zeta}. Since in general iterated integrals of cusp forms are needed to make iterated Eisenstein integrals transform correctly under SL$(2,\mathbb Z)$, we discuss their appearance in section~\ref{sec:4.2}, along with two classes of new periods that arise at modular depth three. Moreover, a generating-series description of equivariant iterated Eisenstein integrals at arbitrary modular depth including cusp forms is proposed in (\ref{uplift}).

The final part~\ref{sec:7} of this section repackages the information of section \ref{sec:4.2} into solutions of inhomogeneous Laplace equations at modular depth three. In this setting, the two classes of new periods arise as the coefficients of solutions to the associated homogeneous Laplace equations.

\subsection{The holomorphic frame}
\label{sec:4.1}

The goal of this section is to rebuild the generating series in (\ref{select.1})
such that $\mathbb B^{\rm sv}$ no longer depends on $\tau$ and $\mathbb J_{\pm}$
become (anti-)meromorphic in $\tau$, respectively. As we will see, both properties
are accomplished by conjugation with the ${\rm SL}_2$ transformation
\beq
\Unew(\tau)
\coloneqq \exp \bigg( {-} \frac{\eee_0^{\vee}}{4y} \bigg) \exp(2\pi i \bar \tau \eee_0)
\in  {\rm SL}_2
\label{eq:Utau}\,.
\eeq
By slight abuse of notation, we employ the same symbol $\Unew(\tau)$ for the ${\rm SL}_2$ transformation $ \exp( {-} \frac{\epsilon_0^{\vee}}{4y} ) \exp(2\pi i \bar \tau \epsilon_0)$ acting on Tsunogai's derivations $\ep_k^{(j)}$ instead of the free Lie-algebra generators $\eee_k^{(j)}$. Accordingly, the definitions in this section pave the way for the inclusion of cusp forms into equivariant iterated Eisenstein integrals.

\subsubsection{Setting up (anti-)holomorphic $(1,0)$ and $(0,1)$-forms}

As a convenient starting point for the construction of (anti-)meromorphic generating series
conjugate to $\mathbb J_{\pm}$, we introduce the following
$(1,0)$-form in $\tau$ that gathers all the relevant integration kernels
\begin{align}
\label{mt.06}
\mathbb{A}_+(\eee_k;\tau)  &\coloneqq \sum_{k=4}^\infty\sum_{j=0}^{k-2}(-1)^j \frac{  (k{-}1)}{j!} \Big[\nuker{j}{k}{\tau} \eee_k^{(j)} +\sum_{\Delta_k \in \mathcal{S}_{k}}\nuker{j}{\Delta_k}{\tau}  \eee_{\Delta_k^+}^{(j)} \Big]\\
&\phantom{:}=  \sum_{k=4}^\infty  \sum_{j=0}^{k-2}(-1)^j
\frac{(k{-}1)}{j!} \, (2\pi i)^{1+j} \, \tau^{j}\, \Big[(2\pi i )^{-k}{\rm  G}_k(\tau) \eee_k^{(j)}+\sum_{\Delta_k \in \mathcal{S}_{k}}\Delta_k(\tau)  \eee_{\Delta_k^+}^{(j)}\Big]\,\dd \tau\,,
\nn
\end{align}
with $\mathcal{S}_{k}$ the vector space of holomorphic cusp forms of weight $k$, and
where $\nuker{j}{k}{\tau}$ and $\nuker{j}{\Delta_k}{\tau}$ are defined
in (\ref{nuker}) and (\ref{eq:nudelta}), respectively.

As an antiholomorphic analogue of the holomorphic $(1,0)$ form $\mathbb{A}_+$ in (\ref{mt.06}),
we shall introduce the following $(0,1)$-form
\begin{align}
\label{mt.07}
\mathbb{A}_-(\eee_k;\tau)  &\coloneqq \sum_{k=4}^\infty\sum_{j=0}^{k-2} \frac{  (k{-}1)}{j!} \Big[\overline{\nuker{j}{k}{\tau} }\eee_k^{(j)} +\sum_{\Delta_k \in \mathcal{S}_{k}}\overline{\nuker{j}{\Delta_k}{\tau}}  \eee_{\Delta_k^-}^{(j)} \Big]\\
&\phantom{:}= - \sum_{k=4}^\infty  \sum_{j=0}^{k-2}(-1)^j
\frac{(k{-}1)}{j!} \, (2\pi i)^{1+j} \, \bar\tau^{j}\, \Big[(2\pi i )^{-k}\overline{{\rm  G}_k(\tau)} \eee_k^{(j)}+\sum_{\Delta_k \in \mathcal{S}_{k}}\overline{\Delta_k(\tau) } \eee_{\Delta_k^-}^{(j)}\Big]\,\dd \bar \tau\,.
\notag
\end{align}
We emphasise that $\mathbb{A}_-$ is not the complex conjugate of $\mathbb{A}_+$ since it differs by a factor $(-1)^j$ in the definition and also by the letters $\eee^{(j)}_{\Delta_{k}^+}$ versus $\eee^{(j)}_{\Delta_{k}^-}$ accompanying the cusp forms.
These letters are non-commmutative variables spanning $\mf{sl}_2$ multiplets according to
\beq
\eee_{\Delta_k^{\pm}}^{(j)} = {\rm ad}_{\eee_0}^j \eee_{\Delta_k^{\pm}} \, , \ \ \ \ \ \
\eee_{\Delta_k^{\pm}}^{(k-1)} =0 
\, , \ \ \ \ \ \
{\rm ad}_{\eee_0^\vee} \eee_{\Delta_k^{\pm}} = 0\,,
\eeq
such that $\eee_{\Delta_k^{\pm}}$ are lowest-weight vectors. In this work,
the letters $\eee_{\Delta_k^{\pm}}$ are not independent from the Eisenstein letters 
$\eee_k$, but are expressible through nested commutators involving at least one Pollack combination $P_w^d$ of section~\ref{sec:rev.2}, for example
\begin{align}
    \eee_{\Delta_{12}^\pm} & = \frac{1}{\Lambda(\Delta_{12},11)} \left( \Lambda(\Delta_{12},13) P_{16}^3 + \Lambda(\Delta_{12},15) P_{20}^5 \right) \label{1stedelta}\\*
    &\quad \quad \mp \frac{1}{\Lambda(\Delta_{12},10)} \left( \Lambda(\Delta_{12},12) P_{14}^2 + \Lambda(\Delta_{12},14) P_{18}^4\right) + \ldots\,,
    \nn
\end{align}
where the ellipsis refers to terms of modular depth $\geq 3$ and \degree\ $\geq 22$, see section \ref{sec:noedelta} below for details. 

In the context of more general classes of iterated integrals, it could be interesting to promote
variants of the
letters $\eee_{\Delta_k^{\pm}}$ to be independent of $\eee_k$. Upon augmenting
the connection (\ref{mt.06}) by weakly holomorphic modular forms, it may be possible
to generalise the non-holomorphic modular forms of \cite{brown2017a} with poles
at the cusp to higher modular depth and to take advantage of independent
letters $\eee_{\Delta_k^{\pm}}$ in a formulation via generating series.\footnote{Since
the letters  $\eee_{\Delta_k^{\pm}}$ in this work reduce to Eisenstein letters
$\eee_k$, one cannot read off the coefficient of $\eee_{\Delta_{k}^\pm}$ from generating
series such as (\ref{eq:Iup}) below. It is tempting to approach
higher-depth analogues of the non-holomorphic modular forms of \cite{brown2017a}
via generating-series methods where the modular completions are obtained
by reading off the coefficients of independent letters associated with
holomorphic cusp forms.}

The occurrence of the Pollack combinations $P_w^d$ ensures that the letters 
$\eee_{\Delta_k}^{\pm}$ vanish when replacing $\eee_k$ by $\ep_k$. 
As indicated by the notation on the left-hand sides of~\eqref{mt.06} and~\eqref{mt.07},  throughout this paper the symbol $\eee_{\Delta_{k}}^\pm$ is always understood as a combination of Eisenstein letters $\eee_k$.
We will spell out additional contributions to $\eee_{\Delta_k^\pm}$ at low modular depth
in (\ref{edeltaex}) below, generalising (\ref{1stedelta}) up to and including \degree\ $20$.

As will become clear in the following, the distinction between
$\eee_{\Delta_k^{+}}$ and $\eee_{\Delta_k^{-}}$ is necessary for instance to ensure that the
modular invariants ${\rm F}_{m,k}^{+(s)}$ and ${\rm F}_{m,k}^{-(s)}$
to be extracted in later sections can be even or odd under the upper-half plane involution $\tau \rightarrow - \bar \tau$. This is
in contrast to the holomorphic Eisenstein series ${\rm E}_k$ in (\ref{eq:Cab})
and (\ref{eq:beqv1}) that are always even by the relative plus sign between
the contributing $\beta_+$ and $\beta_-$, so one has to employ the same
letter $\eee_k$ in $\mathbb J_+$ and $\mathbb J_-$.

\subsubsection{Iterated integrals from path-ordered exponentials}

Given the $(1,0)$- and $(0,1)$-forms $\mathbb{A}_{+}$ and $ \mathbb{A}_{-}$
in (\ref{mt.06}) and (\ref{mt.07}), respectively, we shall next define
generating series of (anti-)holomorphic iterated integrals from their
path-ordered exponentials,
\begin{align}
\mathbb I_{+}(\eee_k;\tau) &= \Pexp \bigg(\int \limits_\tau^{i \infty} \mathbb{A}_{+}(\eee_k;\tau_1)\bigg)
= 1{+}  \int\limits_{\tau}^{i \infty} \mathbb{A}_{+}(\eee_k;\tau_1)
 {+} \int\limits_\tau^{i \infty} \bigg(  \int\limits_{\tau_1}^{i \infty} \mathbb{A}_{+}(\eee_k;\tau_2) \bigg) \mathbb{A}_{+}(\eee_k;\tau_1) +\ldots
\,,  \label{eq:Pexp} \\
\widetilde{\mathbb I}_{-}(\eee_k;\tau) &= \Ptexp \bigg(\int\limits_{\bar \tau}^{-i \infty} \mathbb{A}_-(\eee_k;\tau_1)\bigg)
= 1 {+}  \int\limits_{\bar \tau}^{-i \infty} \mathbb{A}_-(\eee_k;\tau_1)
{+}    \int\limits_{\bar \tau}^{-i \infty} \mathbb{A}_-(\eee_k;\tau_1) \!  \int\limits_{\bar \tau_1}^{-i \infty} \mathbb{A}_-(\eee_k;\tau_2) 
+\ldots 
\,,  \notag
\end{align}
with integrals over $\geq 3$ powers of $\mathbb{A}_{\pm}$ in the ellipses.
In our conventions for $\Pexp$, the integration variable $\tau_1$
closest to the lower endpoint $\tau$ is associated with the rightmost factor of $\mathbb{A}_+$
with respect to the ordering of the $\eee_k,\eee_{\Delta_k^{\pm}}$. The extra tilde in the
notation $\Ptexp$ entering $\widetilde{\mathbb I}_{-}$ instructs us to reverse the 
ordering convention as exemplified in the
second-order term $\sim \mathbb{A}_-(\tau_1) \mathbb{A}_-(\tau_2)$. In terms of
the iterated Eisenstein integrals (\ref{MMVsec.07}) and their cuspidal analogues
such as (\ref{eq:nudelta}), the expansion of the first path-ordered exponential in
(\ref{eq:Pexp}) yields 
\begin{align}
&\Ip(\eee_k;\tau) = 1 
+ \sum_{k_1=4}^{\infty}\sum_{j_1=0}^{k_1-2} (-1)^{j_1} \frac{(k_1{-}1)}{j_1!}
\left\{ 
\ee{j_1}{k_1}{\tau} \eee_{k_1}^{(j_1)}
+ \sum_{\Delta_{k_1}\in \mathcal{S}_{k_1}} \ee{j_1}{\Delta_{k_1}}{\tau} \eee_{\Delta_{k_1}^+}^{(j_1)}\right\}
\label{MMVsec.06}\\
&\ \ + \sum_{k_1=4}^{\infty}\sum_{j_1=0}^{k_1-2}\sum_{k_2=4}^{\infty}\sum_{j_2=0}^{k_2-2} (-1)^{j_1+j_2} \frac{(k_1{-}1)(k_2{-}1)}{j_1!j_2!}\Bigg\{
\ee{j_1 & j_2}{k_1 & k_2}{\tau} \eee_{k_1}^{(j_1)}\eee_{k_2}^{(j_2)}
+
\!\!\!\!\! \sum_{\Delta_{k_2}\in\mathcal{S}_{k_2}} \!\!\!\!\!
\ee{j_1 & j_2}{k_1 & \Delta_{k_2}}{\tau} \eee_{k_1}^{(j_1)}\eee_{\Delta_{k_2}^+}^{(j_2)} \notag\\
&\hspace{14mm} + \sum_{\Delta_{k_1}\in \mathcal{S}_{k_1}}\!\!\!\! \ee{j_1 & j_2}{\Delta_{k_1} & k_2}{\tau} \eee_{\Delta_{k_1}^+}^{(j_1)}\eee_{k_2}^{(j_2)}  + \sum_{\Delta_{k_1}\in\mathcal{S}_{k_1}}\sum_{\Delta_{k_2}\in\mathcal{S}_{k_2}}\!\!\!\!
\ee{j_1 & j_2}{\Delta_{k_1} & \Delta_{k_2}}{\tau} \eee_{\Delta_{k_1}^+}^{(j_1)}\eee_{\Delta_{k_2}^+}^{(j_2)}\Bigg\} +\ldots \, , \notag
\end{align}
with iterated integrals of modular depth $\geq 3$ over holomorphic modular forms in the ellipsis.
The analogous expansion of $\widetilde{ \mathbb I}_-$ reads
\begin{align}
& \widetilde{\mathbb I}_-(\eee_k;\tau) = 1 
+ \sum_{k_1=4}^{\infty}\sum_{j_1=0}^{k_1-2}  \frac{(k_1{-}1)}{j_1!} \left\{\overline{\ee{j_1}{k_1}{\tau} }\eee_{k_1}^{(j_1)}
+ \sum_{\Delta_{k_1}\in \mathcal{S}_{k_1}} \overline{\ee{j_1}{\Delta_{k_1}}{\tau} }\eee_{\Delta_{k_1}^-}^{(j_1)}\right\}
\label{minusees}\\
&\quad + \sum_{k_1=4}^{\infty}\sum_{j_1=0}^{k_1-2}\sum_{k_2=4}^{\infty}\sum_{j_2=0}^{k_2-2}  \frac{(k_1{-}1)(k_2{-}1)}{j_1!j_2!} \Bigg\{ \overline{\ee{j_2 & j_1}{k_2 & k_1}{\tau}} \eee_{k_1}^{(j_1)}\eee_{k_2}^{(j_2)} + \sum_{\Delta_{k_2}\in\mathcal{S}_{k_2}} \overline{\ee{j_2 & j_1}{\Delta_{k_2} & k_1}{\tau}} \eee_{k_1}^{(j_1)}\eee_{\Delta_{k_2}^-}^{(j_2)} \notag \\
&\hspace{12mm} \ \ \ \ \ \ + \sum_{\Delta_{k_1}\in\mathcal{S}_{k_1}} \overline{\ee{j_2 & j_1}{k_2 & \Delta_{k_1}}{\tau}} \eee_{\Delta_{k_1}^-}^{(j_1)}\eee_{k_2}^{(j_2)}
  + \sum_{\Delta_{k_1}\in\mathcal{S}_{k_1}} \sum_{\Delta_{k_2}\in\mathcal{S}_{k_2}} \overline{\ee{j_2 & j_1}{\Delta_{k_2} &\Delta_{k_1}}{\tau}} \eee_{\Delta_{k_1}^-}^{(j_1)}\eee_{\Delta_{k_2}^-}^{(j_2)} \Bigg\}  +\ldots \notag\, ,
\end{align}
where the factors of $(-1)^{j_i}$ seen in (\ref{MMVsec.06}) are absent by our conventions for
$\mathbb{A}_-$ in (\ref{mt.07}), and the ellipsis again refers to iterated integrals of modular depth $\geq 3$.
As mentioned above, our notation ${\mathbb I}_+(\eee_k;\tau)$ and $\widetilde{\mathbb I}_-(\eee_k;\tau)$
for the arguments anticipates that the letters $\eee_{\Delta_{k}^{\pm}}$ will eventually be 
expressible in terms of nested brackets of $\eee_k$ (see (\ref{edeltaex}) below) and do not enter the
construction of this work as independent variables. Nevertheless, the double-integral contributions spelt out
in (\ref{MMVsec.06}) and (\ref{minusees}) illustrate that the reversal from the
$\Ptexp$ operation in (\ref{eq:Pexp}) treats $\eee_{\Delta_{k}^{\pm}}$ as a single letter and 
is more conveniently thought of as a reversal of the integration order of the ${\cal E}[\ldots]$.

\subsubsection{$\mathrm{SL}_2$ transformation of generating series}
\label{sec:csv}

We shall now make contact with the generating series $\mathbb J_{\pm}$ and $\mathbb B^{\rm sv}$ 
of iterated Eisenstein integrals $\beta_{\pm}$ and single-valued MZVs reviewed in 
section \ref{sec:rev.6}. Both of them are related to simpler generating series
by conjugation with the SL$_2$ transformation 
$\Unew$ in (\ref{eq:Utau}). The exponentials can be evaluated as a power series
\begin{align}
& \Unew(\tau) \, \mathbb X\, \Unew(\tau)^{-1} = \sum_{m,n=0}^{\infty} \frac{1}{m! n!} \, \bigg( {-} \frac{1}{4y} \bigg)^m
(2\pi i \bar \tau)^n ({\rm ad}_{\eee_0^\vee})^m ({\rm ad}_{\eee_0})^n \mathbb X\,,
\label{expU}
\end{align}
when acting on $\eee_k$-valued expressions $\mathbb X$. 
For contributions to $\mathbb X$ at fixed \degree, 
the infinite sums in (\ref{expU}) truncate to finitely many terms 
since a given modular-depth-$\ell$ word $\eee_{k_1}^{(j_1)}\ldots \eee_{k_\ell}^{(j_\ell)}$ is annihilated 
at the latest by $1{+}\sum_{i=1}^\ell (k_i{-}2)$ powers of ${\rm ad}_{\eee_0^\vee}$ or ${\rm ad}_{\eee_0}$.

We prove in appendix \ref{sec:Uconj} that the conjugation (\ref{expU}) by the SL$_2$ 
transformation $\Unew$ relates the combination of kernels $\nuker{j}{k}{\tau_1}$ in the
expressions (\ref{mt.06}) and (\ref{mt.07}) for $\mathbb{A}_{\pm}$ to the alternative kernels $\ompm{j}{k}{\tau,\tau_1} $ in section \ref{sec:altker},
\begin{align}
\Unew(\tau) \bigg(\sum_{j=0}^{k-2} \frac{(-1)^j}{j!} \eee_k^{(j)} \nuker{j}{k}{\tau_1} \bigg) \Unew(\tau)^{-1} &=  
\sum_{j=0}^{k-2} \frac{(-1)^j}{j!} \eee_k^{(j)} \omplus{j}{k}{\tau,\tau_1}\, ,
 \label{Unumu}
 \\
\Unew(\tau) \bigg(\sum_{j=0}^{k-2} \frac{1}{j!} \eee_k^{(j)} \overline{\nuker{j}{k}{\tau_1}}  \bigg) \Unew(\tau)^{-1} &=  
 \sum_{j=0}^{k-2} \frac{(-1)^j}{j!} \eee_k^{(j)} \omminus{j}{k}{\tau,\tau_1} 
 \, . \notag
\end{align}
Since the generating series $\mathbb J_{\pm}$ in (\ref{gser.01}) and (\ref{MGFtoBR22})
can be written as path-ordered exponentials involving $\omega_{\pm}$, the conjugations
in (\ref{Unumu}) in fact apply to the full series $\mathbb I_{\pm}$ ,
\begin{align}
\Unew(\tau) \, \mathbb{I}_+(\eee_k;\tau) \, \Unew(\tau)^{-1} &=  R\big[ \mathbb{J}_+(\eee_k;\tau)\big]  \, ,  \label{ItoJ} \\
\Unew(\tau) \,\widetilde{\mathbb{I}}_-(\eee_k;\tau) \, \Unew(\tau)^{-1} &=  R\big[ \widetilde{\mathbb{J}}_-(\eee_k;\tau)\big] 
  \,. \notag
\end{align}
The reflection operator $R$ on the right-hand sides acts via
\beq
R\big[\eee_{k_1}^{(j_1)} \ldots \eee_{k_\ell}^{(j_\ell)}  \big] = \bigg( \prod_{i=1}^\ell \frac{j_i! }{(k_i{-}j_i{-}2)!}\bigg)\,
\eee_{k_\ell}^{(k_\ell-j_\ell-2)} \cdots \eee_{k_1}^{(k_1-j_1-2)} \,,
\label{MMVsec.02}
\eeq
i.e.\ it reverses the ordering of the letters $\eee_{k_i}^{(j_i)}$ and acts on the $\mf{sl}_2$ 
multiplets through the switch operation\footnote{This corresponds to the Weyl reflection (\ref{eq:Weyl}) up to sign.} $\eee_{k}^{(j)}/j! \rightarrow  \eee_{k}^{(k-j-2)}/(k{-}j{-}2)!$.

Comparing the definitions of the two generating series $\mathbb{I}_\pm$ and $\mathbb{J}_\pm$, we see that for a given coefficient function the order of the letters is opposite in the two series, explaining the reversal part of $R$. The switch part instead is due to~\eqref{Unumu} and  the definition of the coefficients of $\mathbb{J}_\pm$ in~\eqref{gser.01} compared with~\eqref{MMVsec.06}.

The reflection operator satisfies the following identities when acting on series $\mathbb{X}$, $\mathbb{Y}$ in the letters $\ep_k^{(j)}$:
\begin{align}
\label{eq:Rop1}
R \left[ \mathbb{X} \,\mathbb{Y} \right] &= R \left[ \mathbb{Y} \right]R \left[ \mathbb{X} \right]\,,\\
R \left[\mathbb{M}^{\rm sv}(z_i)   \mathbb{X}(\ep_k) \mathbb{M}^{\rm sv}(z_i)^{-1} \right] &=  \mathbb{M}^{\rm sv}(z_i)^{-1}  R \big[  \mathbb{X}(\ep_k)\big] \mathbb{M}^{\rm sv}(z_i)\,. \nn
\end{align}
The reflection property in the second line can be seen by expressing the $\ep_k$ produced by $[z_w,\epsilon_m^{(j)}]$ via nested $t^{d}$-operations (\ref{eq:tdpq}) and observing that the sum of the superscripts $d$ has the right parity to attain the negative sign in $R\big[ [ z_w , \ep_m^{(j)} ] \big] = - \big[ z_w , R[ \ep_m^{(j)} ] \big]$. 
We note that while the first identity in~\eqref{eq:Rop1} holds also for series in $\eee_k^{(j)}$, the second one involving the change of alphabet~\eqref{defphisv} is only valid for series $\mathbb{X}$ in $\ep_k^{(j)}$ and we have emphasised this in the notation.
In the later section~\ref{sec:4.2}, we will extend the change of alphabet to a similar formula involving the uplifted generators $\hat{z}_w$ and $\eee^{(j)}_k$, see~\eqref{uplift}, for which the second identity also holds.

It is important to observe that (\ref{ItoJ}) introduces an uplift of the series $\mathbb J_{\pm}(\ep_k;\tau)$
in section \ref{sec:rev.6} since we are now employing the
free-Lie-algebra generators $\eee_k^{(j)}$ instead of $\ep_k^{(j)}$. This
is consistent with the absence of cusp forms in (\ref{gser.01}) since all the $\eee_{\Delta_k^{\pm}}$
are taken to vanish under the restriction $\eee_k^{(j)} \rightarrow \ep_k^{(j)}$.

The SL$_2$ transformation $\Unew$ in (\ref{expU}) transforms the
(anti-)meromorphic generating series $\mathbb I_{\pm}$ into 
path-ordered exponentials $\mathbb J_{\pm}$ of the non-holomorphic
integration kernels $\omega_{\pm}$ that have much cleaner modular properties, see
(\ref{modompm}). Accordingly, the series $\mathbb I_{\pm}$ will henceforth be
referred to as in the {\it holomorphic frame} of $\mf{sl}_2$ whereas
$\mathbb J_{\pm}$ will be said to be in the {\it modular frame}. 

The same SL$_2$ transformation and terminology applies to the
generating series $\mathbb B^{\rm sv}$ of single-valued MZVs.
All the $\tau$-dependence in the definition (\ref{defbsv}) of $\mathbb B^{\rm sv}$
can be identified as an artifact of the SL$_2$ transformation $\Unew$ in (\ref{expU}),
and $\mathbb B^{\rm sv}$ turns out to be conjugate (see (\ref{MMVsec.02}) for the reversal operator $R$)
\begin{align}
\Unew(\tau) \, \CSV\, \Unew(\tau)^{-1} =  R \big[ \mathbb B^{\rm sv}(\eee_k;\tau) \big]\,,
\label{CtoB}
\end{align}
to the following series in real constants $\ccsv{j_1 &\ldots &j_\ell}{k_1 &\ldots &k_\ell}$:
\begin{align}
\CSV &\coloneqq 
 \sum_{P \in \big\{ \begin{smallmatrix} j \\ k \end{smallmatrix} \big\}^\times} \word[P] c^{\rm sv}[P] 
 \label{defcsv} \\
&\phantom{:}= 1 
+  \sum_{k_1=4}^{\infty}\sum_{j_1=0}^{k_1-2}   (-1)^{j_1} \frac{(k_1{-}1)}{j_1!} \ccsv{j_1 }{k_1} \eee_{k_1}^{(j_1)}  
\notag \\
&\phantom{:}\quad
+ \sum_{k_1=4}^{\infty}\sum_{j_1=0}^{k_1-2}\sum_{k_2=4}^{\infty}\sum_{j_2=0}^{k_2-2}
 (-1)^{j_1+j_2} \frac{(k_1{-}1)(k_2{-}1)}{j_1!j_2!}
 \ccsv{j_1 &j_2 }{k_1 &k_2} \eee_{k_1}^{(j_1)}\eee_{k_2}^{(j_2)} + \ldots\,.
\notag 
\end{align}
The ellipsis refers to contributions of 
modular depth $\ell \geq 3$, and we have introduced the shorthand notation
\beq
\word[P] = \word \! \left[\begin{smallmatrix}  j_1 &j_2 &\ldots &j_\ell \\ k_1 &k_2 &\ldots &k_\ell \end{smallmatrix}\right]\coloneqq (-1)^{j_1+\ldots+j_\ell} \frac{(k_1{-}1) (k_2{-}1)\ldots (k_\ell{-}1)}{j_1! j_2!\ldots j_\ell!}  \eee_{k_1}^{(j_1)} \eee_{k_2}^{(j_2)} \ldots \eee_{k_\ell}^{(j_\ell)}\,,
\label{defword}
\eeq
for words in the $\eee_{k_i}^{(j_i)}$. Note that this notation is related to the one in~\eqref{MGFtoBR22} by
\begin{align}
    R\big[\word[P]\big]\Big|_{\eee_k\to \ep_k} = \ep[P] \,.
\end{align}
By analogy with (\ref{ItoJ}), the series $\mathbb C^{\rm sv}$ and $\mathbb B^{\rm sv}$ are
said to be in the holomorphic and modular frame, respectively. Similar to (\ref{MGFtoBR22}),
the summation range $P \in \big\{ \begin{smallmatrix} j \\ k \end{smallmatrix} \big\}^\times$
in the first line of (\ref{defcsv})
gathers all words in $j_i,k_i$ subject to $k_i\geq 4$ and $0\leq j_i\leq k_i{-}2$ with
no separate reference to $\Delta_{k_i}$ as in (\ref{MMVsec.06}).
Upon dressing with derivations $\ep_k^{(j)}$ that satisfy the Pollack relations rather than free generators $\eee_k^{(j)}$, the real constants $\ccsv{j_1 &\ldots &j_\ell}{k_1 &\ldots &k_\ell}$
in (\ref{defcsv}) are projected to single-valued MZVs of \cite{Dorigoni:2022npe}
according to theorem 7.2 in Brown's work \cite{Brown:2017qwo2}. However, we
shall see in section \ref{sec:4.2} that generic $\ccsv{j_1 &\ldots &j_\ell}{k_1 &\ldots &k_\ell}$ 
at modular depth $\ell \geq 3$ and \degree\ $\geq 18$ involve admixtures of new
numbers beyond MZVs with combinations of Pollack combinations $P_w^d$ as coefficients.

Note that alternatively, the change of frame via $\Unew$ can be  performed by introducing
commutative bookkeeping variables $X_i,Y_i$ for each $\eee_{k_i}^{(j_i)}$.
Passing from coefficients of $X_i^a Y_i^b$ (with $a,b\in \mathbb N_0$) to those of 
$(X_i{-}\tau Y_i)^a, \, (X_i{-}\bar \tau Y_i)^b$ as explained in section 7 of \cite{Brown:2017qwo}
introduces the same type of $\tau,\bar \tau$ dependent transformations
as the exponentials in $\Unew$ acting on an $\mf{sl}_2$ multiplet.
This mechanism has also been used in section 2.2 of \cite{Dorigoni:2022npe}
to generate the integration kernels $\omega_{\pm}$ from an alternative
expansion of $(X{-}\tau_1 Y)^{k-2} {\rm G}_k(\tau_1)$. A detailed discussion
with focus on modular depth three will be given in section \ref{sec:8}.

\subsubsection{Equivariant iterated Eisenstein integrals in the holomorphic frame}
\label{sec:holfrm}

Equipped with the previous considerations that apply to generating series in the free Lie algebra generators $\eee_k$, we shall now consider an uplift of the generating series $\mathbb J^{\rm eqv}(\ep_k)$ of~\eqref{select.1} of equivariant iterated Eisenstein integrals to $\mathbb J^{\rm eqv}(\eee_k)$ that will also include cusp forms. The starting point of this uplift is to define a generating series $\mathbb{I}^{\rm eqv}(\ep_k)$ in the holomorphic frame 
\begin{align}
\label{eq:Ieqep}
\mathbb{I}^{\rm eqv}(\ep_k;\tau) = (\phi^{\rm sv})^{ -1}\big(\, \Imint(\ep_k; \tau) \big)
\, \mathbb{C}^{\rm sv}(\ep_k)\, \Ip(\ep_k;\tau) \,,
\end{align}
in terms of~\eqref{MMVsec.06}, \eqref{minusees} and~\eqref{defcsv} with the specialisation $\eee_k\to\ep_k$. 
Here, we have used the inverse of the change of alphabet $\phi^{\rm sv}$ defined in~\eqref{defphisv} that acts on series in $\ep_k$ by $(\phi^{\rm sv})^{ -1} (\mathbb X) = \mathbb M^{\rm sv}( z_i)^{-1} \mathbb X \mathbb M^{\rm sv}( z_i)$.
The generating series (\ref{eq:Ieqep}) is related to  $\mathbb J^{\rm eqv}(\ep_k)$ of~\eqref{select.1} by the change of frame
\begin{align}
R\big[ \Jeqv(\ep_k;\tau) \big]
&= \Unew(\tau)  \,\Ieqv(\ep_k;\tau)\, \Unew(\tau)^{-1} \, . \label{eq:JtoIep} 
\end{align}
In the above generating series~\eqref{eq:Ieqep} and~\eqref{eq:JtoIep} all Tsunogai relations are satisfied. 

All the objects on the right-hand side of~\eqref{eq:Ieqep} have been defined for $\eee_k$ (see~\eqref{MMVsec.06}, \eqref{minusees} and~\eqref{defcsv}), with the exception of the change of alphabet $(\phi^{\rm sv})^{-1}$ that was defined in~\eqref{defphisv} only for series in the generators $\ep_k$ satisfying the Tsunogai relations. We shall postulate an extended change of alphabet $\hat\psi^{\rm sv}$ that operates on series in $\eee_k$: 
\begin{align}
\hat \psi^{\rm sv}\big(\, \Imint(\eee_k; \tau)  \big) &\coloneqq 
\mathbb M^{\rm sv}(\hat z_i)^{-1} \, \Imint(\eee_k; \tau)  \, \mathbb M^{\rm sv}(\hat z_i) + \ldots \, .
\label{defpsisv}
\end{align}
The change of alphabet involves a conjugation 
by the arithmetic parts $\hat{z}_i$ of the uplifted zeta generators that
were introduced in section~\ref{sec:rev.up}. However, as will be explained in sections \ref{sec:ngeozvarpi} and \ref{sec:6} below, the modular 
properties of the uplifted version $\mathbb{I}^{\rm eqv}(\eee_k;\tau)$ of (\ref{eq:Ieqep}) necessitate terms of modular depth $\geq 4$ in the ellipsis that are also due to new periods, see in particular~\eqref{uplift}.  The terms specified on the right-hand side of~\eqref{defpsisv} are complete up to and including modular depth three.
Similar to the change of alphabet $ \phi^{\rm sv}$ for the $\ep_k^{(j)}$ of the antiholomorphic
generating series reviewed in section \ref{subsec:2.4.2}, the $\mf{sl}_2$-invariant map $ \hat \psi^{\rm sv}$  acts on the free-Lie-algebra generators $\eee_k^{(j)}$ and we shall postulate that the conjugation property extends implying $\hat{\psi}^{\rm sv}(\mathbb{X}\mathbb{Y})= \hat{\psi}^{\rm sv}(\mathbb{X})\hat{\psi}^{\rm sv}(\mathbb{Y})$.
By~\eqref{eq:Rop1}, we know that in the specialisation $\eee_k^{(j)} \rightarrow \ep_k^{(j)}$, the change of alphabet (\ref{defpsisv}) reduces to the inverse of $\phi^{\rm sv}$ in (\ref{defphisv}), i.e.
\begin{equation}
\hat \psi^{\rm sv}(\mathbb X) \Big\vert_{\eee_k^{(j)}\to \ep_k^{(j)}}=  (\phi^{\rm sv})^{-1} (\mathbb X) = \mathbb M^{\rm sv}( z_i)^{-1} \mathbb X\, \mathbb M^{\rm sv}( z_i)\,.
\end{equation}

With this extended change of alphabet $\hat \psi^{\rm sv}$, we can now define 
\begin{align}
\label{eq:Iup}
\Ieqv(\eee_k;\tau)&= \hat \psi^{\rm sv}\big(\, \Imint(\eee_k; \tau) \big)
\, \CSV\, \Ip(\eee_k;\tau)\,,
\end{align}
as the uplift of~\eqref{eq:Ieqep} in the holomorphic frame as well as
\begin{align}
\label{eq:JtoII}
R\big[ \Jeqv(\eee_k;\tau) \big]
&= \Unew(\tau)  \,\Ieqv(\eee_k;\tau)\, \Unew(\tau)^{-1}\,,
\end{align}
as the uplift of the equivariant iterated Eisenstein integrals in the modular frame.
Section~\ref{sec:6} is devoted to proving the equivariance of the right-hand 
side of \eqref{eq:JtoII} as well as showing how this relates to the generating series of 
equivariant iterated Eisenstein integrals in \eqref{select.1}.

The uplifted generating series $\mathbb{I}^{\rm eqv}(\eee_k)$ in the holomorphic frame has the following properties (relative to its modular counterpart $\mathbb{J}^{\rm eqv}(\eee_k)$):
\begin{itemize}
\item $\mathbb C^{\rm sv}$ is a generating series involving only real constants, unlike $\mathbb B^{\rm sv}$ that has a rational dependence on $\tau$, $\bar \tau$;
\item the iterated integrals of $\mathbb I_{+}, \Imint$ are (anti-)meromorphic in $\tau$, unlike the $\beta_{\pm}$
entering $\mathbb J_{\pm}$.
\end{itemize}

By analogy with the expansion (\ref{select.1}) of $\Jeqv$ in terms of modular
forms $\beta^{\rm eqv}$, we expand the analogous series $\Ieqv$ in the
holomorphic frame in words (\ref{defword}) composed of the independent generators $\eee_k^{(j)}$,
\beq
\Ieqv(\eee_k;\tau) =
\sum_{P \in \big\{ \begin{smallmatrix} j \\ k \end{smallmatrix} \big\}^\times}  \word[P] {\cal E}^{\rm eqv}[P;\tau] \, .
\label{eeqv.01}
\eeq
As an initial step towards the study of the expansion coefficients ${\cal E}^{\rm eqv}$, we can at first consider only the contribution to~\eqref{eq:Iup} coming simply from  $\hat \psi^{\rm sv} \rightarrow 1$. Restricting to the Eisenstein part this becomes
\beq
\eeqvtau{j_1 &\ldots &j_\ell}{k_1 &\ldots &k_\ell}{\tau} = \! \! \sum_{ 0 \leq p\leq q \leq \ell} \! \! 
(-1)^{j_1+\ldots+ j_p} \overline{ \ee{j_p &\ldots &j_2 &j_1}{ k_p &\ldots &k_2 &k_1 }{\tau} }
\ccsv{j_{p+1} &\ldots &j_q}{k_{p+1} &\ldots &k_q} 
\ee{ j_{q+1} &\ldots &j_\ell }{k_{q+1} &\ldots &k_\ell}{\tau} + \ldots\, ,
\label{eeqv.02}
\eeq
where the ellipsis refers to both  admixtures of MZVs and certain new periods from 
$c^{\rm sv},\hat \psi^{\rm sv}$ and  iterated integrals $\eeno{\ldots &j &\ldots}{\ldots &\Delta_k &\ldots}$
involving cusp forms from the expansion (\ref{MMVsec.06}) and (\ref{minusees}) 
of $\mathbb I_{\pm}$. Examples of these extra terms and further details on their 
extraction from the generating series will be given in section \ref{sec:4.2} below and appendix \ref{app:expeqv.2}.

Since the change of frame between $\mathbb B^{\rm sv} \leftrightarrow \mathbb C^{\rm sv} $ and 
$\mathbb J^{\rm eqv} \leftrightarrow \mathbb I^{\rm eqv} $ follows the same conjugations
(\ref{CtoB}) and (\ref{eq:JtoII}) by $\Unew$, the relation (\ref{defbsv}) between the coefficients
$b^{\rm sv} \leftrightarrow c^{\rm sv}$ can be readily adapted to the equivariant
iterated Eisenstein integrals,
\begin{align}
\beqv{\ldots &j &\ldots}{\ldots &k &\ldots} &=  \sum_{p=0}^{k-2-j} \sum_{\ell=0}^{j+p}
\binom{k{-}j{-}2}{p} \binom{j{+}p}{\ell} \frac{({-}2\pi i \bar \tau)^\ell}{(4y)^p} \,
\eeqv{\ldots &j-\ell+p &\ldots}{\ldots &k &\ldots}\, ,
\label{eeqv.05}
\end{align}
where the double sum again applies independently to each column 
$ \begin{smallmatrix} j_i \\ k_i \end{smallmatrix} $.
As one can anticipate from the powers of $\bar \tau$ and $y^{-1}$ in 
the expansion coefficients of (\ref{eeqv.05}), the ${\cal E}^{\rm eqv}$
in the holomorphic frame do not transform as modular forms.
Still, they enjoy \textit{equivariant} transformation properties in the sense that
\begin{align}
\eeqvtau{j_1 &\ldots &j_\ell }{k_1 &\ldots &k_\ell}{\tau{+}1} &=
\sum_{p_1=0}^{j_1} \ldots \sum_{p_\ell=0}^{j_\ell} \bigg( \prod_{r=1}^\ell (2\pi i)^{j_r-p_r} \binom{ j_r}{p_r}  \bigg) \eeqvtau{p_1 &\ldots &p_\ell }{k_1 &\ldots &k_\ell}{\tau}\,, 
\label{eeqv.06} \\
\eeqvtau{j_1 &\ldots &j_\ell }{k_1 &\ldots &k_\ell}{{-}\tfrac{1}{\tau}} &=
\bigg(  \prod_{r=1}^\ell  (-1)^{j_r} (2\pi i)^{2+2j_r-k_r} \bigg) \eeqvtau{k_1-j_1-2 &\ldots &k_\ell-j_\ell-2}{k_1 &\ldots &k_\ell}{\tau} \,,
\notag
\end{align}
is homogeneous in modular depth.\footnote{After a suitable rescaling, the $\mathcal{E}^{\rm eqv}$ can be viewed as being in a matrix representation of $\SLtwoZ$ such that they form components of a vector-valued modular form.}
More importantly, equivariant iterated Eisenstein
integrals ${\cal E}^{\rm eqv}$ in the holomorphic frame share the modular $T$ and $S$ transformations
(\ref{eq:nukerT}) and (\ref{eq:nukerS}) of the kernels $\nuker{j}{k}{\tau}$ in each column and transform 
in the same way as the formal tensor products
\beq
\eeqvtau{j_1 &j_2 &\ldots &j_\ell }{k_1 &k_2 &\ldots &k_\ell}{\tau} \leftrightarrow  
\nuker{j_1}{k_1}{\tau} \otimes \nuker{j_2}{k_2}{\tau} \otimes \ldots \otimes \nuker{j_\ell}{k_\ell}{\tau}  \, .
\label{liketens}
\eeq
In particular, (\ref{eeqv.06}) takes the same form as the contributions
to $\ee{j}{k}{\tau{+}1}$ and $\ee{j}{k}{-\tfrac1{\tau}}$ in (\ref{eettrf}) 
and (\ref{strfd1}) that preserve modular depth one. Following the original
idea in Brown's construction of equivariant iterated Eisenstein
integrals \cite{Brown:mmv, Brown:2017qwo, Brown:2017qwo2}, it is the joint 
effect of the series $\mathbb C^{\rm sv}$, the extended change of alphabet $\hat \psi^{\rm sv}$ and the cuspidal iterated integrals in
$\mathbb I_{\pm}$ in (\ref{eq:Iup}) that eliminates any term of lower modular depth from the ${\rm SL}(2,\mathbb Z)$ transformations (\ref{eeqv.06}).
Note that the $\eeqv{j_1 &j_2}{k_1 &k_2}$ are $\mathbb Q[2\pi i]$ multiples of the coefficients of 
$X_1^{k_1-2-j_1} Y_1^{j_1}  X_2^{k_2-2-j_2}Y_2^{j_2}$  
in the equivariant double integrals $M_{k_2,k_1}$ in section 9 of \cite{Brown:2017qwo}
whose explicit examples and relations to the modular forms $\beqv{j_1 &j_2}{k_1 &k_2}$
can be found in section 4.3 of \cite{Dorigoni:2022npe}.
We will elaborate more on this approach via commutative formal variables $(X_i,Y_i)$ in section~\ref{sec:8}.


\subsubsection{Differential equations}
\label{sec:dJeqv}

The construction of the generating series in this section from path-ordered exponentials
results in simple differential equations with respect to\ $\tau$ and $\bar \tau$. The definition
(\ref{eq:Pexp}) of $\mathbb{I}_+$ and $\widetilde{\mathbb{I}}_-$ together with the
$(1,0)$- and $(0,1)$-forms $\mathbb{A}_+$ and $\mathbb{A}_-$ in (\ref{mt.06}) and (\ref{mt.07}) readily imply that
\begin{align}
 \partial_{\tau}  \mathbb{I}_+(\eee_k;\tau) \,\dd \tau&= -  \mathbb{I}_+(\eee_k;\tau) 
 \mathbb A_+(\eee_k;\tau) \,,\label{diffeq.01} \\
\partial_{\bar \tau} \widetilde{\mathbb{I}}_-(\eee_k;\tau)\, \dd \bar \tau &= - \mathbb A_-(\eee_k;\tau)  \widetilde{\mathbb{I}}_-(\eee_k;\tau) \,, \notag \\
\partial_{\bar \tau}  \mathbb{I}_+(\eee_k;\tau) &= 
\partial_{\tau} \widetilde{\mathbb{I}}_-(\eee_k;\tau) = 0\,.
\notag
\end{align}
By virtue of (\ref{eq:Iup}), the generating series $\Ieqv$ of equivariant 
iterated Eisenstein integrals in the holomorphic frame share the $\tau$-derivative of (\ref{diffeq.01}) but involve the
$\hat \psi^{\rm sv}$-image of $\mathbb A_-$ in the $\bar \tau$-derivative:
\begin{align}
 \partial_\tau \Ieqv(\eee_k;\tau)\,\dd \tau  &= - \Ieqv(\eee_k;\tau)  \mathbb A_+(\eee_k;\tau) \, ,
\label{diffeq.02} \\
  \partial_{\bar \tau} \Ieqv(\eee_k;\tau)\,\dd \bar \tau &=
-  \hat \psi^{\rm sv} \big(  \mathbb A_-(\eee_k;\tau) \big)
 \Ieqv(\eee_k;\tau) \, .
\notag
\end{align}
For the second equation we have used that $\hat{\psi}^{\rm sv}$ factors on generating series and have assumed that it does not introduce any extra $\tau$-dependence, so that we can differentiate inside the argument.
By the discussion around (\ref{defpsisv}), the contributions to $\hat \psi^{\rm sv} (  \mathbb A_-(\eee_k;\tau) )$ 
of modular depth $\leq 3$ conincide with $\Msv(\zetaE_i)^{-1}  \mathbb A_-(\eee_k;\tau) \Msv(\zetaE_i)$.
Moreover, the restriction $\eee_k^{(j)} \rightarrow \ep_k^{(j)}$ of the $\bar \tau$-derivative 
in (\ref{diffeq.02}) simplifies to 
\beq
 \partial_{\bar \tau} \Ieqv(\ep_k;\tau)\,\dd \bar \tau 
= - \Msv(\zetaT_i)^{-1}  \mathbb A_-(\ep_k;\tau) \Msv(\zetaT_i)
 \Ieqv(\ep_k;\tau)\,,
\label{simpbartau}
\eeq 
 at arbitrary modular depth, reducing $\hat \psi^{\rm sv} $ to arithmetic parts of zeta generators.

The reflected version of the generating series $\mathbb J^{\rm eqv}$ in the
modular frame in (\ref{eq:JtoII}) receives additional contributions
to its $\tau$-derivative from the SL$_2$ transformation in (\ref{eq:Utau}): As
a consequence of $2\pi i (\tau{-}\bar \tau)^2 \partial_\tau \Unew(\tau) = - \eee_0^\vee \Unew(\tau)$,
we have
\begin{align}
-2\pi i &(\tau{-}\bar \tau)^2\partial_\tau R \big[ \Jeqv ( \eee_k;\tau) \big]
= \Big[ \eee_0^\vee , R \big[ \Jeqv ( \eee_k;\tau) \big] \Big]
\label{diffeq.03} \\
 &+ R \big[ \Jeqv ( \eee_k;\tau) \big] \sum_{m=4}^\infty \frac{(m{-}1)}{(m{-}2)!} \, (\tau {-} \bar \tau)^m
\bigg\{
{\rm G}_m(\tau) \eee_m^{(m-2)} + \sum_{\Delta_m \in {\cal S}_m} (2\pi i)^m \Delta_m(\tau) 
\eee_{\Delta_m^+}^{(m-2)}
\bigg\}\, .
\notag
\end{align}
To remove the reflection operator $R$ defined in (\ref{MMVsec.02}) from the left-hand side we simply need to apply it once more on both sides. On the right-hand side this operation replaces in the first line the commutator with $ \eee_0^\vee$ by $ {\rm ad}_{\eee_0}$, while in the second line it exchanges highest- and lowest-weight
vectors via $\frac{1}{(m-2)!}\eee_m^{(m-2)} \rightarrow 
\eee_m$, therefore leading to the simplified result
\begin{align}
-2\pi i &(\tau{-}\bar \tau)^2\partial_\tau   \Jeqv ( \eee_k;\tau) 
=  {\rm ad}_{\eee_0} \Jeqv ( \eee_k;\tau)  
\label{diffeq.04} \\
 &+\sum_{m=4}^\infty  (m{-}1)  \, (\tau {-} \bar \tau)^m
\bigg\{
{\rm G}_m(\tau) \eee_m + \sum_{\Delta_m \in {\cal S}_m} (2\pi i)^m \Delta_m(\tau) 
\eee_{\Delta_m^+} \bigg\}
  \Jeqv ( \eee_k;\tau) \, .
 \notag
\end{align}
The modular forms ${\rm G}_m$ and $\Delta_m$ on the right-hand side of 
(\ref{diffeq.03}) and (\ref{diffeq.04}) illustrate that the construction of $\mathbb J_{\pm}$
and therefore $\mathbb J^{\rm eqv}$ from path-ordered exponentials is consistent
with the modular transformations (\ref{modbeqv}) of the coefficients $\beta^{\rm eqv}$.

Since the letters $\eee_{\Delta^{\pm}}$ associated with the cusp-form contributions to $\mathbb J_{\pm}(\eee_k;\tau)$ are built from Pollack combinations, specialisation to 
$\eee_k \rightarrow \epsilon_k$ results in a holomorphic differential equation
with only Eisenstein series \cite{Brown:2017qwo2, Dorigoni:2022npe}:
\begin{align}
-2\pi i &(\tau{-}\bar \tau)^2\partial_\tau \Jeqv ( \epsilon_k;\tau) 
= {\rm ad}_{\epsilon_0}  \Jeqv( \epsilon_k;\tau)
 \label{diffeq.05} 
+ \sum_{m=4}^\infty
(m{-}1) (\tau{-}\bar \tau)^m {\rm G}_m(\tau) \epsilon_m 
\Jeqv( \epsilon_k;\tau) \, .
\end{align}
The resulting differential equation
for the components $\beta^{\rm eqv}$ reads with $\nabla=2i (\Im \tau)^2 \partial_\tau$:
\begin{align}
-4\pi \nabla
\beqvtau{j_1 &j_2& \ldots &j_\ell}{k_1 &k_2 &\ldots &k_\ell}{\tau} &= \sum_{i=1}^\ell (k_i{-}j_i{-}2)\beqvtau{j_1 & \ldots &j_i+1 &\ldots &j_\ell}{k_1 &\ldots &k_i &\ldots &k_\ell}{\tau}\label{dbetaeqv} \\
&\quad  - \delta_{j_\ell,k_\ell-2} (\tau{-}\bar \tau)^{k_\ell} \GG_{k_\ell}(\tau) 
\beqvtau{j_1 &j_2& \ldots &j_{\ell-1}}{k_1 &k_2 &\ldots &k_{\ell-1}}{\tau} + \ldots \, ,
\notag
\end{align} 
where the ellipsis refers to holomorphic cusp forms multiplying $\beta^{\rm eqv}$ of modular depth $\leq \ell{-}2$. The $\bar\tau$-derivative of $\Jeqv(\eee_k;\tau)$  is not as clean as~\eqref{diffeq.04} as can be anticipated from~\eqref{diffeq.02}: 
Apart from a term ${\rm ad}_{\eee_0^\vee} \Jeqv ( \eee_k;\tau)$ that is completely analogous to ${\rm ad}_{\eee_0} \Jeqv ( \eee_k;\tau)$ in the holomorphic derivative $\partial_\tau \Jeqv ( \eee_k;\tau)$, the $\bar\tau$-derivative also contains contaminations of lower modular depth due to the change of alphabet encoded in $\hat\psi^{\rm sv}$.


\subsection{Modular graph forms from zeta generators}
\label{sec:4zeta}

One of the key results of this work concerns the interplay of the change of alphabet $\hat \psi^{\rm sv}$ and the $\tau$-independent series $\CSV$ within the generating series (\ref{eq:Iup}) of equivariant
iterated Eisenstein integrals. In the specialisation $\eee_k \rightarrow \ep_k$ relevant to MGFs,
the uplifted zeta generators $\hat \zetaT_i, \hat \sigmaT_i$ reduce to the $\zetaT_i, \sigmaT_i$ determined by
the methods of the companion paper \cite{Dorigoni:2023part1} and reviewed in section~\ref{sec:rev.zeta}. 
One of our central claims is that the middle product in the specialisation
\beq
\Ieqv(\ep_k;\tau)=  \Msv(\zetaT_i)^{-1} \,\Imint(\ep_k; \tau)\, \mathbb \Msv(\zetaT_i) 
\, \CSVep\, \Ip(\ep_k;\tau)\,,
\label{spec317}
\eeq
of~\eqref{eq:JtoII} can  be presented in a unified way 
\beq
 \Msv(\zetaT_i)\, \CSVep =  \Msv(\sigmaT_i)\,,
 \label{zcvssigma}
 \eeq
through the same type of group-like series as
in (\ref{Msvz}) with $\sigmaT_i$ in the place of $\zetaT_i$:
\begin{align}
\mathbb M^{\rm sv}(\sigmaT_i) &= \sum_{\ell=0}^{\infty} \sum_{\substack{i_1,i_2,\ldots, i_\ell \\ \in 2\mathbb N+1}} \sigmaT_{i_1}\sigmaT_{i_2}\ldots \sigmaT_{i_\ell} \rho^{-1} \big( {\rm sv}(f_{i_1} f_{i_2} \ldots f_{i_\ell}) \big) 
\label{Msvsigma} \\
&= 1 +2 \! \! \! \sum_{i_1 \in 2\mathbb N+1} \! \! \! \sigmaT_{i_1} \zeta_{i_1}
+ 2 \! \! \!  \!  \! \! \sum_{i_1,i_2 \in 2\mathbb N+1} \! \! \! \! \! \! \sigmaT_{i_1} \sigmaT_{i_2}  \zeta_{i_1} \zeta_{i_2} 
+ \! \! \! \! \! \!  \! \!  \sum_{i_1,i_2,i_3 \in 2\mathbb N+1} \! \! \! \! \! \! \! \!
\sigmaT_{i_1} \sigmaT_{i_2} \sigmaT_{i_3}  \rho^{-1}\big( {\rm sv}(f_{i_1} f_{i_2}f_{i_3} ) \big) +\ldots \, .
\notag
\end{align}

This leads to our main result for the
generating series (\ref{eq:Iup}) of those equivariant iterated Eisenstein integrals which do not feature any holomorphic cusp forms
\beq
\Ieqv(\epsilon_k;\tau) = 
\Msv(\zetaT_i)^{-1} \Imint(\epsilon_k;\tau)
\mathbb \Msv(\sigmaT_i)  \Ip(\epsilon_k;\tau)
\, .
\label{mainiequv}
\eeq
The resulting expressions for
\begin{align}
 \mathbb B^{\rm sv}(\ep_k;\tau) &= R\big[
 \Unew(\tau) \,  \mathbb M^{\rm sv}(\zetaT_i)^{-1} \, \mathbb M^{\rm sv}(\sigmaT_i) 
 \, \Unew(\tau)^{-1} 
 \big]\, ,
 \label{finalBsv} \\
 R\big[ \Jeqv(\ep_k;\tau) \big] &= \Unew(\tau) \Msv(\zetaT_i)^{-1} \Imint(\epsilon_k;\tau)
\mathbb \Msv(\sigmaT_i)  \Ip(\epsilon_k;\tau) \Unew^{-1}(\tau)\,,
\notag
\end{align}
are another major step in making the key ingredients $(\mathbb B^{\rm sv},\phi^{\rm sv})$
of Brown's construction \cite{Brown:2017qwo2} explicit. With the change of alphabet
$\phi^{\rm sv}$ in (\ref{finalphisv}) determined in \cite{Dorigoni:2022npe} 
and the new result (\ref{finalBsv}) of this work for $\mathbb B^{\rm sv}$, the
equivariant iterated Eisenstein integrals entering MGFs are available to the 
modular depths and \degree s where we have computational control over the
zeta generators. There is no conceptual bottleneck in obtaining explicit expansions
of $[\zetaT_w,\ep_k]$ and $\sigmaT_i$ in terms of $\ep_{k_i}^{(j_i)}$ to arbitrary 
\degree\ and modular depth by the methods in the companion paper \cite{Dorigoni:2023part1}. 
Generalisations of the closed-form results (\ref{zetgen.07}) and (\ref{zetgen.08})
beyond modular depth two can be found in the reference as well as appendix \ref{app:F} and the ancillary files of this work.

Note that the action (\ref{MMVsec.02}) of the reflection operator $R$ in (\ref{finalBsv})
is understood to be unchanged under $\eee_k^{(j)} \rightarrow \ep_k^{(j)}$.
However, (\ref{finalBsv}) cannot be uplifted to the larger generating series
$\Jeqv(\eee_k;\tau)$ including cuspidal iterated integrals while preserving the
form (\ref{Msvsigma}) of $\mathbb M^{\rm sv}(\sigmaT_i)$. As we will
see in section \ref{sec:4.2}, the zeta generators $\sigmaT_w \rightarrow \sigmaE_w$
adapted to the free-Lie-algebra generators $\eee_k^{(j)}$ need to be
augmented by additional generators $\sigmaE_\pern$
associated with new periods, $\pern$, beyond MZVs. Moreover, modularity requires these
$\sigmaE_\pern$ to involve arithmetic parts $\zetaE_\pern$ which prevent
us from reducing $\hat \psi^{\rm sv}( \Imint(\eee_k; \tau) ) $ in (\ref{eq:Iup}) to $ \mathbb M^{\rm sv}(\hat z_i)^{-1}  \Imint(\eee_k; \tau)  \mathbb M^{\rm sv}(\hat z_i)$ (though modular depths $\leq 3$ 
are insensitive to the difference). Accordingly, we shall content ourselves
with series in Tsunogai's derivations $\ep_k^{(j)}$ for the rest of this subsection \ref{sec:4zeta}
where the modular completion of $\Imint(\ep_k; \tau) \mathbb I_{+}(\ep_k;\tau)$ can be found in terms
of MZVs.

Similar to our comment below (\ref{finalphisv}), it is not manifest term-by-term that
the expressions (\ref{finalBsv}) for $\mathbb B^{\rm sv}$ and $\Jeqv$
admit an expansion solely in terms of $\ep_k^{(j)}$. In both cases, the
arithmetic parts of the zeta generators from the left-multiplicative series
$ \Msv(\zetaT_i)^{-1}$ can be used to ``drain out'' all the $\zetaT_i$-contributions
to $ \Msv(\sigmaT_i)$. As will be detailed in section \ref{sec:4zeta.2}, this requires iterative use of $\ep_k^{(j)} \zetaT_{w} 
= \zetaT_w  \ep_k^{(j)} - {\rm ad}_{\ep_0}^j [ \zetaT_w , \ep_k]$ and the fact that the
brackets on the right-hand side are expressible via nested commutators 
of $\ep_{k_i}^{(j_i)}$, i.e., that the $\zetaT_w$ normalise the Tsunogai derivation algebra of the $\ep_k$.


\subsubsection{All information from Riemann zeta values}
\label{sec:4zeta.1}

In the new formulation (\ref{finalBsv}), all MZVs entering
the generating series of equivariant iterated Eisenstein integrals arise from
the group-like series $\mathbb M^{\rm sv}(\zetaT_i)$ and
$\mathbb M^{\rm sv}(\sigmaT_i)$ in (\ref{Msvz}) and (\ref{Msvsigma}).
As an important implication of the structure of these $\mathbb M^{\rm sv}$,
the operator-valued coefficients $\zetaT_w$, $\sigmaT_w$ of 
odd Riemann zeta values determine the coefficients of all the remaining
(single-valued) MZVs composed of two or more letters in the $f$-alphabet. 
After reading off the overall coefficient of $\ep_6 \ep_4 \ep_4$ from $\mathbb J^{\rm eqv}$,
for instance, the appearance of $\zeta_3^2, \, \zeta_3 \zeta_5$
and $\zeta_{3,3,5}^{\rm sv}$ in
\begin{align}
\beqv{2 &2 &4}{4 &4 &6}  \big|_{{\rm mod}\,{\rm depth}\; 1}  &= 
\bigg( \frac{ 2 (i \pi \bar \tau)^5   \zeta_3}{14175} 
+ \frac{ (i \pi \bar \tau)^3 \zeta_5}{ 675} 
+ \frac{2 \zeta_3 \zeta_5 }{15} \bigg) \bminusno{2}{4}
+\frac{2 \zeta_3^2}{9}
\bplusno{4}{6} \,,
\label{beqvex446}\\
\beqv{2 &2 &4}{4 &4 &6}  \big|_{{\rm mod}\,{\rm depth}\; 0}  &=
 - \frac{ (i \pi \bar \tau)^8 \zeta_3 }{4536000}
-  \frac{ (i \pi \bar \tau)^6 \zeta_5 }{ 364500}
 - \frac{ 2 (i \pi \bar \tau)^5 \zeta_3^2 }{42525}
 - \frac{ (i \pi \bar \tau)^3 \zeta_3 \zeta_5 }{2025} 
 -\frac{14573 \zeta_{11}}{21600} + \frac{ \zeta_{3,3,5}^{\rm sv}}{225}\,,
 \notag
\end{align}
can be traced back to the coefficients of $\zeta_3$ and $\zeta_5$ in $\beta^{\rm eqv}$
of modular depth $\leq 2$ which in turn follow from low-order terms in (\ref{zetgen.08}). Representations of $\zeta_{3,3,5}^{\rm sv}$ in terms of MZVs and the $f$-alphabet can be found in (\ref{appMZV.g9}) and (\ref{rhoinvlist}), respectively.

More generally, the appearance of higher-depth MZVs or products of arbitrary
MZVs -- say $\rho^{-1}({\rm sv}(f_{i_1} f_{i_2}\ldots))$ in the $f$-alphabet -- 
is interlocked with that of $\zeta_w$ through compositions such as
$\zetaT_{i_1}\zetaT_{i_2}\ldots$ and $\sigmaT_{i_1} \sigmaT_{i_2}\ldots$ in (\ref{Msvz}) and (\ref{Msvsigma}). We reiterate the warning from the end of section~\ref{sec:223} that modular depth zero does not correspond to the Laurent polynomial in $y$ that governs the asymptotics at the cusp.

The same phenomenon was observed in \cite{Schlotterer:2012ny} for a basis of disk
integrals in tree-level amplitudes of massless open-string 
states \cite{Mafra:2022wml}. Instead
of derivations $\zetaT_w$, $\sigmaT_w$, the coefficients of Riemann zeta
values in the low-energy expansion of disk integrals are square matrices $M_w$
whose entries are degree-$w$
polynomials in certain dimensionless kinematic variables with rational coefficients. 
The compositions of operators $\zetaT_w$, $\sigmaT_w$ we
encounter at genus one are then replaced by matrix products $M_{i_1} M_{i_2}\ldots$ 
in the coefficients of MZVs $\rho^{-1}(f_{i_1} f_{i_2}\ldots)$ in the tree-level setting.
Similar matrices and the same type of correlations 
between the coefficients of arbitrary MZVs and those of Riemann zeta values 
can also be found in more general configuration-space integrals at
genus zero \cite{Abreu:2017enx, Abreu:2017mtm, Britto:2021prf}.

Through the single-valued map of MZVs, the sphere integrals in closed-string 
tree-level amplitudes inherit the group-like series structure (\ref{Msvsigma})
with the above $M_w$ in the place of $\sigmaT_w$ \cite{Schlotterer:2012ny, 
Stieberger:2013wea, Stieberger:2014hba, Schlotterer:2018abc, Vanhove:2018mzv, 
Brown:2019wna}. 
The asymptotic expansion as $\tau \rightarrow i\infty$ of generic MGFs is
not yet available in terms of the known low-energy expansion of sphere integrals, 
though the solutions for the two-point genus-one examples in terms of four-point
tree-level integrals \cite{DHoker:2019xef, Zagier:2019eus, Gerken:2020yii} suggest that an 
explicit map can also be found at higher multiplicity. Nevertheless, (\ref{Msvsigma}) and (\ref{finalBsv}) encode the 
modular completion of the iterated Eisenstein integrals in all MGFs, regardless of the 
multiplicity in their configuration-space-integral representation. On these grounds,
the results of this section already incorporate the correlations between MZVs in
sphere integrals into MGFs, even if their explicit connection is not yet worked
out beyond two points.

Still, it remains a pressing problem to deduce the $\tau \rightarrow i\infty$
asymptotics of MGFs from low-energy expansions of sphere integrals
or other single-valued quantities. It is a notorious gap in the literature
to prove the conjecture of \cite{Zerbini:2015rss, DHoker:2015wxz} that the Laurent 
polynomials in $y$ describing the $\tau \rightarrow i\infty$ asymptotics of MGFs 
only have single-valued MZVs as coefficients. From the iterated-integral representation
of the generating series of MGFs in \cite{Gerken:2020yii}, it is not yet rigorously established that
MGFs can only involve $\beta^{\rm eqv}$ with single-valued MZVs as coefficients.
In other words, it remains to rule out that the initial conditions $\tau \rightarrow i\infty$
of the generating series in the reference at $n\geq 3$ points yield
coefficients outside the ring of single-valued MZVs.


\subsubsection{Implications for $c^{\rm sv}$}
\label{sec:4zeta.2}

In order to study the implications of (\ref{zcvssigma}) for the constant coefficients $c^{\rm sv}$ of $\mathbb C^{\rm sv}(\epsilon_k)$ in (\ref{defcsv}), we rewrite the equation as
\begin{align}
\mathbb C^{\rm sv} (\epsilon_k) &=
 \mathbb M^{\rm sv}(\zetaT_i)^{-1} \, \mathbb M^{\rm sv}(\sigmaT_i)
\label{zinvsigma} \\
&= 1 
+ 2 \sum_{i_1 \in 2\mathbb N+1} \zeta_{i_1}  \sigmaT^{\rm g}_{i_1} 
+ 2 \sum_{i_1,i_2 \in 2\mathbb N+1} \zeta_{i_1} \zeta_{i_2} 
\Big( \sigmaT^{\rm g}_{i_1} \sigmaT^{\rm g}_{i_2} + [ \sigmaT^{\rm g}_{i_1} , \zetaT_{i_2} ] \Big) \notag \\
&\quad 
+ \! \! \! \! \! \! \! \! \sum_{i_1,i_2,i_3 \in 2\mathbb N+1} \! \! \! \! \! \! \! \! \rho^{-1}\big( {\rm sv}(f_{i_1} f_{i_2} f_{i_3})  \big)
\Big(
\sigmaT^{\rm g}_{i_1} \sigmaT^{\rm g}_{i_2}   \sigmaT^{\rm g}_{i_3} 
{+}  [ \sigmaT^{\rm g}_{i_1}   \sigmaT^{\rm g}_{i_2}, \zetaT_{i_3}  ]
{+}  [ \sigmaT^{\rm g}_{i_1} , \zetaT_{i_2}  ]\sigmaT^{\rm g}_{i_3}
{+}  \big[  [\sigmaT^{\rm g}_{i_1} ,  \zetaT_{i_2}  ] , \zetaT_{i_3} \big]   \Big)
+\ldots \, ,
\notag
\end{align}
where the geometric contributions
to the zeta generators are denoted by
\beq
\sigmaT_w^{\rm g} = \sigmaT_w - \zetaT_w
\, .
\label{defgeo}
\eeq
The geometric part $\sigmaT^{\rm g}_w$ is valued in Tsunogai's derivation algebra, i.e., a $\mathbb{Q}$-linear combination of nested brackets of the $\ep_k^{(j)}$.

We shall illustrate how different terms in (\ref{zinvsigma}) led us to anticipate (\ref{zcvssigma}) or equivalently (\ref{finalBsv}) and how~\eqref{zinvsigma} yields concrete predictions for the MZVs beyond depth one in various~$c^{\rm sv}$.

\paragraph{First evidence from modular depth one:}

A first indication for the representation (\ref{finalBsv}) of $\mathbb B^{\rm sv}$ in terms of zeta generators came from the following observation on the coefficients of Riemann zeta values: The coefficients of $ \zeta_{w}$ in the constants $c^{\rm sv}$ entering $\mathbb B^{\rm sv}$ in (\ref{defbsv}) turn out to match the expansion coefficients of the zeta generators $\sigmaT_{w}$ in (\ref{zetgen.08}) 
\begin{align}\label{eq:sigmafromcsv}
	\sigmaT_{w} &= \zetaT_{w} + \frac{1}{2} \sum_{k_1=4}^{\infty}\sum_{j_1=0}^{k_1-2} (-1)^{j_1} \frac{(k_1{-}1)}{j_1!} \, \ccsv{j_1}{k_1}\, \Big|_{\rho^{-1}(f_w) }\, \epsilon_{k_1}^{(j_1)} \\
		    &\quad + \frac{1}{2} \sum_{k_1=4}^{\infty}\sum_{j_1=0}^{k_1-2}\sum_{k_2=4}^{\infty}\sum_{j_2=0}^{k_2-2} (-1)^{j_1+j_2} \frac{(k_1{-}1)(k_2{-}1)}{j_1!j_2!}\ccsv{j_1 & j_2}{k_1 & k_2}\, \Big|_{\rho^{-1}(f_w)}\, \epsilon_{k_1}^{(j_1)}\epsilon_{k_2}^{(j_2)}  + \ldots  , \notag
\end{align}
where the coefficients of $\frac{1}{2}$ on the right-hand side stem from
$\zeta_{w}^{\rm sv}=2 \zeta_{w}$.
 For the isomorphism $\rho$ from
(motivic) MZVs to the $f$-alphabet (see appendix \ref{app:A.1} for a brief review) we employ here the preferred choice described in the companion paper \cite{Dorigoni:2023part1} which for instance fixes the ambiguity of $\sigmaT_{11}$ by rational multiples of $[\sigmaT_{3}, [\sigmaT_{3}, \sigmaT_{5}]]$ and all its higher-weight analogues. We have tested (\ref{eq:sigmafromcsv}) to hold for all contributions to $\sigma_{w\leq 15}$ subject to $\sum_i (k_i{-}j_i{-}1) \leq 5$ and for~$\sigma_{17}$ up to $\sum_i (k_i{-}j_i{-}1)=3$.

\paragraph{Additional evidence from maximum modular depth:}

The next piece of evidence for (\ref{finalBsv}) stems from section 4.1.4 of \cite{Dorigoni:2022npe} where the instances of $c^{\rm sv}$ at the maximal value $j_i = k_i{-}2$ in each column were observed to line up with 
\beq
 \ccsv{k_1{-}2 &k_{2}{-}2 &\ldots &k_\ell{-}2 }{ k_1 & k_2 &\ldots &k_\ell} 
= \bigg( \prod_{i=1}^\ell \frac{1}{1{-}k_i}\bigg) \,\rho^{-1} \big( {\rm sv}(f_{k_1-1} f_{k_2-1}
\ldots f_{k_\ell-1}) \big)  +  \ldots \,.
\label{csvf.01}
\eeq
The ellipsis refers to words in the $f$-alphabet with at most $ (\ell{-}2)$ letters as exemplified in the reference, but we will additionally find periods beyond MZVs at \degree\ $k_1{+}\ldots{+}k_\ell \geq 18$ in section \ref{sec:4.2} below. The right-hand side of (\ref{csvf.01}) without the sv map also features in Saad's formula for the (motivic version of the) MMVs $\MMV{0 &0 &\ldots &0}{k_1 &k_2 &\ldots &k_\ell}$, again modulo contributions from  $\leq \ell{-}2$ letters in the $f$-alphabet \cite{Saad:2020mzv}.

Up to and including modular depth three, the terms in (\ref{csvf.01}) are readily seen to stem from the geometric parts of the zeta generators in (\ref{zinvsigma}) by extracting their contribution of lowest modular depth $\sigmaT_w^{\rm g} \rightarrow 
- \frac{1}{(w{-}1)!} \epsilon_{w+1}^{(w-1)}$.\footnote{Geometric contributions to (\ref{zinvsigma}) at higher modular depth or the commutators involving $\zetaT_w$ are incompatible with the property $j_i = k_i{-}2$ of the $c^{\rm sv}$ in (\ref{csvf.01}).}
At modular depth $\ell=1$ and $\ell=2$, the right-hand side
of (\ref{csvf.01}) in fact gives an exact expression for
\begin{align}
&\rho \big( \ccsv{k{-}2}{ k} \big)
= \frac{{\rm sv}(f_{k-1})}{1{-}k}  \, , \ \ \ \
\rho\big(\ccsv{k_1-2 &k_2-2}{ k_1 & k_2} \big) = \frac{{\rm sv}(f_{k_1-1} f_{k_2-1})}{(1{-}k_1)(1{-}k_2)}    \, .
\label{csvf.02}
\end{align}

\paragraph{Predicting MZVs $\zeta_{i_1} \zeta_{i_2}$ in $c^{\rm sv}$ at modular depth two:}
There are two contributions $\sigmaT^{\rm g}_{i_1} \sigmaT^{\rm g}_{i_2} $ and $[ \sigmaT^{\rm g}_{i_1} , \zetaT_{i_2} ]$ along with the MZVs $\zeta_{i_1} \zeta_{i_2}$ in the generating series (\ref{zinvsigma}). While the geometric terms give rise to the expressions $\sim \zeta_{k_1-1} \zeta_{k_2-1}$ for $\ccsv{k_1-2 &k_2-2}{ k_1 & k_2}$ in (\ref{csvf.02}), the commutator multiplying $\zeta_{i_1} \zeta_{i_2}$ in (\ref{zinvsigma}) introduces contributions to $\ccsv{j_1 &j_2}{ k_1 & k_2}$ at non-maximal values $j_i<k_i{-}2$. The modular-depth-two contributions $[\zetaT_w,\epsilon_k] = \frac{ \BF_{w+k-1} }{\BF_k } t^{w+1}(\ep_{w+1},\ep_{w+k-1})+ \ldots$ evaluated using~\eqref{eq:tdpq} and~\eqref{zetgen.07} give rise to the second line of
\begin{align}
\mathbb C^{\rm sv} (\epsilon_k) \, \Big|_{{\rm sv}(f_{i_1}f_{i_2})}^{ \epsilon_{k_1}^{(j_1)}  \epsilon_{k_2}^{(j_2)} } 
&= \frac{1}{(i_1{-}1)!(i_2{-}1)!}  \epsilon_{i_1+1}^{(i_1-1)} \epsilon_{i_2+1}^{(i_2-1)}
\label{seri.31} \\
&\quad 
+ \frac{ \BF_{i_1+i_2} }{ (i_1{+}i_2{-}2)! \BF_{i_1+1}}
\sum_{\ell=0}^{i_2-1}(-1)^\ell [ \epsilon_{i_2+1}^{(\ell)} , \epsilon_{i_1+i_2}^{(i_1+i_2-2-\ell)}]\,,
\notag 
\end{align}
which specialises as follows at low values of $w=3,5,\ldots$ and $k=4,6,\ldots\,$:
\begin{itemize}
\item the $\zeta_3^2$ terms $c^{\rm sv}\big[\smallmatrix j_1 &j_2 \\ 4 &6 \endsmallmatrix \big]$ at $j_1{+}j_2=4$ such as $c^{\rm sv}\big[\smallmatrix 0 &4 \\ 4 &6 \endsmallmatrix \big] =
-\frac{1}{315}  \zeta_3^2\,$;
\item the $\zeta_3 \zeta_5$ terms $c^{\rm sv}\big[\smallmatrix j_1 &j_2 \\ 4 &8 \endsmallmatrix \big]$ at $j_1{+}j_2=6$ such as $c^{\rm sv}\big[\smallmatrix 2 &4 \\ 4 &8 \endsmallmatrix \big] =
-\frac{1}{6300}  \zeta_3 \zeta_5\,$;
\item the $\zeta_3 \zeta_7$ terms $c^{\rm sv}\big[\smallmatrix j_1 &j_2 \\ 4 &10 \endsmallmatrix \big]$ at $j_1{+}j_2=8$ such as $c^{\rm sv}\big[\smallmatrix 0 &8 \\ 4 &10 \endsmallmatrix \big] =
  -\frac{ 5}{2673} \zeta_3 \zeta_7\,$;
\item the $\zeta_3 \zeta_5$ terms $c^{\rm sv}\big[\smallmatrix j_1 &j_2 \\ 6 &8 \endsmallmatrix \big]$ at $j_1{+}j_2=6$ such as $c^{\rm sv}\big[\smallmatrix 1 &5 \\ 6 &8 \endsmallmatrix \big] =
-\frac{1}{176400}  \zeta_3 \zeta_5\,$.
\end{itemize}


\paragraph{Predicting MZVs $\rho^{-1}( {\rm sv}(f_{i_1} f_{i_2} f_{i_3}))$ in $c^{\rm sv}$ at modular depth three:}

The same reasoning allows us to predict the simplest contributions  to the $c^{\rm sv}$ from irreducible single-valued MZVs of
depth greater or equal than three. The contributions from $\zetaT_w$ to the coefficients of $\rho^{-1}( {\rm sv}(f_{i_1} f_{i_2} f_{i_3}))$ in (\ref{zinvsigma}) extend (\ref{csvf.01}) to non-maximal values $j_i<k_i{-}2$. The modular-depth-two contributions $[\zetaT_w,\epsilon_k]
\rightarrow \frac{ \BF_{w+k-1} }{\BF_k } t^{w+1}(\ep_{w+1},\ep_{w+k-1})$ already suffice to derive
\begin{align}
\mathbb C^{\rm sv} (\epsilon_k) \, \Big|_{ {\rm sv}(f_{i_1}f_{i_2}f_{i_3})}^{ \epsilon_{k_1}^{(j_1)}  \epsilon_{k_2}^{(j_2)} \epsilon_{k_3}^{(j_3)} } 
&=  - \frac{1}{(i_1{-}1)!(i_2{-}1)!(i_3{-}1)!}  \epsilon_{i_1+1}^{(i_1-1)} \epsilon_{i_2+1}^{(i_2-1)} \epsilon_{i_3+1}^{(i_3-1)}
\label{seri.32} \\
&\quad 
- \frac{ \BF_{i_1+i_2} }{ (i_1{+}i_2{-}2)! \BF_{i_1+1}}
\sum_{\ell=0}^{i_2-1}(-1)^\ell [ \epsilon_{i_2+1}^{(\ell)} , \epsilon_{i_1+i_2}^{(i_1+i_2-2-\ell)}]
\frac{  \epsilon_{i_3+1}^{(i_3-1)} }{(i_3{-}1)!} \notag  \\
&\quad 
- \frac{  \BF_{i_1+i_3} }{(i_1{+}i_3{-}2)! \BF_{i_1+1}}
\sum_{\ell=0}^{i_3-1}(-1)^\ell [ \epsilon_{i_3+1}^{(\ell)} , \epsilon_{i_1+i_3}^{(i_1+i_3-2-\ell)}]
\frac{  \epsilon_{i_2+1}^{(i_2-1)} }{(i_2{-}1)!} \notag  \\
&\quad 
- \frac{  \epsilon_{i_1+1}^{(i_1-1)} }{(i_1{-}1)!}  \frac{  \BF_{i_2+i_3} }{(i_2{+}i_3{-}2)! \BF_{i_2+1}}
\sum_{\ell=0}^{i_3-1}(-1)^\ell [ \epsilon_{i_3+1}^{(\ell)} , \epsilon_{i_2+i_3}^{(i_2+i_3-2-\ell)}]
\notag \\
&\quad 
- \frac{ \BF_{i_1+i_2} }{ (i_1{+}i_2{-}2)! \BF_{i_1+1}}
\sum_{\ell=0}^{i_2-1} (-1)^\ell \big[ \zetaT_{i_3},  [ \epsilon_{i_2+1}^{(\ell)} , \epsilon_{i_1+i_2}^{(i_1+i_2-2-\ell)}] \big]\,,
\notag 
\end{align}
which for instance predicts
the exact combinations of $\zeta_3^2 \zeta_5$ and $\zeta_{3,3,5}^{\rm sv}$ 
in $c^{\rm sv}\big[\smallmatrix j_1 &j_2 &j_3 \\ 4 &6 &6 \endsmallmatrix \big]$ and $c^{\rm sv}\big[\smallmatrix j_1 &j_2 &j_3 \\ 4 &4 &8 \endsmallmatrix \big]$ with $j_1{+}j_2{+}j_3 = 8$ such as 
$c^{\rm sv}\big[\smallmatrix 0 &4 &4 \\ 4 &6 &6 \endsmallmatrix \big]=  \frac{ \zeta^{\rm sv}_{3,3,5} }{15750} + \frac{ 2\zeta_3^2 \zeta_5}{1575} - \frac{ 232 \zeta_{11}}{23625}$
(cf.\ section 4.1.4 of \cite{Dorigoni:2022npe}). Note that the last term in (\ref{seri.32}) is understood to be rewritten
in terms of $\ep_{k_i}^{(j_i)}$ by means of the Jacobi identity 
\begin{align}
    [ \zetaT_{i_3},  [ \epsilon_{i_2+1}^{(\ell)} , \epsilon_{i_1+i_2}^{(i_1+i_2-2-\ell)}] ]
= [  [ \zetaT_{i_3}, \epsilon_{i_2+1}^{(\ell)}] , \epsilon_{i_1+i_2}^{(i_1+i_2-2-\ell)}] 
+ [ \epsilon_{i_2+1}^{(\ell)} ,  [ \zetaT_{i_3},  \epsilon_{i_1+i_2}^{(i_1+i_2-2-\ell)}] ]\,,
\end{align}
followed by another import of the depth-two terms of $[\zetaT_w,\epsilon_k]
\rightarrow \frac{ \BF_{w+k-1} }{\BF_k } t^{w+1}(\ep_{w+1},\ep_{w+k-1})$.


\subsubsection{Fixing ambiguities in $\mathbb B^{\rm sv} ,\phi^{\rm sv}$ and $\mathbb J^{\rm eqv}$}
\label{sec:4zeta.6}

As detailed in sections 7 and 8 of \cite{Brown:2017qwo2} as well as section 4.1.2 of \cite{Dorigoni:2022npe}, the main ingredients $\mathbb B^{\rm sv} ,\phi^{\rm sv}$ of $\mathbb J^{\rm eqv}$
are only well-defined up to 
\beq
\mathbb B^{\rm sv}(\ep_k;\tau) \rightarrow \mathbb B^{\rm sv}(\ep_k;\tau)  \mathbbm a^{-1}(\ep_k) \, , 
\ \ \ \ 
\phi^{\rm sv}(\mathbb X) \rightarrow \mathbbm a(\ep_k) \phi^{\rm sv}(\mathbb X) \mathbbm a^{-1}(\ep_k)\,,
\label{svieis.01}
\eeq
for some $\mf{sl}_2$-invariant and group-like series $\mathbbm a(\ep_k)$ with single-valued
MZVs as coefficients in the case of iterated Eisenstein integrals~\cite{Brown:2017qwo2}. 
With the
description of $\mathbb B^{\rm sv} $ and $\phi^{\rm sv}$ via zeta generators, the
redefinition (\ref{svieis.01}) amounts to 
\beq
\mathbb M^{\rm sv}(\zetaT_i) \rightarrow \mathbbm a(\ep_k) \mathbb M^{\rm sv}(\zetaT_i) 
\, , \ \ \ \ 
\mathbb M^{\rm sv}(\sigmaT_i) \rightarrow \mathbb M^{\rm sv}(\sigmaT_i)\, ,
\label{svieis.02}
\eeq
which composes well with the reflection property (\ref{eq:Rop1}) since $R[\mathbbm a(\ep_k)]=\mathbbm a(\ep_k)$.\footnote{The reflection property $R[\mathbbm a(\ep_k)]=\mathbbm a(\ep_k)$ can be seen by expressing the admissible $\ep_k^{(j)}$ in its expansion via nested $t^{d}$-operations (\ref{eq:tdpq}) and observing that the sum of the superscripts $d$ has the right parity for eigenvalue $+1$ rather than $-1$ under $R[\cdot]$.}
These redefinitions amount to redistributing $\mf{sl}_2$-invariant nested brackets of $\ep_{k}^{(j)}$
between the two contributions $\sigmaT_w = \sigmaT_w^{\rm g} + \zetaT_w$
to the zeta generators. At modular depth three, such $\mf{sl}_2$-invariants are given (up to scalar multiples) by
\begin{align}
\!\!\!\! I_{k_1,k_2,k_3} \!&=\! \sum_{j=0}^{k_3-2} \sum_{i=0}^{d-2}\sum_{a=0}^j (-1)^{i+j} \binom{j}{a}  \frac{(k_1{-}2{-}i)! (k_2{-}d{+}i)!}{i! (d{-}2{-}i)!} 
\big[ \epsilon_{k_1}^{(i+a)}, [ \epsilon_{k_2}^{(d-2-i+j-a)} ,  \epsilon_{k_3}^{(k_3-2-j)}]\big]\, , \label{svieis.00} 
\end{align}
with $d=\tfrac12( k_1{+}k_2{-}k_3{+}2)$ and $[\ep_0, I_{k_1,k_2,k_3} ]= [\ep_0^{\vee}, I_{k_1,k_2,k_3}] = 0$. The simplest
redistribution of terms within $\sigmaT_w$ compatible
with $\mf{sl}_2$-invariance of $\zetaT_w$ is to shift $\sigmaT_7^{\rm g}  \rightarrow \sigmaT_7^{\rm g}  - q   I_{4,4,6}$ and $ \zetaT_7 \rightarrow \zetaT_7 + q I_{4,4,6}$ with $q\in \mathbb Q$, leaving the total zeta generator $\sigmaT_7$ unaffected. As detailed
in the companion paper \cite{Dorigoni:2023part1}, we fix a unique choice of $\zetaT_w$ by imposing the
associated geometric part $\sigmaT^{\rm g}_w$ to have no $\mf{sl}_2$-singlet
at key \degree\ $\sum_i k_i = 2w$. Different choices of the above coefficient $q$
of $I_{4,4,6}$ will affect the $\mf{sl}_2$-singlet of $\sigmaT_7^{\rm g} $
at its key degree $k_1{+}k_2{+}k_3=14$. Hence, the criterion of \cite{Dorigoni:2023part1} to
select a canonical splitting $\sigmaT_7=\sigmaT^{\rm g}_7 + \zetaT_7$
no longer allows for redistributions of $I_{4,4,6}$ between $\sigmaT^{\rm g}_7$
and $\zetaT_7$.

By the expansion (\ref{eq:sigmafromcsv}) of $\sigmaT_w$, their 
geometric parts are determined in terms of the constants $c^{\rm sv}$.
The ancillary files of \cite{Dorigoni:2022npe} provide the explicit form
of  $\ccsv{j_1 &j_2 &j_3}{4 &4 &6}$ at $j_1{+}j_2{+}j_3=4$
and $\ccsv{j_1 &j_2 &j_3}{4 &6 &6}$ at $j_1{+}j_2{+}j_3=5$
with one-parameter ambiguities $c_{446},c_{466} \in \mathbb Q$.
The first one $c_{446}$ accounts for redefinitions of $\zetaT_7$
by $I_{4,4,6}$ and is fixed to $c_{446}=0$ by the choice of \cite{Dorigoni:2023part1}.
The second parameter in the $\ccsv{j_1 &j_2 &j_3}{4 &6 &6}$ of \cite{Dorigoni:2022npe}
is already fixed to $c_{466}=0$ by the group-like form (\ref{zinvsigma}) of
the generating series $\mathbb C^{\rm sv}$.

The redefinitions (\ref{svieis.01}) of $\mathbb B^{\rm sv} $ and $\phi^{\rm sv}$
affect Brown's generating series of equivariant iterated
Eisenstein integrals via
\beq
\mathbb J^{\rm eqv}(\ep_k;\tau) \rightarrow \mathbb J^{\rm eqv}(\ep_k;\tau)  \mathbbm a^{-1}(\ep_k)\,,
\label{svieis.03}
\eeq
which shift individual modular forms in its coefficients by products of single-valued MZVs and $\beta^{\rm eqv}$ of lower modular depth.
The description of $\mathbb B^{\rm sv} $ and $\phi^{\rm sv}$ in terms
of zeta generators and the criterion of \cite{Dorigoni:2023part1} to fix their arithmetic parts $\zetaT_i$ 
selects the preferred choice of $\mathbb J^{\rm eqv}(\ep_k;\tau)$ in (\ref{finalBsv})
where the ambiguity (\ref{svieis.03}) is resolved.


\subsubsection{Single-valued iterated Eisenstein integrals and analogies with genus zero}
\label{sec:4zeta.8}

Brown defines single-valued (as opposed to equivariant) iterated Eisenstein integrals through their generating
series \cite{Brown:2017qwo2}
\beq
\mathbb J^{\rm sv}(\epsilon_k;\tau) \coloneqq \mathbb J^{\rm eqv}(\epsilon_k;\tau)  \mathbb B^{\rm sv}(\epsilon_k;\tau)^{-1} =
 \mathbb J_{+} ( \ep_k ;\tau ) \mathbb B^{\rm sv}(\epsilon_k;\tau)
 \phi^{\rm sv}\big( \widetilde{\mathbb J_{-}} ( \ep_k ;\tau )  \big)
 \mathbb B^{\rm sv}(\epsilon_k;\tau)^{-1}\,,
 \label{svieis.04}
\eeq
which is invariant under the redefinition (\ref{svieis.01}), i.e.\ well defined
even without the guiding principles for $\zetaT_i$ in section \ref{sec:4zeta.6}. 
The invariance of single-valued iterated Eisenstein integrals under  (\ref{svieis.01})
comes at the cost of their modular properties -- components of (\ref{svieis.04})
do not transform as modular forms under $\SLtwoZ$ in any $\mf{sl}_2$ frame for the $\ep_k^{(j)}$.

In view of (\ref{finalBsv}), the right-multiplication by $\mathbb B^{\rm sv}(\epsilon_k;\tau)^{-1} $ 
in (\ref{svieis.04}) completes the generating series $\mathbb M^{\rm sv}$ in single-valued
MZVs to involve the full zeta generators $\sigmaT_w$,
\beq
 R\big[ \mathbb J^{\rm sv}(\epsilon_k;\tau) \big] = \Unew(\tau) \Msv(\sigmaT_i)^{-1} \Imint(\epsilon_k;\tau)
\mathbb \Msv(\sigmaT_i)  \Ip(\epsilon_k;\tau) \Unew^{-1}(\tau)\,,
 \label{svieis.05}
\eeq
instead of the standalone appearance of the arithmetic parts
$\Msv(\zetaT_i)^{-1} $ in the analogous expression for 
$ R[ \mathbb J^{\rm eqv}(\epsilon_k;\tau) ]$ in the second line of (\ref{finalBsv}).

Converting back to holomorphic frame, (\ref{svieis.05}) is equivalent to the
following generating series of Brown's single-valued iterated Eisenstein integrals 
\beq
\mathbb I^{\rm sv}(\epsilon_k;\tau) = 
\Msv(\sigmaT_i)^{-1} \Imint(\epsilon_k;\tau)
\mathbb \Msv(\sigmaT_i)  \Ip(\epsilon_k;\tau)\,,
\label{svieis.06}
\eeq
which only differs from its equivariant analogue $\Ieqv(\epsilon_k;\tau)$
in (\ref{mainiequv}) by having $\Msv(\sigmaT_i)^{-1}$  in the place
of $\Msv(\zetaT_i)^{-1} $ as its leftmost factor on the right-hand side.
Accordingly, the antiholomorphic differential equation (\ref{simpbartau}) of $\Ieqv(\epsilon_k;\tau)$
carries over to $\mathbb I^{\rm sv}(\epsilon_k;\tau)$ with a modified conjugation
of $\mathbb A_-(\ep_k;\tau)$,
\begin{align}
\, \partial_\tau \mathbb I^{\rm sv}(\ep_k;\tau)\, \dd \tau  &= - \mathbb I^{\rm sv}(\ep_k;\tau)  \mathbb A_+(\ep_k;\tau)\, ,
\label{svieis.07} \\
\partial_{\bar \tau} \mathbb I^{\rm sv}(\ep_k;\tau)\, \dd \bar \tau   &=
- \Msv(\sigmaT_i)^{-1}  \mathbb A_-(\ep_k;\tau) \Msv(\sigmaT_i) \mathbb I^{\rm sv}(\ep_k;\tau)\, .
\notag
\end{align} 
While the coefficients of $\mathbb J^{\rm eqv}(\epsilon_k;\tau) $ are denoted by $\beta^{\rm eqv}$ in (\ref{select.1}), the coefficients of $\mathbb J^{\rm sv}(\epsilon_k;\tau) $ or $\mathbb I^{\rm sv}(\epsilon_k;\tau)$ do \textit{not} obey any simple relation to the building blocks $\beta^{\rm sv}$ for MGFs in (\ref{dictsveqv}). The arrangement of zeta generators in (\ref{svieis.05}) gives rise to non-trivial $T$ transformations at modular depth greater than or equal to two which is incompatible with the $T$-invariance of general $\beta^{\rm sv}$, also see section 4.2.3 of \cite{Dorigoni:2022npe}.
 
As a particular virtue of the representation (\ref{svieis.06}) of
single-valued iterated Eisenstein integrals, it manifests the
close analogy with the construction of single-valued polylogarithms
at genus zero \cite{svpolylog}. By the description of $\mathbb B^{\rm sv} $ and 
$\phi^{\rm sv}$ in terms of zeta generators, (\ref{svieis.06})
takes the same form as the generating series \cite{Frost:2023stm}
\beq
 \mathbb G^{\rm sv}_{\{0,1\}}(x_i;z) =
\Msv(M_i)^{-1} 
 \overline{ \widetilde{\mathbb G}_{\{0,1\}}(x_i;z) }
\Msv(M_i)
  \mathbb G_{\{0,1\}}(x_i;z)\,,
\label{svieis.08}
\eeq 
of single-valued multiple polylogarithms in one variable: The rightmost
series generates meromorphic multiple polylogarithms
\beq
 \mathbb G_{\{0,1\}}(x_i;z) = \sum_{w=0}^\infty \sum_{a_1,\ldots, a_w   = 0,1}
 G(a_1,a_2,\ldots,a_w;z) x_{a_w} \ldots x_{a_2} x_{a_1}\,,
 \label{svieis.09} 
 \eeq
which are recursively defined by ($a_i,z \in \mathbb C$)
\beq
G(a_1,a_2,\ldots,a_w;z) = \int^z_0 \frac{ \dd t}{t{-}a_1} \,G(a_2,\ldots,a_w;t)\,,
\label{svieis.10} 
\eeq
with $G(\emptyset;z)=1$ and shuffle regularisation based on $G(0;z) = \log(z)$. The non-commutative $x_0, x_1$ generate the Lie algebra of the fundamental group of $\mathbb{P}^1\setminus\{0,1,\infty\}$, i.e. the sphere minus three points, and appear as the braid operators in the one-variable version of the
Knizhnik--Zamolodchikov equation
\beq
\partial_z  \mathbb G_{\{0,1\}}(x_i;z)
=  \mathbb G_{\{0,1\}}(x_i;z) \bigg( \frac{ x_0 }{z} + \frac{ x_1 }{z{-}1} \bigg) \,,
\label{svieis.11} 
\eeq
which can be viewed as the genus-zero analogue of the 
Knizhnik--Zamolodchikov--Bernard equation (\ref{diffeq.01})
of the series $\mathbb I_+(\epsilon_k;\tau)$.
Just like the generating series $\Imint(\epsilon_k;\tau)$
of complex-conjugate iterated Eisenstein integrals in (\ref{svieis.06}),
the antimeromorphic polylogarithms in its genus-zero counterpart (\ref{svieis.08})
enter the complex-conjugate series $ \widetilde{\mathbb G}_{\{0,1\}}(x_i;z)$ with a reversed
integration order $ \overline{G(a_1,\ldots,a_w;z)} \rightarrow  \overline{G(a_w,\ldots,a_1;z)}$.

Finally, the series (\ref{svieis.08}) in single-valued polylogarithms 
introduces single-valued MZVs through the same type of series $\mathbb M^{\rm sv}$
in zeta generators encountered in our genus-one setting. The argument
of $\mathbb M^{\rm sv}(M_i)$ instructs us to replace the $\zetaT_i$ in
(\ref{Msvz}) by the genus-zero incarnation $M_i$ of zeta generators that act as
derivations on the braid operators $x_0,x_1$ \cite{Ihara:stable, Furusho2000TheMZ},
e.g.\ $[x_0,M_w]=0$ and $[x_1,M_3] = [[[x_0,x_1],x_0{+}x_1],x_1]$. Hence,
in the same way as the zeta generators $\sigma_w$ (or their arithmetic parts $z_w$) normalise the $\ep_k^{(j)}$, the zeta generators $M_w$ at genus zero normalise the free algebra ${\rm Lie}[x_0,x_1]$. Accordingly,
the stability of the alphabet $\ep_k^{(j)}$ of the generating series $\tilde {\mathbb I}_-(\epsilon_k;\tau)$ under conjugation by $\Msv(\sigmaT_i)$ or $\Msv(\zetaT_i)$ is a genus-one analogue of the fact that $\Msv(M_i)^{-1}  \overline{ \widetilde{\mathbb G}_{\{0,1\}}(x_i;z) }
\Msv(M_i)$ is expressible in terms of $x_0,x_1$.

The original construction of single-valued polylogarithms \cite{svpolylog}
introduces single-valued MZVs through the Drinfeld 
associator, a series in $(x_0,x_1)$. The reformulation in (\ref{svieis.08}) 
relies on the more abstract generating series $\Msv(M_i)$ in zeta 
generators \cite{Frost:2023stm} which no longer exposes the geometric origin 
of its MZVs from the monodromies of the Knizhnik--Zamolodchikov 
equation (\ref{svieis.11}). Instead, the motivation to employ a Lie-algebraic 
reformulation of the single-valued map at genus zero stems from the following features:

First, all known $\mathbb Q$-relations 
among single-valued MZVs are automatically incorporated by the use of the 
$f$-alphabet in (\ref{Msvz}) and $\mathbb M^{\rm sv}(M_i)$. Second, the coefficients of the MZVs $\rho^{-1} 
( {\rm sv}(f_{i_1} f_{i_2} \ldots f_{i_\ell}) ) $ beyond depth one 
(i.e.\ for $\ell \geq 2$) are deduced from those of the primitives $\rho^{-1}(f_{w})$
as discussed in section \ref{sec:4zeta.1}. Third and most
importantly, upon comparison of (\ref{svieis.06}) with (\ref{svieis.08}), 
the use of zeta generators brings the explicit realisation of the 
single-valued map into the same form at genus zero and genus one.
Single-valued polylogarithms at genus zero in (\ref{svieis.06}) are constructed
from the same type of conjugation by $\Msv$ of the antiholomorphic 
contributions as seen in (\ref{svieis.08}) for Brown's single-valued 
iterated Eisenstein integrals at genus one. The dictionary between the
respective ingredients at genus zero and genus one is summarised in table \ref{dict01}.

\begin{table}
\centering
\begin{tabular}{c||c}
genus zero&genus one
\\\hline\hline
$ \displaystyle \Bigg. \frac{\dd z}{z{-}a}\, , \ \ \ a \in \{0,1 \}$ 
&$\ \ \  \tau^j {\rm G}_k(\tau)\, \dd \tau \,, \ \ \ k\geq 4 \, , \ \ \ 0 \leq j \leq k{-}2 \ \ \ $
\\\hline
$\bigg.  \mathbb G_{\{0,1\}} \mbox{ series in } x_0,x_1$ normalised by $M_{w}\Big.$ 
&$\Big.  \mathbb I_{\pm} \mbox{ series in } \ep_{k\geq 4}^{(j)}$ normalised by $\sigma_{w}\Big.$
\\\hline
$\bigg. \ \ \ \mathbb G^{\rm sv}_{\{0,1\}}=
\Msv(M_i)^{-1} \, 
 \overline{ \widetilde{\mathbb G}_{\{0,1\}}} \,
\Msv(M_i) \,
  \mathbb G_{\{0,1\}} \ \ \ $
  &$\mathbb I^{\rm sv}  = \Msv(\sigmaT_i)^{-1} \, \Imint  \,
\mathbb \Msv(\sigmaT_i) \, \Ip $
\end{tabular}
\caption{\label{dict01}\textit{Dictionary between the constructions of single-valued polylogarithms in one variable (genus zero) and single-valued iterated Eisenstein integrals (genus one).}}
\end{table}

More specifically, the relations among the $\ep_k$ imply that the construction
applies to those iterated Eisenstein integrals in $\mathbb I_+(\ep_k;\tau)$ 
which admit a representation
via configuration space integrals, so-called elliptic MZVs \cite{Broedel:2015hia, LMSonEMZV}.
In order to relate MGFs in the expansion of $I^{\rm eqv}(\ep_k;\tau)$ 
to single-valued elliptic MZVs, it remains to adjust the left-multiplicative
series in zeta generators. The single-valued map 
$\mathbb I_+(\ep_k;\tau) \rightarrow \mathbb I^{\rm sv}(\ep_k;\tau)$ 
to Brown's single-valued iterated Eisenstein integrals differs from the generating series $\mathbb I^{\rm eqv}(\ep_k;\tau)$ of MGFs by a product of two different
series $\Msv$, i.e.
\beq
\mathbb I^{\rm sv}(\ep_k;\tau) =
\Msv(\sigmaT_i)^{-1}
\Msv(\zetaT_i) \mathbb I^{\rm eqv}(\ep_k;\tau) \, .\ \ \ \
\eeq


\subsection{Reinstating iterated integrals of holomorphic cusp forms}
\label{sec:4.2}

The previous section \ref{sec:4zeta} was dedicated to the generating
series of MGFs obtained from the specialisation $\eee_k \rightarrow \ep_k$
in the generating series $ \Jeqv ( \eee_k;\tau) $ or $ \Ieqv ( \eee_k;\tau) $ of 
more general non-holomorphic modular forms. In particular, the specialisation
$\eee_k \rightarrow \ep_k$ is essential for deriving the analogy with genus zero as discussed in section \ref{sec:4zeta.8} where we showed how the constructions of 
single-valued polylogarithms and single-valued iterated Eisenstein integrals rest on the same type of conjugation by series in single-valued MZVs
and zeta generators.

The goal of the present section is to describe new features of the generating series
$ \Jeqv ( \eee_k;\tau) ,  \Ieqv ( \eee_k;\tau) $ in the free algebra of
$\eee_k^{(j)}$. The Pollack combinations $P_w^d$ of $\eee_k^{(j)}$ introduced in section 
\ref{sec:rev.2} govern the appearance of holomorphic cusp forms
and more general classes of periods beyond MZVs. We shall use these $P_w^d$ to construct the 
letters $\eee_{\Delta_k^{\pm}}$
accompanying the cuspidal integration kernels in the one-forms $\mathbb A_{\pm}$
of (\ref{mt.06}) and (\ref{mt.07}) that vanish under $\eee_k \rightarrow \ep_k$.
Both $\eee_{\Delta_k^{\pm}}$ and the constants 
$c^{\rm sv}$ in the expansion (\ref{zinvsigma}) of the series $ \CSV $ in general 
involve periods beyond MZVs and L-values which drop out upon specialising to
$\eee_k \rightarrow \ep_k$ and therefore do not appear in the MGF world.

Based on a comprehensive study of non-holomorphic modular forms 
accompanying all of $\eee_{k_1}^{(j_1)} \eee_{k_2}^{(j_2)} \eee_{k_3}^{(j_3)}$
at \degree\ $k_1{+}k_2{+}k_3\leq 20$, we here encounter the following new phenomena
at modular depth three:
\begin{itemize}
\item in section \ref{sec:hybint}: 
(anti-)holomorphic double integrals involving an Eisenstein series and a holomorphic cusp form;
\item in section \ref{sec:noedelta}: new periods beyond ratios of 
L-values accompanying depth-one integrals of holomorphic cusp forms;
\item in section \ref{sec:sivarpi}: new periods beyond MZVs and L-values 
entering $c^{\rm sv}$ and hence $ \CSV $.
\end{itemize}
All of these phenomena are inferred from the modular completion of
triple Eisenstein integrals. The Pollack combinations of $\eee_k^{(j)}$
in their generating-series description are determined by imposing
modularity (\ref{modbeqv}) of all the coefficients $\beqv{j_1 &j_2 &j_3}{k_1 &k_2 &k_3}$
in $\Jeqv ( \eee_k;\tau)$ at \degree\ $k_1{+}k_2{+}k_3\leq 20$. Finally, section
\ref{sec:ngeozvarpi} provides a qualitative discussion of yet another new feature at
modular depth four whose explicit study at the level of non-holomorphic modular
form is left for the future.

\subsubsection{Double integrals of Eisenstein series and cusp forms}
\label{sec:hybint}

A decomposition of $\Jeqv ( \eee_k;\tau)$ into (anti-)meromorphic iterated
integrals of modular forms, a change of alphabet $\hat \psi^{\rm sv}$ and a series $ \CSV$ in constants were already given in
(\ref{eq:Iup}) and (\ref{eq:JtoII}). We now focus on the letters $\eee_{\Delta^{\pm}}$ 
accompanying the (anti-)holomorphic cusp forms in the one-forms (\ref{mt.06}) and (\ref{mt.07}) for the
(anti-)meromorphic series
$\mathbb I_{\pm}$ and their images $\mathbb J_{\pm} $ in the modular frame in (\ref{ItoJ}). 

At modular depth two, the admixtures of cuspidal integrals $\bpmno{j }{\Delta_{2s}}$ in (\ref{cuspat3.02}) entering the coefficients $\beqv{j_1 &j_2}{k_1 &k_2}$ of $\mathbb J^{\rm eqv}$
are known from \cite{Dorigoni:2022npe}, see the ancillary files of the reference
for all examples up to and including $k_1{+}k_2=24$. These admixtures
are informed by the modular invariants ${\rm F}^{\pm (s)}_{m,k}$ mentioned in 
section \ref{sec:rev.4} and discussed in more detail in section~\ref{sec:7}. They  uniquely specify the appearance of
$\bpmno{j }{\Delta_{2s}}$ in the coefficients of $\eee_{k_1}^{(j_1)}\eee_{k_2}^{(j_2)}$
of $\mathbb J_+$ and $ \widetilde{\mathbb J}_-$ at modular depth two.
Since the ${\rm F}^{\pm (s)}_{m,k}$ are respectively even and odd under
$\tau \rightarrow -\bar \tau$, one encounters both relative signs between
$\bplusno{j }{\Delta_{2s}} \pm \bminusno{j }{\Delta_{2s}}$ in $\mathbb J^{\rm eqv}$
which necessitates two different letters
\beq
\eee_{\Delta_{2s}^+}  = \eee_{\Delta_{2s}^{\rm ev}}   - \eee_{\Delta_{2s}^{\rm odd}}\, , \ \ \ \ \ \
\eee_{\Delta_{2s}^-}   = \eee_{\Delta_{2s}^{\rm ev}}  + \eee_{\Delta_{2s}^{\rm odd}} \,,
 \label{edelta.00}
\eeq
in $\mathbb J_+$ and $ \widetilde{\mathbb J}_-$.
This is to be contrasted
with non-holomorphic Eisenstein series in (\ref{eq:Cab}) and (\ref{eq:beqv1}) where
$\bplusno{j }{k }$ and $\bminusno{j}{k}$ only occur with a relative
plus sign at modular depth one and therefore descend from the
same letter $\eee_k$ in both of $\mathbb J_+$ and $ \widetilde{\mathbb J}_-$.

As indicated by the superscripts in (\ref{edelta.00}), the
contributions $\eee_{k_1}^{(j_1)}\eee_{k_2}^{(j_2)}$ of modular depth two to
$ \eee_{\Delta_{2s}^{\rm ev}} $ ($ \eee_{\Delta_{2s}^{\rm odd}}$) 
are engineered to reproduce the cusp-form contributions to the 
even functions ${\rm F}^{+(s)}_{m,k}$ (odd functions ${\rm F}^{-(s)}_{m,k}$).
In both cases, the modular weight of the holomorphic cusp form $\Delta_{2s}$ is
twice the Laplace eigenvalue of the underlying ${\rm F}^{\pm(s)}_{m,k}$,
and the \degree\ of the associated contribution to $\eee_{k_1}^{(j_1)}\eee_{k_2}^{(j_2)}$ is
given by $k_1{+}k_2= 2m{+}2k$.

The simplest two examples ${\rm F}^{-(6)}_{2,5}$ and ${\rm F}^{+(6)}_{2,6}$
of the modular invariants with cuspidal contributions \cite{Dorigoni:2021ngn} for instance 
imply the following cuspidal parts (denoted by $\big|_{\Delta_{2s}}$) \cite{Dorigoni:2022npe}:
\begin{align}
\beqv{0 &5}{4 &10}\,  \big|_{\Delta_{2s}}&= 
\beqv{1 &4}{4 &10}\,  \big|_{\Delta_{2s}}=
\beqv{2 &3}{4 &10}\,  \big|_{\Delta_{2s}}=\frac{ \Lambda(\Delta_{12},12)  \big(\, \bminusno{5 }{\Delta_{12}} - \bplusno{5 }{\Delta_{12}} \big)  }{  122472000 \, \Lambda(\Delta_{12},10) }\, ,
\notag \\
 \beqv{0 &6}{4 &12}\,  \big|_{\Delta_{2s}}&= -  \beqv{2 &4}{4 &12}\,  \big|_{\Delta_{2s}}
 =  \frac{ \Lambda(\Delta_{12},13)  \big(\, \bplusno{5 }{\Delta_{12}} + \bminusno{5 }{\Delta_{12}}  \big)  }{ 5746356000\, \Lambda(\Delta_{12},11) } \, ,\notag \\
  \beqv{1 &5}{4 &12}\,  \big|_{\Delta_{2s}} &= 0\, ,
   \label{edelta.ex}
 \end{align}
with similar results on $\beqv{j_1 &j_2}{6 &8}\,  \big|_{\Delta_{2s}}$ from ${\rm F}^{-(6)}_{3,4}$ and
$\beqv{j_1 &j_2}{6 &10}\,  \big|_{\Delta_{2s}},\,  \beqv{j_1 &j_2}{8 &8}\,  \big|_{\Delta_{2s}}$ from ${\rm F}^{+(6)}_{3,5},{\rm F}^{+(6)}_{4,4}$.
This needs to be matched with the
contributions $\eeno{j}{\Delta_{12}} \eee_{\Delta_{12}^+}^{(j)}$ to (\ref{MMVsec.06})
and $\overline{\eeno{j}{\Delta_{12}} }\eee_{\Delta_{12}^-}^{(j)}$ to (\ref{minusees}) translated into the modular-frame series
$\mathbb J_{\pm}(\eee_k;\tau)$ via (\ref{ItoJ}). 

Since MGFs obtained from the specialisation $\eee_k \rightarrow \ep_k$ are free of
cusp forms, the contributions to $ \eee_{\Delta_{12}^{\rm odd}} $ and 
$ \eee_{\Delta_{12}^{\rm even}}$ encoded in (\ref{edelta.ex}) have to be proportional
to the simplest Pollack combinations $P_w^d$ of section \ref{sec:rev.2},
\beq
 \eee_{\Delta_{12}^{\rm odd}} = \frac{\Lambda(\Delta_{12}, 12) }{\Lambda(\Delta_{12}, 10)} \, P_{14}^2 + \ldots\, ,  \ \ \ \
 \eee_{\Delta_{12}^{\rm ev}} = \frac{\Lambda(\Delta_{12}, 13) }{\Lambda(\Delta_{12}, 11)} \, P_{16}^3 + \ldots \, ,
 \label{edelta.01}
 \eeq
with an infinite series of nested brackets of $\eee_k^{(j)}$ of \degree\ $\geq 18$ and modular depth $\geq 3$ in the ellipsis. 
Here and below, we fix normalisation conventions for $P_w^d$ such
that they enter $ \eee_{\Delta_{2s}^{\rm odd}} $ and 
$ \eee_{\Delta_{2s}^{\rm even}}$ in (\ref{edelta.01}) with unit coefficients besides the
ratios of L-values (\ref{defLfct}). This is always possible since
the defining property $P_w^d(\ep_k) = 0$ holds in any normalisation, and
each $P_w^d$ is uniquely associated with cusp forms $\Delta_{w-2d+2}$ 
and \degree\ $w$. The large integers in the denominators of (\ref{edelta.ex}) 
are then responsible for the similarly unwieldy normalisation factors in the expressions
(\ref{P182}) and (\ref{P203}) for $P_{14}^2$ and $P_{16}^3$. 

In fact, inspection of the cuspidal contributions
to $\beqv{0 &5}{4 &10}$ and $\beqv{0 &6}{4 &12}$ is already enough to
determine these normalisations. The appearance of the cuspidal $\bpmno{5 }{\Delta_{12}} $ in
$\beqv{1 &4}{4 &10}$, $\beqv{2 &3}{4 &10}$ and $\beqv{1 &5}{4 &12}$, $\beqv{2 &4}{4 &12}$
in (\ref{edelta.ex}) is interlocked with that in $\beqv{0 &5}{4 &10}$ and $\beqv{0 &6}{4 &12}$
by the rewriting of ${\rm ad}_{\eee_0}^j( P_{14}^2 )$ and ${\rm ad}_{\eee_0}^j( P_{16}^3)$ in the 
cuspidal sector of $\mathbb J_{\pm}$ in terms of $\eee_{k_1}^{(j_1)}  \eee_{k_2}^{(j_2)}$.
Similarly, the cuspidal contributions to $\beqv{j_1 &j_2}{6 &8}$,
$\beqv{j_1 &j_2}{6 &10}$ and $\beqv{j_1 &j_2}{8 &8}$ are completely
fixed by the results on $\beqv{j_1 &j_2}{4 &10}$ and
$\beqv{j_1 &j_2}{4 &12}$ in (\ref{edelta.ex}) since they follow from the known 
relative factors of $t^2_{4,10}, t^2_{6,8}$ 
in $P_{14}^2$ and $t_{4,12}^3, t_{6,10}^3, t_{8,8}^3$ in $P_{16}^3$ (see (\ref{P182}) 
and (\ref{P203})) which are in fact determined by the period 
polynomial of $\Delta_{12}$ \cite{Pollack}.

By the same logic, the $\beqv{j_1 &j_2}{k_1 &k_2}$ informed by 
${\rm F}^{\pm (s)}_{m,k}$ at higher Laplace eigenvalues and higher \degree s $k_1{+}k_2 = 2m{+}2k$ determine the ratios of L-values in
	\begin{align}
		\eee_{\Delta_{12}^{\rm odd}} &= \frac{\Lambda(\Delta_{12}, 12) }{\Lambda(\Delta_{12}, 10)} \,
		P_{14}^2 +  \frac{\Lambda(\Delta_{12}, 14) }{\Lambda(\Delta_{12}, 10)} \, P_{18}^4+\ldots\, ,
		&\eee_{\Delta_{16}^{\rm odd}} &=  \frac{\Lambda(\Delta_{16}, 16)}{ \Lambda(\Delta_{16}, 14)} \,P_{18}^2+\ldots \, ,  \label{pre.edex} \\
		\eee_{\Delta_{12}^{\rm ev}} &= \frac{\Lambda(\Delta_{12}, 13)}{\Lambda(\Delta_{12}, 11)} \,  P_{16}^3 +\frac{\Lambda(\Delta_{12}, 15)}{\Lambda(\Delta_{12}, 11)} \, P_{20}^5 +\ldots\, ,
		&\eee_{\Delta_{16}^{\rm ev}} &= \frac{\Lambda(\Delta_{16}, 17)}{ \Lambda(\Delta_{16}, 15)} \, P_{20}^3+\ldots  \, ,
		\notag \\
		&&\eee_{\Delta_{18}^{\rm odd}} &= \frac{\Lambda(\Delta_{18}, 18)}{\Lambda(\Delta_{18}, 16)} \,P_{20}^2
		+\ldots \, ,\notag
	\end{align}
with \degree\ $\geq 18$ and modular depth $\geq 3$ in the ellipsis.
Moreover, the detailed prefactors of the $\bpmno{j }{\Delta_{2s}}$ in 
these $\beqv{j_1 &j_2}{k_1 &k_2}$ fix the normalisations of the
$P_{w}^d$ in (\ref{P182}) to (\ref{Patd5}) that are compatible with
unit coefficients in (\ref{pre.edex}). 

By the construction of the generating series $\mathbb J^{\rm eqv}$ from
path-ordered exponentials in (\ref{eq:Iup}) and (\ref{eq:JtoII}), our systematic handle on the 
modular-depth-two contributions to $ \eee_{\Delta_{2s}^{\pm}}$ also
has implications for $\beta^{\rm eqv}$ at modular depth three. More 
specifically, expressions like (\ref{pre.edex}) at modular depth two
translate into modular-depth-three contributions
$\eeno{j_1 & j_2}{k_1 & \Delta_{k_2}} \eee_{k_1}^{(j_1)}\eee_{\Delta_{k_2}^+}^{(j_2)} $
and $\eeno{j_1 & j_2}{\Delta_{k_1} & k_2} \eee_{\Delta_{k_1}^+}^{(j_1)}\eee_{k_2}^{(j_2)} $ to (\ref{MMVsec.06}) as well as
$\overline{\eeno{j_2 & j_1}{\Delta_{k_2} & k_1}} \eee_{k_1}^{(j_1)}\eee_{\Delta_{k_2}^-}^{(j_2)} $ and $\overline{\eeno{j_2 & j_1}{k_2 & \Delta_{k_1}}} \eee_{\Delta_{k_1}^-}^{(j_1)}\eee_{k_2}^{(j_2)}$ to (\ref{minusees}). These terms inject double integrals involving one holomorphic 
Eisenstein series and one cusp forms into various $\beqv{j_1 &j_2 &j_3}{k_1 &k_2 &k_3}$.
Similarly, contributions like (\ref{pre.edex}) to $ \eee_{\Delta_{2s}^{\pm}}$
determine the products of holomorphic and antiholormophic integrals of modular depth one 
involving one Eisenstein series and one cusp form that enter generic $\beqv{j_1 &j_2 &j_3}{k_1 &k_2 &k_3}$. These mechanisms lead to terms of the schematic form
\begin{align}
\beqv{j_1 &j_2 &j_3}{k_1 &k_2 &k_3} &= \ldots + \bigg\{  \frac{\Lambda(\Delta_{2s}, 2s {+} 2n)}{\Lambda(\Delta_{2s}, 2s{-}2)}   \ {\rm or} \  \frac{\Lambda(\Delta_{2s}, 2s {+} 2n{+}1)}{\Lambda(\Delta_{2s}, 2s{-}1)}  \bigg\} \label{schem:d3} \\
&\quad  \times \Big\{ \bpmno{ \ell_1 &\ell_2 }{k &\Delta_{2s}}
\ {\rm or} \  \bpmno{ \ell_1 &\ell_2 }{\Delta_{2s} &k} 
\ {\rm or} \ \bplusno{ \ell_1}{k } \bminusno{ \ell_2}{ \Delta_{2s} }
\ {\rm or} \ \bminusno{ \ell_1}{k } \bplusno{ \ell_2}{ \Delta_{2s} }
\Big\}\,,
\notag
\end{align}
with $n\in \mathbb N_0$ and $\ell_1,\ell_2$ distinct from $j_1,j_2,j_3$. One
can see from the differential equations in $\tau$ that the contributions to
$\beqv{j_1 &j_2 &j_3}{k_1 &k_2 &k_3}$ in (\ref{schem:d3}) are necessary
to attain the desired modular properties (\ref{modbeqv}). For instance, 
translating (\ref{diffeq.04}) into components identifies contributions
\beq
2\pi i(\tau{-}\bar \tau)^2 \partial_\tau  \beqv{j_1 &j_2 &j_3}{k_1 &k_2 &k_3}= 
  - \delta_{j_3,k_3-2} (\tau{-}\bar \tau)^{k_3} \GG_{k_3}(\tau) 
 \beqv{j_1 &j_2}{k_1 &k_2}+\ldots \, ,
\eeq
where the modular completions $\bpmno{j}{ \Delta_{2s} }$ of the double Eisenstein integrals present in the $\beqv{j_1 &j_2}{k_1 &k_2}$
on the right-hand side, can only come from terms of the form $\bplusno{j &k_3{-}2}{\Delta_{2s} &k_3 }$ and $\bplusno{k_3{-}2}{k_3 } \bminusno{j }{\Delta_{2s}}$  in the $\beqv{j_1 &j_2 &j_3}{k_1 &k_2 &k_3}$ on the left-hand side.

Since the expansion of $ \eee_{\Delta_{2s}^{\pm}}$ in terms 
of brackets of $\eee_k^{(j)}$
starts at \degree\ $2s{+}2$ (i.e.\ at degrees $\geq 14$), the mixing (\ref{schem:d3}) of Eisenstein and cuspidal integrals in
$\beqv{j_1 &j_2 &j_3}{k_1 &k_2 &k_3}$ begins at \degree\ $k_1{+}k_2{+}k_3=18$, for instance
\begin{align}
\beqv{0 &0 &0}{4 &4 &10}\,  \big|_{\Delta_{2s}} &= \frac{\Lambda(\Delta_{12},12)}{122472000\Lambda(\Delta_{12},10)} \bigg(   \bminusno{0 & 0}{4& \Delta_{12}}  
 -  \bplusno{0 & 0}{4& \Delta_{12}}  \label{ex4410} \\*
 &\quad \quad\quad \quad\quad  \quad\quad  
 -  \bminusno{0}{4} \bplusno{0}{\Delta_{12}} 
  + \frac{\zeta_3}{24y^2} \bplusno{0}{\Delta_{12}}\bigg) \, . \notag
  \end{align}

\subsubsection{New periods and depth-one integrals of cusp form}
\label{sec:noedelta}

Apart from the contributions (\ref{schem:d3}) mixing holomorphic 
Eisenstein series and cusp forms, generic $\beqv{j_1 &j_2 &j_3}{k_1 &k_2 &k_3}$
will involve additional terms $\bpmno{j}{ \Delta_{2s} }$ without any admixtures
of Eisenstein series. A simple class of cuspidal contributions with coefficients $\sim \zeta_{k-1} 
\frac{\Lambda(\Delta_{2s}, 2s {+} 2n)}{\Lambda(\Delta_{2s}, 2s{-}2)} $
or $\sim \zeta_{k-1} \frac{\Lambda(\Delta_{2s}, 2s {+} 2n{+}1)}{\Lambda(\Delta_{2s}, 2s{-}1)}$
exemplified by the last term of (\ref{ex4410})
arises from combining the coefficients $\sim \zeta_{k-1}$ of $\eee_k^{(j)}$ in $\mathbb B^{\rm sv}$
with terms of modular depth two in the $\eee_{\Delta_{2s}^{\pm}}$
of $\mathbb J_{\pm}$.\footnote{More specifically, the relevant terms of modular depth one are given by 
\[
\mathbb B^{\rm sv}(\eee_k;\tau) = 1
-2 \sum_{k=4}^{\infty} \sum_{j=0}^{k-2} \frac{(-1)^j \zeta_{k-1} \eee_k^{(j)} }{j! (4y)^j} + \ldots \, .
\] } However, our main interest in this section is on the terms of
schematic form $[\eee_{k_1}^{(j_1)}, [\eee_{k_2}^{(j_2)},\eee_{k_3}^{(j_3)}]]$ 
entering the letters $\eee_{\Delta_{2s}^{\pm}}$ whose contributions to
$\beqv{j_1 &j_2 &j_3}{k_1 &k_2 &k_3}$ involve neither an Eisenstein series nor any MZVs.

Contributions to $\eee_{\Delta_{2s}^{\pm}}$ of modular depth three can be anticipated
from the fact that generic Pollack combinations $P_w^d$ at $d\geq 3$ mix
nested brackets of $\eee_k^{(j)}$ of different modular depth $2,3\ldots,d$. In the simplest
case of $P_{16}^3$, terms of modular depth two such as $t_{4,12}^3, t_{6,10}^3, t_{8,8}^3$
coexist with $ [ \eee_4, [ \eee_4, \eee_8 ] ]$ and $ [ \eee_6, [ \eee_6, \eee_4 ] ]$
of modular depth three, see (\ref{P203}) for their coefficients. The contributions
$ [ \eee_4, [ \eee_4, \eee_8 ] ]$ and $ [ \eee_6, [ \eee_6, \eee_4 ] ]$ to $P_{16}^3$
are the unique modular-depth-three completion of $t_{4,12}^3, t_{6,10}^3, t_{8,8}^3$
to attain a vanishing combination in Tsunogai's derivation algebra.
In this way, the expansion of (\ref{eq:JtoII}) together with (\ref{edelta.01}) predicts
contributions $\sim  \bplusno{j}{ \Delta_{12} } + \bminusno{j}{ \Delta_{12} }$
to various $\beqv{j_1 &j_2 &j_3}{4 &6 &6}$ and $\beqv{j_1 &j_2 &j_3}{4 &4&8}$ 
with $j_1{+}j_2{+}j_3=j$, for instance
\begin{align}
\beqv{0 &1 &2}{4 &4 &8} \,  \big|_{\Delta_{2s}} &= - \frac{ 11 \Lambda(\Delta_{12}, 13)}{ 25075008000 \Lambda(\Delta_{12}, 11)}\, \Big( \, \bplusno{3}{ \Delta_{12} } + \bminusno{3}{ \Delta_{12} } \Big)
 \, ,
\\
\beqv{2 &3 &2}{4 &6 &6} \,  \big|_{\Delta_{2s}} &= -
\frac{ 23 \Lambda(\Delta_{12}, 13)}{ 16716672000 \Lambda(\Delta_{12}, 11)}
\, \Big( \, \bplusno{7}{ \Delta_{12} } + \bminusno{7}{ \Delta_{12} } \Big)
\, . \notag
\end{align}
However, the analogous completion of generic $P_w^d$ at $d\geq 3$ and $w \geq 18$
is ambiguous. Only the terms $t^d_{p,q}$ of modular depth two 
and \degree\ $p{+}q=w$ are canonical and determined by period
polynomials of $\Delta_{2s}$. The additional nested brackets
at modular depth $\geq 3$ such as $[\eee_{k_1}^{(j_1)}, [\eee_{k_2}^{(j_2)},\eee_{k_3}^{(j_3)}]]$  
needed  to attain $P_w^d \rightarrow 0$ as $ \eee_{k}^{(j)} \rightarrow  \ep_{k}^{(j)}$ can typically be shifted by
brackets of $\eee_k^{(j)}$ with simpler Pollack combinations while
maintaining their defining properties. For instance, $P_{18}^4$ 
in (\ref{Patd4}) can be shifted by multiples of 
$t^3(\eee_4,P_{14}^2) \sim t^3(\eee_4,t^2(\eee_4,\eee_{10}))
- 3 t^3(\eee_4,t^2(\eee_6,\eee_{8}))$ while preserving
\begin{itemize}
\item[(i)] the lowest-weight-vector condition $[\eee_0^\vee , P_{18}^4]=0$, see (\ref{PwdHW});
\item[(ii)] \degree\ $18$; 
\item[(iii)] total number $4$ of $\eee_{k\geq 0}$.
\end{itemize}
Similar ambiguities arise for other $P_{w}^d$ in (\ref{P203}) to (\ref{Patd5}) with
one free parameter for $P_{20}^3$ and three free parameters for 
$P_{20}^5$,
\begin{align}
P_{18}^4 &\leftrightarrow t^3(\eee_4,P_{14}^2)  \, , \ \ \ \ 
P_{20}^3 \leftrightarrow t^2(\eee_6,P_{14}^2) \, , \notag \\
P_{20}^5 &\leftrightarrow t^4(\eee_6,P_{14}^2) , \, t^3(\eee_4,P_{16}^3) , \, [\zetaE_3, P_{14}^2]\, ,
\label{ambigee}
\end{align}
all of which preserve the vanishing of $P_{w}^d$ under $\eee_{k}^{(j)} \rightarrow  \ep_{k}^{(j)}$, 
the lowest-weight vector property $[\eee_0^\vee , P_{w}^d]=0$, \degree\ $w$ and the total number
$d$ of $\eee_{k\geq 0}$.

Hence, it is a priori unclear how to generalise the contribution
$\sim \frac{\Lambda(\Delta_{12}, 14) }{\Lambda(\Delta_{12}, 10)}  P_{18}^4$ 
generating the modular-depth-two terms of
$\eee_{\Delta_{12}^{\rm odd}}$ in (\ref{pre.edex}) to modular depth three.
We consider a one-parameter ansatz of possible deformations
$P_{18}^4 \rightarrow P_{18}^4 + c \,t^3(\eee_4,t^2(\eee_4,\eee_{10}))
- 3 c \,t^3(\eee_4,t^2(\eee_6,\eee_{8}))$ to $\eee_{\Delta_{12}^{\rm odd}}$
and then impose the modular properties
(\ref{modbeqv}) on the resulting $\beqv{j_1 &j_2 &j_3}{4 &4 &10}$
and $\beqv{j_1 &j_2 &j_3}{4 &6 &8}$ to determine $c \in \mathbb R$.
Remarkably, there is no rational solution for $c$. Instead,
we obtain the expression (with modular depth $\geq 4$ and \degree\ $\geq 22$ in the
ellipsis)
\begin{align}
 \eee_{\Delta_{12}^{\rm odd}} &= \frac{1}{\Lambda(\Delta_{12}, 10)} \, \Big\{\Lambda(\Delta_{12}, 12) P_{14}^2 + \Lambda(\Delta_{12}, 14)P_{18}^4+\Lambda^{3,2}_{4,14} t^3 ( \eee_4, P_{14}^2)
 \Big\}+\ldots \,, \label{prevex}
 \end{align}
involving a new period $\Lambda^{3,2}_{4,14} \in \mathbb R$ that we did not manage to express in terms
of MZVs and (critical or non-critical) L-values (\ref{defLfct}).
By numerical and analytical studies of the modularity properties
of $\beqv{j_1 &j_2 &j_3}{k_1&k_2 &k_3}$ at $k_1{+}k_2{+}k_3=18$,
the new number is available both as an approximation 
$\Lambda^{3,2}_{4,14} = 0.0001108864\dots$ to 300 digits and
as a combination of MMVs (\ref{eq:mmvdef}) and (\ref{MMVsec.17b}) in an ancillary file.

More generally, we insert free parameters along with the ambiguities
in (\ref{ambigee}) and impose the modular properties
(\ref{modbeqv}) on all $\beqv{j_1 &j_2 &j_3}{k_1 &k_2 &k_3}$
at $k_1{+}k_2{+}k_3 \leq 20$. This process gives rise to
additional new periods $\Lambda^{d_1,d_2}_{k,w} \in \mathbb R$ that cannot be
expressed in terms of MZVs or L-values and
that we label according to the accompanying $t^{d_1}(\eee_k,P_{w}^{d_2})$ in
\begin{align}
 \eee_{\Delta_{12}^{\rm odd}} &= \frac{1}{\Lambda(\Delta_{12}, 10)} \, \Big\{\Lambda(\Delta_{12}, 12) P_{14}^2 + \Lambda(\Delta_{12}, 14)P_{18}^4+\Lambda^{3,2}_{4,14} t^3 ( \eee_4, P_{14}^2)
 \Big\} +\ldots\, , \notag \\
\eee_{\Delta_{12}^{\rm ev}} &= \frac{1}{\Lambda(\Delta_{12}, 11)} \, \Big\{ \Lambda(\Delta_{12}, 13) P_{16}^3 +\Lambda(\Delta_{12}, 15)P_{20}^5\notag \\
&\quad +\Lambda^{4,2}_{6,14} \big( t^4(\eee_6, P_{14}^2) -\tfrac{364}{15} [\zetaE_3,P_{14}^2]\big) + \Lambda^{3,3}_{4,16}  t^3(\eee_4, P_{16}^3) \Big\}+\ldots \, ,
\notag \\
\eee_{\Delta_{16}^{\rm odd}} &=  \frac{\Lambda(\Delta_{16}, 16)}{ \Lambda(\Delta_{16}, 14)} P_{18}^2+\ldots \, ,  \notag \\
\eee_{\Delta_{16}^{\rm ev}} &= \frac{1}{ \Lambda(\Delta_{16}, 15)}\, \Big\{ \Lambda(\Delta_{16}, 17) P_{20}^3+ \Lambda^{2,2}_{6,14} t^2 (\eee_6, P_{14}^2)\Big\}+\ldots \, , \notag \\
\eee_{\Delta_{18}^{\rm odd}} &= \frac{\Lambda(\Delta_{18}, 18)}{\Lambda(\Delta_{18}, 16)} P_{20}^2
+\ldots \, . \label{edeltaex}
 \end{align}
 The ellipsis refers to $\eee_{k}$ of \degree\ $\geq 22$, and the expansions of 
 $\eee_{\Delta_{18}^{\rm ev}} $ as well as $\eee_{\Delta_{2s}^{\rm even}},
 \eee_{\Delta_{2s}^{\rm odd}}$ with $2s\geq20$
start at \degree\ $\geq 22$. We have normalised the new periods
 $\Lambda^{d_1,d_2}_{k,w}$ such that their products with
$ t^{d_1}(\eee_k,P_{w}^{d_2})$ enter the curly brackets of 
 (\ref{edeltaex}) with unit coefficients. The coefficients of $ [\zetaE_3,P_{14}^2]$
in our ansatz for ambiguities of $P_{20}^5$ in $\eee_{\Delta_{12}^{\rm ev}} $
turned out to be a rational multiple of $\Lambda^{4,2}_{6,14} $ along with $  t^4(\eee_6, P_{14}^2)$,
hence, there was no need to extend the notation
beyond $\Lambda^{d_1,d_2}_{k,w} \leftrightarrow t^{d_1}(\eee_k,P_{w}^{d_2})$.
It would be interesting to investigate the origin of the ratio $-\tfrac{364}{15}$
in this expression for $\eee_{\Delta_{12}^{\rm ev}} $.
 
The new period  $\Lambda^{3,2}_{4,14}$ at \degree\ $k_1{+}k_2{+}k_3=18$
and the new periods $\Lambda^{4,2}_{6,14},\Lambda^{3,3}_{4,16},$ $\Lambda^{2,2}_{6,14}$
at $k_1{+}k_2{+}k_3=20$ are only well defined up to rational multiples of certain
non-critical L-values. For instance, in the combination
$ \Lambda(\Delta_{12}, 14)P_{18}^4+\Lambda^{3,2}_{4,14} t^3 ( \eee_4, P_{14}^2)$
entering $\eee_{\Delta_{12}^{\rm odd}}$ in (\ref{prevex}), redefining $P_{18}^4 \rightarrow P_{18}^4
+ c\, t^3 ( \eee_4, P_{14}^2)$ with $c \in \mathbb Q$ amounts to the redefinition
$\Lambda^{3,2}_{4,14} \rightarrow \Lambda^{3,2}_{4,14} +  c\,\Lambda(\Delta_{12}, 14)$.
Accordingly, we propose to assign the transcendental weight $n{+}1$ of the
non-critical L-value $\Lambda(\Delta_{2s},2s{+}n)$ in the potential redefinition
to each $\Lambda^{d_1,d_2}_{k,w}$. Table \ref{newLambdas} summarises the ambiguity, transcendental weight and numerical value of the new
periods encountered in (\ref{edeltaex}) up to $k_1{+}k_2{+}k_3=20$, also see an ancillary file for representations of $\Lambda^{d_1,d_2}_{k,w}$ in terms of MMVs. 


\begin{table}
\centering
\begin{tabular}{c||c|c|c}
new period&well defined up to &transcendental weight &approx. numerical value
\\\hline\hline
$\Big. \Lambda^{3,2}_{4,14} \Big.$ &$\Lambda(\Delta_{12}, 14)$ &3 &$0.0001108864\dots$
\\\hline
$\Big. \Lambda^{4,2}_{6,14} \Big.$ &$\Lambda(\Delta_{12}, 15)$ &4 &$-0.0001258966\dots$
\\\hline
$\Big. \Lambda^{3,3}_{4,16} \Big.$ &$\Lambda(\Delta_{12}, 15)$ &4 &$-0.0009720209\dots$
\\\hline
$\Big. \Lambda^{2,2}_{6,14} \Big.$ &$\Lambda(\Delta_{16}, 17)$ &2 &$0.00006694828\dots$  
\end{tabular}
\caption{\label{newLambdas}\textit{New periods $\Lambda^{d_1,d_2}_{k,w}$ encountered in the contributions to $\eee_{\Delta_{2s}^{\rm even}},
 \eee_{\Delta_{2s}^{\rm odd}}$ at modular depth three and \degree\ $k_1{+}k_2{+}k_3 = 18,20$.}}
\end{table}

\subsubsection{New periods in $c^{\rm sv}$ and Laurent polynomials}
\label{sec:sivarpi}

Upon specialising $\eee_k \rightarrow \ep_k$, the zeta generators $\sigmaT_i$ provided a unified description (\ref{zcvssigma}) of 
the ingredients $\Msv( \zetaT_i)$ and $\mathbb C^{\rm sv}(\ep_k)$
in the expression (\ref{eq:Iup}) for $\Ieqv(\ep_k;\tau)$ where all
coefficients are $\mathbb Q$-linear combinations of (single-valued) MZVs. 
Also in the free algebra of $\eee_k^{(j)}$, the MZV-contributions 
to $\ccsv{j_1 &j_2 &j_3}{k_1 &k_2 &k_3}$ at $k_1{+}k_2{+}k_3 \leq 20$
are still governed by the group-like series $\Msv( \sigmaE_i )$ in uplifted zeta generators
given by (\ref{Msvz}) with $\hat \sigma_i$ in place of $z_i$. We have performed 
comprehensive checks up to \degree\ 20 that the coefficients of single-valued MZVs $\zeta_{i_1} \zeta_{i_2}$ and
$ \rho^{-1}( {\rm sv}(f_{i_1} f_{i_2}f_{i_3} ) ) $ in $\mathbb C^{\rm sv}(\eee_k)$ follow
from the uplift $(\sigmaT_i , \zetaT_i )\rightarrow (\sigmaE_i , \zetaE_i)$ 
of (\ref{zinvsigma}). It is important to use the form of zeta generators 
in (\ref{zetgen.08}) and the data in appendix \ref{app:exd3} as well as the brackets $[\zetaE_w,\eee_k]$ of appendix \ref{app:zep3} without adding any Pollack combinations.

However, in the free algebra of $\eee_k^{(j)}$, the 
modular completions of iterated Eisenstein integrals to several $\beqv{j_1 &j_2 &j_3}{k_1 &k_2 &k_3}$
at $k_1{+}k_2{+}k_3\geq 18$ give rise to periods beyond MZVs and L-values.
A first class of new numbers $\Lambda^{d_1,d_2}_{k,w}$ entering the
letters $\eee_{\Delta^{\pm}}$ for cusp forms was discussed in the previous section.
We shall here describe a second class of new periods $\varpi^{d_1,d_2}_{k,w} \in \mathbb R$
with similar labels $d_1,d_2,k,w \in \mathbb N$ to be explained below. As a defining property
of the new periods $\varpi^{d_1,d_2}_{k,w}$, they appear in the constants
$\ccsv{j_1 &j_2 &j_3}{k_1 &k_2 &k_3}$ that enter $\beqv{j_1 &j_2 &j_3}{k_1 &k_2 &k_3}$ 
of \degree\ $k_1{+}k_2{+}k_3 \geq 18$ without any accompanying iterated integrals. 

At the level of generating series, the new periods $\varpi^{d_1,d_2}_{k,w}$ from $c^{\rm sv}$
enter $\Jeqv(\eee_k;\tau)$ on a similar footing as the MZVs in $\Msv( \sigmaE_i )$.
This leads us to associate additional generators
$\sigmaE_{\pern} = \hat \sigma^{d_1,d_2}_{k,w}$ which
aim to generalise the way that $\sigmaE_m$ track the appearance of $\zeta_m$ in
the series $\Msv( \sigmaE_i )$. 
Following this analogy, the generators
$\hat \sigma^{d_1,d_2}_{k,w}$ are taken to have arithmetic terms
$\zetaE_{\pern} = \zetaE^{d_1,d_2}_{k,w}$ similar to $\sigmaE_{i}$. 
Generators of the type $\sigmaE_{\pern}$ have also appeared in~\cite{Brown2019} as modular elements associated with primitive periods of the moduli space $\mathcal{M}_{1,1}$ beyond MZVs, starting with non-critical L-values of holomorphic cusp forms.
The modular elements of the reference have a similar decomposition into geometric and arithmetic parts.

In this setting, we propose the following uplift of (\ref{zcvssigma}) and (\ref{finalBsv})
to the free algebra of $\eee_k^{(j)}$ instead of the~$\ep_k^{(j)}$,
\begin{align}
&\hat \psi^{\rm sv}(\mathbb X) =   \mathbb M^{\rm sv}(\hat z_i,\zetaE_{\pern})^{-1} \, \mathbb X  \, \mathbb M^{\rm sv}(\hat z_i,\zetaE_{\pern})
\, , \ \ \
 \Msv( \zetaE_i, \zetaE_{\pern}) \CSV =  \Msv( \sigmaE_i,\sigmaE_{\pern})\, ,
 \label{uplift} \\
  &R\big[ \Jeqv(\eee_k;\tau) \big] = \Unew(\tau) \Msv( \zetaE_i,\zetaE_{\pern})^{-1} \Imint(\eee_k;\tau)
\mathbb \Msv(  \sigmaE_i, \sigmaE_{\pern})  \Ip(\eee_k;\tau) \Unew^{-1}(\tau)\, ,
\notag
\end{align}
see (\ref{eq:Iup}) and (\ref{defpsisv}) for the change of alphabet $\hat \psi^{\rm sv}$. Both of
\beq
\Msv( \sigmaE_i,\sigmaE_{\pern})= \Msv(  \sigmaE_i)+\ldots \, , \ \ \ \
\Msv( \zetaE_i,\zetaE_{\pern})= \Msv(  \zetaE_i)+\ldots 
\label{extendMM}
\eeq
refer to extensions of the series (\ref{Msvz}) beyond MZVs starting with pairs
$\varpi^{d_1,d_2}_{k,w} \hat \sigma^{d_1,d_2}_{k,w}$ and
$\varpi^{d_1,d_2}_{k,w} \hat z^{d_1,d_2}_{k,w}$. 
The first periods beyond MZVs in $\CSV$ and in $ \Msv( \sigmaE_i,\sigmaE_{\pern})$
are encountered at modular depth three and \degree\ $k_1{+}k_2{+}k_3 =18$ and $20$. 
After a comprehensive
study of the relevant $\beqv{j_1 &j_2 &j_3}{k_1 &k_2 &k_3}$, we find two and four
contributions at leading and subleading order to the uplift of $\Msv(  \sigmaE_i)$
in the generating series (\ref{uplift}),
\begin{align}
 \Msv(  \sigmaE_i, \sigmaE_{\pern})
 &= \Msv(  \sigmaE_i)
 + \varpi^{2,2}_{4,14} \hat  \sigma^{2,2}_{4,14} 
+ \varpi^{4,2}_{4,14} \hat  \sigma^{4,2}_{4,14}
\label{svieis.14}  \\
 &\quad
    + \varpi^{2,2}_{6,14}\hat  \sigma^{2,2}_{6,14} 
       + \varpi^{4,2}_{6,14}\hat  \sigma^{4,2}_{6,14}
 + \varpi^{6,2}_{6,14}\hat   \sigma^{6,2}_{6,14} 
  + \varpi^{3,3}_{4,16}\hat  \sigma^{3,3}_{4,16} 
     + \ldots\, .
 \notag
\end{align}
Just like the zeta generators, the new generators $\sigmaE_\varpi \rightarrow \hat  \sigma^{d_1,d_2}_{k,w}$
are expected to involve infinite series in nested brackets of $\eee_{k}^{(j)}$,
starting with the following highest-weight vectors 
identified by our computations at modular depth three,
\begin{align}
\hat   \sigma^{2,2}_{4,14} &= s^{2}(\eee_4,P^2_{14}) +  \ldots
  \, , \ \ \ \
&\hat \sigma^{4,2}_{4,14} &= s^{4}(\eee_4,P^2_{14}) + \ldots \, ,
\label{svieis.15}  \\
\hat  \sigma^{2,2}_{6,14} &= s^{2}(\eee_6,P^2_{14}) + \ldots
  \, , \ \ \ \       
&\hat \sigma^{4,2}_{6,14} &= s^{4}(\eee_6,P^2_{14}) + \ldots \, ,
\notag \\
 \hat \sigma^{6,2}_{6,14} &= s^{6}(\eee_6,P^2_{14}) + \ldots
  \, , \ \ \ \ 
 &\hat \sigma^{3,3}_{4,16} &= s^{3}(\eee_4,P^3_{16}) + \ldots  \, .
 \notag
\end{align}
In both (\ref{svieis.14}) and (\ref{svieis.15}) the ellipses refer to terms
of \degree\ $\geq 22$ or modular depth $\geq 4$. While the notation for the
new periods $\Lambda^{d_1,d_2}_{k,w}$ in the letters $\eee_{\Delta^{\pm}}$
is adapted to the associated lowest-weight vectors $t^{d_1}(\eee_k,P^{d_2}_{w})$,
the labels of the new periods $\varpi^{d_1,d_2}_{k,w}$ in this section refer
to highest-weight vectors $s^{d_1}(\eee_k,P^{d_2}_{w})$ in (\ref{svieis.15}) describing their
lowest-order occurrence in $\Msv( \sigmaE_i,\sigmaE_{\pern})$ or $\mathbb C^{\rm sv}(\eee_k)$ (see (\ref{uplift})).

An independent realisation of the six new periods
in (\ref{svieis.14}) is provided by
\begin{align}
\ccsv{2 &2 &8}{4 &4 &10} &\notag =
-\tfrac{2}{81} \,  
\rho^{-1} (2 f_{3} f_{3} f_{9}  +  f_{3} f_{9} f_{3}  +  f_{9} f_{3} f_{3} ) 
+ \tfrac{1}{4041576000} \varpi^{2,2}_{4,14}\\*
&\phantom{=} +\tfrac{26869796704014139979194459442197511 }{21407683345986402107516651097196800}\zeta_{15} \, ,
\notag \\
\ccsv{2&2&6}{4& 4&10} &= 
\tfrac{5}{112266}  \,  
\rho^{-1} ( f_{3} f_{7} f_{3}  +  f_{7} f_{3} f_{3} ) 
 + \tfrac{1}{181870920000}  \varpi^{4,2}_{4,14} - \tfrac{12816078754315007321  }{2004437855486547204570000}\zeta_{13}\,, \notag\\
\ccsv{2& 2& 8}{4& 4& 12}  &\label{svieis.31}  = 
\tfrac{691}{30405375} \,  
\rho^{-1} ( f_{3} f_{9} f_{3}  +  f_{9} f_{3} f_{3} ) 
   + \tfrac{1}{5688892440000} \varpi^{3,3}_{4,16} \\
   &\notag \phantom{=} + 
   \tfrac{9052954611948991353652521347483 }{3794998411333953100877951785412160000 } \zeta_{15} \, , 
 \\
\ccsv{4& 4& 6}{6& 6& 8} &=
-\tfrac{2}{175}  \,  
\rho^{-1} (2 f_{5} f_{5} f_{7}  +  f_{5} f_{7} f_{5}  +  f_{7} f_{5} f_{5} ) 
- \tfrac{1}{2910600000} \varpi^{2,2}_{6,14} \notag\\
&\notag \phantom{=}-\scalebox{1.0}{$\frac{957793720722761645418983400734693368126541316174567357528329392131303221923951350785293}{46407385633150935787854910109345468165086540982881530827257255703301703974682993546875}$} \zeta_{17}\, ,
   \notag \\
\ccsv{4& 4& 4}{6& 6& 8} &= 
 - \tfrac{1}{130977000000} \varpi^{4,2}_{6,14} -\tfrac{4199  }{1343304000}\zeta_{15} \, ,
  \notag \\
\ccsv{4 &4 &2}{6 &6 &8} &= 
-\tfrac{1}{1102500} \,  
\rho^{-1} ( f_{3} f_{5} f_{5}  +  f_{5} f_{3} f_{5} ) 
-\tfrac{ 1 }{611226000000} \varpi^{6,2}_{6,14}+ \tfrac{ 108119242521250682513 }{166672044710760167555760000 }\zeta_{13}\, ,   \notag
\end{align}
and we have determined both numerical values of these $\varpi^{d_1,d_2}_{k,w}$ 
(typically to 300 digits)
and analytical representations in terms of MMVs by imposing the modular properties
(\ref{modbeqv}) on the associated $\beqv{j_1 &j_2 &j_3}{k_1 &k_2 &k_3}$, see also the ancillary file for the representation of these numbers in terms of MMVs.  The increasingly large rational coefficients of $\zeta_{w}$ are due to our choice of $f$-alphabet discussed in appendix~\ref{app:A} and \cite{Dorigoni:2023part1}.

From the $\ccsv{j_1 &j_2 &j_3}{k_1 &k_2 &k_3}$ that are expressible in terms of MZVs we can read directly that the transcendental weight is $j_1{+}j_2{+}j_3{+}3$. Hence, 
(\ref{svieis.31}) leads us to assign transcendental
weight 13 to $\varpi^{4,2}_{4,14}$, $\varpi^{6,2}_{6,14}$, 
weight 15 to $ \varpi^{2,2}_{4,14}$, $\varpi^{4,2}_{6,14}$, $\varpi^{3,3}_{4,16}$ and 
17 to $ \varpi^{2,2}_{6,14} $. This is consistent with the total number of $\eee_{k\geq 0}$
in the associated $\hat \sigma^{d_1,d_2}_{k,w}$ in (\ref{svieis.14}) and
(\ref{svieis.15}) which more generally assigns
transcendental weight $k{+}w{-}d_1{-}d_2{+}1$ to $\varpi^{d_1,d_2}_{k,w}$.

However, the new periods $\varpi^{d_1,d_2}_{k,w}$ are only well-defined 
up to adding rational multiples of $\zeta_{k{+}w{-}d_1{-}d_2{+}1}$. Such redefinitions
simply shift the associated zeta generator $\sigmaE_{k{+}w{-}d_1{-}d_2{+}1}$
by rational multiples of the highest-weight vectors $s^{d_1}(\eee_k,P^{d_2}_{w})$
at the leading order of the new generators $\hat \sigma^{d_1,d_2}_{k,w}$ in (\ref{svieis.15}).
At modular depth three, these shifts by $s^{d_1}(\eee_k,P^{d_2}_{w})$ are
ambiguities in the zeta generators $\hat \sigma_{k{+}w{-}d_1{-}d_2{+}1}$ since
\begin{itemize}
\item[(i)] as highest-weight vectors, the $s^{d_1}(\eee_k,P^{d_2}_{w})$ preserve
the key property $[\hat{N},\sigmaE_m]=0$ at modular depth three, with $\hat{N}$ given by (\ref{zetgen.06}),
\item[(ii)] the Pollack combination $P^{d_2}_{w}$ vanishes under $\eee_k \rightarrow \ep_k$, so
the image of $s^{d_1}(\eee_k,P^{d_2}_{w})$ in the Tsunogai algebra annihilates
the free-Lie-algebra generators $a,b$ of (\ref{TSaction}) and cannot be detected from
the methods in \cite{Dorigoni:2023part1}.
\end{itemize}
Table \ref{newvarpis} summarises the possible redefinition, transcendental weight 
and numerical value of the new periods encountered in (\ref{svieis.14}).

Eventually, the modular completion of iterated Eisenstein integrals to $\beqv{j_1 &j_2 &\ldots &j_\ell}{k_1 &k_2 &\ldots &k_\ell}$ 
at modular depth $\ell \geq 4$ should inform the general structure of the extension
$\Msv( \sigmaE_i,\sigmaE_{\pern})= \Msv(  \sigmaE_i)+\ldots$
involving words in both zeta generators and (one or more) 
$\sigmaE^{d_1,d_2}_{k,w}$. The wealth of new periods expected
as the coefficients of such words in $\sigmaE_i$ and $\sigmaE^{d_1,d_2}_{k,w}$
may admit a classification into primitive ones (akin to $\zeta_k$) and `composite' ones (akin
to MZVs $\rho^{-1}( {\rm sv}(f_{i_1} f_{i_2}\ldots f_{i_\ell} ) )$ at $\ell \geq 2$).
The systematics of $\Msv( \sigmaE_i,\sigmaE_{\pern})$ is beyond the reach of the studies in this
work, and the mathematical understanding of the vast system of periods and 
associated $(\sigmaE_i,\sigmaE_\varpi)$-generators in
$\Msv( \sigmaE_i,\sigmaE_{\pern})$ is expected to crucially build
upon Brown's work \cite{Brown2019}.


\begin{table}
\centering
\begin{tabular}{c||c|c|c}
new period&well defined up to &transcendental weight &approx. numerical value
\\\hline\hline
$\Big. \varpi^{4,2}_{4,14} \Big.$ &$\zeta_{13}$ &13 & $-3.370063\ldots\times10^6$
\\\hline
$\Big. \varpi^{2,2}_{4,14} \Big.$ &$\zeta_{15}$ &15 &$-5.066084\ldots\times10^9$
\\\hline
$\Big. \varpi^{6,2}_{6,14} \Big.$ &$\zeta_{13}$ &13 &$-19555.45\ldots$
\\\hline
$\Big. \varpi^{4,2}_{6,14}\Big.$ &$\zeta_{15}$ &15 &$506391.8\ldots$
\\\hline
$\Big. \varpi^{3,3}_{4,16}\Big.$ &$\zeta_{15}$ &15 &$-8.585493\ldots\times10^7$
\\\hline
$\Big. \varpi^{2,2}_{6,14}\Big.$ &$\zeta_{17}$ &17 &$-6.007268\ldots \times10^{10}$
\end{tabular}
\caption{\label{newvarpis}\textit{New periods $\varpi^{d_1,d_2}_{k,w}$ encountered in the
constants $c^{\rm sv}[\ldots]$ at $k_1{+}k_2{+}k_3 = 18,20$ and 
hence in the simplest extensions $ \Msv(  \sigmaE_i, \sigmaE_{\pern})$ of the series 
in zeta generators in (\ref{svieis.14}).}}
\end{table}

 \subsubsection{Arithmetic parts of the new generators}
\label{sec:ngeozvarpi}

On the one hand, the arithmetic parts $\zetaE^{d_1,d_2}_{k,w}$ expected for the new generators
$\sigmaE^{d_1,d_2}_{k,w}$ are inaccessible to our explicit studies of $\beqv{j_1 &j_2 &j_3}{k_1 &k_2 &k_3}$
at modular depth $\leq 3$. On the other hand, the necessity for $\zetaE^{d_1,d_2}_{k,w}$ can be easily inferred
from modular $T$ invariance at modular depth $\geq 4$. We will see in the later discussion around (\ref{mt.22})
that the series $\mathbb \Msv(  \sigmaE_i, \sigmaE_{\pern}) $ has to commute with $\hat{N}$ in (\ref{zetgen.06})
which implies
\beq
0 = [\hat{N} , \sigmaE^{d_1,d_2}_{k,w} ] = 
- [\eee_0 , \sigmaE^{d_1,d_2}_{k,w} ] +  \sum_{\ell=4}^{\infty } (\ell{-}1) \BF_\ell [\eee_\ell , \sigmaE^{d_1,d_2}_{k,w} ]  \, .
\label{zvarpi.01}
\eeq
This condition will be studied in a decomposition $ \sigmaE^{d_1,d_2}_{k,w} = \zetaE^{d_1,d_2}_{k,w}  +s^{d_1}(\eee_k, P^{d_2}_w) 
+ \hat \rho^{d_1,d_2}_{k,w} $ with $ \hat \rho^{d_1,d_2}_{k,w} $ referring to all geometric
modular-depth $\geq 4$ terms in the new generator. The contribution to (\ref{zvarpi.01}) at
modular depth three vanishes by the highest-weight-vector condition $[\eee_0,s^{d_1}(\eee_k, P^{d_2}_w)]=0$,  assuming
that the expansion of $[\hat N , \zetaE^{d_1,d_2}_{k,w} ]$ with $\mf{sl}_2$ invariant 
$\zetaE^{d_1,d_2}_{k,w} $ starts at modular depth four. Validity of (\ref{zvarpi.01}) at
modular depth four in turn imposes
\beq
 [\eee_0 ,  \hat \rho^{d_1,d_2}_{k,w}]  =  \sum_{\ell=4}^{\infty } (\ell{-}1) \BF_\ell
 \Big( [\eee_\ell , s^{d_1}(\eee_k, P^{d_2}_w) ] 
 +  [\eee_\ell ,\zetaE^{d_1,d_2}_{k,w} ]   \Big) + \ldots  \, ,
\label{zvarpi.02}
\eeq
where the ellipsis instructs us to discard terms of modular depth $\geq 5$.
In absence of $\zetaE^{d_1,d_2}_{k,w}$, (\ref{zvarpi.02}) could only hold
if $[\eee_\ell , s^{d_1}(\eee_k, P^{d_2}_w) ]$ was expressible as an ${\rm ad}_{\eee_0}$ image for all $\ell\geq 4$.

However, as soon as the $\mf{sl}_2$-module with $\eee_\ell$ as its lowest-weight vector is bigger than the $\mf{sl}_2$-module with  $s^{d_1}(\eee_k, P^{d_2}_w) $ as its highest-weight vector, this is impossible.\footnote{Note that $P^{d_2}_w$ is the lowest-weight vector in a $(w{-}2d_2{+}1)$-dimensional $\mf{sl}_2$-module and  $s^{d_1}(\eee_k, P^{d_2}_w)$ is the highest-weight vector in a  $(k{+}w{-}2d_1{-}2d_2{+}3)$-dimensional $\mf{sl}_2$-module.} 
The reason is that $\ad_{\eee_0}$ raises the $\ad_\hhh$-eigenvalue determined by (\ref{eq:ade0e0chH}), and the one of $[\eee_\ell , s^{d_1}(\eee_k, P^{d_2}_w) ]$ is negative for sufficiently large $\ell$. 
However, the dimension of the $\ad_\hhh$-eigenspaces in the tensor product of the two representations grows strictly monotonically with the eigenvalue for negative eigenvalues. Therefore $\ad_{\eee_0}$ cannot be surjective in this range of $\ad_\hhh$-eigenvalues and it is a quick check that $[\eee_\ell , s^{d_1}(\eee_k, P^{d_2}_w) ]$ for sufficiently large $\ell$ is not in the image since $\eee_\ell$ is a lowest-weight vector of the bigger representation.
Hence, for an infinity of values $\ell > k{+}w{-}2d_1{-}2d_2{+}4$, the arithmetic terms in $ [\eee_\ell ,\zetaE^{d_1,d_2}_{k,w} ] $ in (\ref{zvarpi.02}) have to contribute non-trivially such that the right-hand side is expressible as
 $ [\eee_0 ,  \hat \rho^{d_1,d_2}_{k,w}] $ for a suitable choice of the modular-depth-four
 contributions to $\hat \rho^{d_1,d_2}_{k,w}$. 
 
The $\mf{sl}_2$-invariance $[ \eee_0, \zetaE^{d_1,d_2}_{k,w}] = 0$ of the arithmetic parts used in the above arguments will be justified when
analysing the modular properties of $\Jeqv ( \eee_k;\tau)$ in section \ref{sec:6}.
As we will soon show, SL$_2$-equivariance of the expression (\ref{uplift}) for $R[ \Jeqv(\eee_k;\tau)] $ hinges on
\beq
[ \eee_0, \Msv( \zetaE_i,\zetaE_{\pern}) ] = [ \eee_0^\vee, \Msv( \zetaE_i,\zetaE_{\pern}) ] = 0\, .
\label{zvarpi.03}
\eeq
Similar to the discussion of zeta generators in section \ref{sec:4zeta.6}, the arithmetic parts $\zetaE_{\pern}$ in the decomposition $\hat \sigma_{\pern}= \zetaE_{\pern}+\sigma^{\rm g}_{\pern}$ (with geometric terms $\sigma^{\rm g}_{\pern}$ built from nested brackets of $\eee_k^{(j)}$) are not yet fixed by (\ref{zvarpi.03}). One could still redistribute $\mf{sl}_2$ invariant terms between $\zetaE_{\pern}$ and $\sigma^{\rm g}_{\pern}$ without altering the overall generator $\hat \sigma_{\pern}$. By (\ref{uplift}), the net effect of such redefinitions of $\zetaE_{\pern}$ is to add $\mathbb Q$ multiples of the associated period $\varpi$ to some of the $c^{\rm sv}$ and $\beta^{\rm eqv}$ that fall into singlet representations of $\mf{sl}_2$.

In order to fix these ambiguities, we follow the logic applied to zeta generators in section \ref{sec:4zeta.6}. We single out a preferred choice of $\hat z_{\pern}$ by imposing that $\sigma^{\rm g}_{\pern}$ does not contain any $\mf{sl}_2$ singlet. In this way, $\hat z_{\pern}$ captures the entire $\mf{sl}_2$ singlet of $\hat \sigma_{\pern}$, and one is led to canonically defined $c^{\rm sv}$ and $\beta^{\rm eqv}$ from the coefficients in (\ref{uplift}). However, the preferred choices of $\hat z_i$ and $\hat z_{\pern}$ as the $\mf{sl}_2$ singlets of $\hat \sigma_i$ and $\hat \sigma_{\pern}$ do not eliminate the possibility to redefine $\varpi^{d_1,d_2}_{k,w}$ by adding rational multiples of $\zeta_{k{+}w{-}d_1{-}d_2{+}1}$ which amounts to shifting $\hat \sigma_{k{+}w{-}d_1{-}d_2{+}1}$ by $\hat \sigma^{d_1,d_2}_{k,w}$. 

\subsection{Reformulation in terms of Laplace equations}
\label{sec:7}

At modular depth two, the space of modular invariants $\beqv{j_1&j_2}{k_1&k_2}$ can be re-expressed through irreducible modular-depth-two functions and products of modular-depth-one functions~\cite{Dorigoni:2021jfr,Dorigoni:2021ngn}. More specifically, the modular invariant functions satisfy $2j_1{+}2j_2=k_1{+}k_2{-}4$ and can be expressed through solutions $\mathrm{F}_{m,k}^{\pm(s)}$ to the inhomogeneous Laplace equations (with $(2m,2k) = (k_1,k_2)$)
\begin{align}
\big( \Delta - s(s{-}1) \big) \Fp{m}{k}{s} &=  \EE_m \EE_k\,,  \label{eq:Lap2} \\*
\big( \Delta - s(s{-}1) \big) \Fm{m}{k}{s} &=  \frac{(\nabla \EE_m) (\overline{\nabla} \EE_k) - (\nabla \EE_k) (\overline{\nabla} \EE_m) }{2 (\Im\tau)^2}\,.
\notag
\end{align}
Assuming without loss of generality that $m\leq k$, the spectrum of the irreducible modular-depth-two functions is given by
\begin{align}
\Fp{m}{k}{s}&: \quad& s &\in \{ k{-}m{+}2, \, k{-}m{+}4, \, \ldots, \, k{+}m{-}4, \, k{+}m{-}2\}\,,\nn\\
\Fm{m}{k}{s}&: \quad& s &\in \{ k{-}m{+}1, \, k{-}m{+}3, \, \ldots, \, k{+}m{-}3, \, k{+}m{-}1\}\,,
\label{countfs}
\end{align}
with multiplicity one.  The superscripts $+/-$ denote whether the functions are even/odd under the involution $\tau\mapsto -\bar\tau$.

We can use this basis to count the number of modular invariants functions $\beqv{j_1&j_2}{2m&2k}$ at given $m$ and $k$. For $m <k$, there are $2m{-}1$ modular invariants among  $\beqv{j_1&j_2}{2m&2k}$, which ties in with the $2m{-}1$ irreducible functions $\mathrm{F}_{m,k}^{\pm(s)}$ due to the counting in (\ref{countfs}). Products $\nabla^\ell\EE_m  \overline{\nabla}^\ell\EE_k$ or $\overline{\nabla}^\ell\EE_m \nabla^\ell \EE_k$ with $\ell< m$ in turn exhaust the
symmetric combinations or \textit{shuffles} 
\begin{align}
\beqv{j_1&j_2}{2m&2k}+ \beqv{j_2&j_1}{2k&2m} = \beqv{j_1}{2m}  \beqv{j_2}{2k} \,,
\end{align}
and thus the reversely ordered $\beqv{j_2&j_1}{2k&2m}$.
Note that, up to shuffles, the functions $\mathrm{F}_{m,k}^{\pm(s)}$ belong to $\mathfrak{sl}_2$ multiplets of dimensions $2s{-}1$ which also agrees with the tensor-product decomposition~\eqref{eq:sl2tens}.
For $m=k$, there are no odd functions $\Fm{m}{m}{s}$ and the number of modular invariants reduces to $m{-}1$ up to shuffles which is in agreement with antisymmetric part of the tensor product~\eqref{eq:sl2tens}.

The $\mathfrak{sl}_2$ multiplet here is defined by the number of $\nabla$ or $\overline{\nabla}$ derivatives that can be applied to $\mathrm{F}_{m,k}^{\pm(s)}$  before the Cauchy--Riemann equation becomes a sum of products of modular-depth-one functions~\cite{Dorigoni:2021jfr,Dorigoni:2021ngn}. 
This notion of $\mathfrak{sl}_2$ multiplets in terms of $\nabla$ and $\overline\nabla$ is related to the one in terms of the generators $\ad_{\eee_0}$ and $\ad_{\eee_0^\vee}$ introduced in section~\ref{sec:rev.2} as follows:
In the holomorphic derivative $\nabla \Jeqv ( \eee_k;\tau)$ there is a  contribution $\sim {\rm ad}_{\eee_0} \Jeqv ( \eee_k;\tau)$ from the differential equation (\ref{diffeq.04}) such that the  $\mathfrak{sl}_2$ raising operator $\eee_0$ on the generating series captures the action of $\nabla$ on its components $\beta^{\rm eqv}$ in (\ref{dbetaeqv}). There is an analogous structure in the antiholomorphic derivative $\overline{\nabla} \Jeqv ( \eee_k;\tau)$ with $\eee_0^\vee$ such that $\overline\nabla$ takes the role of the lowering operator of $\mathfrak{sl}_2$.\footnote{Strictly speaking,
the lowering operator on modular forms $\beta^{\rm eqv}$ of modular weight $(0,\bar{w})$ is $y^{-\bar w}\overline{\nabla} y^{\bar w}$ as done in
\eqref{eq:CRLap} with $\bar w=-2s$.}
The leftover terms in (\ref{diffeq.04}) involve factors of ${\rm G}_m$ or $\Delta_m$ that are not relevant to the $\mathfrak{sl}_2$ multiplet structure and similarly for the antiholomorphic derivative. The interplay of Laplace equations, Cauchy--Riemann equations and $\mathfrak{sl}_2$ multiplets  at modular depth three is further discussed in section \ref{sec:7.1.new} below.

By abuse of terminology, we will refer to the functions in~\eqref{eq:Lap2} as Laplace eigenfunctions and $s$ as their eigenvalue, even though the equation they satisfy is inhomogeneous.
Given the usefulness of the equivalent basis of Laplace eigenfunctions at modular depth two~\cite{Dorigoni:2021jfr,Dorigoni:2021ngn}, we will now discuss to what extent an analogous basis can be set up at modular depth three. 

\subsubsection{Counting of irreducible functions at modular depth three}
\label{sec:7.1}

In the quest for constructing similar modular invariants $\mathrm{F}_{m,k,\ell}^{\pm(s)}$ at modular depth three, a first step is to
investigate what source terms can be expected on the right-hand sides of the candidate inhomogeneous Laplace equations.
Let us consider first integers $2\leq m< k< \ell$, so that all subscripts are different. Up to shuffles there are the two independent arrangements of columns in the equivariant modular-depth-three Eisenstein integrals, for instance
\begin{align}
\label{eq:dep3F}
\beqv{j_1&j_2&j_3}{2m & 2k & 2\ell} 
\quad\quad \text{and}\quad\quad
\beqv{j_2&j_1&j_3}{2k & 2m & 2\ell} \,.
\end{align}
The counting of the number of modular invariants is more intricate in this case,
 and we begin with analysing the triple tensor product using associativity and the 
 notation of~\eqref{eq:sl2tens},
\begin{align}
\label{eq:trip2}
V(\eee_{2m}) \otimes V(\eee_{2k}) \otimes V(\eee_{2\ell}) 
&= V(\eee_{2m}) \otimes \!\!\!\bigoplus_{p=\ell-k+1}^{\ell+k-1} \!\!\! V(\eee_{2p})
= \bigoplus_{p=\ell-k+1}^{\ell+k-1}   \bigoplus_{q=|m-p|+1}^{m+p-1} V(\eee_{2q})\,.
\end{align}
Each of the $(2q{-}1)$-dimensional representations $V(\eee_{2q})$ above contains one modular invariant and, in view of~\eqref{eq:dep3F}, they occur with multiplicity two for the shuffle-irreducible modular-depth-three functions with $2\leq m< k< \ell$. 

A main difference to the modular-depth-two system~\eqref{eq:Lap2} is that there are now multiplicities involved in the tensor product, e.g.\ for $(m,k,\ell)=(2,3,5)$:
\begin{align}
\label{eq:s235}
V(\eee_4) \otimes V(\eee_6)\otimes V(\eee_{10}) &= V(\eee_4) \oplus  2{\times} V(\eee_{6}) \oplus 3{\times} V(\eee_8) \oplus 3{\times} V(\eee_{10})\nn\\
&\hspace{10mm} \oplus3{\times} V(\eee_{12}) \oplus 2{\times} V(\eee_{14}) \oplus V(\eee_{16}) \, ,
\end{align}
where we recall that $V(\eee_{2m})$ denotes the $(2m{-1})$-dimensional representation of $\mathfrak{sl}_2$ and that it has Casimir eigenvalue $m(m{-}1)$, see (\ref{caseig}). 

For $\ell\geq m{+}k$, the number of invariants is $2{\times}(2m{-}1)(2k{-}1)$, but for lower values of $\ell$ the counting is more restricted. We also note that, at modular depth three, the one-dimensional representation $V(\eee_2) $ can appear in the tensor product even after having removed shuffles. 
The phenomenon happens for instance for $\ell=m{+}k{-}1$, and the explicit form of the
$\mf{sl}_2$ singlets can be obtained from the uplift $\ep_k^{(j)} \rightarrow \eee_k^{(j)}$ of
the $I_{k_1,k_2,k_3}$ given by (\ref{svieis.00}).

When some of $(m,k,\ell)$ coincide, the number of shuffle irreducibles, giving rise to new irreducible modular-depth-three functions, has to be adapted.
The general pattern is 
\begin{itemize}
\item 
All three of $(m,k,\ell)$ distinct. The shuffle irreducibles are given by \textit{twice} the triple tensor product $V(\eee_{2m})\otimes V(\eee_{2k})\otimes V(\eee_{2\ell})$ as detailed above.
\item 
Exactly two of $(m,k,\ell)$ are identical. The shuffle irreducibles are given by the triple tensor product $V(\eee_{2m})\otimes V(\eee_{2k})\otimes V(\eee_{2\ell})$ (i.e.\ without the doubling of the previous case), for instance
\begin{align}
\!\! \! \! \! V(\eee_4) \otimes V(\eee_4)\otimes V(\eee_{6}) &= V(\eee_2) \oplus 2{\times} V(\eee_4) \oplus 3{\times} V(\eee_6) \oplus 2{\times} V(\eee_8) \oplus V(\eee_{10}) 
\, , \label{exaab} \\
\!\! \! \! \! V(\eee_4) \otimes V(\eee_4)\otimes V(\eee_{8}) &= V(\eee_4) \oplus 2{\times} V(\eee_6) \oplus 3{\times} V(\eee_8) \oplus 2{\times} V(\eee_{10}) \oplus V(\eee_{12}) 
\, , \notag \\
\!\! \! \! \!  V(\eee_4) \otimes V(\eee_6)\otimes V(\eee_{6}) &= V(\eee_2) \oplus 3{\times} V(\eee_4) \oplus 3{\times} V(\eee_6) \oplus 3{\times} V(\eee_8) \oplus 2{\times} V(\eee_{10}) \oplus V(\eee_{12}) 
\, , \notag
\end{align}
where the $\mf{sl}_2$ singlet $V(\eee_2)$ in the first line is realized through the quantity $I_{4,4,6}$ discussed below (\ref{svieis.00}).
\item 
All three of $(m,k,\ell)$ are the same. The shuffle irreducibles are given by \textit{half} of the `mixed symmetry projection' of the triple tensor product:
\begin{align}
\frac12\Big[\big(V(\eee_{2m})\otimes V(\eee_{2m})\otimes V(\eee_{2m})\big) \ominus \text{Sym}^3 \big(V(\eee_{2m}) \big) \ominus \text{Alt}^3 \big(V(\eee_{2m}) \big)\Big]\,.
\label{exaaa}
\end{align}
Here, $\text{Sym}^3$ and $\text{Alt}^3$ denote the fully symmetrised and antisymmetrised triple tensor product of representations. The removal of $\text{Alt}^3 (V(\eee_{2m}) )$ can be understood from the vanishing of totally antisymmetrised nested brackets $[\eee_{2m}^{(j_1)},[\eee_{2m}^{(j_2)},\eee_{2m}^{(j_3)}]]$ by Jacobi identities. The factor of $1/2$ is meaningful since all representations appear with even multiplicity. In the simplest examples, (\ref{exaaa}) gives rise to shuffle irreducibles in
\begin{align}
(m,k,\ell)=(2,2,2) \ \ &\Rightarrow \ \  V(\eee_{4}) \oplus V(\eee_{6}) \, ,
\\
(m,k,\ell)=(3,3,3) \ \ &\Rightarrow \ \  V(\eee_{4}) \oplus 2{\times} V(\eee_{6})\oplus V(\eee_{8})\oplus V(\eee_{10})\oplus V(\eee_{12})\,,
 \notag
\end{align}
after for instance removing the totally antisymmetric and symmetric parts $V(\eee_2)$ and $V(\eee_8) \oplus V(\eee_4)$ from 
$V(\eee_4) \otimes V(\eee_4)\otimes V(\eee_{4}) = V(\eee_{2})  \oplus 3{\times} V(\eee_4)  \oplus 2{\times} V(\eee_6)\oplus    V(\eee_8)$.
\end{itemize}

\subsubsection{Examples of Laplace equations at modular depth three}
\label{sec:7.1.2}

The tensor-product decomposition~\eqref{eq:trip2} also suggests the types of sources that can arise in the Laplace equations for a given eigenvalue $s$, corresponding to the various factors in the decomposition. These source terms can be constructed out of products of modular-depth-two functions with modular-depth-one functions (and up to first derivatives) as well as triple products of modular-depth-one functions. 

Representative equations are
\begin{align}
(\Delta {-}2 ) \mathrm{F}_{2,2,2}^{+(2)} &=  \frac16 \EE_2^3 + \EE_2 \Fp{2}{2}{2}\,,
\label{repexs}\\
(\Delta {-}6 ) \mathrm{F}_{2,2,2}^{-(3)} &=  \frac{ (\nabla\EE_2) \overline{\nabla} \Fp{2}{2}{2}-  (\overline{\nabla}\EE_2) {\nabla} \Fp{2}{2}{2}}{2 (\Im\tau)^2}\,,\nn\\
(\Delta {-}2 ) \mathrm{F}_{2,2,3}^{-(2a)}&= \EE_2 \Fm{2}{3}{2} + \frac{ (\nabla \EE_2) \overline{\nabla}\Fp{2}{3}{3}- (\overline{\nabla} \EE_2) {\nabla}\Fp{2}{3}{3}}{2 (\Im \tau)^2}\,,\nn\\
(\Delta {-}2 ) \mathrm{F}_{2,2,3}^{-(2b)}&= 2\EE_2 \Fm{2}{3}{2} + \frac{ (\nabla \EE_3) \overline{\nabla}\Fp{2}{2}{2}- (\overline{\nabla} \EE_3) {\nabla}\Fp{2}{2}{2}}{2 (\Im \tau)^2}\,,\nn\\
(\Delta {-}6 ) \mathrm{F}_{2,2,3}^{+(3a)}&=
 \EE_2^2 \EE_3 +2\EE_3 \Fp{2}{2}{2} +12  \EE_2 \Fp{2}{3}{3}
\, , \nn \\
(\Delta {-}30 ) \mathrm{F}_{2,2,5}^{-(6a)}&= \EE_2 \Fm{2}{5}{6} - \frac{(\nabla\EE_5)\overline{\nabla}\Fp{2}{2}{2}  - (\overline{\nabla}\EE_5){\nabla}\Fp{2}{2}{2} }{10(\Im\tau)^2}\,,\nn\\
(\Delta {-}42) \mathrm{F}_{2,2,5}^{+(7)}&= 2 \EE_2^2 \EE_5+ \frac{(\nabla\EE_2) \overline{\nabla}\Fm{2}{5}{6} - (\overline{\nabla}\EE_2) {\nabla}\Fm{2}{5}{6}}{2(\Im\tau)^2}  - \frac{\EE_2 (\nabla \EE_2)\overline{\nabla}\EE_5+ \EE_2 (\overline{\nabla} \EE_2){\nabla}\EE_5}{4(\Im\tau)^2} \, .\notag
\end{align}
The examples above were chosen to illustrate the possible types of source terms. We will comment below what fixes the right-hand sides of these equations and the relative factors therein. 
When there is a multiplicity to an eigenvalue, we have made a choice of basis and labelled the corresponding eigenfunctions by $a,b,\ldots$ along
with the superscript $(s)$. 
The ancillary file contains a list of all the irreducible functions ${\rm F}_{m,k,\ell}^{\pm(s)}$ of degree $2m+2k+2\ell \leq 20$ in terms of $\beta^{\rm eqv}$ along with an explanation of the choice of basis $a,b, \ldots$.

\subsubsection{$\mathfrak{sl}_2$ structure of the Laplace eigenfunctions}
\label{sec:7.1.new}

The form of the Laplace equations (\ref{repexs}) at modular depth three is more complicated compared to the modular-depth-two case~\eqref{eq:Lap2}.
The precise form of the Laplace equations is dictated by the fact that, up to shuffles, the modular invariant functions with eigenvalue $s(s{-}1)$ belong to an $\mathfrak{sl}_2$ multiplet of dimension $2s{-}1$.
The association of ${\rm F}_{m,k,\ell}^{\pm(s)}$ with $(2s{-}1)$-dimensional multiplets follows from their Cauchy--Riemann equations: 
By the differential equations discussed in section~\ref{sec:dJeqv} and the comments just before section~\ref{sec:7.1}, the Maa\ss{} operators $\nabla$ and $\overline\nabla$ in (\ref{mops}) can be thought of as raising and lowering operators, dropping contributions proportional to holomorphic modular forms.
While $\nabla^s \mathrm{F}^{\pm(s)}$ and $\overline{\nabla}^s \mathrm{F}^{\pm(s)}$ of order $s$ simplify to quantities of lower modular depth, their lower-order Cauchy--Riemann derivatives ($\nabla^\ell \mathrm{F}^{\pm(s)}$ and $\overline{\nabla}^\ell \mathrm{F}^{\pm(s)}$ with $\ell<s$) are still indecomposable modular-depth-three objects.
For example,
\begin{align}
\label{eq:CRF3}
\nabla^2 \mathrm{F}_{2,2,2}^{+(2)} &= \frac 12 (\nabla\EE_2) (  \nabla \Fp{2}{2}{2} )\,,\\
\nabla^3 \mathrm{F}_{2,2,2}^{-(3)} &= -\frac12 \EE_2 (\nabla \EE_2) (\nabla^2 \EE_2) + \frac12 (\nabla^3 \EE_2) \Fp{2}{2}{2} + 2 (\nabla^2 \EE_2) (\nabla \Fp{2}{2}{2}) + \frac{21\zeta_3}{200} (\nabla^3 \EE_3)\,,
\nn
\end{align} 
leave a triplet $(\nabla\mathrm{F}_{2,2,2}^{+(2)}, \,\mathrm{F}_{2,2,2}^{+(2)}, \, \overline{\nabla}\mathrm{F}_{2,2,2}^{+(2)})$ of quantities which cannot be expressed in terms of ${\rm E}_k$, $\mathrm{F}_{m,k}^{\pm(s)}$ and their Cauchy--Riemann derivatives. In $\nabla^3 \mathrm{F}_{2,2,2}^{-(3)}$, the occurrence of the last source term with an explicit $\zeta_3$ might appear surprising, but is in line with similar homogeneous terms arising at modular depth two~\cite{Dorigoni:2021jfr} and it can be traced back to the term $  t^2(\eee_4,t^3(\eee_4,\eee_4))$ in the expression (\ref{zwep3.51}) for $[\zetaE_3,\eee_{6}]$.

The fact that the $s^{\rm th}$ Cauchy--Riemann derivative of $\mathrm{F}^{\pm(s)}$ can be expressed through product functions is tantamount to being  outside of the $\mathfrak{sl}_2$ multiplet of irreducible functions. More generally, the interpretation of $\nabla$ and $\overline\nabla$ as raising and lowering operators is useful to group generic $\beta^{\rm eqv}$ subject to the Cauchy--Riemann equations (\ref{dbetaeqv}) into multiplets of $\mathfrak{sl}_2$ and to make contact with the functions ${\rm F}_{m,k,\ell}^{\pm(s)}$ at leading modular depth.
At depth one, for instance, \eqref{eq:EGk} relates $\EE_k$ to a multiplet of dimension $2k{-}1$ since $(\Im\tau)^{2k} \GG_{2k}$ (and its complex conjugate) generate submodules of the infinite-dimensional non-unitary principal series of SL$_2(\mathbb{R})$ with spherical vector $\EE_k$ whose quotient space is of dimension $2k{-}1$, see for instance~\cite{Fleig:2015vky}.
At higher modular depth, one finds submodules spanned by shuffles, i.e.\
the functions appearing on the right-hand sides of~\eqref{eq:CRF3} are not lowest-weight vectors of $\mathfrak{sl}_2$
but they  do not yield any shuffle irreducible functions at modular depth three under the action with a lowering operator $\overline\nabla$.

Equipped with this understanding of the $\mathfrak{sl}_2$-multiplets at higher modular depth, we can now explain how the interplay between the Laplace equation and the Cauchy--Riemann equation  fixes the right-hand sides in~\eqref{repexs}.
The fact that the $s^{\rm th}$ Cauchy--Riemann derivative must be expressible through (products of) derivatives of functions of lower modular depth leads to an ansatz for $\nabla^s \mathrm{F}_{m,k,\ell}^{\pm(s)}$ in terms of known objects by selecting objects built from the constituent $\mathfrak{sl}_2$ representations as well as homogeneous terms (such as the last term in the second line of~\eqref{eq:CRF3}). 
Furthermore, the identity
\begin{align}
\label{eq:CRLap}
    \pi \overline{\nabla} \left( y^{-2s} (\pi \nabla)^s \mathrm{F}_{m,k,\ell}^{\pm(s)} \right) = y^{-2(s-1)} (\pi \nabla)^{s-1} \big( \Delta -s(s{-}1)\big) \mathrm{F}_{m,k,\ell}^{\pm(s)}\,,
\end{align}
imposes a non-trivial condition on the ansatz for the Laplace equation of $\mathrm{F}_{m,k,\ell}^{\pm(s)}$. Solving this equation constrains the Laplace equation and the Cauchy--Riemann equation at the same time and leads to solution spaces of the dimension discussed after~\eqref{eq:s235}.\footnote{Both ans\"atze can be simplified by considering the effect of redefining $\mathrm{F}_{m,k,\ell}^{\pm(s)}$ itself by products of functions of lower modular depth.}
In particular, this fixes the linear combination $\frac16 \EE_2^3+ \EE_2 \mathrm{F}_{2,2}^{+(2)}$ on the right-hand side of the first equation in~\eqref{repexs} since for no other linear combination of these terms (excluding an overall scaling) can one find a Cauchy--Riemann equation---corresponding to a triplet of $\mathfrak{sl}_2$---that is consistent with the Laplace equation in the sense of~\eqref{eq:CRLap}. 

As a final comment, we emphasise that it is possible to obtain modular invariant solutions to  inhomogeneous Laplace eigenvalue problems more general than \eqref{eq:Lap2} and \eqref{repexs}. In particular, using spectral analysis for square-integrable functions on the upper half-plane quotiented by $\SLtwoZ$ \cite{Klinger-Logan:2018sjt}, we can construct solutions to \eqref{eq:Lap2} whose eigenvalues lie outside of the spectrum \eqref{countfs} or possess source terms given by $\EE_m\EE_k$ with $m,k\in \mathbb{C}$. Particularly important is the case where the source terms involve half-integral non-holomorphic Eisenstein series, see e.g.\ \cite{Green:2014yxa,Fedosova:2023cab,Alday:2023pet}. 
Although these more general inhomogeneous Laplace equations do give rise to well-defined modular invariant functions, these objects will not be expressible in terms of the equivariant iterated integrals considered here since their Cauchy--Riemann equations will not have the proper factorised form discussed above. 

\subsubsection{Laplace equations modulo lower modular depth}
\label{sec:7.1.3}

As another method to determine the admissible source terms in (\ref{repexs}),
one can evaluate the Laplacian of the modular-depth-three contributions to modular
invariant $\beqv{j_1 &j_2  &j_3}{k_1 &k_2 &k_3}$ in closed form: Based on the
first-order equations (\ref{dtaubeta}) and (\ref{dtaubarbeta}) of $\beta_{\pm}$ in
the terms (\ref{beqvdell}) of leading modular depth, it is not hard to derive (see section 3.1 of \cite{Dorigoni:2021jfr})
\begin{align}
\Delta \beqv{j_1 &j_2  }{k_1 &k_2 } \, &\big|_{j_1+j_2 =\frac{1}{2}(k_1+k_2-4)} =
\big(  (k_1{-}j_1{-}2)(j_1{+}1)+ (k_2{-}j_2{-}2)(j_2{+}1) \big) \beqv{j_1 &j_2 }{k_1 &k_2 }  \notag \\
&\quad + j_1(k_2{-}j_2{-}2)  \beqv{j_1-1 &j_2+1  }{k_1 &k_2 }
+ j_2(k_1{-}j_1{-}2)  \beqv{j_1+1 &j_2-1 }{k_1 &k_2 }  \notag \\
  &\quad - \delta_{j_2,k_2-2} (\tau {-} \bar \tau)^{k_2} {\rm G}_{k_2}
 j_1  \beqv{j_1-1 }{k_1  }  
  - \delta_{j_1,0} \, \frac{ \overline{ {\rm G}_{k_1} } }{(2\pi i)^{k_1}}    \, 
  (k_2{-}j_2{-}2)  \beqv{j_2+1  }{k_2 }   \notag \\
&\quad   +    \delta_{j_1,0} \,\delta_{j_2,k_2-2} (\tau {-} \bar \tau)^{k_2} {\rm G}_{k_2}    \, \frac{ \overline{ {\rm G}_{k_1} } }{(2\pi i)^{k_1}}   \ \textrm{mod modular depth $\leq 1$}\,,
 \label{lapmd2}
\end{align}
and its generalisation to modular depth three,
\begin{align}
&\Delta \beqv{j_1 &j_2  &j_3}{k_1 &k_2 &k_3} \, \big|_{j_1+j_2 + j_3=\frac{1}{2}(k_1+k_2+k_3-6)} =
\bigg( \sum_{i=1}^3 (k_i{-}j_i{-}2)(j_i{+}1) \bigg) \beqv{j_1 &j_2  &j_3}{k_1 &k_2 &k_3}  \notag \\
&\quad\quad\quad + j_1(k_2{-}j_2{-}2)  \beqv{j_1-1 &j_2+1  &j_3}{k_1 &k_2 &k_3}
+ j_2(k_1{-}j_1{-}2)  \beqv{j_1+1 &j_2-1  &j_3}{k_1 &k_2 &k_3}  \notag \\
&\quad\quad\quad + j_1(k_3{-}j_3{-}2)  \beqv{j_1-1 &j_2  &j_3+1}{k_1 &k_2 &k_3}
+ j_3(k_1{-}j_1{-}2)  \beqv{j_1+1 &j_2  &j_3-1}{k_1 &k_2 &k_3}  \notag \\
&\quad\quad\quad + j_2(k_3{-}j_3{-}2)  \beqv{j_1 &j_2-1  &j_3+1}{k_1 &k_2 &k_3}
+ j_3(k_2{-}j_2{-}2)  \beqv{j_1 &j_2+1  &j_3-1}{k_1 &k_2 &k_3} \notag \\
&\quad\quad\quad - \delta_{j_3,k_3-2} (\tau {-} \bar \tau)^{k_3} {\rm G}_{k_3}
  \big( j_1  \beqv{j_1-1 &j_2 }{k_1 &k_2 } +  j_2  \beqv{j_1 &j_2-1 }{k_1 &k_2 } \big) \notag \\
&\quad\quad\quad - \delta_{j_1,0} \, \frac{ \overline{ {\rm G}_{k_1} } }{(2\pi i)^{k_1}}    \, 
     \big(  (k_2{-}j_2{-}2)  \beqv{j_2+1 &j_3 }{k_2 &k_3 } 
     + (k_3{-}j_3{-}2)  \beqv{j_2 &j_3+1 }{k_2 &k_3 }  \big) \notag \\
&\quad\quad\quad +    \delta_{j_1,0} \, \delta_{j_3,k_3-2} \,(\tau {-} \bar \tau)^{k_3} {\rm G}_{k_3}    \frac{ \overline{ {\rm G}_{k_1} } }{(2\pi i)^{k_1}}  \,  \beqv{j_2 }{k_2 } \ \textrm{mod modular depth $\leq 2$} \, .
 \label{lapmd3}
\end{align}
In both of (\ref{lapmd2}) and (\ref{lapmd3}), the disclaimer mod modular depth $\leq \ell$ indicates that
$\beta_{\pm}$ of modular depth $\leq \ell$ are discarded. One can easily check via
(\ref{lapmd3}) as well as ${\rm E}_2 = - 6 \beqv{1 }{4 } $ and ${\rm F}_{2,2}^{+(2)} = 18 \beqv{2 &0 }{4 &4 } $
that the expressions
\begin{align}
\mathrm{F}_{2,2,2}^{+(2)} &= -54  \beqv{2 &1 &0}{4 &4 &4} 
\, ,
\label{exf222} \\
\mathrm{F}_{2,2,2}^{-(3)} &=
 54 (\beqv{1 &2 &0}{4 &4 &4} 
 - \beqv{2 &0 &1}{4 &4 &4} ) 
 - \frac{63}{20} \zeta_3 \beqv{2}{6} \,,
 \notag
\end{align}
are consistent with the first two Laplace equations in (\ref{repexs}) (the last contribution $\sim \zeta_3 \beqv{2}{6}$ can be fixed from the Cauchy--Riemann equation (\ref{eq:CRF3})). Note that the even modular function in (\ref{exf222}) is related to the combination ${\rm E}_{2,2,2} = - C_{2,2,1,1}+\ldots$ of MGFs (with a three-loop MGF $C_{2,2,1,1}=(\frac{\Im \tau}{\pi})^6 \sum_{p_1,p_2,p_3,p_4 \neq 0} \frac{\delta(p_1{+}p_2{+}p_3{+}p_4)}{|p_1|^4 |p_2|^4 |p_3|^2 |p_4|^2 }$ and graphs involving $\leq 2$ loops in the ellipsis) via $\mathrm{F}_{2,2,2}^{+(2)}= \frac{1}{4} {\rm E}_{2,2,2}$, see \cite{Broedel:2018izr, Gerken:2020yii} for different integral representations of ${\rm E}_{2,2,2}$.

Similar to the construction of $\mathrm{F}_{m,k}^{\pm(s)}$ in \cite{Dorigoni:2021jfr},
the guiding principle for $\beqv{j_1 &j_2  &j_3}{k_1 &k_2 &k_3}$-representations
of $\mathrm{F}_{m,k,\ell}^{\pm(s)}$ is to minimize the value of $j_3$ and to maximize the
value of $j_1$ while preserving the modular-invariance condition $j_1{+}j_2 {+} j_3=\frac{1}{2}(k_1{+}k_2{+}k_3{-}6)$. 
In this way, the occurrence of holomorphic Eisenstein series ${\rm G}_{k_3}$ and $\overline{{\rm G}_{k_1}}$ in Laplace or Cauchy--Riemann equations is delayed, and the source terms in the Laplace equations involve at most one $\nabla, \overline{\nabla}$-derivative of ${\rm E}_s$ or $\mathrm{F}_{m,k}^{\pm(s)}$. By the same reasoning, ${\rm F}_{2,2}^{+(2)} $ is constructed from a single-term $\beqv{2 &0 }{4 &4 }$ and not from the combination $\frac{1}{2}(\beqv{2 &0 }{4 &4 }- \beqv{0 &2 }{4 &4 })$ which is annihilated by $\nabla^2$ modulo terms involving ${\rm G}_4$.

We reiterate that, thinking of $\nabla$ and $\overline\nabla$ as raising and lowering operators and dropping contributions proportional to (anti-)holomorphic modular forms, the differential equation (\ref{dbetaeqv})
can be used to group the $\beta^{\rm eqv}$ into multiplets of $\mathfrak{sl}_2$. These multiplets are useful for relating the functions ${\rm F}_{m,k,\ell}^{\pm(s)}$ to the $\beta^{\rm eqv}$ at leading modular depth.

Note that the involution $\tau\mapsto -\bar\tau$ that maps $\mathrm{F}_{m,k}^{\pm(s)} \mapsto \pm \mathrm{F}_{m,k}^{\pm(s)}$ and $\mathrm{F}_{m,k,\ell}^{\pm(s)} \mapsto \pm \mathrm{F}_{m,k,\ell}^{\pm(s)}$ acts on the leading-depth contributions to the modular forms $\beta^{\rm eqv}$ via
\begin{align}
&\beqv{j_1 &j_2 &\ldots &j_\ell}{k_1 &k_2 &\ldots &k_\ell} \, \big|_{\tau\mapsto -\bar\tau} =
(4y)^{2\ell+2(j_1+j_2+\ldots+j_\ell)-k_1-k_2-\ldots-k_\ell}   \label{ccbeqv} \\*
&\quad \times \beqv{k_\ell-j_\ell-2 &\ldots &k_2-j_2-2  &k_1-j_1-2}{k_\ell &\ldots &k_2 &k_1} \, \textrm{mod modular depth $\leq \ell{-}1$} \,, \notag
\end{align}
which explains the two-term combination in the second line of (\ref{exf222}).

\subsubsection{New periods in Laurent polynomials}
\label{sec:7.2}

The new periods $\varpi^{d_1,d_2}_{k,w} $ of section \ref{sec:sivarpi} appear in the Laurent polynomials of the irreducible modular-depth-three functions $\mathrm{F}_{m,k,\ell}^{+(s)}$ as coefficients of the term $y^{1-s}$ in the Laurent polynomial. This homogeneous solution to the Laplace equation is the only power where periods different from those already contained in the source terms on the right-hand sides of the Laplace equation can arise. The coefficients of this homogeneous solution $y^{1-s}$ cannot be fixed from the differential equation but are fixed by modularity.

Examples covering all six new periods listed in table~\ref{newvarpis} are
\begin{align}
{\rm F}^{+(5c)}_{2,2,5} \, \big|_{\text{LP},y^{-4}} &= 
   \left( \frac{\varpi^{4,2}_{4,14}}{41472000} + \frac{235431 \zeta_{13}}{1408000} \right) y^{-4} 
\,,\nn\\
{\rm F}^{+(7)}_{2,2,5} \, \big|_{\text{LP},y^{-6}} &= 
	\bigg( - 
	\frac{\varpi^{2,2}_{4,14}}{221184000}+\frac{35}{512} \rho^{-1}\big( {\rm sv}(f_3 f_3 f_9) \big) - \frac{161}{2048} \rho^{-1}\big({\rm sv}(f_{3} f_9 f_3 )\big)\nn\\
 &\hspace{9mm} -\tfrac{26869796704014139979194459442197511 }{1171581836689117404633480344371200} \zeta_{15}\bigg) y^{-6}\,, \nn\\
{\rm F}^{+(6a)}_{2,2,6} \, \big|_{\text{LP},y^{-5}} &= 
\left(-\frac{\varpi^{3,3}_{4,16}}{116751360000}
+\frac{6973601 \zeta_{15}}{31841280000}   \right)y^{-5}\,,\nn\\
{\rm F}^{+(8)}_{3,3,4} \, \big|_{\text{LP},y^{-7}} &=
\bigg( \frac{13 \varpi^{2,2}_{6,14}}{1032192000} + \frac{125}{2048} \rho^{-1}\big({\rm sv}(f_5 f_5 f_7 )\big) - \frac{19}{2048} \rho^{-1}\big({\rm sv}(f_5 f_7 f_5 )\big)\nn\\
&\hspace{-15mm}
+\scalebox{1.0}{$\frac{12451318369395901390446784209551013785645037110269375647868282097706941885011367560208809}{16457545589037769089788200156526323602094759468907597425839456221707686535091042560000}$}\zeta_{17}
   \bigg)y^{-7} \,, \nn\\
{\rm F}^{+(6e)}_{2,3,5} \, \big|_{\text{LP},y^{-5}} &= 
\left(  \frac{13 \varpi^{4,2}_{6,14}}{1013760000 }-\frac{615627961 \zeta_{15}}{11208130560} \right) y^{-5}\,,
 \nn\\
{\rm F}^{+(4e)}_{2,3,5} \, \big|_{\text{LP},y^{-3}} &= 
   \left(-\frac{\varpi^{6,2}_{6,14}}{8064000}+\frac{21240581 \zeta_{13}}{1341204480}\right) y^{-3} \, ,
\end{align}
where the letters in the superscripts of ${\rm F}^{+(5c)}_{2,2,5}$ and similar functions are again due to the non-trivial multiplicities of the relevant Laplace eigenspaces. In these equations we have restricted ourselves to writing the term in the Laurent polynomials which corresponds to the homogeneous solution $y^{1-s}$, as all remaining terms can be uniquely fixed from the Laplace equations \eqref{repexs}.

Similarly, indecomposable single-valued MZVs beyond depth one firstly appear through contributions of the homogeneous solution $y^{1-s}$ to Laurent polynomials. For instance, the Laurent polynomial of the modular invariant $\mathrm{F}_{2,2,3}^{+(5)}$ in the largest $\mf{sl}_2$-module $V(\eee_{10}) $ of the relevant tensor product in the first line of (\ref{exaab}) has a contribution $\sim \zeta^{\rm sv}_{3,3,5}/y^4$.

Finally, although as stressed the coefficient of the homogeneous term $y^{1-s}$ cannot be fixed directly from the Laplace equation, this can nonetheless be reconstructed by analysing the behaviour as $y\to0$ of the series of contributions $(q\bar{q})^n = \exp(-4\pi n y)$, with $n\in \mathbb{N}$, exponentially suppressed at the cusp $y\gg1$.
This fact is a consequence of modularity which deeply intertwines the two expansion regimes $y\gg1$ and $y\to0$ for any modular invariant function. In \cite{Dorigoni:2022bcx,Dorigoni:2023nhc}, this phenomenon was manifested for the modular-depth-two functions $\mathrm{F}_{m,k}^{+(s)}$ defined in \eqref{eq:Lap2} by exploiting a Mellin transform argument and spectral theory. It would be extremely interesting to understand how these new periods appearing in the Laurent polynomials of $\mathrm{F}_{m,k,\ell}^{\pm(s)}$ can be recovered from the small-$y$ behaviour of the series of contributions $(q\bar{q})^n$.

\subsubsection{New periods multiplying cusp forms}
\label{sec:7.3}

In the same way that the Laurent polynomials contain the new periods of table~\ref{newvarpis}, the new periods $\Lambda^{d_1,d_2}_{k,w}$ of section \ref{sec:noedelta} and table~\ref{newLambdas} appear as the coefficients of iterated integrals of cusp forms. For the irreducible modular-depth-three functions this can happen when the eigenvalue parameter $s$ is half the modular weight of a holomorphic cusp form, $\Delta_{2s}$, since in this case the accompanying $\bpmno{s-1}{\Delta_{2s}}$ are solutions of the homogeneous Laplace equation.

Four explicit examples covering all four new periods of table~\ref{newLambdas} are
\begin{align}
{\rm F}^{-(6)}_{2,2,5} \, \big|^{\Delta_{2s}}_{\text{mod\, depth 1}} &= 
\left(
-\frac{8599 \Lambda(\Delta_{12}, 14)}{  10945440 \Lambda(\Delta_{12}, 10)} - \frac{\Lambda^{3, 2}_{4, 14}}{\Lambda(\Delta_{12}, 10)}
\right)
\frac{\bplusno{5}{\Delta_{12}}\!{-} \bminusno{5}{\Delta_{12}}}{90000}
+\ldots \, ,
\notag \\
{\rm F}^{+(6a)}_{2,2,6} \, \big|^{\Delta_{2s}}_{\text{mod\, depth 1}} &= 
\left(
\frac{13934837129 \Lambda(\Delta_{12}, 15)}{1082489265000 \Lambda(\Delta_{12},11)} 
  + \frac{\Lambda^{3, 3}_{4, 16}}{  \Lambda(\Delta_{12}, 11)}
\right)
\frac{\bplusno{5}{\Delta_{12}}\!{+} \bminusno{5}{\Delta_{12}}}{10365000}
+\ldots\, ,
\notag 
\\
{\rm F}^{+(8)}_{2,3,5} \, \big|^{\Delta_{2s}}_{\text{mod\, depth 1}} &=  - \frac{13 \Lambda^{2, 2}_{6, 14}}{
 12600 \Lambda(\Delta_{16}, 15)}
\big(\bplusno{7}{\Delta_{16}}\!{+} \bminusno{7}{\Delta_{16}} \big) +\ldots\, ,
\notag
\\
{\rm F}^{+(6)}_{2,3,5} \, \big|^{\Delta_{2s}}_{ \text{mod\, depth 1}} &= 
\left(\frac{13929 \Lambda(\Delta_{12}, 15)}{425656 \Lambda(\Delta_{12}, 11)}+ \frac{13 \Lambda^{4,2}_{6,14}}{
 \Lambda(\Delta_{12}, 11)}\right)
\frac{\bplusno{5}{\Delta_{12}}\!{+} \bminusno{5}{\Delta_{12}}}{90000} + \ldots\, .
\end{align}
The notation $|^{\Delta_{2s}}_{\text{mod\, depth 1}} $ indicates that we are only selecting the cuspidal contributions at modular depth one, while mixed integrals
$ \bpmno{ j_1 &j_2 }{k &\Delta_{2s}} $, $ \bpmno{ j_1 &j_2 }{\Delta_{2s} &k} $ or
$ \bpmno{ j_1 }{k }   \bmpno{j_2 }{\Delta_{2s} } $
are not tracked on the right-hand sides.

\section{Modular properties}
\label{sec:6}

In this section we show that the generating series $\Jeqv $ introduced in \eqref{uplift} indeed generates equivariant iterated integrals.
Given the modular properties (\ref{modbeqv}) satisfied by $\beta^{\rm eqv}$,
equivariance can be recast at the level of the generating series \eqref{uplift} as
\begin{align}
&R\big[ \Jeqv (\eee_k;\tau{+}1)\big] = R\big[ \Jeqv (\eee_k;\tau)\big] \, ,\label{mt.04} \\
&R\big[\Jeqv \big(\eee_k;{-}\,\tfrac{1}{\tau}\big)\big] \,=  \bar \tau^{-[\eee_0 , \eee_0^\vee]}  R\big[\Jeqv (\eee_k;\tau) \big]\bar \tau^{[\eee_0 , \eee_0^\vee]}\,.\label{mt.04S}
\end{align}
This can be easily seen by realising that the desired modular properties \eqref{modbeqv} imply invariance under $T$-transformation, $T\cdot\tau \coloneqq \tau{+}1$, of the corresponding generating series.
While under $S$-transformation, $S\cdot \tau\coloneqq -1/\tau$, we note that given a word $P=\left[\begin{smallmatrix} j_1 & ... & j_\ell \\ k_1 & ... & k_\ell \end{smallmatrix}\right]$ equation \eqref{modbeqv} is equivalent to
\begin{align}
R\big(\eee[P] \beta^{\rm eqv}\! \left[P;-\tfrac{1}{\tau}\right]\big)
&= \bar \tau^{-[\eee_0 , \eee_0^\vee]}  R\big(\eee[P] \beta^{\rm eqv}\! \left[P;\tau\right]\big) \bar\tau^{[\eee_0 , \eee_0^\vee]} \,,
\end{align} 
where we used the definition \eqref{MGFtoBR22}, adapted to free-algebra generators
\beq
\eee[P] \coloneqq
\bigg( \prod_{i=1}^\ell \frac{ (-1)^{j_i}(k_i{-}1) }{(k_i{-}j_i{-}2)!} \bigg)
\eee^{(k_\ell-2-j_\ell)}_{k_\ell} \cdots \eee^{(k_2-2-j_2)}_{k_2}\eee^{(k_1-2-j_1)}_{k_1}
\, , \label{adaptdef}
\eeq
combined with the $R$-operation defined in \eqref{MMVsec.02}, and the following eigenvalues under the SL$_2$ Cartan element $[\eee_0 , \eee_0^\vee]$
\beq
[\eee_0 , \eee_0^\vee] \eee_k^{(j)} = (2{+}2j{-}k) \eee_k^{(j)}
\ \ \Rightarrow \ \
[\eee_0 , \eee_0^\vee] \eee\!\left[\begin{smallmatrix} j_1 & ... & j_\ell \\ k_1 & ... & k_\ell \end{smallmatrix}\right]
= \sum_{i=1}^\ell (2{+}2j_i{-}k_i) \eee\!\left[\begin{smallmatrix} j_1 & ... & j_\ell \\ k_1 & ... & k_\ell \end{smallmatrix}\right]
\,.
\label{mt.03}
\eeq
Note that for the rest of the section we will consider general words $P$, where the entries can be both Eisenstein series and holomorphic cusp forms. For this reason we keep the derivations $\eee_k$ and $\eee_{\Delta^{\pm}}$ as independent symbols for part of the computation and eventually substitute the relations (\ref{edeltaex}) expressing $\eee_{\Delta^{\pm}}$ in terms of $\eee_k$. Along the road, we will comment on the key differences between this more general case and the special case where we restrict to Tsunogai's derivations, i.e.\ $\eee_k\to\epsilon_k$ and $\eee_{\Delta^{\pm}}\to0$.

The remainder of this section is devoted to showing that the right-hand side of \eqref{uplift} satisfies the equivariance properties \eqref{mt.04}.
To this end let us show separately $T$-equivariance and $S$-equivariance.

\subsection{\texorpdfstring{$T$-equivariance}{T-equivariance}}
\label{sec:Tequiv}
We want to show that the right-hand side of \eqref{uplift} is equivariant under $T$-transformation $T\cdot \tau = \tau{+}1$.
Given the definition \eqref{eq:Utau} of the ${\rm SL}_2$ transformation $\Unew(\tau)$, it is immediate to conclude that
\begin{equation}
\Unew(\tau{+}1) = \Unew(\tau) e^{2\pi i \eee_0}\,.\label{UT}
\end{equation}

Let us now turn our attention towards obtaining the $T$-transformation for the generating series $\mathbb{I}_\pm$ using their representations via path-ordered exponentials \eqref{eq:Pexp} constructed using the $(1,0)$-form $\mathbb{A}_+(\eee_k;\tau_1)$ and the $(0,1)$-form $\mathbb{A}_-(\eee_k;\tau_1)$, respectively.

Starting with $\Ip$, a simple application of the $T$-transformation \eqref{eq:nukerT} for the $\nu$-kernels combined with the $\mathfrak{sl}_2$ relations \eqref{eq:ade0e0chH} shows that $\mathbb{A}_+(\eee_k;\tau_1)$ defined in \eqref{mt.06} transforms as
\beq
\mathbb{A}_+(\eee_k;\tau_1{+}1) =
e^{-2\pi i \eee_0}
\mathbb{A}_+(\eee_k;\tau_1)  e^{2\pi i \eee_0}\,.
\label{mt.08}
\eeq
The path-ordered exponential then transforms both through the
$T$-action \eqref{mt.08} on the integrand and through the transformation
of the integration domain 
\begin{align}
\Ip(\eee_k;\tau{+}1) &=\Pexp \bigg( \int_{\tau{+}1}^{i\infty} \mathbb{A}_+(\eee_k;\tau_1) \bigg)
=\Pexp \bigg( \int_{\tau}^{i\infty-1} \mathbb{A}_+(\eee_k;\tau_1{+}1) \bigg)
\label{mt.09} \\
&=\Pexp \bigg( \int_{i\infty}^{i\infty-1} \mathbb{A}_+(\eee_k;\tau_1{+}1) \bigg)
\Pexp \bigg( \int_{\tau}^{i\infty} \mathbb{A}_+(\eee_k;\tau_1{+}1) \bigg)
\notag \\
&=  e^{-2\pi i \eee_0}\Pexp \bigg( \int_{i\infty}^{i\infty-1} \mathbb{A}_+(\eee_k;\tau_1) \bigg)
 \Ip(\eee_k;\tau) e^{2\pi i \eee_0} \notag\\
&=  e^{-2\pi i \eee_0}
\mathbb{T}_+(\eee_k)^{-1}
 \Ip(\eee_k;\tau) e^{2\pi i \eee_0} \, .\notag
\end{align}
It remains to simplify the $\tau$-independent path-ordered exponential 
\begin{equation}
\mathbb T_+(\eee_k)\coloneqq  \Pexp \bigg( \int_{i\infty-1}^{i\infty} \mathbb{A}_+(\eee_k;\tau_1) \bigg)
\,,\label{eq:Tcocyle}
\end{equation} i.e.\ the $T$-cocycle at infinity.
This integral can be evaluated via tangential-basepoint
regularisation (see in particular section 6 of \cite{Brown:mmv}) effectively selecting
contributions only from Eisenstein series and in particular from their zero-mode ${\rm G}_k(\tau) = 2 \zeta_k + { O}(e^{2\pi i \tau})$.
This means that to evaluate the $T$-cocycle at infinity we can just focus on the zero-mode part of $\mathbb{A}_+(\eee_k;\tau_1)$ which, using the identity $2 \zeta_k = - \BF_k (2\pi i)^k$ with $\BF_k= {\rm B}_k/k!$, reduces to
\beq
\mathbb{A}_+^{i\infty}(\eee_k;\tau_1) = - 2\pi i \sum_{k=4}^\infty (k{-}1) \BF_k
 \sum_{j=0}^{k-2} \frac{1}{j!} \, (-2\pi i \tau_1)^j \eee_k^{(j)}\,.
\label{mt.10}
\eeq
The  inverse $T$-cocycle at infinity can then be computed from the zero modes as 
\begin{align}
\mathbb T_+(\eee_k)^{-1}
&=  \Pexp \bigg( \int_{i\infty}^{i\infty-1} \mathbb{A}^{i\infty}_+(\eee_k;\tau_1) \bigg)
\label{mt.11} \\
&= 1+ \sum_{r=1}^{\infty} \sum_{j_1,j_2,\ldots,j_r=0}^{\infty}
\bigg( \prod_{i=1}^r \frac{ (2\pi i)^{j_i+1} }{j_i! \sum_{m=i}^r (j_m{+}1)} \bigg)   \hat N_+^{(j_1)} 
\hat N_+^{(j_2)} \ldots  \hat N_+^{(j_r)} \,,
\notag
\end{align}
where we introduced the
variant
\beq
\hat{N}_+ \coloneqq  \sum_{k=4}^{\infty} (k{-}1) \BF_k \eee_k \, , \ \ \ \ \ \
\hat{N} = \hat{N}_+  - \eee_0\,,
\label{mt.12}
\eeq
of the generator $\hat{N}$ in \eqref{zetgen.06} subject to $[\hat N,\sigmaE_w]=0$.

It is particularly useful to express the $T$-cocycle at infinity via \eqref{mt.11} as it allows us to prove the following lemma
which should be equivalent to the results in section 6 of \cite{Brown:mmv}.

\begin{lemma} 
\label{lem2}
The expression (\ref{mt.11}) for the inverse $T$-cocycle at infinity admits the generating-series
representation
\beq
\mathbb T_+(\eee_k)^{-1} = e^{2\pi i \eee_0} 
 e^{2\pi i (\hat N_+ - \eee_0)} = e^{2\pi i \eee_0} 
 e^{2\pi i  \hat N } \,.
 \label{mt.13}
\eeq
\end{lemma}
The proof of this lemma can be found in appendix \ref{app:Tcocyc}.

Finally, by combining (\ref{mt.09}) with (\ref{mt.13}), we arrive at the
modular $T$-transformation
\beq
\Ip(\eee_k;\tau{+}1) = e^{2\pi i  \hat N} \Ip(\eee_k;\tau) e^{2\pi i \eee_0}\, .
 \label{mt.14}
\eeq
Note that, since cusp forms do not contribute to the $\tau_1 \rightarrow i\infty$
regime in \eqref{mt.10}, the $T$-cocycle at infinity \eqref{mt.11} takes the same form
upon restriction to Eisenstein series and Tsunogai's derivations, i.e.\ $\eee_k\to \epsilon_k\,,\,\eee_{\Delta^{\pm}}\to 0 $.
Hence, the same modular $T$-transformation \eqref{mt.14} holds for $\Ip(\epsilon_k;\tau)$, i.e. 
\beq
\Ip(\ep_k;\tau{+}1) = e^{2\pi i N} \Ip(\ep_k;\tau) e^{2\pi i \ep_0}\, ,
 \label{mt.14Tsu}
\eeq
where $N$ in (\ref{zetgen.04}) is the specialisation of $\hat N$ to $\eee_k\to \epsilon_k$.

A similar argument can be repeated for $\Imint$ starting from the path-ordered exponential representation \eqref{eq:Pexp} in terms of the $(0,1)$-form $\mathbb{A}_-(\eee_k;\tau_1)$ defined in \eqref{mt.07}.
Given the immediate analogue of (\ref{mt.08}),
\beq
\mathbb{A}_-(\eee_k;\tau_1{+}1) =
e^{-2\pi i \eee_0}
\mathbb{A}_-(\eee_k;\tau_1)  e^{2\pi i \eee_0}\,,
\label{mt.00}
\eeq
the steps in (\ref{mt.09}) can be adapted with the reversal prescription to arrive at
\begin{align}
\Imint(\eee_k;\tau{+}1) 
&= e^{-2\pi i \eee_0}\, \Imint(\eee_k;\tau) \,
 \Ptexp \bigg( \int_{-i\infty}^{-i\infty-1} \mathbb{A}_-(\eee_k;\tau_1) \bigg) e^{2\pi i \eee_0}\,. \label{mt.15}
\end{align}
A computation similar to (\ref{mt.11}) yields the inverse $T$-cocycle at infinity 
\beq
\widetilde{\mathbb T}_-(\eee_k)^{-1}\coloneqq \Ptexp \bigg( \int_{-i\infty}^{-i\infty-1} \mathbb{A}_-(\eee_k;\tau_1) \bigg) 
= e^{- 2\pi i (\hat N_+ -\eee_0)} e^{- 2\pi i \eee_0}= e^{- 2\pi i \hat N} e^{- 2\pi i \eee_0} \,.
 \label{mt.18}
\eeq
 Inserting this expression in (\ref{mt.15}), we conclude that
\beq
\Imint(\eee_k;\tau{+}1) = e^{-2\pi i \eee_0} \Imint(\eee_k;\tau) e^{- 2\pi i \hat N}\,.
 \label{mt.19}
\eeq
The same modular $T$-transformation with $ \eee_0 \rightarrow  \ep_0$ and $\hat N \rightarrow N$
 applies to the generating series of Tsunogai's derivations, i.e. $\eee_k\to\epsilon_k\,,\,\eee_{\Delta^{\pm}}\to 0$, by the arguments below (\ref{mt.14}).

We are now in position of computing the $T$-transformation of \eqref{uplift}.
By inserting the $T$-transformations (\ref{UT}), (\ref{mt.14}) and
(\ref{mt.19}) into that of $R[\Jeqv]$ in \eqref{uplift}, it follows that
\begin{align}
&\nn \!\! R\big[\Jeqv(\eee_k;\tau{+}1)\big] = \Unew(\tau{+}1) \Msv(\zetaE_i, \zetaE_\varpi)^{-1} \Imint(\eee_k;\tau{+}1) 
\mathbb \Msv(\sigmaE_i,\sigmaE_\pern)  \Ip(\eee_k;\tau{+}1) \Unew^{-1}(\tau{+}1)\\
 &\!\!= \Unew(\tau) e^{2\pi i \eee_0} \Msv(\zetaE_i , \zetaE_\varpi)^{-1} e^{-2\pi i \eee_0}  \Imint(\eee_k;\tau )  e^{-2\pi i \hat N}
\Msv(\sigmaE_i,\sigmaE_\pern) e^{2\pi i  \hat N} \Ip(\eee_k;\tau ) \Unew^{-1}(\tau )\,,
 \label{mt.21}
\end{align}
which manifests that $T$-invariance of $\Jeqv(\eee_k;\tau)$ is ensured by the following conditions:
\beq
e^{2\pi i \eee_0} \mathbb M^{\rm sv}(\zetaE_i, \zetaE_\varpi)^{-1} e^{-2\pi i \eee_0} =  \mathbb M^{\rm sv}(\zetaE_i, \zetaE_\varpi)^{-1}
\, , \ \ \ \  
e^{-2\pi i \hat N}\mathbb M^{\rm sv}(\sigmaE_i,\sigmaE_\pern) e^{2\pi i \hat N}= 
\mathbb M^{\rm sv}(\sigmaE_i,\sigmaE_\pern)  \, .
 \label{mt.22}
\eeq
The first equation is implied by
imposing SL$_2$-invariance of
$\zetaE_i,\zetaE_\varpi$
as already anticipated in~\eqref{zvarpi.03}.
The second equation of (\ref{mt.22}) in turn is satisfied if $[\hat{N},\sigmaE_i]=[\hat{N},\sigmaE_\pern]=0$.

We conclude by noting again that the above relations are still valid after we impose Tsunogai's relations, i.e.\ even after having substituted $\eee_k\to \epsilon_k$ and having set to zero all cusp-form contributions, $\eee_{\Delta^{\pm}}\to 0 $ and $\sigmaE_\pern, \zetaE_\varpi \to0$. In case of the single-valued series $ R[ \mathbb J^{\rm sv}(\epsilon_k;\tau)]$ in (\ref{svieis.05}), however, the analogue of (\ref{mt.21}) 
\begin{align}
 R\big[\mathbb J^{\rm sv}(\ep_k;\tau{+}1)\big] &=  \Unew(\tau) e^{2\pi i \ep_0} \Msv(\sigma_i )^{-1} e^{-2\pi i \ep_0}  \Imint(\ep_k;\tau )  e^{-2\pi i  N}
\Msv(\sigma_i) e^{2\pi i N} \Ip(\ep_k;\tau ) \Unew^{-1}(\tau )
 \label{mt.21sv}
\end{align}
does not produce $R[\mathbb J^{\rm sv}(\ep_k;\tau)]$ on the right-hand side since
$e^{2\pi i \ep_0} \Msv(\sigma_i )^{-1} e^{-2\pi i \ep_0} \neq \Msv(\sigma_i )^{-1} $. That is why the coefficients of $R[\mathbb J^{\rm sv}(\ep_k;\tau)]$ cannot possibly match the $T$-invariant $\beta^{\rm sv}$ as discussed below (\ref{svieis.07}).

\subsection{\texorpdfstring{$S$-equivariance}{S-equivariance}}
\label{sec:sequiv}
The analysis of the modular $S$ properties of the constituents of $\Ieqv,\Jeqv$
proceeds in a similar fashion. We start from the definition \eqref{eq:Utau} 
to conclude that
\begin{equation}
\Unew\big({-}\tfrac{1}{\tau}\big) = \bar \tau^{-[\eee_0 , \eee_0^\vee]}  \, \Unew(\tau)\, \weyl\, (2\pi i)^{-[\eee_0 , \eee_0^\vee]} \,,\label{US}
\end{equation}
where $\weyl$ denotes the Weyl reflection $\weyl=e^{\eee_0^\vee} e^{-\eee_0} e^{\eee_0^\vee}$.
This equation can be checked more easily in any faithful representation of SL$_2$, e.g.\ by considering the case $k=4$.

Upon conjugation by $\Unew\big({-}\tfrac{1}{\tau}\big)$, we see that it is sensible to define the ``S-modular'' transformation on the derivation $\eee_k^{(j)}$ as
\begin{align}
\label{mt.04a}
\eee_k^{(j)}\big|_{S} \coloneqq  \weyl \left[ (2\pi i)^{-[\eee_0 , \eee_0^\vee]}  \eee_k^{(j)} (2\pi i)^{[\eee_0 , \eee_0^\vee]} \right] \weyl^{-1}\,,
\end{align}
which, given the Weyl reflection \eqref{eq:Weyl} and the $\mathfrak{sl}_2$ relations \eqref{eq:ade0e0chH}, can be rewritten as
\beq
\eee_k^{(j)} \big|_{S} = (-1)^j (2\pi i)^{k-2-2j} \frac{ j! }{(k{-}j{-}2)!} \eee_k^{(k-j-2)} \,.
\label{mmvtob.02}
\eeq

Note that since the above relations are only sensitive to the $\mathfrak{sl}_2$ 
structure of the derivations $\eee_k$, (\ref{mmvtob.02}) holds in identical form for the cuspidal variables $\eee_{\Delta^{\pm}}$ and for Tsunogai's derivations, i.e.\ if we substitute  $\eee_0 \to \epsilon_0\,,\,\eee_0^\vee \to\epsilon_0^\vee$ and $\eee_k^{(j)}\to \epsilon_k^{(j)}$.

We can now turn our attention towards the $S$-transformations of the generating series~$\mathbb{I}_\pm$.
Starting again with $\Ip$, we can use the $S$-transformation of the $\nu$ kernels \eqref{eq:nukerS} and the definition \eqref{mt.06} of $\mathbb{A}_+$ to deduce 
\begin{align}
\mathbb{A}_+(\eee_k;-\tfrac{1}{\tau}) & = \mathbb{A}_+(\eee_k;\tau)\big\vert_S\,,
\end{align}
where the $S$-transformation on the right-hand side only acts on the derivations $\eee_k^{(j)}$, as given in \eqref{mmvtob.02} or equivalently \eqref{mt.04a}.

Computing the $S$-transformation of the path-ordered exponential \eqref{eq:Pexp} yields
\begin{align}
\Ip(\eee_k;-\tfrac{1}{\tau}) &\nn=\Pexp \bigg( \int_{-\frac{1}{\tau}}^{i\infty} \mathbb{A}_+(\eee_k;\tau_1) \bigg)
=\Pexp \bigg( \int_{\tau}^{0} \mathbb{A}_+(\eee_k;-\tfrac{1}{\tau_1}) \bigg) \\
&=\Pexp \bigg( \int_{i\infty}^{0} \mathbb{A}_+(\eee_k;-\tfrac{1}{\tau_1}) \bigg)
\Pexp \bigg( \int_{\tau}^{i\infty} \mathbb{A}_+(\eee_k;-\tfrac{1}{\tau_1}) \bigg)
\notag \\
&=   \big( \mathbb{S}_+(\eee_k)^{-1} \Ip(\eee_k;\tau)  \big)\big|_S\,,\label{eq:IpS1}
\end{align}
where we defined the tangentially regulated $S$-cocycle at infinity as
\begin{equation}
 \mathbb{S}_+(\eee_k) \coloneqq \Pexp \bigg( \int_{0}^{i\infty} \mathbb{A}_+(\eee_k;\tau_1) \bigg)\,,
\label{spluscocyc}
\end{equation} 
or equivalently the generating series of all MMVs \eqref{eq:mmvdef} and \eqref{MMVsec.17b}
\begin{align}
&\mathbb S_{+}(\eee_k) =  1 + \sum_{k_1=4}^{\infty}\sum_{j_1=0}^{k_1-2} (-1)^{j_1} \frac{(k_1{-}1)}{j_1!}
( 2\pi i)^{j_1+1-k_1} \bigg\{ \MMV{j_1}{k_1} \eee_{k_1}^{(j_1)} + \!\!\sum_{\Delta_{k_1}\in \mathcal{S}_{k_1}}\!\! \MMV{j_1}{\Delta_{k_1}}\eee_{\Delta^{+}_{k_1}}^{(j_1)} \bigg\} \label{mmvtob.23}   \\
&\quad  + \sum_{k_1=4}^{\infty}\sum_{j_1=0}^{k_1-2}\sum_{k_2=4}^{\infty}\sum_{j_2=0}^{k_2-2} (-1)^{j_1+j_2} \frac{(k_1{-}1)(k_2{-}1)}{j_1!j_2!}
 ( 2\pi i)^{j_1+j_2+2-k_1-k_2} 
  \bigg\{ \MMV{j_1 & j_2}{k_1 & k_2}  \eee_{k_1}^{(j_1)}{\eee_{k_2}^{(j_2)}}
  \notag \\
&\quad \quad \quad\quad \quad \quad\quad \quad \quad \quad \quad \quad \notag 
 + \sum_{\Delta_{k_2}\in \mathcal{S}_{k_2}} \MMV{j_1 & j_2}{k_1 & \Delta_{k_2}} \eee_{k_1}^{(j_1)}\eee_{\Delta^{+}_{k_2}}^{(j_2)}
+\sum_{\Delta_{k_1}\in \mathcal{S}_{k_1}} \MMV{j_1 & j_2}{\Delta_{k_1} & k_2}
\eee_{\Delta^{+}_{k_1}}^{(j_1)}
\eee_{k_2}^{(j_2)} \\
&\quad \quad \quad\quad \quad \quad\quad \quad \quad\quad \quad \quad  \notag + \sum_{\Delta_{k_1}\in \mathcal{S}_{k_1}}\sum_{\Delta_{k_2}\in \mathcal{S}_{k_2}} \MMV{j_1 & j_2}{\Delta_{k_1} & \Delta_{k_2}}
\eee_{\Delta^{+}_{k_1}}^{(j_1)}
\eee_{\Delta^{+}_{k_2}}^{(j_2)}
\bigg\} + \ldots \,,\notag
 \end{align}
with terms of modular depth greater than or equal to three in the ellipsis.
In contrast to the $T$-cocycle at infinity in \eqref{mt.11}, the
$S$-cocycle at infinity receives contributions from both Eisenstein series and cusp forms
and considerably simplifies when we specialise to Tsunogai's derivations, i.e.\ $\eee_k\to\epsilon_k$ and $\eee_{\Delta^{\pm}}\to0$. In this case, the $S$-cocycle at infinity reduces to the generating series of the 
MMVs \eqref{eq:mmvdef} from Eisenstein series,
\begin{align}
 \mathbb{S}_+(\epsilon_k) &= 1 + \sum_{k_1=4}^\infty \sum_{j_1=0}^{k_1-2} \frac{(-1)^{j_1} (k_1{-}1) }{j_1!}
 (2\pi i)^{j_1+1-k_1}  \MMV{j_1}{k_1} \epsilon_{k_1}^{(j_1)}    \label{mmvtob.05}  \\
 &\quad + \sum_{k_1,k_2=4}^\infty \sum_{j_1=0}^{k_1-2}   \sum_{j_2=0}^{k_2-2} \frac{(-1)^{j_1+j_2} (k_1{-}1)(k_2{-}1) }{j_1! j_2!}
 (2\pi i)^{j_1+j_2+2-k_1-k_2}  \MMV{j_1 &j_2}{k_1 &k_2} \epsilon_{k_1}^{(j_1)} \epsilon_{k_2}^{(j_2)}
 +\ldots \,.
\notag
\end{align}
Note that we can find an alternative representation of \eqref{eq:IpS1} using the identity
\begin{equation}
\MMV{k_1{-}j_1{-}2 &k_2{-}j_2{-}2 & \ldots & k_\ell{-}j_\ell{-2}}{k_1&k_2& \ldots&k_\ell} = (-1)^{\sum_{i=1}^\ell (j_i+1)} \MMV{j_\ell & \ldots &j_2 &j_1}{k_\ell&\ldots &k_2& k_1}\,,\label{reflmmv}
\end{equation}
and similar generalisations upon replacing Eisenstein integration kernels with 
holomorphic cusp forms. This identity translates into 
 \begin{equation}
\mathbb S_+(\eee_k) \big|_S = \mathbb S_+(\eee_k)^{-1}\,,
 \label{mt.29}
 \end{equation}
at the level of generating series which leads to the
equivalent form for the $S$-transformation \eqref{eq:IpS1} of $\Ip$ given by
\beq
\Ip\big(\eee_k;{-}\tfrac{1}{\tau} \big) =\mathbb S_+(\eee_k) \big( \Ip(\eee_k;\tau) \big|_S \big)\,.
 \label{mt.28}
\eeq
Similar arguments apply to the $S$-transformation of $\Imint$. 
Firstly, from \eqref{eq:nukerS} and the definition \eqref{mt.07}, we deduce 
\begin{align}
\mathbb{A}_-(\eee_k;-\tfrac{1}{\tau}) & = \mathbb{A}_-(\eee_k;\tau)\big\vert_S\,.
\end{align}
From this equation we can proceed as above and write
\begin{align}
\Imint(\eee_k;-\tfrac{1}{\tau}) 
&= \bigg(\,\Imint(\eee_k;\tau) \,
 \Ptexp \bigg( \int_{-i\infty}^{0} \mathbb{A}_-(\eee_k;\tau_1) \bigg) \bigg) \Big\vert_S\notag \\
 &= \Big(\,\Imint(\eee_k;\tau) \,
\mathbb{S}_-(\eee_k)\Big) \Big\vert_S\,,  \label{mt.15new}
\end{align}
where in the last line we introduced the $S$-cocycle at infinity\footnote{The careful reader may be confused as to why our definition (\ref{eq:IpS1}) for $\mathbb{S}_+$ involves the path ordered exponential of $\int_{0}^{i\infty} \mathbb{A}_+(\eee_k;\tau_1)$ while the present definition (\ref{mt.15new}) for $\mathbb{S}_-$ is written in terms of $\int_{-i\infty}^{0} \mathbb{A}_-(\eee_k;\tau_1)$. The reason for this apparent discrepancy lies in the two different conventions for the path ordered exponentials, $\Pexp$ and the reverse ordering $\Ptexp$, as defined in \eqref{eq:Pexp}. Upon expanding (\ref{mt.15new}), the coefficient of a generic term $ \eee_{k_1}^{(j_1)}\cdots \eee_{k_\ell}^{(j_\ell)}$ is given by an iterated integral over the domain $0<\bar{\tau}_\ell<...<\bar{\tau}_1 <-i\infty$. From \eqref{eq:mmvdef} we see that this is precisely the correct structure to produce the combination $ \eee_{k_1}^{(j_1)}\cdots \eee_{k_\ell}^{(j_\ell)}\overline{ \MMV{j_1  &\ldots &j_\ell}{k_1  &\ldots &k_\ell} }$.}
\begin{equation}
\mathbb S_- (\eee_k)\coloneqq \Ptexp \bigg( \int_{-i\infty}^{0} \mathbb{A}_-(\eee_k;\tau_1) \bigg)\,.
\label{defscoc}
\end{equation}
Based on tangential-basepoint regularisation and the reality properties
\beq
\overline{ \MMV{j_1 &j_2 &\ldots &j_\ell}{k_1 &k_2 &\ldots &k_\ell} } =
(-1)^{\sum_{i=1}^\ell (j_i+1)}\MMV{j_1 &j_2 &\ldots &j_\ell}{k_1 &k_2 &\ldots &k_\ell}\,,
\eeq
of MMVs \eqref{eq:mmvdef} and \eqref{MMVsec.17b} (which hold in
identical form for $k_i \rightarrow \Delta_{k_i}$), the expansion of
(\ref{defscoc}) takes a form analogous to (\ref{mmvtob.23}),
\begin{align}
&\mathbb S_{-}(\eee_k) =  1 + \!\sum_{k_1=4}^{\infty}\sum_{j_1=0}^{k_1-2} (-1)^{j_1} \frac{(k_1{-}1)}{j_1!}
(- 2\pi i)^{j_1+1-k_1} \!\bigg\{ \MMV{j_1}{k_1} \eee_{k_1}^{(j_1)} + \!\!\sum_{\Delta_{k_1} \in \mathcal{S}_{k_1}} \!\!\!
\MMV{j_1}{\Delta_{k_1}} 
 \eee_{\Delta^{-}_{k_1}}^{(j_1)}\bigg\} \label{mmvtob.23m} \\
 &\notag+ \sum_{k_1=4}^{\infty}\sum_{j_1=0}^{k_1-2}\sum_{k_2=4}^{\infty}\sum_{j_2=0}^{k_2-2} (-1)^{j_1+j_2} \frac{(k_1{-}1)(k_2{-}1)}{j_1!j_2!}
 (- 2\pi i)^{j_1+j_2+2-k_1-k_2} \bigg\{ \MMV{j_1 & j_2}{k_1 & k_2}  \eee_{k_1}^{(j_1)}\eee_{k_2}^{(j_2)} \notag \\
&\notag \quad \quad \quad\quad \quad \quad\quad \quad \quad \quad \quad \quad + \sum_{\Delta_{k_2}\in \mathcal{S}_{k_2}}  \MMV{j_1 & j_2}{k_1 & \Delta_{k_2}} \eee_{k_1}^{(j_1)}\eee_{\Delta^{-}_{k_2}}^{(j_2)} +\sum_{\Delta_{k_1}\in\mathcal{S}_{k_1}}   \MMV{j_1 & j_2}{\Delta_{k_1} & k_2}
\eee_{\Delta^{-}_{k_1}}^{(j_1)} 
\eee_{k_2}^{(j_2)} \\
& \quad \quad \quad\quad \quad \quad\quad \quad \quad \quad \quad \quad +\sum_{\Delta_{k_1}\in\mathcal{S}_{k_1}} \sum_{\Delta_{k_2}\in\mathcal{S}_{k_2}}  \MMV{j_1 & j_2}{\Delta_{k_1} & \Delta_{k_2}}
\eee_{\Delta^{-}_{k_1}}^{(j_1)} 
\eee_{\Delta^{-}_{k_2}}^{(j_2)} \bigg\}+ \ldots \, . \notag
\end{align}

Once more, we can easily specialise to Tsunogai's derivations by substituting $\eee_k\to \epsilon_k$ and $\eee_{\Delta^\pm}\to0$. All MMVs involving cusp forms drop out, and we retrieve the modified generating series of Eisenstein MMVs \eqref{eq:mmvdef}
\begin{align}
\mathbb S_{-} (\epsilon_k)&= 1 + \sum_{k_1=4}^\infty \sum_{j_1=0}^{k_1-2} \frac{(-1)^{j_1} (k_1{-}1) }{j_1!}
 (-2\pi i)^{j_1+1-k_1}  \MMV{j_1}{k_1} \epsilon_{k_1}^{(j_1)}    \label{mmvtob.06} \\
 &\quad + \!\! \sum_{k_1,k_2=4}^\infty \sum_{j_1=0}^{k_1-2}   \sum_{j_2=0}^{k_2-2} \frac{(-1)^{j_1+j_2} (k_1{-}1)(k_2{-}1) }{j_1! j_2!}
 (-2\pi i)^{j_1+j_2+2-k_1-k_2}  \MMV{j_1 &j_2}{k_1 &k_2} \epsilon_{k_1}^{(j_1)} \epsilon_{k_2}^{(j_2)}
 +\ldots \,.
\notag
\end{align}
We can again use the reflection identities (\ref{reflmmv}) between MMVs to deduce
\begin{equation}
\mathbb{S}_-(\eee_k) \big\vert_S = \mathbb{S}_-(\eee_k)^{-1}\,,
\end{equation}
and upon substitution into \eqref{mt.15new} arrive at
\begin{equation}
\Imint(\eee_k;-\tfrac{1}{\tau}) =
\Big(\,\Imint(\eee_k;\tau) \,
 \Big\vert_S\Big)\,\mathbb{S}_-(\eee_k)^{-1}\,.\label{ImintS}
\end{equation} 
We can now assemble the modular $S$-transformations of $\Jeqv$
in \eqref{uplift} from those of its constituents presented in
 \eqref{US}, \eqref{mt.28} and \eqref{ImintS}:
\begin{align}
 R\big[ \Jeqv\big(\eee_k;{-}\tfrac{1}{\tau} \big)\big] 
&= \bar \tau^{-[\eee_0 , \eee_0^\vee]}  \Unew(\tau)  \mathbb M^{\rm sv}(\zetaE_i, \zetaE_\varpi)^{-1} \Imint(\eee_k;\tau) 
\notag \\
&\quad \times \Big( \mathbb S_-(\eee_k)^{-1} \mathbb M^{\rm sv}(\sigmaE_i,\sigmaE_\pern) \mathbb S_+(\eee_k)
\Big) \Big|_{S} \, \Ip(\eee_k;\tau)  \Unew^{-1}(\tau) \bar \tau^{[\eee_0 , \eee_0^\vee]}  \notag \\
&=\bar \tau^{-[\eee_0 , \eee_0^\vee]} R \big[\Jeqv(\eee_k;\tau)\big]\bar \tau^{[\eee_0 , \eee_0^\vee]}\,.
\label{sequvprf}
\end{align}
Here we used the fact that the generating series $\mathbb M^{\rm sv}(\zetaE_i,\zetaE_\varpi)$ of arithmetic derivations $\zetaE_i,\zetaE_\varpi$ is unaffected by $\vert_S$. More crucially, in passing to the last line of (\ref{sequvprf}), we have conjectured that Brown's formula in section 7.2 of \cite{Brown:2017qwo2}
\beq
\mathbb S_-(\epsilon_k)^{-1} \mathbb M^{\rm sv}(\sigmaT_i) \mathbb S_+(\epsilon_k) = \mathbb M^{\rm sv}(\sigmaT_i)\, \big|_S\,,
\eeq
proven to be valid for the Tsunogai's derivations $\epsilon_k$ and the associated zeta generators $\sigma_i$, 
can be extended to hold on the more general derivations $\eee_k\,,\,\eee_{\Delta^\pm}, \, \sigmaE_i,\, \sigmaE_\pern$ by 
\beq
\mathbb S_-(\eee_k)^{-1} \mathbb M^{\rm sv}(\sigmaE_i,\sigmaE_\pern) \mathbb S_+(\eee_k) = \mathbb M^{\rm sv}(\sigmaE_i,\sigmaE_\pern)\, \big|_S\,.
\label{beyond1708}
\eeq
This conjecture is established up to and including modular depth three and \degree\ $20$ since we have confirmed the modular properties (\ref{modbeqv}) of all the $\beqv{j_1 &j_2  &j_3}{k_1 &k_2 &k_3}$ up to these modular depths and \degree. In section \ref{sec:gen}, we prove the existence of a series $\mathbb M^{\rm sv}(\sigmaE_i,\sigmaE_\pern)$ subject to (\ref{beyond1708}) at arbitrary degree and modular depth. However, as discussed at the end of section \ref{sec:corr}, this proof does not determine the number-theoretic structure of $\mathbb M^{\rm sv}(\sigmaE_i,\sigmaE_\pern)$. In other words, the methods of section \ref{sec:gen} do not predict the new periods at a given degree and modular depth. Moreover, the relations between the coefficients of $\rho^{-1} 
( {\rm sv}(f_{i_1} f_{i_2} \ldots f_{i_\ell}) )$ and those of $\zeta_{i_k}$ as implied by the expansion (\ref{Msvz}) of the MZV sector $\mathbb M^{\rm sv}(\sigmaE_i)$ remain conjectural.

\section{\texorpdfstring{Reformulation in terms of commutative $X_i, Y_i$}{Reformulation in terms of commutative X, Y}}
\label{sec:8}

In this section, we combine various iterated integrals with the help of auxiliary commutative variables $(X_i,Y_i)$ for each $1 \leq i \leq r$, where $r$ is the modular depth. Each pair of commutative variables transforms as a doublet under $\SLtwoZ$:
\begin{align}
\begin{pmatrix} X_i\\Y_i\end{pmatrix} 
    \mapsto 
    \begin{pmatrix}
         a X_i+ b Y_i\\ c X_i +d Y_i
    \end{pmatrix}
    \quad\quad\text{with}\quad
    \gamma=\begin{pmatrix}a&b \\c &d\end{pmatrix}\in \SLtwoZ \,.
    \label{trfxandy}
\end{align}
Thus the combination
\begin{align}
\label{eq:XYtau}
    X_i {-} \tau Y_i \mapsto (c\tau{+}d)^{-1} (X_i {-}\tau Y_i)\,,
\end{align}
transforms with modular weight $(-1,0)$. Combining powers of $(X_i{-}\tau Y_i)$ with holomorphic modular forms in the integration kernels will then produce {\it equivariant} expressions, i.e.\ modular invariants after taking the transformation (\ref{trfxandy}) of $X_i$ and $Y_i$ into account. These are the building blocks of the equivariant integrals introduced by Brown~\cite{Brown:2017qwo,Brown:2017qwo2} (see also~\cite{Drewitt:2021,Dorigoni:2022npe}) and that represent a different organisation of the same iterated integrals introduced in~\eqref{MMVsec.07}.

Factors $(X_i{-}\tau Y_i)^{k_i-2}$ of weight $(2{-}k_i,0)$ are homogeneous polynomials in the $X_i$ and $Y_i$ of degree $k_i{-}2$. Since the variables $(X_i,Y_i)$ are commuting and transform in a doublet, the transformation of $(X_i,Y_i)$ in $(X_i{-}\tau Y_i)^{k_i-2}$ under $\SLtwoZ$ is that of a $(k{-}1)$-dimensional representation isomorphic to $V(\eee_{k_i})$ in section~\ref{sec:rev.2.1}. 
Products $(X_i{-}\tau Y_i)^{k_i-2} (X_j{-}\tau Y_j)^{k_j-2}$ with distinct bookkeeping variables $(X_i,Y_i)$ and $(X_j,Y_j)$ then yield polynomials transforming in the tensor product $V(\eee_{k_i}) \otimes V(\eee_{k_j})$. The projection to the various irreducible components of the tensor product can be achieved by applying certain differential operators in $(X_i,Y_i)$ and $(X_j,Y_j)$ \cite{Brown:2017qwo} that are reviewed below.

In this section we review such a  construction at modular depths up to three and show in particular in section~\ref{sec:modcomp3} that a completion of elementary integrals leads to a completely equivalent description of the modular forms discussed in the previous sections. Moreover, we will argue in section~\ref{sec:gen} that this construction generalises to arbitrary modular depth.

\subsection{Equivariant iterated integrals at modular depths one and two}
  \label{sec:mk.1}

In this section, we will explain the connection between the work of the previous sections and equivariant triple integrals. The construction of our triple integrals will involve functions produced from the double-integral case and follows a similar procedure (originally given by Brown in \cite{Brown:2017qwo} and also recapped in \cite{Dorigoni:2022npe}).

\subsubsection{Modular depth one}
  
A first step is to combine the integration kernels $\nuker{j}{k}{\tau}$  defined in (\ref{nuker}) to the $(1,0)$-form valued polynomial in the commutative bookkeeping variables $X$ and $Y$, 
 \begin{align}
\underline{\rm G}_k[X,Y;\tau] &\coloneqq \frac{ (k{-}1)! }{2 (2\pi i)^{k-1} } \, (X{-}\tau Y)^{k-2} \, {\rm G}_k(\tau)\, \dd \tau 
\label{defundG}\\
&=  \frac{1}{2}\,(k{-}1)! \sum_{j=0}^{k-2} (-1)^j \binom{k{-}2}{j} \, X^{k-2-j} \bigg( \frac{Y}{2\pi i}\bigg)^j \, \nuker{j}{k}{\tau}\, .
\notag
\end{align}
In view of~\eqref{eq:XYtau}, this $(1,0)$-form is equivariant under $\SLtwoZ$, i.e.\ invariant when transforming both $(X,Y)$ and $\tau$. 

We wish to use this $(1,0)$-form and its complex conjugate as integration kernels for a line integral. In order for the integral to be path-independent, the forms must be closed.
For a one-form of the type $\varphi(\tau) \coloneqq \varphi^+(\tau) + \varphi^-(\tau) \coloneqq \varphi^\tau(\tau)  {\rm d}\tau + \varphi^{\bar\tau}(\tau)  {\rm d}\bar\tau$, the closure condition $\dd \varphi=0$ translates into the following partial differential equations for its components $\varphi^{\tau}, \varphi^{\bar\tau}$ or constituent $(1,0)$- and $(0,1)$-forms $\varphi^{\pm}$:
\begin{equation}
\partial_{\bar\tau} \varphi^\tau = \partial_\tau \varphi^{\bar\tau} 
\quad\quad \Longleftrightarrow \quad\quad
\partial_{\bar\tau} \varphi^+ \wedge 2\dd\bar\tau = 2\dd \tau \wedge \partial_\tau \varphi^-\,.
    \label{clphi}
\end{equation}
Since $\underline{\GG}_k[X,Y;\tau]$ is holomorphic, closure is trivially satisfied,  and this form and its complex conjugate 
can be used to reorganise the depth-one modular forms $\beta^{\rm eqv}$ in (\ref{eq:Cab}) and (\ref{eq:beqv1}) as follows:
\begin{align}
    M_{k}[X,Y;\tau]  & \coloneqq -\frac{1}{2} \int_\tau^{i \infty} \underline{\rm G}_k[X,Y;\tau_1] 
- \frac{1}{2} \int_{\bar \tau}^{-i \infty} \overline{ \underline{\rm G}_k[X,Y;\tau_1]} 
+ \frac{(k{-}2)!  \zeta_{k-1}}{2(2\pi i)^{k-2}}  \, Y^{k-2} \notag \\
& = -  \frac{1}{4} (k{-}1)! 
\sum_{j=0}^{k-2}
 \frac{ \binom{k-2}{j}  }{(-4y)^{j}}
 \beqvtau{j}{k}{\tau}  
(X{-}\tau Y)^{j} (X{-}\bar \tau Y)^{k-2-j} \, .
\label{mkdepth1}
\end{align}
The last term  $\sim \zeta_{k-1} Y^{k-2}$ of the first line is engineered to attain the equivariant transformation law under $(\begin{smallmatrix}
    a &b \\ c &d
\end{smallmatrix}) \in \SLtwoZ$,
\beq
 M_{k}\bigg[aX{+}bY,cX{+}dY;\frac{a\tau{+}b}{c\tau{+}d} \bigg] = M_{k}[X,Y;\tau] \, ,
 \label{equivdpt1}
\eeq
i.e.\ to cancel the cocycle under the modular $S$-transformation of the first two terms of (\ref{mkdepth1}). By the modular weights $(-1,0)$ and $(0,-1)$ of $(X{-}\tau Y)$ and $(X{-}\bar \tau Y)$, the equivariance property (\ref{equivdpt1}) identifies the coefficients proportional to $ \beqvtau{j}{k}{\tau}/y^{j} $ of $(X{-}\tau Y)^{j} (X{-}\bar \tau Y)^{k-2-j}$ as  modular forms of weight $(j,\, k{-}j{-}2)$, reproducing the known modular properties of $ \beqvtau{j}{k}{\tau}$, namely $\beqvtau{j}{k}{\tfrac{a\tau+b}{c\tau+d}} =
(c\bar \tau{+}d)^{k-2j-2} \beqvtau{j}{k}{\tau}$. 
From expression~\eqref{mkdepth1} it is straightforward to check the differential equations
\begin{align}
\label{eq:dM1}
         2  \,  \partial_{\tau} M_k[\tau] \, {\rm d} \tau &= \underline{\rm G}_k[\tau]\, ,&
    2 \,  \partial_{\bar \tau}  M_k[\tau] \, {\rm d}\bar \tau &=  \overline{\underline{\rm G}_k[\tau]} \, .
\end{align}

In the remainder of this section, we will use the analogous fact that equivariant functions in several pairs of commutative variables $(X_i,Y_i), \ i=1,2,\ldots,r$ yield modular forms of weights $\sum_{i=1}^r (j_i,\, k_i{-}j_i{-}2)$ upon rewriting
\beq
X = \frac{ \tau (X{-} \bar \tau Y)}{\tau {-} \bar \tau} - \frac{ \bar \tau (X{-}  \tau Y)}{\tau {-} \bar \tau} \, , \ \ \ \ 
Y = \frac{ (X{-} \bar \tau Y)}{\tau {-} \bar \tau} - \frac{ (X{-}  \tau Y)}{\tau {-} \bar \tau}
\label{tauto}\,,
\eeq
and taking their coefficients of $\prod_{i=1}^r (X_i{-}\tau Y_i)^{j_i} (X_i{-}\bar \tau Y_i)^{k_i-2-j_i}$ \cite{Brown:2017qwo}. In the language
of section \ref{sec:4.1}, converting the coefficients of $Y_i^{j_i} X_i^{k_i-j_i-2}$ to those of $(X_i{-}\tau Y_i)^{j_i} (X_i{-}\bar \tau Y_i)^{k_i-2-j_i}$ amounts to passing from the holomorphic frame to the modular frame. In other words, the rewriting (\ref{tauto}) of expansion variables is an alternative way of implementing the SL$_2$ transformation $\Unew(\tau)$ given by (\ref{eq:Utau}).

\subsubsection{Towards equivariant integrals at modular depth two}

In \cite{Brown:2017qwo}, a crucial ingredient for the construction of the equivariant double integrals was an equivariant closed one-form of the type
 \beq
     D_{k_1,k_2}[X_1,Y_1,X_2,Y_2;\tau] \coloneqq -\frac{1}{2} \left(\underline{\rm G}_{k_1}[X_1,Y_1;\tau] M_{k_2}[X_2,Y_2;\tau]
+M_{k_1}[X_1,Y_1;\tau] \overline{ \underline{\rm G}_{k_2}[X_2,Y_2;\tau] } \right)\, ,
\label{dk1k2}
\eeq
which now depends on four commutative bookkeeping variables $X_1,Y_1,X_2,Y_2$, and where $\underline{\rm G}_{k_i}[X_i,Y_i;\tau]$ and $M_{k_j}[X_j,Y_j;\tau]$ are given by (\ref{defundG}) and (\ref{mkdepth1}), respectively.
Closure of $D_{k_1,k_2}$ in the sense of~\eqref{clphi} can be verified using~\eqref{defundG} and~\eqref{eq:dM1}.
As a closed one-form, $D_{k_1,k_2}$ in (\ref{dk1k2}) yields a well-defined function of $\tau$ upon integration,
\begin{align}
K_{k_1,k_2}[\tau] &\coloneqq 
\frac{1}{4}  \, \int_{\tau}^{i\infty}  \underline{\rm G}_{k_1}[\tau_1]  \!\int_{\tau_1}^{i\infty} \underline{\rm G}_{k_2}[\tau_2]  
   \, + \frac{1}{4} \, \int_{\tau}^{i\infty}   \underline{\rm G}_{k_1}[\tau_1]  
 \times \! \int_{\bar \tau}^{-i\infty} \overline{ \underline{\rm G}_{k_2}[\tau_2] } 
\label{Kdepthtwo} \\
 &\quad + \frac{1}{4} \, \int_{\bar \tau}^{-i\infty} \overline{ \underline{\rm G}_{k_2}[\tau_2] }   \!
\int_{\bar \tau_2}^{-i\infty} \overline{ \underline{\rm G}_{k_1}[\tau_1] }
 +  \frac{1}{2}\, c_{k_2}  \int_{\tau}^{i\infty}  \underline{\rm G}_{k_1}[\tau_1]  
  \, +  \frac{1}{2}\, c_{k_1}  \int_{\bar \tau}^{-i\infty} \overline{ \underline{\rm G}_{k_2}[\tau_2] }\,, 
  \notag 
\end{align}
where we employ the following shorthand in the last two terms,
\beq
c_{k_i} = c_{k_i}[Y_i] \coloneqq  - \frac{(k_i{-}2)! \zeta_{k_i-1}}{2(2\pi i)^{k_i-2}} \, Y_i^{k_i-2} \, .
\label{def:ck}
\eeq
To avoid cluttering (especially in later parts), from now on we will no longer display the dependence of $K_{k_1, k_2}$ and related objects on the commutative bookkeeping variables $X_i, Y_i$. We choose to keep the square brackets, however, to emphasise the connection to these variables. Furthermore, whenever this notation is used, the correct $X_i, Y_i$ can always be inferred by the values assigned to the subscript $k_i$, for example, $K_{k_1,k_2}[\tau]=K_{k_1,k_2}[X_1,Y_1,X_2,Y_2;\tau]$.

By the alternative expansions
\begin{align}
 \underline{\rm G}_{k}[X,Y;\tau_1]  &= \frac{1}{2}\, (k{-}1)! \sum_{j=0}^{k-2} \binom{k{-}2 }{ j } \, \frac{1}{(-4y)^j} \, (X{-}\tau Y)^j (X{-}\bar \tau Y)^{k-2-j} \omplus{j}{k}{\tau,\tau_1}
\, , \notag \\
\overline{ \underline{\rm G}_{k}[X,Y;\tau_1] } &= \frac{1}{2}\, (k{-}1)! \sum_{j=0}^{k-2} \binom{k{-}2}{j } \, \frac{1}{(-4y)^j} \, (X{-}\tau Y)^j (X{-}\bar \tau Y)^{k-2-j} \omminus{j}{k}{\tau,\tau_1}\,, \label{altgkexp} 
\end{align}
of the holomorphic $(1,0)$-form (\ref{defundG}) and its complex conjugate, the expression (\ref{Kdepthtwo}) for $K_{k_1,k_2}$ can be rewritten in terms of the iterated integrals $\beta_{\pm}$ in (\ref{cuspat3.03}) \cite{Dorigoni:2022npe}\footnote{In the first line of (\ref{kdepth2}), we fixed a minus-sign mistake concerning the coefficient of $\bplusno{j_1}{k_1} \bminusno{j_2}{k_2} $ in (4.45) of \cite{Dorigoni:2022npe}.} 
\begin{align}
K_{k_1,k_2}&[\tau] =
\frac{(k_1{-}1)!(k_2{-}1)!}{16}\sum_{j_1=0}^{k_1-2}\sum_{j_2=0}^{k_2-2}
 \frac{ \binom{k_1{-}2}{j_1}\binom{k_2{-}2}{j_2} }{ (-4y)^{j_1+j_2}}
\Big(
 \bplusno{j_2 &j_1}{k_2 &k_1} + \bplusno{j_1}{k_1} \bminusno{j_2}{k_2}  + \bminusno{ j_1 &j_2}{k_1 &k_2} \Big) \notag\\
&\quad \quad \quad \times    (X_1{-}\tau Y_1)^{j_1}(X_1{-} \bar{\tau} Y_1)^{k_1-2-j_1}
 (X_2{-}\tau Y_2)^{j_2}(X_2{-} \bar{\tau} Y_2)^{k_2-2-j_2} \label{kdepth2}  \\
& \ - \frac{(k_2{-}2)!(k_1{-}1)! \zeta_{k_2-1}}{8 (2\pi i)^{k_2-2}}\sum_{j=0}^{k_1-2}\binom{k_1{-}2}{j}\frac{({-}1)^j}{(4y)^j}\bplusno{j}{k_1}  (X_1{-}\tau Y_1)^j(X_1{-}\bar{\tau}Y_1)^{k_1{-}2{-}j}  Y_2^{k_2-2} \nonumber\\
& \ - \frac{(k_1{-}2)!(k_2{-}1)! \zeta_{k_1-1}}{8 (2\pi i)^{k_1-2}}\sum_{j=0}^{k_2-2}\binom{k_2{-}2}{j}\frac{(-1)^j}{(4y)^j}\bminusno{j}{k_2} Y_1^{k_1{-}2} (X_2{-}\tau Y_2)^j(X_2{-}\bar{\tau}Y_2)^{k_2{-}2{-}j} \, .\nonumber
\end{align}

\subsubsection{Equivariant completion at modular depth two}
\label{subsubsec:5.1.1}

In spite of its equivariant total differential (\ref{dk1k2}), the function (\ref{kdepth2}) itself fails to be equivariant by a cocycle: evaluating the $\SLtwoZ$ transformation 
\begin{align}
    (X_i,Y_i,\tau) \rightarrow \bigg(aX_i{+}bY_i,cX_i{+}dY_i,\frac{a\tau{+}b}{c\tau{+}d}  \bigg)
\end{align}
of $K_{k_1,k_2}[\tau]$ introduces an inhomogeneous term independent on $\tau$ and polynomial in $X_i,Y_i$. Still, there is a method to restore equivariance by adding simple classes of terms \cite{Brown:2017qwo} with compensating cocycles. We start by introducing the equivariant projector 
\begin{equation}\label{deltaj}
    \delta^\ell \coloneqq    m \circ \left( \dfrac{\partial}{\partial X_1} \otimes \dfrac{\partial}{\partial Y_2} - \dfrac{\partial}{\partial Y_1}\otimes \dfrac{\partial }{\partial X_2} \right)^\ell,
    \end{equation}
defined for $\ell \geq 0$, and where $m:\mathbb{Q}[X_1,Y_1]\otimes \mathbb{Q}[X_2,Y_2] \to \mathbb{Q}[X_1,Y_1]$ is the multiplication map setting $X_2=X_1$ and $Y_2=Y_1$ after evaluating the derivatives in (\ref{deltaj}). We then define the function 
\beq
 K^{(\ell)}_{k_1,k_2}[X_1,Y_1;\tau] \coloneqq \frac{(i\pi)^\ell}{(\ell!)^2} \, \delta^\ell \Big( K_{k_1,k_2}[\tau] \Big) \,,
\eeq
in the normalisation conventions of \cite{Brown:2017qwo} which is equivariant up to a cocycle that only depends on one pair $X_1,Y_1$ of bookkeeping variables. It transforms in an $\mf{sl}_2$-representation of dimension $k_1{+}k_2{-}2\ell{-}3$, showing that only a finite range of $\ell$-values have to be considered.

By the Eichler--Shimura theorem \cite{Eichler:1957, Shimura:1959} we can express cocycles of $K^{(\ell)}_{k_1,k_2}$ in terms of a coboundary and linear combinations of cocycles of modular forms. In other words, there is a systematic completion $M^{(\ell)}_{k_1,k_2}$ of $K^{(\ell)}_{k_1,k_2}$ which produces an equivariant function of the form 
\begin{align}\label{eq:Mj}
M^{(\ell)}_{k_1,k_2}[X_1,Y_1; \tau] = \ & K^{(\ell)}_{k_1,k_2}[X_1,Y_1;\tau]  - c^{(\ell)}_{k_1,k_2}[X_1,Y_1]
\\* &  - \frac{1}{2}
\bigg\{ 
\int_{\tau}^{i\infty}  \underline{\rm f}^{\,(\ell)}_{k_1,k_2}[X_1,Y_1;\tau_1]
+\int_{\bar \tau}^{-i\infty} \!   \overline{\underline{\rm g}^{(\ell)}_{k_1,k_2}[X_1,Y_1;\tau_1]}
 \bigg\}\, , \nonumber
\end{align}
where the $\tau$-independent quantity $c^{(\ell)}_{k_1,k_2}[X_1,Y_1]$ in the first line is a polynomial in $X_1,Y_1$ of homogeneity degree $k_1{+}k_2{-}2\ell{-}4$.
In the second line of (\ref{eq:Mj}), $\underline{\rm f}^{\,(\ell)}_{k_1,k_2}[X_1,Y_1;\tau_1]$ is an equivariant $(1,0)$-form composed of a holomorphic cusp form of weight $k_1{+}k_2{-}2\ell{-}2$
multiplying $(X_1{-}\tau_1 Y_1)^{k_1{+}k_2{-}2\ell{-}4}$,
and $\overline{\underline{\rm g}^{(\ell)}_{k_1,k_2}[X_1,Y_1;\tau_1]}$  is an equivariant $(0,1)$-form composed of an antiholomorphic modular form of weight $k_1{+}k_2{-}2\ell{-}2$ multiplying $(X_1{-}\bar \tau_1 Y_1)^{k_1{+}k_2{-}2\ell{-}4}$. As a consequence, holomorphic cusp forms of modular weight $2s$ can occur in $\underline{\rm f}^{\,(\ell)}_{k_1,k_2},  \overline{\underline{\rm g}^{(\ell)}_{k_1,k_2}}$ at degrees $k_1{+}k_2\geq 2s{+}2$ and values $\ell=\frac{1}{2}(k_1{+}k_2){-}s{-}1$. 
We note that, while $\underline{\rm f}_{k_1,k_2}^{\,(\ell)}$ involves a cusp form, the form entering $\underline{\rm g}_{k_1,k_2}^{\,(\ell)}$ is a general holomorphic modular form and so can be a combination of a cuspidal and Eisenstein part. 

We note that the application of the Eichler--Shimura theorem relies on the presence of only one pair of bookkeeping variables $X_i,Y_i$, which explains the need to introduce the projector $\delta^\ell$.
Fortunately, we can study $M^{(\ell)}_{k_1,k_2}$ and their components at each level $\ell=0,1,\ldots,{\rm min}(k_1,k_2){-}2$ and, as was demonstrated in \cite{Dorigoni:2022npe}, reconstruct equivariant functions $M_{k_1,k_2}[\tau]= M_{k_1,k_2}[X_1,Y_1,X_2,Y_2;\tau]$. Such functions can be represented in the same format as in \eqref{eq:Mj}, but they can also be described simply in terms of modular-depth-two $\beta^{\rm eqv}$ (see appendix \ref{app:expeqv.3} for a detailed comparison of the two representations of $M_{k_1,k_2}[\tau]$ in (\ref{eq:Mbeqv})):
\begin{align}\label{eq:Mbeqv}
M_{k_1,k_2}[\tau] & \coloneqq  K_{k_1,k_2}[\tau] - c_{k_1,k_2} - \frac{1}{2}
\bigg\{ 
\int_{\tau}^{i\infty}  \underline{\rm f}_{k_1,k_2}[\tau_1]
+\int_{\bar \tau}^{-i\infty} \!   \overline{\underline{\rm g}_{k_1,k_2}[\tau_1]}
 \bigg\}
\\[2mm]
& =\frac{1}{16} (k_1{-}1)! (k_2{-}1)!
\sum_{j_1=0}^{k_1-2} \sum_{j_2=0}^{k_2-2} 
 \frac{ \binom{k_1-2}{j_1} \binom{k_2-2}{j_2}    }{(-4y)^{j_1+j_2}}
 \beqv{j_2&j_1}{k_2&k_1} \nonumber
 \\
& \qquad \times
(X_1{-}\tau Y_1)^{j_1} (X_1{-}\bar \tau Y_1)^{k_1-2-j_1}  
(X_2{-}\tau Y_2)^{j_2}  (X_2{-}\bar \tau Y_2)^{k_2-2-j_2} \, , \nonumber
\end{align}
such that
\begin{align}
    \dfrac{(i\pi)^\ell}{(\ell!)^2} \, \delta^\ell \big(M_{k_1,k_2}\big) &= M^{(\ell)}_{k_1,k_2}\, , &    \dfrac{(i\pi)^\ell}{(\ell!)^2} \, \delta^\ell \big(c_{k_1,k_2}\big) &= c^{\,(\ell)}_{k_1,k_2} \,  ,
    \label{md2proj} \\
   \dfrac{(i\pi)^\ell}{(\ell!)^2} \, \delta^\ell \big(\underline{\rm f}_{k_1,k_2}\big) &= \underline{\rm f}^{\,(\ell)}_{k_1,k_2}
   \, , \qquad    &\dfrac{(i\pi)^\ell}{(\ell!)^2} \, \delta^\ell \big(\overline{\underline{\rm g}_{k_1,k_2}}\big) &=\overline{\underline{\rm g}^{(\ell)}_{k_1,k_2}} \, .
   \notag
\end{align}
For fixed values of $(k_1,k_2)$, the $\beqv{j_2&j_1}{k_2&k_1}$ in (\ref{eq:Mbeqv}) enter the generating series $\Jeqv(\eee_k;\tau)$ with words $\eee_{k_1}^{(j_1)}\eee_{k_2}^{(j_2)}$ in the $\mf{sl}_2$ tensor product $V(\eee_{k_1})\otimes V(\eee_{k_2})$ as coefficients. The highest-dimensional module
$V(\eee_{k_1+k_2-2})$ in this tensor product is isolated by $M^{(0)}_{k_1,k_2}$ in (\ref{md2proj}) resulting from the projector (\ref{deltaj}) at $\ell=0$. The operator $\delta^{0}$ preserves the combined homogeneity degree in $X_i,Y_i$ which is $k_1{+}k_2{-}4$ in both $M^{(0)}_{k_1,k_2}$ and the unprojected $M_{k_1,k_2}$. In the same way as the differential operator entering $\delta^\ell$ in (\ref{deltaj}) reduces this homogeneity degree by $2\ell$, the projected equivariant integrals in $M^{(\ell)}_{k_1,k_2}$ at generic $\ell=0,1,\ldots,{\rm min}(k_1,k_2){-}2$ fall into the $\mf{sl}_2$ irreducible $V(\eee_{k_1+k_2-2-2\ell})$. 

By isolating the contributions to (\ref{eq:Mbeqv}) at modular depth zero, the $\tau$-independent quantities $c_{k_1,k_2}$  can be assembled from the expansion coefficients $c^{\rm sv}$ in (\ref{defbsv}) or (\ref{defcsv}) \cite{Dorigoni:2022npe}
\begin{align}
c_{k_1,k_2} &= - \frac{1}{16} (k_1{-}1)!(k_2{-}1)!
\sum_{j_1=0}^{k_1-2} \sum_{j_2=0}^{k_2-2}
(-1)^{j_1+j_2}\binom{ k_1{-}2 }{ j_1 } \binom{ k_2{-}2 }{ j_2 } \notag \\
&\quad \times 
\bigg( \frac{  Y_1 }{2\pi i} \bigg)^{j_1}  
X_1^{k_1-2-j_1} 
 \bigg( \frac{  Y_2 }{2\pi i} \bigg)^{j_2}
  X_2^{k_2-2-j_2} \,  \ccsv{j_2  &j_1 }{ k_2 &k_1}\, .
 \label{de2csv}
\end{align}
The $c^{\rm sv}$ on the right-hand side are rational multiples of $\zeta_{2n_1+1}$ or $\zeta_{2n_1+1}\zeta_{2n_2+1}$ with $n_i \in \mathbb N$ that can be explicitly determined in all degrees from (\ref{zetgen.08}) and (\ref{seri.31}).

The Eisenstein contributions to $\overline{\underline{\rm g}_{k_1,k_2}}$ in \eqref{eq:Mbeqv} are known in closed form from (4.49) of \cite{Dorigoni:2022npe}. In the setting of this work, the expressions $\overline{\underline{\rm g}_{k_1,k_2}} \sim \zeta_{{\rm min}(k_1,k_2)-1} \overline{ \underline{\rm G}_{|k_1-k_2|+2} } $ of the reference can be derived from the terms $\sim [\eee_{k_1}^{(j_1)},\eee_{k_2}^{(j_2)}]$
in $ \Msv( \zetaE_i,\zetaE_{\pern})^{-1} \Imint(\eee_k;\tau)
\mathbb \Msv(  \zetaE_i, \zetaE_{\pern})$ and their contributions to the $\beqv{j_2&j_1}{k_2&k_1}$ in \eqref{eq:Mbeqv}.
Similarly, the cusp forms entering $\underline{\rm f}_{k_1,k_2}$ and $\overline{\underline{\rm g}_{k_1,k_2}}$ have rational multiples of $\frac{\Lambda(\Delta_{2s}, 2s {+} 2n)}{\Lambda(\Delta_{2s}, 2s{-}2)}$ or $\frac{\Lambda(\Delta_{2s}, 2s {+} 2n{+}1)}{\Lambda(\Delta_{2s}, 2s{-}1)} $ as coefficients ($n \in \mathbb N_0$), see section 4.3.4 of \cite{Dorigoni:2022npe} for explicit examples. The exact form of these expressions can be reproduced from the terms $\sim [\eee_{k_1}^{(j_1)},\eee_{k_2}^{(j_2)}]$ in the letters $\eee_{\Delta_{2s}^{\pm}}$ of section \ref{sec:4.2} and their contributions to the $\beqv{j_2&j_1}{k_2&k_1}$ in \eqref{eq:Mbeqv}.

From expression~\eqref{eq:Mbeqv} along with~\eqref{dk1k2} and~\eqref{eq:dM1} it can be verified that
\begin{align}
\label{eq:dM2}
    2  \,  \partial_{\tau} M_{k_1,k_2}[\tau]\, {\rm d}\tau  &=  \underline{\rm G}_{k_1}[\tau]M_{k_2}[\tau] + \underline{\rm f}_{k_1,k_2}[\tau]\, , &
     2  \,  \partial_{\bar \tau}  M_{k_1,k_2}[\tau] \, {\rm d}\bar \tau &=  M_{k_1}[\tau]\,\overline{\underline{\rm G}_{k_2}[\tau]} + \overline{\underline{\rm g}_{k_1,k_2}[\tau]}\, .
\end{align}
With the expansions (\ref{mkdepth1}) and (\ref{eq:Mbeqv}) of $ M_k[\tau]$ and $M_{k_1,k_2}[\tau] $ in terms of the modular forms $ \beqv{j_1&j_2}{k_1&k_2}$, the differential equations~\eqref{eq:dM1} and~\eqref{eq:dM2} are an alternative way of projecting the differential equation (\ref{diffeq.02}) of the equivariant series $\mathbb I^{\rm eqv}$ to modular depth one and two. The projectors (\ref{deltaj}) to irreducible representations of $\mf{sl}_2$ then isolate those combinations of $\beta^{\rm eqv}$ from (\ref{eq:Mbeqv}) which form an $\mf{sl}_2$-multiplet under the raising and lowering operators $\nabla$ and~$\overline{\nabla}$.

\subsection{Modular completion at modular depth three}
\label{sec:modcomp3}

We will now repeat  at modular depth three the analysis of the preceding section. The first step is to find an equivariant closed form involving triple integrals and the second step is to find a completion that is modular.

\subsubsection{Towards equivariant integrals at modular depth three}

As an analogue of $D_{k_1,k_2}[\tau]$ in (\ref{dk1k2}) at modular depth three, we aim to construct a closed and equivariant one-form that comprises differentials of triple Eisenstein integrals. A first guess compatible with equivariance and the desired triple integrals is $\underline{\rm G}_{k_1}[\tau] M_{k_2,k_3}[\tau]
+M_{k_1,k_2}[\tau] \overline{ \underline{\rm G}_{k_3}[\tau] }$, where we recall the convention that $M_{k_i,k_j}[\tau] \coloneqq M_{k_i,k_j}[X_i,Y_i,X_j,Y_j;\tau]$. However, starting from this initial candidate one-form we see that closure requires the addition of two further terms, leading us to the definition
\begin{align}
 D_{k_1,k_2,k_3}[\tau]\coloneqq -\frac{1}{2} \Big(&\underline{\rm G}_{k_1}[\tau] M_{k_2,k_3}[\tau]
+M_{k_1,k_2}[\tau] \overline{ \underline{\rm G}_{k_3}[\tau] } \label{dk1k2k3} \\
+\,&\underline{\rm f}_{k_1,k_2}[\tau]M_{k_3}[\tau] + M_{k_1}[\tau]\overline{\underline{\rm g}_{k_2,k_3}[\tau]}\Big)\, . \quad \nonumber
\end{align}
Closure of $D_{k_1,k_2,k_3}$ in the sense of~\eqref{clphi} can be checked using~\eqref{eq:dM1} and~\eqref{eq:dM2}.

Since all of $M_{k_i,k_j}[\tau]$, $\underline{\rm f}_{k_i,k_j}[\tau]$ and $\overline{\underline{\rm g}_{k_i,k_j}[\tau]}$ depend on $X_i,Y_i,X_j,Y_j$, the closed and equivariant one-form $D_{k_1,k_2,k_3}[\tau]$ depends on three pairs $X_i,Y_i$ at $i=1,2,3$. Considering its integral then gives the following lengthy expression
\begin{align}
& \nonumber K_{k_1,k_2,k_3}[\tau]=  
-\frac{1}{8}  \int\limits_{\tau}^{i\infty}  \underline{\rm G}_{k_1}[\tau_1]  
 \int\limits_{\tau_1}^{i\infty} \underline{\rm G}_{k_2}[\tau_2] 
 \int\limits_{\tau_2}^{i\infty}  \underline{\rm G}_{k_3}[\tau_3] 
 -\frac{1}{8}  \int\limits_{\bar\tau}^{-i\infty}  \overline{\underline{\rm G}_{k_3}[\tau_3]  }
 \int\limits_{\bar\tau_3}^{-i\infty}  \overline{\underline{\rm G}_{k_2}[\tau_2]  }
 \int\limits_{\bar\tau_2}^{-i\infty}  \overline{\underline{\rm G}_{k_1}[\tau_1]  }  
\\
& \nonumber \quad
-\frac{1}{8}  \int\limits_{\tau}^{i\infty} \underline{\rm G}_{k_1}[\tau_1] 
  \times \! \int\limits_{\bar\tau}^{-i\infty} \overline{\underline{\rm G}_{k_3}[\tau_3] }
 \int\limits_{\bar\tau_3}^{-i\infty} \overline{  \underline{\rm G}_{k_2}[\tau_2]}
-\frac{1}{8}   \int\limits_{\bar\tau}^{-i\infty} \overline{\underline{\rm G}_{k_3}[\tau_3]  }
 \times \! \int\limits_{\tau}^{i\infty} \underline{\rm G}_{k_1}[\tau_1] 
 \int\limits_{\tau_1}^{i\infty} \underline{\rm G}_{k_2}[\tau_2]
\\
& \nonumber \quad
 - \frac{1}{4} \,c_{k_1}
 \int\limits_{\bar\tau}^{-i\infty}  \overline{\underline{\rm G}_{k_3}[\tau_3]  } 
 \int\limits_{\bar\tau_3}^{-i\infty}  \overline{\underline{\rm G}_{k_2}[\tau_2]} 
 - \frac{1}{4} \, c_{k_2} 
 \int\limits_{\tau}^{i\infty} \underline{\rm G}_{k_1}[\tau_1]   \times \!\!\!
 \int\limits_{\bar\tau}^{-i\infty}  \overline{\underline{\rm G}_{k_3}[\tau_3]} 
  - \frac{1}{4} \, c_{k_3} 
 \int\limits_{\tau}^{i\infty}  \underline{\rm G}_{k_1}[\tau_1]  
 \int\limits_{\tau_1}^{i\infty}  \underline{\rm G}_{k_2}[\tau_2] \\
&  \nonumber \quad
  +\frac{1}{4}  \int\limits_{\bar\tau}^{-i\infty}  \overline{\underline{\rm G}_{k_3}[\tau_3]} \times 
 \int\limits_{\tau}^{i\infty} \underline{\rm f}_{k_1,k_2}[\tau_1] \, + \frac{1}{4}
 \int\limits_{\tau}^{i\infty}  \underline{\rm G}_{k_1}[\tau_1] \int\limits_{\tau_1}^{i\infty} \underline{\rm f}_{k_2,k_3}[\tau_2]
 \, +\frac{1}{4}\int\limits_{\tau}^{i\infty} \underline{\rm f}_{k_1,k_2} [\tau_1]\int\limits_{\tau_1}^{i\infty}  \underline{\rm G}_{k_3}[\tau_3]
 \\
 & \nonumber  \quad  + \frac{1}{4} \int\limits_{\bar\tau}^{-i\infty} \overline{\underline{\rm g}_{k_2,k_3}[\tau_2]} \int\limits_{\bar\tau_2}^{-i\infty}  \overline{\underline{\rm G}_{k_1}[\tau_1] }  + \frac{1}{4} \int\limits_{\tau}^{i\infty} \underline{\rm G}_{k_1}[\tau_1] \times \!\!\!
 \int\limits_{\bar\tau}^{-i\infty} \overline{\underline{\rm g}_{k_2,k_3}[\tau_2]} 
 +\frac{1}{4} \int\limits_{\bar\tau}^{-i\infty}  \overline{\underline{\rm G}_{k_3}[\tau_3]}
 \int\limits_{\bar\tau_3}^{-i\infty} \overline{\underline{\rm g}_{k_1,k_2}[\tau_1]}
 \\
 &\quad + \frac{1}{2} \, c_{k_3}\int\limits_{\tau}^{i\infty} \underline{\rm f}_{k_1,k_2}[\tau_1]   + \frac{1}{2}\, c_{k_1}\int\limits_{\bar\tau}^{-i\infty} 
 \overline{\underline{\rm g}_{k_2,k_3}[\tau_2]}
 + \dfrac{1}{2} \, c_{k_2,k_3}\int\limits_{\tau}^{i\infty} \underline{\rm G}_{k_1}[\tau_1]  + \frac{1}{2} \, c_{k_1,k_2}\int\limits_{\bar\tau}^{-i\infty} 
 \overline{\underline{\rm G}_{k_3}[\tau_3]}
\, .
 \label{k123expr}
\end{align}
The simple monomials $c_{k_i} \sim \zeta_{k_i-1} Y_i^{k_i-2}$ are given by (\ref{def:ck}), and the explicit form of $c_{k_i,k_j}$, $\underline{\rm f}_{k_i,k_j}$ and $\overline{\underline{\rm g}_{k_i,k_j}}$ defined by (\ref{eq:Mbeqv}) is discussed below (\ref{md2proj}).

\subsubsection{Equivariant completion at modular depth three}
\label{sec:eqv3}

Similar to the case of modular depth two, the combinations $K_{k_1,k_2,k_3}$ in (\ref{k123expr}) are not equivariant in spite of their construction by integrating an equivariant one-form (\ref{dk1k2k3}). Still, it is possible to find an equivariant completion of $K_{k_1,k_2,k_3}$ by solely adding terms of modular depth zero and one. The existence of the equivariant completion once more follows from the Eichler--Shimura theorem whose application relies on a generalisation of the projector $\delta^j$ in (\ref{deltaj}) to triple integrals. More specifically, this projector needs to be designed to handle polynomials in three pairs of bookkeeping variables $X_i$, $Y_i$ (i.e.\ for $i=1,2,3$, and not just for $i=1,2$). With this in mind, we first introduce the following generalisation of $\delta^\ell$ defined in~\eqref{deltaj} (for $\ell \geq 0$),
\begin{equation}
\delta^\ell_{a,b} \coloneqq  m_{a,b} \circ \bigg( \dfrac{\partial}{\partial X_a} \otimes \dfrac{\partial}{\partial Y_b} - \dfrac{\partial}{\partial Y_a}\otimes \dfrac{\partial }{\partial X_b} \bigg)^\ell\,,
\label{defdelab}
\end{equation}
where $m_{a,b}$ is the multiplication map $\mathbb{Q}[X_a,Y_a]\otimes \mathbb{Q}[X_b,Y_b] \to \mathbb{Q}[X_a,Y_a]$ setting $X_b=X_a$ and $Y_b=Y_a$ after evaluating the derivatives in (\ref{defdelab}). Using a pair of these projectors we  can construct an equivariant projector $\delta^{\ell_1,\ell_2}$ via 
\beq
\delta^{\ell_1,\ell_2} \coloneqq \delta^{\ell_1}_{1,2} \circ \delta^{\ell_2}_{2,3} \, ,
\label{defdelab.2}
\eeq
where we note that $\delta^\ell_{1,2}$ is simply the original $\delta^\ell$ in (\ref{deltaj}). For example we have
\begin{align}
    \delta^{0,0}(X_1 X_2 X_3 Y_1 Y_2 Y_3) & = \delta^0_{1,2} \big( \delta^0_{2,3}(X_1 X_2 X_3 Y_1 Y_2 Y_3) \big)
   =\delta^0_{1,2} \big(m_{2,3}(X_1 X_2 X_3 Y_1 Y_2 Y_3) \big)  \nonumber
    \\ & = \delta^0_{1,2}(X_1X_2^2Y_1Y_2^2)
    = m_{1,2}(X_1X_2^2Y_1Y_2^2)  
    =   X_1^3 Y_1^3 \, ,
\notag \\
\delta^{0,1} (X_1^{a_1} X_2^{a_2} X_3^{a_3} Y_1^{b_1} Y_2^{b_2} Y_3^{b_3}) &= (a_2 b_3 {-} a_3 b_2) X_1^{a_1+a_2+a_3-1} Y_1^{b_1+b_2+b_3-1} \, ,
\label{exdeltaij} \\
\delta^{1,0} (X_1^{a_1} X_2^{a_2} X_3^{a_3} Y_1^{b_1} Y_2^{b_2} Y_3^{b_3}) &= \big(a_1 (b_2{+}b_3) {-} (a_2{+}a_3) b_1) X_1^{a_1+a_2+a_3-1} Y_1^{b_1+b_2+b_3-1}
\notag
\end{align}
and the last two cases here presented illustrate that generically $\delta^{\ell_1,\ell_2} \neq \delta^{\ell_2,\ell_1}$. The homogeneity degrees with respect to the $X_i,Y_i$ translate into degrees $k_{1}{+}k_2{+}k_3=6+\sum_{i=1}^3(a_i{+}b_i)$ of the accompanying iterated Eisenstein integrals in (\ref{k123expr}). The order of writing the projectors as in~\eqref{defdelab.2} corresponds to a specific choice of putting parentheses when evaluating the triple tensor product~\eqref{eq:trip2} of $V(\eee_{k_i})$. We will comment more on the group-theoretic interpretation of $\delta^{\ell_1,\ell_2}$ in section~\ref{sec:deltasl2}.

Using the projector in (\ref{defdelab}) and (\ref{defdelab.2}), we can define the new function
\beq
 K^{(\ell_1,\ell_2)}_{k_1,k_2,k_3}[X_1,Y_1;\tau] \coloneqq \frac{(i\pi)^{\ell_1+\ell_2}}{(\ell_1!)^2(\ell_2!)^2} \, \delta^{\ell_1,\ell_2} \Big( K_{k_1,k_2,k_3}[\tau] \Big) \, ,
 \label{projKs}
\eeq
which, as before, is equivariant up to a cocycle depending on one pair $X_1,Y_1$ of bookkeeping variables. An application of the Eichler--Shimura theorem then gives a systematic completion $M^{(\ell_1,\ell_2)}_{k_1,k_2,k_3}$ of $K^{(\ell_1,\ell_2)}_{k_1,k_2,k_3}$ which results in an equivariant function:
\begin{align} \label{TripleM}
M^{(\ell_1,\ell_2)}_{k_1,k_2,k_3}[X_1,Y_1;\tau]  = & \ K^{(\ell_1,\ell_2)}_{k_1,k_2,k_3}[X_1,Y_1;\tau]  - c^{(\ell_1,\ell_2)}_{k_1,k_2,k_3}[X_1,Y_1] \\*
& \ - \frac{1}{2}
\bigg\{ 
\int_{\tau}^{i\infty}  \underline{\rm f}^{\,(\ell_1,\ell_2)}_{k_1,k_2,k_3}[X_1,Y_1;\tau_1]
+\int_{\bar \tau}^{-i\infty} \!   \overline{\underline{\rm g}^{(\ell_1,\ell_2)}_{k_1,k_2,k_3}[X_1,Y_1;\tau_1]} 
 \bigg\} \, . \nonumber
\end{align}
In perfect analogy with the modular-depth-two case discussed below (\ref{eq:Mj}),
the $\tau$-indepen\-dent  $c^{(\ell_1,\ell_2)}_{k_1,k_2,k_3}[X_1,Y_1]$ are polynomials in the bookkeeping variables $X_1,Y_1$ of homogeneity degree $w{-}2=k_1{+}k_2{+}k_3{-}2\ell_1{-}2\ell_2{-}6$. Moreover, $\underline{\rm f}^{(\ell_1,\ell_2)}_{k_1,k_2,k_3}[X_1,Y_1;\tau_1]$ is an equivariant $(1,0)$-form composed of holomorphic cusp forms of weight $w$
multiplied by $(X_1{-}\tau_1 Y_1)^{w{-}2}$,
and $\overline{\underline{{\rm g}}^{( \ell_1,\ell_2)}_{k_1,k_2,k_3}[X_1,Y_1;\tau_1]}$  is an equivariant $(0,1)$-form composed of antiholomorphic modular forms of weight $w$
multiplied by $(X_1{-}\bar\tau_1 Y_1)^{w{-}2}$.

Similar to our reasoning at modular depth two in section \ref{subsubsec:5.1.1}, we have projected the functions $K_{k_1,k_2,k_3}[X_1,Y_1,X_2,Y_2,X_3,Y_3;\tau]$ in (\ref{k123expr}) to the $K^{(\ell_1,\ell_2)}_{k_1,k_2,k_3}[X_1,Y_1;\tau]$ in (\ref{projKs}) in order to infer the existence of the equivariant completion (\ref{TripleM}) from the Eichler--Shimura theorem. The union of the projected $M^{(\ell_1,\ell_2)}_{k_1,k_2,k_3}[X_1,Y_1;\tau]$ at all $\ell_1,\ell_2 \geq 0$ with $\ell_1{+}\ell_2 \leq \frac{1}{2}(k_1{+}k_2{+}k_3){-}3$ can be used to reconstruct an equivariant function
\beq
M_{k_1,k_2,k_3}[\tau]  =  K_{k_1,k_2,k_3}[\tau] - c_{k_1,k_2,k_3} - \frac{1}{2}
\bigg\{ 
\int_{\tau}^{i\infty}  \underline{\rm f}_{k_1,k_2,k_3}[\tau_1]
+\int_{\bar \tau}^{-i\infty} \!   \overline{\underline{\rm g}_{k_1,k_2,k_3}[\tau_1]}
 \bigg\}\,,
 \label{KtoM3}
\eeq
of three pairs $X_i,Y_i$ of bookkeeping variables
subject to
\beq
M^{(\ell_1,\ell_2)}_{k_1,k_2,k_3}[X_1,Y_1;\tau] = \frac{(i\pi)^{\ell_1+\ell_2}}{(\ell_1!)^2(\ell_2!)^2} \, \delta^{\ell_1,\ell_2} \Big( M_{k_1,k_2,k_3}[\tau] \Big) \, .
\label{unproj3}
\eeq
The same conventions apply to the extraction of the projected $c^{(\ell_1,\ell_2)}_{k_1,k_2,k_3}$, $\underline{\rm f}^{\,(\ell_1,\ell_2)}_{k_1,k_2,k_3}$ and $\overline{\underline{\rm g}^{(\ell_1,\ell_2)}_{k_1,k_2,k_3}}$ from the more general quantities $c_{k_1,k_2,k_3} $, $\underline{\rm f}_{k_1,k_2,k_3}$ and $   \overline{\underline{\rm g}_{k_1,k_2,k_3}}$ in (\ref{KtoM3}).

In perfect analogy with (\ref{eq:Mbeqv}) at modular depth two, the $M_{k_1,k_2,k_3}[\tau]$ in (\ref{KtoM3}) admit an alternative description in terms of the equivariant functions $\beqv{j_3&j_2&j_1}{k_3&k_2&k_1}$,
\begin{align}
&M_{k_1,k_2,k_3}[\tau] = \frac{-(k_1{-}1)! (k_2{-}1)! (k_3{-}1)!}{64} 
\sum_{j_1=0}^{k_1-2} \sum_{j_2=0}^{k_2-2} \sum_{j_3=0}^{k_3-2} 
 \frac{ \binom{k_1-2}{j_1} \binom{k_2-2}{j_2} \binom{k_3-2}{j_3} }{(-4y)^{j_1+j_2+j_3}}
 \,  \beqv{j_3&j_2&j_1}{k_3&k_2&k_1}
\label{olidreams}\\
&\quad \! \! \times 
(X_1{-}\tau Y_1)^{j_1} (X_1{-}\bar \tau Y_1)^{k_1-2-j_1} 
(X_2{-}\tau Y_2)^{j_2} (X_2{-}\bar \tau Y_2)^{k_2-2-j_2}
(X_3{-}\tau Y_3)^{j_3} (X_3{-}\bar \tau Y_3)^{k_3-2-j_3}\, ,
\notag
\end{align}
see appendix \ref{app:expeqv.4} for a detailed proof of equivalence to (\ref{KtoM3}).

In summary, the Eichler--Shimura theorem implies the existence of equivariant integrals $M^{(\ell_1,\ell_2)}_{k_1,k_2,k_3}[X_1,Y_1;\tau]$ at arbitrary degree $k_1{+}k_2{+}k_3$. Their multivariate combinations $M_{k_1,k_2,k_3}[\tau]$ obtained from assembling the contributions from all projectors $\delta^{\ell_1,\ell_2}$ at $0\leq \ell_1{+}\ell_2 \leq \frac{1}{2}(k_1{+}k_2{+}k_3){-}3$ determine the $\beta^{\rm eqv}$ at modular depth three via (\ref{olidreams}). As a consequence, the generating-series construction of modular forms $\beqv{j_3&j_2&j_1}{k_3&k_2&k_1}$ as in section \ref{sec:4} is bound to succeed at arbitrary degree, i.e.\ beyond the explicit checks at $k_1{+}k_2{+}k_3\leq 20$ we have performed. 

\subsubsection{$\delta^{\ell_1,\ell_2}$ versus $\mf{sl}_2$ representations}
\label{sec:deltasl2}

Similar to the decomposition (\ref{eq:Mbeqv}) of $M_{k_1,k_2}[\tau]$ at modular depth two, the projector $\delta^{\ell_1,\ell_2}$ in (\ref{defdelab.2}) isolates irreducible $\mf{sl}_2$ representations of the tensor product $V(\eee_{k_1})\otimes V(\eee_{k_2}) \otimes V(\eee_{k_3})$ comprising the coefficients $\eee_{k_1}^{(j_1)}\eee_{k_2}^{(j_2)}\eee_{k_3}^{(j_3)}$ of $\beta^{\rm eqv}$ in $\Jeqv(\eee_k;\tau)$. 
The particular choice of order in~\eqref{defdelab.2} is parallel to the way the triple tensor product was written in~\eqref{eq:trip2}: The inner projector $\delta_{2,3}^{\ell_2}$ decomposes the product $ V(\eee_{k_2}) \otimes V(\eee_{k_3})$ into irreducibles and the outer operator $\delta^{\ell_1}_{1,2}$ then projects onto irreducibles in the product of these with $V(\eee_{k_1})$. All possible ways of putting parentheses when evaluating the triple tensor correspond to one of the three projectors  $\delta^{\ell_1}_{1,2} \circ \delta^{\ell_2}_{2,3}$, $\delta^{\ell_1}_{1,2} \circ \delta^{\ell_2}_{1,3}$ and $\delta^{\ell_1}_{1,3} \circ \delta^{\ell_2}_{1,2}$. These projectors can be checked to have the correct linear dependencies expected from associativity of the tensor product.  Therefore, it is sufficient to simply choose one ordering---as we have done in~\eqref{defdelab.2}---and to consider all possible values of $\ell_1$ and $\ell_2$.

At fixed $(k_1,k_2,k_3)$, the highest-dimensional representation $V(\eee_{k_1+k_2+k_3-4})$ in (\ref{eq:trip2}) occurs with multiplicity one and it is isolated by selecting the $\ell_1=\ell_2=0$ instance of $\delta^{\ell_1,\ell_2}$ which preserves the homogeneity degree $k_1{+}k_2{+}k_3{-}6$ of (\ref{olidreams}) in $X_i, Y_i$. The $M^{(\ell_1,\ell_2)}_{k_1,k_2,k_3}[X_1,Y_1;\tau] $ in (\ref{unproj3}) obtained from $\delta^{\ell_1,\ell_2}$ at non-zero $\ell_1$ or $\ell_2$ have reduced homogeneity degree given by
$k_1{+}k_2{+}k_3{-}2\ell_1{-}2\ell_2{-}6$. Accordingly, these $M^{(\ell_1,\ell_2)}_{k_1,k_2,k_3}[X_1,Y_1;\tau]$ correspond to $\mf{sl}_2$ modules $V(\eee_{k_1+k_2+k_3-2\ell_1-2\ell_2-4})$ whose reduction in dimensions by $2\ell_1{+}2\ell_2$ matches the shift of homogeneity degree by $\delta^{\ell_1,\ell_2}$.

In contrast to the $\mf{sl}_2$ modules in the tensor products $V(\eee_{k_1})\otimes V(\eee_{k_2})$ at modular depth two, the $V(\eee_{k_1+k_2+k_3-2\ell_1-2\ell_2-4})$ isolated from $M^{(\ell_1,\ell_2)}_{k_1,k_2,k_3}[X_1,Y_1;\tau] $ at modular depth three have non-trivial multiplicities for generic $\ell_1,\ell_2\geq 0$. For instance, the $\mf{sl}_2$ modules $V(\eee_{k_1+k_2+k_3-6})$ and $V(\eee_{k_1+k_2+k_3-8})$ in $V(\eee_{k_1})\otimes V(\eee_{k_2}) \otimes V(\eee_{k_3})$ of subleading and subsubleading dimensions in (\ref{eq:s235})
and (\ref{exaab}) have multiplicities two and three, respectively. The multiplicities $2$ and $3$ in
\beq
V(\eee_{k_1}) \! \otimes \! V(\eee_{k_2}) \! \otimes \! V(\eee_{k_3}) =
V(\eee_{k_1+k_2+k_3-4})\! \oplus \! 2{\times} V(\eee_{k_1+k_2+k_3-6})\! \oplus \! 3{\times}
V(\eee_{k_1+k_2+k_3-8})\oplus\ldots
\label{triptp}
\eeq
(with $V(\eee_{k})$ at $k \leq k_1{+}k_2{+}k_3{-}10$ in the ellipsis) are in one-to-one correspondence with the counting of independent projectors $\delta^{\ell_1,\ell_2}$ at a given sum $\ell_1{+}\ell_2=1,2$. More specifically, the distinct equivariant functions $M^{(0,1)}_{k_1,k_2,k_3}[X_1,Y_1;\tau] $ and $M^{(1,0)}_{k_1,k_2,k_3}[X_1,Y_1;\tau] $ in (\ref{TripleM}) are expressible in terms of those $\beqv{j_3&j_2&j_1}{k_3&k_2&k_1}$ which occur in the projection of $\Jeqv(\eee_k;\tau)$ to the modules $2{\times} V(\eee_{k_1+k_2+k_3-6})$ in (\ref{triptp}). Similarly, the three distinct $M^{(\ell_1,\ell_2)}_{k_1,k_2,k_3}[X_1,Y_1;\tau] $ resulting from 
$\delta^{0,2}, \delta^{1,1}$ and $\delta^{2,0}$ comprise those combinations of $\beqv{j_3&j_2&j_1}{k_3&k_2&k_1}$ in
the projection of $\Jeqv(\eee_k;\tau)$ to $3{\times}
V(\eee_{k_1+k_2+k_3-8})$.

For the projectors $\delta^{\ell_1,\ell_2}$ at higher $N=\ell_1{+}\ell_2 \geq 3$, however, not all of the $N{+}1$ partitions $\delta^{0,N},\delta^{1,N-1},\ldots$, $\delta^{N-1,1}$, $\delta^{N,0}$ lead to linearly independent $M^{(\ell_1,\ell_2)}_{k_1,k_2,k_3}[X_1,Y_1;\tau]$ via (\ref{unproj3}). This corresponds to the fact that the largest representation $V(\eee_{k_1+k_2+k_3-10})$ in the ellipsis of (\ref{triptp})  often appears with multiplicity $\leq 3$, see for instance (\ref{eq:s235}) and (\ref{exaab}). In other words, relations among the $\delta^{0,3},\delta^{1,2},\delta^{2,1},\delta^{3,0}$ actions on $M_{k_1,k_2,k_3}[\tau]$ explain the deviation from multiplicity $4{\times} V(\eee_{k_1+k_2+k_3-10})$ as one may naively expect. The same effect is responsible for analogous multiplicity drops for $V(\eee_{k})$ in (\ref{triptp}) at $k\leq k_1{+}k_2{+}k_3{-}12$.

Nevertheless, by exhausting the $\delta^{\ell_1,\ell_2}$ with all $\ell_1,\ell_2 \geq 0$ subject to the bound $\ell_1{+}\ell_2\leq \frac{1}{2}(k_1{+}k_2{+}k_3){-}3$, one can be sure to capture each $\mf{sl}_2$ module in the triple tensor product (\ref{eq:trip2}) at least once. This is why the reconstruction of $M_{k_1,k_2,k_3}[\tau]$ from
equviariant functions (\ref{TripleM}) will always be possible. On these grounds, the existence of equivariant $M^{(\ell_1,\ell_2)}_{k_1,k_2,k_3}[X_1,Y_1;\tau]$ resulting from the Eichler--Shimura theorem propagates to that of the $M_{k_1,k_2,k_3}[\tau]$ where the complete tensor product (\ref{eq:trip2}) is unified via three pairs $X_i,Y_i$ of bookkeeping variables.

Note that the combinations of $\beta^{\rm eqv}$ in a given $\mf{sl}_2$ irreducible $M^{(\ell_1,\ell_2)}_{k_1,k_2,k_3}[X_1,Y_1;\tau]$ obtained from the projection (\ref{unproj3}) of (\ref{olidreams}) form a multiplet under the raising and lowering operators $\nabla$ and $\overline{\nabla}$.

\subsubsection{Properties and examples of the equivariant completions}

We shall next exploit the representation (\ref{olidreams}) of the equivariant triple integrals $M_{k_1,k_2,k_3}[\tau]$ to further specify the ingredients $ c_{k_1,k_2,k_3}$, $\underline{\rm f}_{k_1,k_2,k_3}[\tau_1]$ and $\overline{\underline{\rm g}_{k_1,k_2,k_3}[\tau_1]}$ of the equivariant completion in (\ref{KtoM3}). Similar to (\ref{de2csv}) at modular depth two, the $\tau$-independent $ c_{k_1,k_2,k_3}$ are easily determined by truncating (\ref{olidreams}) to terms of modular depth zero,
\begin{align}
c_{k_1,k_2,k_3} &=  \frac{1}{64} (k_1{-}1)!(k_2{-}1)! (k_3{-}1)!
\sum_{j_1=0}^{k_1-2} \sum_{j_2=0}^{k_2-2}  \sum_{j_3=0}^{k_3-2}
(-1)^{j_1+j_2+j_3} \binom{ k_1{-}2 }{ j_1 } \binom{ k_2{-}2}{ j_2 }  \binom{ k_3{-}2}{ j_3 } \notag \\
&\quad \times 
\bigg( \frac{  Y_1 }{2\pi i} \bigg)^{j_1}  
X_1^{k_1-2-j_1} 
 \bigg( \frac{  Y_2 }{2\pi i} \bigg)^{j_2}
  X_2^{k_2-2-j_2} 
   \bigg( \frac{  Y_3 }{2\pi i} \bigg)^{j_3}
  X_3^{k_3-2-j_3} \,  \ccsv{j_3 &j_2  &j_1 }{ k_3 &k_2 &k_1} \, .
 \label{de3csv}
\end{align}
As a consequence, the coefficients of $X_i$ and $\frac{Y_i}{2\pi i}$ in $c_{k_1,k_2,k_3}$ of degree $k_1{+}k_2{+}k_3 \leq 16$ are single-valued MZVs, for instance
\begin{align}
c_{4,4,6}^{(0,0)} &=
\frac{3 Y_1^8 }{2560 \pi^8} \bigg(
 \zeta^{\rm sv}_{3, 3, 5} - \frac{14573}{96} \zeta_{11}
\bigg)
- \frac{ i X_1^3 Y_1^5 \zeta_3 \zeta_5}{7680 \pi^5} 
+ \frac{ i  X_1^5 Y_1^3 \zeta_3^2}{80640 \pi^3}
+ \frac{ X_1^6 Y_1^2 \zeta_5}{1382400 \pi^2}
- \frac{ X_1^8 \zeta_3}{19353600 } \, ,
\end{align}
where the projection $c_{k_1,k_2,k_3}^{(\ell_1,\ell_2)}$ of
$c_{k_1,k_2,k_3}$ is defined in analogy with (\ref{unproj3}). Starting from degree $k_1{+}k_2{+}k_3 = 18$, however, one additionally encounters the new periods $\varpi^{d_1,d_2}_{k,w}$ of section \ref{sec:sivarpi} among the coefficients of $c_{k_1,k_2,k_3}$. From the examples (\ref{svieis.31}) of $c^{\rm sv}$ involving new periods, one can straightforwardly generate instances of $\varpi^{d_1,d_2}_{k,w}$ in $c_{k_1,k_2,k_3}$ via (\ref{de3csv}).

Note that the $c_{k_1,k_2,k_3}^{(\ell_1,\ell_2)}$ in singlet representations of $\mf{sl}_2$ (i.e.\ at $\ell_1{+}\ell_2=\frac{1}{2}(k_1{+}k_2{+}k_3){-}3$) are independent on $X_1,Y_1$ and therefore unconstrained by equivariance. By (\ref{de3csv}), this reflects the freedom to redefine combinations of $\ccsv{j_3 &j_2  &j_1 }{ k_3 &k_2 &k_1}$ in $\mf{sl}_2$ singlets as described in section \ref{sec:4zeta.6} and at the end of section \ref{sec:ngeozvarpi}. In the context of the generating series $\mathbb J^{\rm eqv}$, these ambiguities in $\ccsv{j_3 &j_2  &j_1 }{ k_3 &k_2 &k_1}$ were fixed by imposing certain conditions on the arithmetic parts $\hat z_i,\hat z_{\pern}$ of the generators $\hat \sigma_i,\hat \sigma_{\pern}$. By the relations (\ref{olidreams}) and (\ref{de3csv}), our conventions for $\hat z_i,\hat z_{\pern}$ single out a preferred choice for $c_{k_1,k_2,k_3}$ and thereby for $M_{k_1,k_2,k_3}$.

The Eisenstein part of $\overline{\underline{\rm g}_{k_1,k_2,k_3}[\tau_1]}$ is determined by those terms in $ \Msv( \zetaE_i,\zetaE_{\pern})^{-1} \Imint(\eee_k;\tau)$  $\mathbb \Msv(  \zetaE_i, \zetaE_{\pern})$ where $\Imint$ contributes with modular depth one and the change of alphabet maps the respective letter $\eee_k$ to $[ \eee^{(j_1)}_{k_1},[ \eee^{(j_2)}_{k_2} , \eee^{(j_3)}_{k_3}]]$. These nested brackets can arise from two types of contributions from the change of alphabet via $\Msv$ -- either from $[\hat z_{2n+1} ,\eee_k ]$ with coefficients $\zeta_{2n+1}$ or from $[\hat z_{2n_1+1} ,[  \hat z_{2n_2+1} ,\eee_k ]]$ with coefficients $\zeta_{2n_1+1}\zeta_{2n_2+1}$. Hence, the Eisenstein part of $\overline{\underline{\rm g}^{(\ell_1,\ell_2)}_{k_1,k_2,k_3}[\tau_1]}$ is given by $(0,1)$-forms $\overline{ \underline{\rm G}_{k} [\tau_1]}$ multiplied by $\mathbb Q[(2\pi i)^{\pm 1}]$-linear combinations of $\zeta_{2n+1}$ and $\zeta_{2n_1+1}\zeta_{2n_2+1}$.

Finally, the cusp forms entering
$\underline{\rm f}_{k_1,k_2,k_3}[\tau_1]$ and $\overline{\underline{\rm g}_{k_1,k_2,k_3}[\tau_1]}$ are obtained from $\mathbb I_{+}(\eee_k;\tau)$ and $\Imint(\eee_k;\tau)$ by isolating the contributions $\sim [ \eee^{(j_1)}_{k_1},[ \eee^{(j_2)}_{k_2} , \eee^{(j_3)}_{k_3}]]$ to the expansion of $\eee_{\Delta^{\pm}} = \eee_{\Delta^{\rm even}} \mp \eee_{\Delta^{\rm odd}}$. From the expressions (\ref{edeltaex}) at degrees $k_1{+}k_2{+}k_3 \leq 20$, one can already see that the coefficients of the modular-depth-one integrals over cusp forms not only involve the $\frac{\Lambda(\Delta_{2s}, 2s {+} 2n)}{\Lambda(\Delta_{2s}, 2s{-}2)}$ and $\frac{\Lambda(\Delta_{2s}, 2s {+} 2n{+}1)}{\Lambda(\Delta_{2s}, 2s{-}1)} $ of the $\underline{\rm f}_{k_1,k_2}$ and $\overline{\underline{\rm g}_{k_1,k_2}}$ 
but additionally feature the new periods $\Lambda^{d_1,d_2}_{k,w}$ of section \ref{sec:noedelta}. This is most conveniently exemplified via
\begin{align}
\Big(\underline{\rm f}^{\,( 0,0)}_{8,6,6}[X_1, Y_1; \tau]  &,\, \overline{\underline{\rm g}^{\, (0,0)}_{8,6,6}}[X_1, Y_1; \tau] \Big)  = i \pi \bigg( \frac{2861}{748632192}\frac{\Lambda(\Delta_{16},17)}{\Lambda(\Delta_{16},15)}   -  \frac{9}{385}  \frac{\Lambda^{2,2}_{6,14}}{\Lambda(\Delta_{16},15)} \bigg) \nonumber
    \\
    &\quad \quad\quad\quad\quad \times
    \big((X_1 {-} \tau Y_1)^{14} \Delta_{16}(\tau)\, \dd\tau , \,
    -(X_1 {-} \bar \tau Y_1)^{14} \overline{ \Delta_{16}(\tau) }\, \dd \bar \tau\big)
    \, .
\end{align}
In summary, the new periods in $\beta^{\rm eqv}$ at modular depth three enter their alternative construction via $M_{k_1,k_2,k_3}$ through the equivariant completion of $K_{k_1,k_2,k_3}$ in (\ref{KtoM3}). The two classes $\varpi^{d_1,d_2}_{k,w}$ and $\Lambda^{d_1,d_2}_{k,w}$ of new periods originate from $c_{k_1,k_2,k_3}$ and $\underline{\rm f}_{k_1,k_2,k_3}[\tau_1]$, $\overline{\underline{\rm g}_{k_1,k_2,k_3}[\tau_1]}$, respectively. In both cases, the need for new periods can be traced back to the multiple modular values $\MMV{j_1 & j_2}{k_1 & k_2}$, $\MMV{j_1 & j_2 &j_3}{k_1 & k_2 &k_3}$ and $\MMV{j_1 & j_2}{k_1 & \Delta_{k_2}}$ in the modular $S$-transformation of the iterated integrals in the expression (\ref{k123expr}) for $K_{k_1,k_2,k_3}$.


\subsection{Equivariant integrals at arbitrary modular depth}
\label{sec:gen}

The examples at modular depth up to three suggest a natural generalisation to equivariant iterated Eisenstein integrals at arbitrary modular depth. In these examples, the construction of equivariant integrals $M_{k_1,k_2}[\tau]$ and $M_{k_1,k_2,k_3}[\tau]$ was initiated through the identification of closed and equivariant one-forms $D_{k_1,k_2}[\tau]$ and $D_{k_1,k_2,k_3}[\tau]$ explicitly given in (\ref{dk1k2}) and (\ref{dk1k2k3}), respectively.
One aim of this section is to generalise these formulae to closed and equivariant one-forms $D[\tau]=D_{k_1,k_2,\ldots,k_n}[\tau]$ at arbitrary modular depth $n$. 

Closure is a key feature of these forms, since it is this property that ensures homotopy invariance of their integrals $K[\tau]=K_{k_1,k_2,\ldots,k_n}[\tau]$. Equivariance of the one-forms $D[\tau]$ in turn implies that the associated $K[\tau]$ function is equivariant up to a cocycle. We then used this equivariance to give a modular completion $M[\tau]$ of $K[\tau]$. The second aim of this section is to generalise this modular completion to any modular depth $n$. We will achieve both of the above goals via an inductive proof.

\subsubsection{Setting up the inductive hypothesis and the base case}

We will take the modular-depth-two example to be our base case and show how this lines up with a recursive formula we introduce below. Although it is not strictly necessary, we will often discuss the modular-depth-three case too, in order to familiarise the reader with the recursive definitions.

We now proceed to construct a closed  one-form $D_{k_1,\ldots,k_n}[\tau]$ at arbitrary modular depth~$n$.
For $n=1$, we recall the following functions from section \ref{sec:mk.1}:
\begin{align}
D_k[\tau]&=\underline{{\rm G}}_{k}[\tau] + \overline{\underline{{\rm G}_{k}}[\tau]} \, ,
&\underline{\rm f}_{k}[\tau] &= 0 \, , \label{n1case} \\*
M_k[\tau] &= -\frac{1}{2} \left\{\int_\tau^{i \infty} \underline{\rm G}_k[\tau_1] 
+ \int_{\bar \tau}^{-i \infty} \overline{ \underline{\rm G}_k[\tau_1]} \right\}
+ c_k  \, ,
&\overline{\underline{\rm g}_{k}[\tau]} &= 0 \, ,\notag
\end{align}
where the $\tau$-independent quantities $c_k\sim \zeta_{k-1}Y^{k-2}$ are given by \eqref{def:ck}.

We then recursively define, for $n \geq 2$, the following one-forms:
\beq
D_{k_1,k_2,\ldots,k_n}[\tau] \coloneqq
-\frac{1}{2} \, \Big( 
 D^+_{k_1,k_2,\ldots,k_n}[\tau]
+ D^-_{k_1,k_2,\ldots,k_n}[\tau]
\Big)\,,
\label{alld.01}
\eeq
with
\begin{align}
 D^+_{k_1,k_2,\ldots,k_n}[\tau] &\coloneqq \underline{{\rm G}}_{k_1}[\tau]M_{k_2,k_3,\ldots ,k_n}[\tau]
 + \sum_{r=2}^{n-1} \underline{\rm f}_{k_1,\ldots,k_r}[\tau]M_{k_{r+1},\ldots ,k_n}[\tau]\, ,
\label{alld.02} \\
 D^-_{k_1,k_2,\ldots,k_n}[\tau] &\coloneqq 
 M_{k_1,k_2,\ldots,k_{n-1}}[\tau]\overline{\underline{{\rm G}}_{k_n}[\tau]}
 + \sum_{r=1}^{n-2} M_{k_1,\ldots,k_r}[\tau]\overline{\underline{\rm g}_{k_{r+1},\ldots,k_n}[\tau]} \, .
\notag
\end{align}
where, for $m<n$, all of $M_{k_1,\ldots, k_m}[\tau]$,  $\underline{\rm f}_{k_1,\ldots, k_m}[\tau]$ and  $\overline{\underline{\rm g}_{k_1,\ldots, k_m}[\tau]}$ are some equivariant functions satisfying the following differential equations:
\begin{align}
2 \, \partial_\tau M_{k_1,\ldots, k_m}[\tau] \, {\rm d} \tau &= D^+_{k_1,\ldots,k_{m}}[\tau] +
\underline{\rm f}_{k_1,\ldots,k_{m}}[\tau] \, ,
\label{alld.04} \\
2  \, \partial_{\bar \tau} M_{k_1,\ldots, k_m}[\tau]  \, {\rm d}\bar\tau &= 
D^-_{k_1,\ldots,k_{m}}[\tau] 
+ \overline{\underline{\rm g}_{k_1,\ldots,k_m}[\tau]} \, , \nonumber
\end{align}
which are generalisations of \eqref{eq:dM1} and~\eqref{eq:dM2}.
We wish to show that, for every $n$, the one-forms $D_{k_1,k_2,\ldots,k_n}[\tau]$ are closed and equivariant, and that equivariant functions $M_{k_1,\ldots, k_n}[\tau]$ satisfying \eqref{alld.04} exist.

The instances of \eqref{alld.01} at $n=2$ and $n=3$ are easily seen to reproduce the expressions (\ref{dk1k2}) and (\ref{dk1k2k3}) for $D_{k_1,k_2}[\tau]$ and $D_{k_1,k_2,k_3}[\tau]$, respectively.
Similarly, for these $n$, equivariant $M[\tau]$, $\underline{\rm f}[\tau]$ and $\overline{\underline{\rm g}[\tau]}$ satisfying \eqref{alld.04} have already been shown to exist via the modular completions \eqref{eq:Mbeqv} and \eqref{KtoM3}.

To summarise, at the base case $n=2$, we have the closure and equivariance of \eqref{alld.01} and the existence of equivariant functions
satisfying \eqref{alld.04}.
For $n\geq3$, we will now assume, as our inductive hypothesis, the closure and equivariance of $D_{k_1,k_2\ldots,k_{m}} [\tau]$ \eqref{alld.01} and the existence of 
equivariant $M[\tau]$, $\underline{\rm f}[\tau]$ and $\overline{\underline{\rm g}[\tau]}$ satisfying \eqref{alld.04} at modular depth $m<n$. In the next section, we will prove, under this inductive hypothesis, that $D_{k_1,\ldots, k_n}$ at modular depth $n$ is closed. In the following section~\ref{sec:ES}, we will then show the existence of equivariant functions $M_{k_1,\ldots,k_n}[\tau]$
satisfying \eqref{alld.04} at modular depth $n$, completing our proof by induction. 

\subsubsection{Proof of equivariance and closure at modular depth $n$}

We recall that $D_{k_1,k_2\ldots,k_n}$ is defined via $D^+_{k_1,k_2\ldots,k_n}$ and $D^-_{k_1,k_2\ldots,k_n}$ according to~\eqref{alld.01} and these definitions, in turn, contain various $M[\tau]$, $\underline{\rm f}[\tau]$ and $\overline{\underline{\rm g}[\tau]}$ at modular depth $n{-}1$ or lower. All of these objects are well defined and equivariant by our induction hypothesis. This observation, coupled with the fact that the $\underline{\rm G}_{k_1}[\tau]$ and $\overline{\underline{\rm G}_{k_n}[\tau]}$ are also equivariant, proves the equivariance of  $D_{k_1,k_2\ldots,k_n}[\tau]$ at modular depth $n$.

The closure condition~\eqref{clphi} that we need to prove at modular depth $n$  is equivalent to proving
\beq
\partial_{\bar \tau}  D^+_{k_1,\ldots,k_{n}}[\tau] \wedge \, 2 {\rm d}\bar \tau  
=2 {\rm d}  \tau \wedge \, \partial_{ \tau}  D^-_{k_1,\ldots,k_{n}}[\tau]  .
\label{allD.05}
\eeq
Using (\ref{alld.02}) and \eqref{alld.04}, the left-hand side can be written as
\begin{align}
&\underline{{\rm G}}_{k_1}[\tau]  \wedge \big(
D^-_{k_2,\ldots,k_{n}}[\tau] 
+ \overline{\underline{{\rm g}}_{k_2,\ldots,k_n}[\tau]}
\big) +
\sum_{r=2}^{n-2} \underline{{\rm f}}_{k_1,\ldots,k_r}[\tau]  \wedge \big(
D^-_{k_{r+1},\ldots,k_{n}}[\tau] 
+ \overline{\underline{{\rm g}}_{k_{r+1},\ldots,k_n}[\tau]}
\big) \notag \\
&\quad +  \underline{{\rm f}}_{k_1,\ldots,k_{n-1}}[\tau]  \wedge \overline{\underline{{\rm G}}_{k_n}[\tau]}\, ,
\label{allD.06}
\end{align}
where we have used the fact that $D^-_{k_n}[\tau]=\overline{\underline{{\rm G}}_{k_n}}[\tau]$ and $\overline{\underline{\rm g}_{k_n}[\tau]}=0$.
Similarly, we can write the right-hand side of (\ref{allD.05}) as
\begin{align}
&\big( D^+_{k_1,\ldots,k_{n-1}}[\tau] +  \underline{{\rm f}}_{k_1,\ldots,k_{n-1}}[\tau]  \big) \wedge \overline{\underline{{\rm G}}_{k_n}[\tau]}
 + \sum_{r=2}^{n-2} 
 \big( D^+_{k_1,\ldots,k_{r}}[\tau] +  \underline{{\rm f}}_{k_1,\ldots,k_{r}}[\tau]  \big) \wedge
 \overline{\underline{{\rm g}}_{k_{r+1},\ldots,k_n}[\tau]} \notag \\
 &\quad + \underline{{\rm G}}_{k_1}[\tau] 
 \wedge \overline{\underline{{\rm g}}_{k_{2},\ldots,k_n}[\tau]} \, .
\label{allD.07}
\end{align}
The terms in (\ref{allD.06}) and (\ref{allD.07}) not containing any $D^{\pm}$ match by inspection. The remaining terms in (\ref{allD.06}) and (\ref{allD.07}) can be shown to agree by inserting (\ref{alld.02}) for the $D^{\pm}$ which, as before, are of lower modular depth and well defined.
 In this setting, the terms involving $D^{\pm}$ in both (\ref{allD.06}) and (\ref{allD.07}) simplify to
\begin{align} 
 &\underline{{\rm G}}_{k_1}[\tau] 
M_{k_{2},\ldots,k_{n-1}}[\tau] 
 \wedge \overline{\underline{{\rm G}}_{k_n}[\tau]}
 +\underline{{\rm G}}_{k_1}[\tau] 
 \wedge \sum_{r=2}^{n-2}
 M_{k_{2},\ldots,k_{r}}[\tau] \overline{\underline{{\rm g}}_{k_{r+1},\ldots,k_n}[\tau]} \label{allD.08} \\
 &\quad
 +  \sum_{r=2}^{n-2}  \underline{{\rm f}}_{k_1,\ldots,k_{r}}[\tau] 
M_{k_{r+1},\ldots,k_{n-1}}[\tau] 
  \wedge \overline{\underline{{\rm G}}_{k_n}[\tau]}
   +  \sum_{2\leq r< s \leq n-2}^{n-2}   \underline{{\rm f}}_{k_1,\ldots,k_{r}}[\tau]   \wedge
   M_{k_{r+1},\ldots,k_{s}}[\tau]  \overline{\underline{{\rm g}}_{k_{s+1},\ldots,k_n}[\tau]}\, .
  \notag
 \end{align} 
 Therefore, equation \eqref{allD.05} holds and $D_{k_1,k_2,\ldots,k_n}$ is closed.

\subsubsection{Proving the existence of a modular completion at depth $n$}
\label{sec:ES}

We have demonstrated the closure and equivariance of $D_{k_1,k_2,\ldots,k_n}[\tau]$ in the previous section and, therefore, the following integral is equivariant up to a cocycle depending on $n$ pairs $X_i,Y_i$ of commutative bookkeeping variables, $i=1,2,\ldots,n$:
\beq \label{eq:kn}
K_{k_1,k_2,\ldots ,k_n}[\tau] =  -\dfrac{1}{2} \, 
\bigg\{ 
\int_{\tau}^{i\infty}  
\!D^+_{k_1,k_2,\ldots,k_{n}}[\tau_1] 
+\int_{\bar \tau}^{-i\infty} \! 
 \!D^-_{k_1,k_2,\ldots,k_{n}}[\tau_1] 
 \bigg\} \, . 
\eeq
The cancellation of its cocycle in multiple pairs $X_i,Y_i$ follows the logic of sections \ref{subsubsec:5.1.1} and \ref{sec:eqv3}. We first introduce projectors analogous to (\ref{defdelab.2}) which reduce the cocycle from the integral over $D^{\pm}_{k_1,k_2,\ldots,k_n}$ to a function of only one pair $X_1,Y_1$. One possible choice for the desired family of projectors is
\beq
\delta^{\ell_1,\ell_2,\ldots,\ell_{n-1}} \coloneqq  \delta^{\ell_1}_{1,2}\circ \delta^{\ell_2}_{2,3}\circ \ldots \circ \delta^{\ell_{n-1}}_{n-1,n}
\label{dptnproj}
\eeq
with $\delta^\ell_{a,b}$ given by (\ref{defdelab}). This choice of projector corresponds to a particular way of putting parentheses when decomposing the $n$-fold tensor product $V(\eee_{k_1})\! \otimes \! V(\eee_{k_2}) \! \otimes \!\ldots\! \otimes\! V(\eee_{k_n})$ into irreducible representations of $\mf{sl}_2$. This definition is a natural generalisation of~\eqref{defdelab.2}. We formally allow for arbitrary non-negative values of the $\ell_i$, but in practice only a finite number of projectors will be non-zero when acting on a polynomial in the $(X_i,Y_i)$.

For a given choice of the integers $\ell_i$ characterising the projector (\ref{dptnproj}), we reduce the function $K_{k_1,k_2,\ldots ,k_n}[\tau]$ of $2n$ bookkeeping variables $X_i, Y_i$ to a function of only $X_1,Y_1$: 
\begin{align}
K^{(\ell_1,\ldots,\ell_{n-1})}_{k_1,k_2,\ldots ,k_n}[X_1,Y_1;\tau] &=  \frac{(i\pi)^{\ell_1+\ldots+\ell_{n-1}} \delta^{\ell_1,\ell_2,\ldots,\ell_{n-1}}}{(\ell_1!)^2 \ldots (\ell_{n-1}!)^2} \, 
K_{k_1,k_2,\ldots ,k_n}[\tau]
\label{dptnprojM} \, ,
\end{align}
which is equivariant up to a cocycle now only depending on one pair $X_1,Y_1$ of bookkeeping variables. An application of the Eichler--Shimura theorem then gives a systematic completion which results in the equivariant function at fixed $\ell_1,\ldots,\ell_{n-1}$:
\begin{align}
M^{(\ell_1,\ldots,\ell_{n-1})}_{k_1,k_2,\ldots ,k_n}[X_1,Y_1;\tau] &= K^{(\ell_1,\ldots,\ell_{n-1})}_{k_1,k_2,\ldots ,k_n}[X_1,Y_1;\tau] - c^{(\ell_1,\ldots,\ell_{n-1})}_{k_1,k_2,\ldots ,k_n}[X_1,Y_1] \label{hereES} \\
&\quad
- \frac{1}{2}
\bigg\{ 
\int_{\tau}^{i\infty}  \underline{\rm f}^{(\ell_1,\ldots,\ell_{n-1})}_{k_1,k_2,\ldots ,k_n}[X_1,Y_1;\tau_1]
+\int_{\bar \tau}^{-i\infty} \!   \overline{\underline{\rm g}^{(\ell_1,\ldots,\ell_{n-1})}_{k_1,k_2,\ldots ,k_n}[X_1,Y_1;\tau_1]} \bigg\} \, , \notag
\end{align}
where $c^{(\ell_1,\ldots,\ell_{n-1})}_{k_1,k_2,\ldots ,k_n}[X_1,Y_1]$ is a polynomial and
\begin{itemize}
\item $ \underline{\rm f}^{(\ell_1,\ldots,\ell_{n-1})}_{k_1,k_2,\ldots ,k_n}[X_1,Y_1;\tau_1]$ is a unique equivariant $(1,0)$-form composed of holomorphic cusp forms of weight $w=k_1{+}\ldots{+}k_n{-} 2(\ell_1{+}1){-} \ldots {-} 2(\ell_{n-1}{+}1)$ 
multiplying $(X_1{-}\tau_1 Y_1)^{w-2}$,
\item $\overline{\underline{\rm g}^{(\ell_1,\ldots,\ell_{n-1})}_{k_1,k_2,\ldots ,k_n}[X_1,Y_1;\tau_1]}$ is a unique equivariant $(0,1)$-form  composed of antiholomorphic modular forms of the same weight $w$ multiplying $(X_1{-}\bar\tau_1 Y_1)^{w-2}$.
\end{itemize}
By constructing these equivariant completions $M^{(\ell_1,\ldots,\ell_{n-1})}_{k_1,k_2,\ldots ,k_n}[X_1,Y_1;\tau]$ for all integers $\ell_i \geq 0$ in the finite range $\sum_{i=1}^{n-1}\ell_i \leq \frac{1}{2}(k_1{+}k_2{+}\ldots{+}k_n){-}n$, one captures all irreducible representation of $\mf{sl}_2$ in the tensor product $V(\eee_{k_1})\! \otimes \! V(\eee_{k_2}) \! \otimes \!\ldots\! \otimes\! V(\eee_{k_n})$ relevant for $M_{k_1,k_2,\ldots,k_n}$ at least once. Hence, it is possible to construct an unprojected equivariant 
 function
\begin{align}\label{eq:mkn}
M_{k_1,k_2,\ldots ,k_n}[\tau] &= K_{k_1,k_2,\ldots ,k_n}[\tau] - c_{k_1,k_2,\ldots ,k_n}  \\
&\qquad
- \frac{1}{2}
\bigg\{ 
\int_{\tau}^{i\infty}  \underline{\rm f}_{k_1,k_2,\ldots ,k_n}[\tau_1]
+\int_{\bar \tau}^{-i\infty} \!   \overline{\underline{\rm g}_{k_1,k_2,\ldots ,k_n}[\tau_1]} 
 \bigg\} \,,\notag
\end{align}
of $n$ pairs of bookkeeping variables subject to
\begin{align}
      M^{(\ell_1,\ldots,\ell_{n-1})}_{k_1,k_2,\ldots ,k_n}[X_1,Y_1;\tau] = \frac{(i\pi)^{\ell_1+\ldots+\ell_{n-1}} \delta^{\ell_1,\ell_2,\ldots,\ell_{n-1}}}{(\ell_1!)^2 \ldots (\ell_{n-1}!)^2} \, M_{k_1,k_2,\ldots ,k_n}[\tau] \,,
      \label{projmn}
\end{align}
and analogous conditions for $c_{k_1,k_2,\ldots ,k_n}$, $\underline{\rm f}_{k_1,k_2,\ldots ,k_n}[\tau_1]$ and $ \overline{\underline{\rm g}_{k_1,k_2,\ldots ,k_n}[\tau_1]} $.
The functions defined in \eqref{eq:mkn} can be seen to satisfy the differential equation \eqref{alld.04}.

To summarise, under our induction hypothesis, we have proven the closure and equivariance of $D_{k_1,k_2,\ldots,k_n}$ and shown the existence of 
 equivariant $M_{k_1,k_2,\ldots,k_n}[\tau]$, $\underline{\rm f}_{k_1,k_2,\ldots,k_n}[\tau]$ and $\overline{\underline{\rm g}_{k_1,k_2,\ldots,k_n}[\tau]}$ satisfying \eqref{alld.04}. Therefore, by induction, $D_{k_1,k_2,\ldots,k_n}$ is equivariant and closed and equivariant $M_{k_1,k_2,\ldots,k_n}$, $\underline{\rm f}_{k_1,k_2,\ldots,k_n}$ and $\overline{\underline{\rm g}_{k_1,k_2,\ldots,k_n}}$ satisfying \eqref{alld.04} exist, for all $n \geq2$.
Furthermore, we can always assume that $M_{k_1,k_2,\ldots,k_n}$ is the modular completion of $K_{k_1,k_2,\ldots,k_n}$ \eqref{eq:kn} and takes the form given by \eqref{eq:mkn}.

\subsubsection{Consequences for $\beta^{\rm eqv}$ at modular depth $n$}
\label{sec:corr}

The equivariant integrals $M_{k_1,k_2}$ and $M_{k_1,k_2,k_3}$ offer
an equivalent description of the modular forms
$\beta^{\rm eqv}$ at modular depths two and three via (\ref{eq:Mbeqv}) and (\ref{olidreams}), respectively. The
construction of equivariant $M_{k_1,k_2,\ldots,k_n}$
at arbitrary modular depth should similarly capture the $\beta^{\rm eqv}$ at all orders in the generating series $\mathbb J^{\rm eqv}$ for suitable choices of the $\tau$-independent $c_{k_1,k_2,\ldots,k_n}$ in the form
\begin{align}
&M_{k_1,k_2,\ldots,k_n}[\tau] = \bigg( {-}\frac{1}{4 } \bigg)^{n} \, (k_1{-}1)! (k_2{-}1)! \ldots (k_n{-}1)!
\sum_{j_1=0}^{k_1-2} \sum_{j_2=0}^{k_2-2} \ldots \sum_{j_n=0}^{k_n-2} 
 \frac{ \binom{k_1-2}{j_1} \binom{k_2-2}{j_2}\ldots \binom{k_n-2}{j_n} }{(-4y)^{j_1+j_2+\ldots+j_n}}
 \notag \\
&\quad \! \! \times 
 \beqv{j_n &\ldots &j_2&j_1}{k_n &\ldots &k_2&k_1}
(X_1{-}\tau Y_1)^{j_1} (X_1{-}\bar \tau Y_1)^{k_1-2-j_1} 
\ldots
(X_n{-}\tau Y_n)^{j_n} (X_n{-}\bar \tau Y_n)^{k_n-2-j_n}\, .
\label{Mbeqvn}
\end{align}
The $\mf{sl}_2$ irreducible representations (\ref{projmn}) obtained from the $\delta^{\ell_1,\ell_2,\ldots,\ell_{n-1}}$ projections of (\ref{Mbeqvn}) then feature combinations of $\beta^{\rm eqv}$ where $\nabla$ and $\overline{\nabla}$ act as raising and lowering operators.

In the same way as the $c^{(\ell_1,\ldots,\ell_{n-1})}_{k_1,k_2,\ldots ,k_n}$ in $\mf{sl}_2$-singlet representations are unconstrained by equivariance, there is an analogous freedom of redefining the modular-depth-zero contributions $\ccsv{j_1 &\ldots &j_\ell}{k_1 &\ldots &k_\ell}$ to $\beqv{j_1 &\ldots &j_\ell}{k_1 &\ldots &k_\ell}$ in an $\mf{sl}_2$-invariant way. As detailed in section \ref{sec:4zeta.6} and at the end of section \ref{sec:ngeozvarpi}, the ambiguities in the constants $c^{\rm sv}$ entering the generating series $\mathbb J^{\rm eqv}$ are fixed by imposing certain conditions on its arithmetic generators $\zetaE_i,\zetaE_{\pern}$. This choice can be propagated to fix ambiguities in $M_{k_1,k_2,\ldots,k_n}$ by imposing the following corollary of (\ref{Mbeqvn}) iteratively in $n$:
\begin{align}
c_{k_1,k_2,\ldots,k_n} &= -\bigg( {-}\frac{1}{4 } \bigg)^{n} \, (k_1{-}1)! (k_2{-}1)! \ldots (k_n{-}1)!
\sum_{j_1=0}^{k_1-2} \sum_{j_2=0}^{k_2-2} \ldots \sum_{j_n=0}^{k_n-2} 
 \frac{ \binom{k_1-2}{j_1} \binom{k_2-2}{j_2}\ldots \binom{k_n-2}{j_n} }{(-4y)^{j_1+j_2+\ldots+j_n}}
 \notag \\
&\quad 
\times \ccsv{j_n &\ldots &j_2&j_1}{k_n &\ldots &k_2&k_1}
\bigg( \frac{  Y_1 }{2\pi i} \bigg)^{j_1} 
X_1^{k_1-2-j_1} 
\ldots \bigg( \frac{  Y_n }{2\pi i} \bigg)^{j_n} 
  X_n^{k_n-2-j_n} 
\, .
\label{ccsvn}
\end{align}
We should highlight another advantage of the generating-series approach to $\beta^{\rm eqv}$ over the alternative construction via (\ref{Mbeqvn}): the structure of the generating series $\mathbb J^{\rm eqv}$ -- in particular the conjectural organisation of MZVs through the group-like series $\mathbb M^{\rm sv}$ in (\ref{uplift}) and (\ref{extendMM}) -- fixes the appearance of arbitrary (products of) MZVs from that of odd Riemann zeta values. This has implications on the MZVs entering all of $c_{k_1,\ldots ,k_n}$, $\underline{\rm f}_{k_1,\ldots ,k_n}[\tau_1]$ and $\overline{\underline{\rm g}_{k_1,\ldots ,k_n}[\tau_1]}$. 

Still, both approaches to the construction of $\beta^{\rm eqv}$ have in common that the new periods at modular depth $n$ (i.e.\ generalisations of the $\Lambda^{d_1,d_2}_{k,w}$ and $\varpi^{d_1,d_2}_{k,w}$ of section  \ref{sec:4.2}) have to be determined from modularity. By imposing equivariance of (\ref{eq:mkn}), the constants $c_{k_1,\ldots ,k_n}$ in non-singlet representations of $\mf{sl}_2$ as well as the real coefficients of the modular forms in $\underline{\rm f}_{k_1,\ldots ,k_n}[\tau_1]$ and $\overline{\underline{\rm g}_{k_1,\ldots ,k_n}[\tau_1]}$ at modular depth $n$ are determined by the MMVs (\ref{eq:mmvdef}) and (\ref{MMVsec.17b}) from the $S$-transformation of $K_{k_1,\ldots ,k_n}[\tau]$. 

In the formulation via $\mathbb J^{\rm eqv}$, the new periods enter via $c^{\rm sv}, \hat \psi^{\rm sv}$ as well as higher-order terms in the expansion of the letters $\eee_{\Delta^{\pm}}$ in (\ref{edeltaex}). Their detailed expressions in terms of MMVs are generated by the $S$-cocycle condition (\ref{beyond1708}). Within the discussion of section \ref{sec:sequiv}, the existence of a series $\mathbb M^{\rm sv}(\sigmaE_i,\sigmaE_\pern)$ subject to (\ref{beyond1708}) was conjectural and relied on order-by-order checks at increasing modular depth and degree. The inductive proof of this section together with the relation (\ref{Mbeqvn}) in turn establishes the existence of a solution to (\ref{beyond1708}) at arbitrary modular depth and degree.

In order to determine the new periods entering $c_{k_1,\ldots ,k_n}$, $\underline{\rm f}_{k_1,\ldots ,k_n}[\tau_1]$ and $\overline{\underline{\rm g}_{k_1,\ldots ,k_n}[\tau_1]}$ at a given modular depth and degree, one would still need additional information on the MMVs entering the $S$-transformation of $K_{k_1,\ldots ,k_n}[\tau]$. From the methods of this work, one would have to test order by order whether the $S$-cocycle of $K_{k_1,\ldots ,k_n}[\tau]$ can be expressed in terms of (sums of products of) known periods and add a minimal set of new ones whenever this is not the case. This kind of case-by-case study for instance leads to the new periods $\Lambda^{d_1,d_2}_{k,w}$ and $\varpi^{d_1,d_2}_{k,w}$ at modular depth three and degree $\leq 20$ described in section \ref{sec:4.2} which were initially derived from $\mathbb J^{\rm eqv}$. 

As another open problem for the future, the procedure of this section to determine $c_{k_1,\ldots ,k_n}$, $\underline{\rm f}_{k_1,\ldots ,k_n}[\tau_1]$ and $\overline{\underline{\rm g}_{k_1,\ldots ,k_n}[\tau_1]}$ from MMVs does not manifest any correlations among the MZVs. More specifically, beyond our explicit checks at modular depth three and degree $\leq 20$ it is still conjectural  that the appearances of $\zeta_{i_k}$ and $\rho^{-1} ( {\rm sv}(f_{i_1} f_{i_2} \ldots f_{i_\ell}))$ in $\mathbb M^{\rm sv}(\sigmaE_i,\sigmaE_\pern)$ are interlocked through the expansion (\ref{Msvz}) of the generating series $\mathbb M^{\rm sv}(\sigmaE_i)$ of its MZV sector.

\section{Conclusion and outlook}
\label{sec:last}

In this work, we have connected different perspectives on non-holomorphic modular forms built from
iterated integrals of holomorphic modular forms. The explicit construction of such modular forms 
was advanced in several directions, and elegant number-theoretic structures have been identified in the coefficients of the iterated integrals.

A first line of results concerns modular graph forms (MGFs) known from the low-energy expansion of genus-one
string amplitudes \cite{DHoker:2015wxz, DHoker:2016mwo}. Following Brown's generating-series 
approach to assemble MGFs from iterated Eisenstein integrals \cite{Brown:2017qwo2},
we give the first explicit description of all contributions from single-valued multiple zeta values (MZVs). 
The key ingredients in our construction are so-called zeta generators (see section~\ref{sec:rev.zeta}) that conspire with the 
non-commutative variables $\epsilon_k$ in the generating series. As a main result of this work, 
we find simple composition rules for zeta generators which interlock products and higher-depth 
instances of MZVs with the appearance of odd Riemann zeta values $\zeta_{2k+1}$ as explained in section~\ref{sec:4zeta}.

A more general class of non-holomorphic modular forms is obtained by augmenting
iterated Eisenstein integrals by holomorphic cusp forms as additional integration kernels. 
The modular completions of double Eisenstein integrals via cusp forms are known to
feature odd Riemann zeta values and L-values of holomorphic cusp forms
\cite{Brown:2017qwo, Dorigoni:2021ngn}. Another main result of this work is the
systematic construction of the analogous modular completions of triple Eisenstein integrals.
Apart from double integrals that mix holomorphic Eisenstein series and cusp forms, 
we find irreducible MZVs of depth $\geq 3$ and two classes of new periods.
This is explained in detail in section~\ref{sec:4} with general arguments related to the modular properties of the generating series in section~\ref{sec:6}.
The existence of similar modular completions is established for 
iterated integrals over an arbitrary number of Eisenstein series by the more direct construction methods in section~\ref{sec:8}.

Our results motivate and support a variety of directions of follow-up research:
\begin{itemize}
\item The use of zeta generators in this work manifests striking parallels between the constructions
of single-valued genus-zero polylogarithms in one variable and modular graph forms at genus one, see section~\ref{sec:4zeta.8}. 
It would be rewarding to pinpoint echoes of these parallels in the multi-variable case and at higher genus.
For instance, the description of single-valued polylogarithms in $n$ variables via zeta generators
\cite{Frost:2023stm} may shed light on the appearance of MZVs in elliptic modular graph forms
\cite{DHoker:2018mys, Dhoker:2020gdz, new:eMGF} depending on $n{-}1$ marked points on a torus. At higher
genus, it remains to find realisations of zeta generators that introduce MZVs into the respective
modular graph forms \cite{DHoker:2017pvk} and tensors \cite{DHoker:2020uid}, starting with the 
apparance of $\zeta_3$ in the
expansion of the Zhang--Kawazumi invariant in \cite{Pioline:2015qha}.
\item In this work, zeta generators are applied to modular completions of iterated Eisenstein integrals
where the integration variables are modular parameters. The tight interplay of zeta generators with 
the fundamental group of the once punctured torus \cite{Dorigoni:2023part1} suggests to reformulate
and extend our results at the level of configuration-space integrals over marked points on genus-one 
surfaces. More specifically, one could try to (i) organize the MZVs in the
$\tau \rightarrow i\infty$ asymptotics of elliptic associators \cite{Enriquez:Emzv} via
zeta generators, (ii) describe the low-energy expansions of the open- and closed-string
integrals in \cite{Mafra:2019ddf, Mafra:2019xms, Gerken:2019cxz} via matrix representations 
of zeta generators, (iii) connect the proposal for a single-valued 
map between open- and closed-string integrals at genus one \cite{Gerken:2020xfv} with the more recent
double-copy formulae for integrals over marked points on a torus \cite{Stieberger:2022lss, Stieberger:2023nol, Bhardwaj:2023vvm, Mazloumi:2024wys}.
\item The new periods encountered in the modular completion of triple Eisenstein integrals call for an organizing principle akin to the $f$-alphabet description of MZVs \cite{Brown:2011mot, Brown:2011ik}. It is tempting to aim for an extension of the group-like series $\mathbb M^{\rm sv}$ in MZVs to accommodate L-values of holomorphic cusp forms and the new periods that start appearing at the level of triple integrals. Such extensions of $\mathbb M^{\rm sv}$ should involve the counterparts of zeta generators for non-critical L-values discussed in \cite{Brown2019}, see section~\ref{sec:4.2} for preliminary suggestions. Moreover, a group-like series featuring L-values and beyond may offer a natural interpretation of the composite letters $\eee_{\Delta^{\pm}}$ accompanying the holomorphic cusp forms in the generating series of this work: The expansions of $\eee_{\Delta^{\pm}}$ in terms of brackets of Eisenstein letters $\eee_k$ could be traced back to conjugations of elementary letters for cusp forms \cite{brown2017a} with suitable extensions or cuspidal analogues of $\mathbb M^{\rm sv}$.
\item The Laplace equations discussed in section \ref{sec:7} simplify the integration of modular-invariant MGFs over the fundamental domain of SL(2,$\mathbb{Z}$) as demonstrated in \cite{Green:1999pv,DHoker:2019mib,DHoker:2021ous,Doroudiani:2023bfw}. Our results on equivariant iterated Eisenstein integrals of arbitrary modular depth in section \ref{sec:gen} are a first step to extend the integration techniques of the references to MGFs of modular depth four and higher. 
Since having the zero-mode of MGFs with respect to $\Re(\tau)$ is enough to integrate them over the fundamental domain \cite{zagier1981rankin,Angelantonj:2011br,Angelantonj:2012gw,Doroudiani:2023bfw}, 
another possible investigation is to
take advantage of the contributions $\sim (q\bar{q})^n$ to the modular invariants in this work exposed by their iterated-integral representations.
With this approach, we can gain valuable insights into the number-theoretic properties of the integrated MGFs and evaluate previously inaccessible coefficients of low-energy interactions $D^{w}R^4$ of type-II superstrings with $w\geq 14$, going beyond the state of the art on $D^{\leq 12}R^4$~\cite{Green:2008uj,Basu:2015ayg,Basu:2016xrt} summarised with a view to transcendentality in \cite{DHoker:2019blr}.
\item Modular forms involving double Eisenstein integrals
were obtained from Poincar\'e series of integrals over a single Eisenstein series
\cite{Dorigoni:2021jfr, Dorigoni:2021ngn}. It would be interesting to generalize
these results to new representations of modular triple integrals in terms of
Poincar\'e series of suitable double integrals and to make contact with the cuspidal Poincar\'e series of \cite{Diamantis:2020}. These connections between triple, double and single integrals should contain key information on the systematics of how Poincar\'e series relate iterated integrals over different numbers of holomorphic modular forms.
\end{itemize}

\section*{Acknowledgements}

We are grateful to Francis Brown, Emiel Claasen, Eric D'Hoker, Benjamin Enriquez, Hadleigh Frost, Michael Green, Deepak Kamlesh, Carlos Rodriguez and Congkao Wen 
for combinations of inspiring discussions, collaboration on related topics and helpful comments 
on a draft version of this work. In particular, we are indebted to Francis Brown for bringing
the zeta generators~$\sigmaT_w$ to our attention.
The authors would like to thank the organisers of the workshops ``Geometries and Special Functions for Physics and Mathematics'' at the BCTP Bonn and ``New connections between physics and number theory'' at the Pollica Physics Centre for creating a stimulating atmosphere.
AK would like to thank the Isaac Newton Institute for Mathematical Sciences, Cambridge, for support and hospitality during the programme ``Black holes: bridges between number theory and holographic quantum information'' where work on this paper was undertaken. 

This research was supported in part by grant NSF PHY-2309135 to the Kavli Institute for Theoretical Physics (KITP). This work was supported by EPSRC grant no EP/R014604/1.
The research of MD was supported by the IMPRS for Mathematical and Physical Aspects of Gravitation, Cosmology and Quantum Field Theory. JD is supported by the Royal Society (Spectral theory of automorphic forms: trace formulas and more). MH, OS and BV are partially supported by the European Research Council under
ERC-STG-804286 UNISCAMP. MH and BV are furthermore supported by the Knut and Alice Wallenberg Foundation under grants KAW 2018.0116 and KAW 2018.0162, respectively. MH is also partially supported by the Swiss National Science Foundation through the NCCR SwissMAP. The research of OS is partially supported by the strength area ``Universe and mathematical physics'' which is funded by the Faculty of Science and Technology at Uppsala University.

\appendix

\section*{Appendix}

\section{\texorpdfstring{Multiple zeta values and $f$-alphabet}{Multiple zeta values and f-alphabet}}
\label{app:A}

This appendix gathers basics of multiple zeta values (MZVs), their $f$-alphabet description and their single-valued map. We follow conventions where MZVs, $\zeta_{n_1,n_2,\ldots,n_r}$, of depth $r$
are defined by nested sums over integers $k_i$
\beq
\zeta_{n_1,n_2,\ldots,n_r} \coloneqq \sum_{1\leq k_1<k_2<\ldots<k_r}^{\infty} k_1^{-n_1}
k_2^{-n_2}\ldots k_r^{-n_r}\,,
\label{appMZV.01}
\eeq
indexed by integers $n_i$ in the range $n_1,\ldots,n_{r-1} \geq 1$ and $n_r\geq 2$. MZVs admit an alternative description in terms of multiple polylogarithms (\ref{svieis.10}) at unit argument $z=1$,
\beq
\zeta_{n_1,n_2,\ldots,n_r} = (-1)^r G(\underbrace{0,\ldots,0}_{n_r-1},1,\ldots,\underbrace{0,\ldots,0}_{n_2-1},1,\underbrace{0,\ldots,0}_{n_1-1},1;1)\, .
\label{appMZV.g1}
\eeq
By combining the properties of their representations via nested sums and iterated integrals, one can generate infinite families of relations among MZVs over $\mathbb Q$. All known $\mathbb Q$-relations among MZVs preserve the weight $w=n_1{+}n_2{+}\ldots{+}n_r$ of $\zeta_{n_1,n_2,\ldots,n_r}$ and can therefore be solved weight by weight. From the joint effort of all currently known $\mathbb Q$-relations up to weight $w=18$, for instance, one can express all MZVs in this range via products of Riemann zeta values $\zeta_n$ (i.e.\ MZVs at depth one) and the following list of conjecturally indecomposable higher-depth MZVs \cite{Blumlein:2009cf}:
\beq
\begin{array}{lc|c||c|c}
&w &{\rm basis} \ {\rm MZVs} &w &{\rm basis} \ {\rm MZVs}\\\hline
&8 &\zeta_{3,5} &14 & \zeta_{3,3,3,5} , \, \zeta_{3,11} , \, \zeta_{5,9}  \\
&10 &\zeta_{3,7} &15 &\zeta_{5,3,7} , \, \zeta_{3,3,9} , \, \zeta_{1,1,3,4,6}\\
&11 &\zeta_{3,3,5}&16 &\zeta_{3,3,3,7} , \, \zeta_{3,3,5,5}
, \, \zeta_{3,13}, \, \zeta_{5,11} , \, \zeta_{1,1,6,8}  \\
&12 &\zeta_{3,9} , \, \zeta_{1,1,4,6} &17 &\zeta_{3,3,3,3,5}, \, \zeta_{1,1,3,6,6} ,\,
\zeta_{5,5,7}, \, \zeta_{3,3,11} ,\, \zeta_{5,3,9} , \, \zeta_{3,5,9} \\
&13 &\zeta_{3,3,7}, \, \zeta_{3,5,5} &18 &\zeta_{3,15} , \, \zeta_{5,13}, \, \zeta_{1,1,6,10},\,
\zeta_{3,5,5,5}, \, \zeta_{5,3,3,7} , \, \zeta_{3,3,3,9} , \, \zeta_{3,5,3,7} , \, \zeta_{1,1,3,3,4,6}
\end{array}
\label{appMZV.03}
\eeq

\subsection{\texorpdfstring{The $f$-alphabet}{The f-alphabet}}
\label{app:A.1}

The systematics of conjecturally indecomposable higher-depth MZVs in (\ref{appMZV.03}) can be conveniently captured by mapping them to certain non-commutative variables $f_k$ known as the $f$-alphabet \cite{Brown:2011mot, Brown:2011ik}. However, in view of the conjectural status of $\mathbb Q$-linear independence results for MZVs and their products, one performs the mapping into the $f$-alphabet at the level of so-called motivic MZVs $\zeta^{\mot}_{n_1,n_2,\ldots,n_r}$. By their definition in the algebraic-geometry literature \cite{Goncharov:2005sla, Brown:2011mot, Brown2014MotivicPA, brown2015notes}, motivic MZVs obey all currently known $\mathbb Q$-relations of their real counterparts in (\ref{appMZV.01}) but are otherwise independent over $\mathbb Q$. 

The key idea behind the $f$-alphabet is to map motivic MZVs to a Hopf-algebra comodule
\beq
{\cal F}\coloneqq  \mathbb Q[f_2] \otimes_{\mathbb Q} \mathbb Q \langle f_3,f_5,f_7,\ldots \rangle\,,
\label{appMZV.g2}
\eeq
spanned by monomials in non-commutative generators $f_3,f_5,f_7,\ldots$ of odd degree and a single commutative generator $f_2$ of even degree. Given an isomorphism $\rho$ mapping $\zeta^{\mot}_{n_1,n_2,\ldots,n_r}$ to non-commutative polynomials in ${\cal F}$, all $\mathbb Q$-relations among motivic MZVs are exposed. Such isomorphisms $\rho$ are taken to obey the normalisation condition on Riemann zeta values
\beq
\rho( \zeta^{\mot}_{n} ) = f_n\, ,
\label{appMZV.g3}
\eeq
(with $f_{2k} \coloneqq  \frac{\zeta_{2k}}{(\zeta_2)^k} f_2^k \in \mathbb Q f_2^k$ in case of even degree) and to translate products of motivic MZVs into shuffles\footnote{The shuffle product in ${\cal F}$ is understood to act trivially on the commutative generator $f_2$,
\begin{equation*}
f_2^a f_{i_1}\ldots f_{i_r} \shuffle f_2^b f_{i_{r+1}}\ldots f_{i_s} = f_2^{a+b} (f_{i_1}\ldots f_{i_r} \shuffle  f_{i_{r+1}}\ldots f_{i_s} )  \, , \ \ \ \ \ \ 
i_j \in 2\mathbb N{+}1\, .
\end{equation*}} (with $\vec{n},\vec{m} \in \mathbb N^\times$),
\beq
\rho( \zeta^{\mot}_{\vec{n}}\cdot \zeta^{\mot}_{\vec{m}} ) = 
\rho( \zeta^{\mot}_{\vec{n}}  )  \shuffle \rho(  \zeta^{\mot}_{\vec{m}} ) 
\, .
\label{appMZV.g4}
\eeq
Most importantly, $\rho$-isomorphisms are required to translate the Goncharov--Brown coaction $\Delta_G$ of motivic MZVs \cite{goncharov2001multiple, Goncharov:2005sla, Brown:2011mot, Brown:2011ik} into the simple deconcatenation coaction $\Delta_{\rm dec}$ in ${\cal F}$,\footnote{We note that $f_2$ here is only mapped to the left factor in the tensor product under the coaction. This choice is not uniform in the literature.}
\begin{align}
\rho \big( \Delta_G( \zeta^{\mot}_{n_1,n_2,\ldots,n_r} ) \big) &=  \Delta_{\rm dec}  \big( \rho  ( \zeta^{\mot}_{n_1,n_2,\ldots,n_r} ) \big) \, , \label{appMZV.g5} \\
\Delta_{\rm dec} \big(f_2^n f_{i_1} f_{i_2}\ldots f_{i_r}\big) &= \sum_{j=0}^r (f_2^n f_{i_1} f_{i_2} \ldots f_{i_j})\otimes
f_{i_{j+1}}\ldots f_{i_r}\, .
\notag
\end{align}
The conditions (\ref{appMZV.g3}) to (\ref{appMZV.g5}) determine the $\rho$-image of motivic MZVs $\zeta^{\mot}_{n_1,n_2,\ldots,n_r}$ of depth $r\geq 2$ up to the coefficient of $f_{n_1+n_2+\ldots+n_r}$.

A canonical choice of these leftover coefficients, i.e.\ a canonical choice of isomorphism $\rho$, is described in a companion paper \cite{Dorigoni:2023part1} and, from now on and everywhere in this work, $\rho$ will denote this preferred isomorphism. As discussed in \cite{Dorigoni:2023part1}, this leads to the following $f$-alphabet images of the simplest indecomposable motivic MZVs in (\ref{appMZV.03})
\begin{align}
\rho( \zeta^\mot_{3,5}) &= - 5 f_3 f_5 + \frac{100471}{35568} f_8 \, , \label{appMZV.g6} \\
\rho( \zeta^\mot_{3,7}) &= - 14 f_3 f_7 - 6 f_5 f_5 + \frac{408872741707}{40214998720} f_{10}\, , \notag \\
\rho( \zeta^\mot_{3,3,5}) &= - 5 f_3 f_3 f_5 - 45 f_2 f_9 - \frac{6}{5} f_2^2 f_7  + \frac{4}{7} f_2^3 f_5 + \frac{1119631493}{14735232} f_{11} \, .
\notag
\end{align}
The large integers appearing in the  denominators of the coefficients of $f_w$ in \eqref{appMZV.g6} are purely artifacts of the choice of $\zeta^\mot_{3,5}$, $\zeta^\mot_{3,7}$ etc.\ as irreducibles (similar artifacts occur in (\ref{rhoinv}) and (\ref{appMZV.g8}) below). In the companion paper \cite{Dorigoni:2023part1} we give a natural method to choose weight-$w$ irreducibles such that the coefficients of the $f_w$ in their image under the canonical map $\rho$ is zero.

Being an isomorphism, $\rho$ is invertible, and one can retrieve arbitrary motivic MZVs from $\mathbb Q$-linear combinations of $\rho^{-1}(f_2^n f_{i_1} f_{i_2}\ldots f_{i_r})$ for suitable choices of $n$ and $i_j$. One can readily check at fixed weight that the indecomposable motivic MZVs obtained from (\ref{appMZV.03}) are in one-to-one correspondence with compositions $f_{i_1}f_{i_2}\ldots \neq f_{i_2}f_{i_1}\ldots$ with $i_j \in 2\mathbb N{+}1$ that cannot be written as shuffles. For instance, (\ref{appMZV.g6}) relates the existence of indecomposable $\zeta^\mot_{3,5}$ and $\zeta^\mot_{3,3,5}$ to $\rho^{-1}(f_3 f_{5}) \neq \rho^{-1}(f_5 f_{3})$ and $\rho^{-1}(f_3 f_3 f_{5})$ being different from $\rho^{-1}(f_3 f_5 f_{3})$ and $\rho^{-1}(f_5 f_3 f_{3})$.
We note that, starting from weights $w=12,15,\ldots$, the minimal depth of some of the motivic MZVs in a $\mathbb Q$-basis necessarily exceeds the number of odd letters in their $f$-alphabet images.
For instance, the two shuffle independent combinations $\rho^{-1}(f_3 f_{9})$ and $\rho^{-1}(f_5 f_{7})$ at $w=12$ necessitate at least one motivic MZV of depth $\geq 4$ such as $\zeta_{1,1,4,6}$ in (\ref{appMZV.03}) \cite{Broadhurst:1996kc, Blumlein:2009cf}.

By a slight abuse of notation, we will drop the motivic superscript when stating explicit $\rho^{-1}$-images in the remainder of this appendix and in the main body of the paper, for instance
\begin{align}
\rho^{-1}(f_3 f_5)&= - \frac{1}{5}  \,  \zeta_{3,5}
 + \frac{100471}{177840} \zeta_8\, ,  \label{rhoinv}  \\
 \rho^{-1}(f_3 f_7) &=
- \frac{1}{14} \, \zeta_{3,7} 
 -\frac{3}{14}  \,\zeta_{5}^2   + \frac{408872741707}{563009982080} \zeta_{10}\, ,
\notag\\
\rho^{-1}(f_3f_3f_5) &= 
\frac{4}{35}\,  \zeta_2^3 \zeta_5
 -\frac{6}{25} \,  \zeta_2^2 \zeta_7
-9  \zeta_2 \zeta_9
+ \frac{1}{5} \zeta_{3,3,5} + \frac{1119631493}{73676160} \zeta_{11}\, . \notag
\end{align}
In other words, $\rho^{-1}$ in (\ref{rhoinv}) and below is understood to automatically comprise the period map $\zeta^{\mot}_{n_1,n_2,\ldots,n_r} \mapsto \zeta_{n_1,n_2,\ldots,n_r}$ which is well-defined in spite of the currently unproven statements for MZVs (see e.g.\ the textbooks \cite{Jianqiang, GilFresan} for an overview). 
Note that (\ref{rhoinv}) determines $\rho^{-1}$ images of permutations in $f_{i_j}$ via shuffle relations (\ref{appMZV.g4}), e.g.\ $\rho^{-1}(f_{i_1}\shuffle f_{i_2})= \zeta_{i_1}\zeta_{i_2}$.

Before spelling out additional $\rho^{-1}$-images relevant to the group-like series $\mathbb M^{\rm sv}$ in (\ref{Msvz}), we shall briefly review the single-valued map sv.

\subsection{Single-valued MZVs}
\label{app:A.2}

In the same way as MZVs are multiple polylogarithms at unit argument, see (\ref{appMZV.g1}), one arrives at so-called single-valued MZVs by evaluating single-valued polylogarithms at $z=1$ \cite{Schnetz:2013hqa, Brown:2013gia}
\beq
\zeta^{\rm sv}_{n_1,n_2,\ldots,n_r} = (-1)^r G^{\rm sv}(\underbrace{0,\ldots,0}_{n_r-1},1,\ldots,\underbrace{0,\ldots,0}_{n_2-1},1,\underbrace{0,\ldots,0}_{n_1-1},1;1)\, .
\label{appMZV.g0}
\eeq
The single-valued polylogarithms, $G^{\rm sv}$, in one variable were firstly constructed in \cite{svpolylog} and generalised to multiple variables in \cite{Broedel:2016kls, DelDuca:2016lad}. With the ingredients reviewed in this work, single-valued polylogarithms $G^{\rm sv}(a_1,\ldots,a_r;z)$ at $a_i \in \{0,1 \}$ can also be obtained from the coefficient of $x_{a_r}\ldots x_{a_1}$ in the series $\mathbb G^{\rm sv}_{\{0,1\}}(x_i;z)$ given by (\ref{svieis.08}) \cite{Frost:2023stm}. These constructions of $G^{\rm sv}$ manifest that single-valued MZVs (\ref{appMZV.g0}) are again expressible as $\mathbb Q$-linear combinations of products of MZVs.

At the level of motivic MZVs, one can define a single-valued map via
\beq
{\rm sv}\, :\, \zeta^{\mot}_{n_1,n_2,\ldots,n_r} \mapsto \zeta^{\rm sv}_{n_1,n_2,\ldots,n_r}\,,
\label{appMZV.g7}
\eeq
which takes a particularly convenient form in the $f$-alphabet \cite{Schnetz:2013hqa, Brown:2013gia}\footnote{By slight abuse of notation, we do not distinguish the map sv in (\ref{appMZV.g7}) from its composition $\rho \circ {\rm sv}\circ \rho^{-1}$ seen in (\ref{appMZV.02}).}
\beq
{\rm sv}(f_2^n f_{i_1} f_{i_2}\ldots f_{i_r}) = \delta_{n,0 }\sum_{j=0}^r (f_{i_j}\ldots f_{i_2} f_{i_1})\shuffle
(f_{i_{j+1}}\ldots f_{i_{r-1}} f_{i_r})\,,
\label{appMZV.02}
\eeq
and preserves the (shuffle) multiplication in the sense that ${\rm sv}( \zeta^{\mot}_{\vec{n}}\cdot \zeta^{\mot}_{\vec{m}} ) = 
{\rm sv}( \zeta^{\mot}_{\vec{n}}  )  {\rm sv}(  \zeta^{\mot}_{\vec{m}} ) $. For Riemann zeta values, one readily finds $\zeta^{\rm sv}_{2n}=0$ and $\zeta^{\rm sv}_{2n+1}= 2 \zeta_{2n+1}$ from 
(\ref{appMZV.02}) at $r=0$ and $r=1$, respectively. For words in $r=2$ and $r=3$ odd-degree generators $f_{i_j}$ with $i_j \in 2\mathbb N{+}1$, (\ref{appMZV.02}) specialises to 
\begin{align}
{\rm sv}(f_{i_1} f_{i_2}) &= 2f_{i_1} f_{i_2} + 2 f_{i_2} f_{i_1} = 2 f_{i_1} \shuffle f_{i_2}\, ,
\label{appMZV.g8} \\
{\rm sv}(f_{i_1} f_{i_2} f_{i_3}) &= 2( f_{i_1} f_{i_2} f_{i_3} + f_{i_2} f_{i_1} f_{i_3} + f_{i_3} f_{i_2} f_{i_1} + f_{i_2} f_{i_3} f_{i_1})
=  {\rm sv}(f_{i_3} f_{i_2} f_{i_1})
\, . \notag
\end{align}
By the first line, the expansion of the group-like $\mathbb M^{\rm sv}$ in (\ref{Msvz}) up to and including the second order in $f_i$ is expressible via Riemann zeta values since $\rho^{-1}({\rm sv}(f_{i_1}))= 2 \zeta_{i_1}$ and $\rho^{-1}({\rm sv}(f_{i_1} f_{i_2}))= 2 \zeta_{i_1} \zeta_{i_2}$. Starting from the third order, the $\rho^{-1}$ images of ${\rm sv}(f_{i_1} f_{i_2} f_{i_3})$ in the second line of (\ref{appMZV.g8}) generically involve (conjecturally indecomposable) single-valued MZVs beyond depth one, for instance
\beq
\zeta^{\rm sv}_{3,3,5} = 2 \zeta_{3,3,5} - 5 \zeta_3^2 \zeta_5 + 90 \zeta_2 \zeta_9 + \frac{12}{5} \zeta_2^2 \zeta_7 - \frac{8}{7} \zeta_2^3 \zeta_5\, .
\label{appMZV.g9}
\eeq
The construction (\ref{uplift}) of the generating series $\mathbb J^{\rm eqv}$ from the series $\mathbb M^{\rm sv}(\hat\sigma_i,\hat\sigma_\varpi)= \mathbb M^{\rm sv}(\hat\sigma_i) +\ldots$ in zeta generators implies that the modular forms $\beqv{j_1 &j_2 &j_3}{k_1 &k_2 &k_3}$ at modular depth three involve the single-valued MZV $\rho^{-1}({\rm sv}(f_{k_1-1} f_{k_2-1}f_{k_3-1}))$. Hence, our investigations up to degree $k_1{+}k_2{+}k_3=20$ lead to all single-valued MZVs $\rho^{-1}({\rm sv}(f_{i_1} f_{i_2}f_{i_3}))$ up to and including weight $i_1{+}i_{2}{+}i_3 = 17$ and three non-commutative letters in the $f$-alphabet. Their representations in terms of conventional MZVs are given by 
\begin{align}
\label{rhoinvlist}
\rho^{-1}\big({\rm sv}(f_3 f_3 f_5)\big) &= -\frac{1}{5} \zeta_{3,3,5}^{\mathrm{sv}} + \frac{1119631493}{36838080} \zeta_{11} \,,\\
\rho^{-1}\big({\rm sv}(f_3 f_5 f_3)\big) &= \frac{2}{5} \zeta_{3,3,5}^{\mathrm{sv}} + 4 \zeta_{3}^2 \zeta_{5} - \frac{1119631493}{18419040} \zeta_{11} \nonumber\,,\\
\rho^{-1}\big({\rm sv}(f_3 f_3 f_7)\big) &= -\frac{1}{14} \zeta_{3,3,7}^{\mathrm{sv}} + \frac{3}{35} \zeta_{3,5,5}^{\mathrm{sv}} + \frac{8607216661079268929}{999844297536624120} \zeta_{13} \nonumber\,,\\
\rho^{-1}\big({\rm sv} (f_3 f_7 f_3)\big) &= \frac{1}{7} \zeta_{3,3,7}^{\mathrm{sv}} - \frac{6}{35} \zeta_{3,5,5}^{\mathrm{sv}} + 4 \zeta_{3}^2 \zeta_{7} - \frac{8607216661079268929}{499922148768312060} \zeta_{13} \nonumber\,,\\
\rho^{-1}\big({\rm sv}(f_3 f_5 f_5)\big) &= \frac{1}{25} \zeta_{3,5,5}^{\mathrm{sv}} 
- \frac{839332307937696179}{39676361013358100} \zeta_{13} \nonumber\,,\\
\rho^{-1}\big({\rm sv}(f_5 f_3 f_5)\big) &= -\frac{2}{25} \zeta_{3,5,5}^{\mathrm{sv}} + 4 \zeta_{3} \zeta_{5}^2 
+ \frac{839332307937696179}{19838180506679050} \zeta_{13} \nonumber\,,\\
\rho^{-1}\big({\rm sv}(f_5 f_3 f_7)\big) &= \frac{1}{70} \zeta_{5,3,7}^{\mathrm{sv}} + \frac{4}{5} \zeta_{5}^3 + \frac{24}{5} \zeta_{3} \zeta_{5} \zeta_{7} 
- \frac{1156747681600394679684642590233}{52156108470099943251903139200} \zeta_{15}\, ,
\nonumber\\
\rho^{-1}\big({\rm sv}(f_3 f_5 f_7)\big) &=
{-} \frac{ 1706 }{68409} \zeta_{3,3,9}^{\mathrm{sv}} - \frac{58001 }{1596210}\zeta_{5,3,7}^{\mathrm{sv}} + \frac{
 144 }{7601} \zeta_{1,1,3,4,6}^{\mathrm{sv}}
+ \frac{
 759436 }{114015} \zeta_{3} \zeta_{5} \zeta_{7} 
+\frac{384 }{38005}\zeta_{3}^5  \notag \\
&\quad + \frac{557516 }{114015} \zeta_{5}^3 + \frac{5024}{691} \zeta_{3}^2 \zeta_{9} + \frac{
 70011715120065369545804422936104641 }{396438580481229668657715761059200 } \zeta_{15} \, ,
\nonumber\\
\rho^{-1}\big({\rm sv}(f_3 f_3 f_9)\big) &=
{-} \frac{ 655}{22803}  \zeta_{3,3,9}^{\mathrm{sv}} - \frac{17203 }{532070} \zeta_{5,3,7}^{\mathrm{sv}} - \frac{
 48 }{7601} \zeta_{1,1,3,4,6}^{\mathrm{sv}}
  - \frac{
 178972 }{38005} \zeta_{3} \zeta_{5} \zeta_{7}
-\frac{128 }{38005} \zeta_{3}^5  \notag \\
&\quad - \frac{88972 }{38005} \zeta_{5}^3 - \frac{5024 }{2073} \zeta_{3}^2 \zeta_{9} + \frac{
 26869796704014139979194459442197511 }{264292386987486445771810507372800 } \zeta_{15} \, ,
\nonumber
\end{align}
and a comprehensive list of all cases up to weight $17$ in the ancillary file. Note that instances of $\rho^{-1}({\rm sv}(f_{i_1} f_{i_2}f_{i_3}))$ with $i_1>i_3$ can be inferred from the symmetry under $i_1 \leftrightarrow i_3$ noted in (\ref{appMZV.g8}), and we reiterate that we have implicitly applied the period map $\zeta^{\mot}_{n_1,n_2,\ldots,n_r} \mapsto \zeta_{n_1,n_2,\ldots,n_r}$ to the images of $\rho^{-1}$ in this section. 

In order to find the numerical values of the above $\rho^{-1}\left(\operatorname{sv}\left(f_a f_b f_c\right)\right)$, we have rewritten the single-valued MZVs on the right-hand sides of (\ref{rhoinvlist}) via (\ref{appMZV.g9}) and
\begin{align}
\zeta_{3,3,7}^{\mathrm{sv}}
&= 2 \zeta_{3,3,7} + 12 \zeta_{3,5} \zeta_{5} + 60 \zeta_{3} \zeta_{5}^2 - 
 \frac{64}{35} \zeta_{2}^3 \zeta_{7} - 14 \zeta_{3}^2 \zeta_{7} + 
 \frac{112}{5} \zeta_{2}^2 \zeta_{9} + 407 \zeta_{2} \zeta_{11} \, , 
\label{svmzvex} \\
\zeta_{3,5,5}^{\mathrm{sv}} 
&= 2 \zeta_{3,5,5} + 10 \zeta_{3,5} \zeta_{5} + 50 \zeta_{3} \zeta_{5}^2 + 
 20 \zeta_{2}^2 \zeta_{9} + 275 \zeta_{2} \zeta_{11}  \, , 
\notag \\
\zeta_{5,3,7}^{\mathrm{sv}} 
&= 2 \zeta_{5, 3, 7} - \frac{192}{385} \zeta_{2}^5 \zeta_{5} - 12 \zeta_{3,7} \zeta_{5} - 
 78 \zeta_{5}^3 - \frac{96}{25} \zeta_{2}^4 \zeta_{7} - 28 \zeta_{3,5} \zeta_{7} - 
 336 \zeta_{3} \zeta_{5} \zeta_{7} \notag \\
 &\quad - \frac{272}{35} \zeta_{2}^3 \zeta_{9} + 
 44 \zeta_{2}^2 \zeta_{11} + 1001 \zeta_{2} \zeta_{13}  \, , 
\notag \\
\zeta_{3,3,9}^{\mathrm{sv}} &= 2 \zeta_{3, 3, 9} + 12 \zeta_{3,7} \zeta_{5} + 72 \zeta_{5}^3 + 
 \frac{144}{175} \zeta_{2}^4 \zeta_{7} + 30 \zeta_{3,5} \zeta_{7} + 
 318 \zeta_{3} \zeta_{5} \zeta_{7} + \frac{232}{35} \zeta_{2}^3 \zeta_{9} \notag \\
 &\quad - 
 27 \zeta_{3}^2 \zeta_{9} + \frac{504}{5} \zeta_{2}^2 \zeta_{11} + 1209 \zeta_{2} \zeta_{13}  \, , 
\notag \\
\zeta_{1,1,3,4,6}^{\mathrm{sv}} &=  2 \zeta_{1,1,3,4,6} 
+ 4 \zeta_{2} \zeta_{3,3,7} - 
 \frac{28}{5} \zeta_{2} \zeta_{3,5,5} 
 +  2 \zeta_{1,1,4,6} \zeta_{3}
 + 8 \zeta_{2}^2 \zeta_{3,3,5}
 - \frac{16}{5} \zeta_{2}^2 \zeta_{3,5} \zeta_{3}  \notag \\
 &\quad - 
 12 \zeta_{2} \zeta_{3,7} \zeta_{3} + \frac{58}{9} \zeta_{3,9} \zeta_{3} + 
 21 \zeta_{2} \zeta_{3,5} \zeta_{5} - \frac{145}{56} \zeta_{3,7} \zeta_{5} - 
 \frac{481}{10} \zeta_{3,5} \zeta_{7} + \frac{
 2903944 \zeta_{2}^6 \zeta_{3}}{716625} \notag \\
 &\quad + \frac{24}{35} \zeta_{2}^3 \zeta_{3}^3 - \frac{
 2 \zeta_{3}^5}{3} - \frac{47270 \zeta_{2}^5 \zeta_{5}}{1617}  - 
 \frac{23}{5} \zeta_{2}^2 \zeta_{3}^2 \zeta_{5} - 16 \zeta_{2} \zeta_{3} \zeta_{5}^2 - \frac{
 13133 \zeta_{5}^3}{56} + \frac{2792 \zeta_{2}^4 \zeta_{7}}{2625} \notag \\
 &\quad - 52 \zeta_{2} \zeta_{3}^2 \zeta_{7} - 
 \frac{12509}{24} \zeta_{3} \zeta_{5} \zeta_{7} + \frac{16837}{63} \zeta_{2}^3 \zeta_{9} - 
 \frac{2927}{9} \zeta_{3}^2 \zeta_{9} - \frac{27199}{30} \zeta_{2}^2 \zeta_{11} - 
 \frac{56717}{60} \zeta_{2} \zeta_{13}
\, , \notag
\end{align}
see the ancillary files for their weight-17 analogues.

The examples of (\ref{rhoinvlist}) illustrate that the modular depth of $\beta^{\rm eqv}$ does not bound the depth $r$ of the MZVs $\zeta_{n_1,\ldots,n_r}$ in its Fourier expansion. Instead, the modular depth of $\beta^{\rm eqv}$ sets an upper bound on the number of non-commutative letters in the $f$-alphabet representation of the associated MZVs, see \cite{Saad:2020mzv} for the analogous statement for MMVs. The MZV representation of a given $\rho^{-1}(f_{i_1} f_{i_2}\ldots f_{i_\ell})$ will in general necessitate representatives of depth $> \ell$
as exemplified by $\zeta_{1,1,3,4,6}^{\mathrm{sv}}$ in the last four lines of (\ref{rhoinvlist}).

\section{Changing from holomorphic to modular frame}
\label{sec:Uconj}

In this section we will prove the validity of \eqref{Unumu} relating the two types of integration kernels $\nuker{j}{k}{\tau_1} $ and $\omplus{j}{k}{\tau,\tau_1}$ through the SL$_2$ transformation $\Unew(\tau)$ in (\ref{eq:Utau}).
To this end we will actually prove a stronger statement.
Let us define the auxiliary operator
\begin{equation}
    \widetilde {\Unew}(a,b)\coloneqq \exp(a \eee_0^{\vee}) \exp(b\eee_0)\,.
\end{equation}
\begin{lemma} From Lemma~\ref{lemma1} we can derive the identity:
\begin{align}
 \widetilde {\Unew}(a,b)\Big[ \sum_{j=0}^{k-2} \frac{(-1)^j}{j!}(2\pi i \tau_1)^j \eee_k^{(j)}\Big] \widetilde {\Unew}(a,b)^{-1} =\sum_{\ell=0}^{k-2}\frac{\eee_k^{(\ell)}}{\ell!} [1+a(b-2\pi i \tau_1)]^{k-2-\ell}(b-2\pi i \tau_1)^{\ell}\,.
 \label{ablemm}
\end{align}
\end{lemma}

Note that given our definition \eqref{eq:Utau}, we simply have
\begin{equation}
\Unew(\tau) = \widetilde {\Unew}\big({-}\tfrac{1}{4y},2\pi i\bar\tau \big)
\, , \quad a= - \tfrac{1}{4y}
\, , \quad b= 2\pi i \bar \tau
\,,\label{eq:UabU}
\end{equation}
such that \eqref{Unumu} follows as a simple corollary of this more general lemma.

\textit{Proof:}
We start by expanding both exponential factors in $\widetilde{\Unew}(a,b)$ on
the left-hand side of (\ref{ablemm}) as a power series:
\begin{align}
&\sum_{\ell_1,\ell_2=0}^\infty \frac{a^{\ell_1} b^{\ell_2}}{\ell_1!\ell_2!} ({\rm ad}_{\eee_0^\vee})^{\ell_1}  \Big[ ({\rm ad}_{\eee_0})^{\ell_2}\Big[\sum_{j=0}^{k-2} \frac{(-1)^j}{j!}(2\pi i \tau_1)^j \eee_k^{(j)} \Big]\Big]
\\
&=\sum_{j=0}^{k-2}\sum_{\ell_2=0}^{k-2-j} \sum_{\ell_1=0}^{j+\ell_2} \frac{a^{\ell_1} b^{\ell_2}}{\ell_1!\ell_2!} \frac{(-1)^j}{j!}(2\pi i \tau_1)^j \frac{(j{+}\ell_2)!\, (k{+}\ell_1{-}2{-}j{-}\ell_2)!}{(j{+}\ell_2{-}\ell_1)!\,(k{-}2{-}j{-}\ell_2)!} \eee_k^{(j+\ell_2-\ell_1)}\,,
\notag
\end{align}
where we used crucially \eqref{eq:adeochIter}.

We then perform the change in summation variables 
$j\to \ell\coloneqq j{+}\ell_2{-}\ell_1$ to arrive at
\begin{align}
 & \widetilde \Unew(a,b)\Big[ \sum_{j=0}^{k-2} \frac{(-1)^j}{j!}(2\pi i \tau_1)^j \eee_k^{(j)}\Big] \widetilde \Unew(a,b)^{-1}\\
&\nn =\sum_{\ell=0}^{k-2} \frac{\eee_k^{(\ell)}}{\ell!} \sum_{\ell_1=0}^{k-2-\ell} \binom{k{-}2{-}\ell}{\ell_1} a^{\ell_1} \sum_{\ell_2=0}^{\ell+\ell_2} \binom{\ell{+}\ell_1}{\ell_2} b^{\ell_2}(-2\pi i \tau_1)^{\ell+\ell_1-\ell_2}\\
&\nn =\sum_{\ell=0}^{k-2} \frac{\eee_k^{(\ell)}}{\ell!} \sum_{\ell_1=0}^{k-2-\ell} \binom{k{-}2{-}\ell}{\ell_1} a^{\ell_1} (b-2\pi i \tau_1)^{\ell+\ell_1}\\
&\nn=\sum_{\ell=0}^{k-2}\frac{\eee_k^{(\ell)}}{\ell!} [1+a(b-2\pi i \tau_1)]^{k-2-\ell}(b-2\pi i \tau_1)^{\ell}\,,
\end{align}
thus concluding our proof of (\ref{ablemm}).

Note that this lemma only relies on the relations \eqref{eq:adeochIter} coming from the $\mathfrak{sl}_2$ action on the module spanned by $\eee_k^{(j)}$ with $0\leq j\leq k{-}2$. In particular, it can be also reformulated in terms of the formal variables $\ep_k^{(j)}$ rather than $\eee_k^{(j)}$.

If we specialise $a,b$ according to \eqref{eq:UabU} we obtain
\begin{equation}
\Unew(\tau) \Big[ \sum_{j=0}^{k-2} \frac{(-1)^j}{j!}(2\pi i \tau_1)^j \eee_k^{(j)}\Big] \Unew(\tau)^{-1} = (2\pi i)^{k-2} \sum_{\ell=0}^{k-2}\frac{(-1)^\ell \eee_k^{(\ell)}}{\ell!}  \left(\frac{\tau{-}\tau_1}{4y}\right)^{k-2-\ell}(\bar\tau{-}\tau_1)^{\ell}\,,
\end{equation}
and upon adding some factors of $(2\pi i )$ we arrive at the kernels $\omega_\pm$ defined in \eqref{cuspat3.01},
\begin{align}
\Unew(\tau) \left[ \sum_{j=0}^{k-2} \frac{(-1)^j}{j!} \eee_k^{(j)} \nuker{j}{k}{\tau_1} \right] \Unew(\tau)^{-1} &=
\sum_{\ell=0}^{k-2} \frac{(-1)^\ell}{\ell!}
\eee_k^{(\ell)} 
\omplus{\ell}{k}{\tau,\tau_1} %
\,,\\
\Unew(\tau) \left[ \sum_{j=0}^{k-2} \frac{1}{j!} \eee_k^{(j)} \overline{\nuker{j}{k}{\tau_1}} \right] \Unew(\tau)^{-1} &= %
\sum_{\ell=0}^{k-2} \frac{(-1)^\ell}{\ell!}
\eee_k^{(\ell)} 
\omminus{\ell}{k}{\tau,\tau_1} \, .%
\notag
\end{align}
This dictionary between the differential forms $\nu$ and $\omega_{\pm}$ also holds for holomorphic cusp forms instead of Eisenstein series, i.e.\ for kernels $\nuker{j}{\Delta_k}{\tau}$ defined in~\eqref{eq:nudelta} such that
\begin{align}
\Unew(\tau) \left[ \sum_{j=0}^{k-2} \frac{(-1)^j}{j!} \eee_{\Delta_k^{+}}^{(j)} \nuker{j}{{\Delta_k}}{\tau_1} \right] \Unew(\tau)^{-1} &=
\sum_{\ell=0}^{k-2} \frac{(-1)^\ell}{\ell!}
\eee_{\Delta_k^{+}}^{(\ell)} \omplus{\ell}{{\Delta_k}}{\tau,\tau_1} \, , \\
\Unew(\tau) \left[ \sum_{j=0}^{k-2} \frac{1}{j!} \eee_{\Delta_k^{-}}^{(j)} \overline{\nuker{j}{{\Delta_k}}{\tau_1}} \right] \Unew(\tau)^{-1} &=
\sum_{\ell=0}^{k-2} \frac{(-1)^\ell}{\ell!}
\eee_{\Delta_k^{-}}^{(\ell)} \omminus{\ell}{{\Delta_k}}{\tau,\tau_1}
\notag \,,
\end{align}
where the only property used is that $\eee_{\Delta_k^{\pm}}^{(j)}$ span an $\mf{sl}_2$ multiplet.

\section{\texorpdfstring{Comparing $M_{k_1,k_2,k_3}$ with combinations of ${\cal E}$ and $c^{\rm sv}$}{Comparing M(k1,k2,k3) with combinations of calE and csv}}
\label{app:expeqv}

A first goal of this appendix is to derive more explicit formulae for the expansion coefficients
$\eeqv{j_1 &\ldots &j_\ell}{k_1 &\ldots &k_\ell}$ of equivariant iterated Eisenstein 
integrals from the generating series $\Ieqv(\eee_k;\tau)$ in (\ref{eq:Iup}) and (\ref{eeqv.01}).
The expressions for $\eeqv{j_1 &\ldots &j_\ell}{k_1 &\ldots &k_\ell}$ obtained in section \ref{app:expeqv.2} will then be 
used to demonstrate the equivalence between the two constructions of the modular forms $\beqv{j_1 &j_2 &j_3}{k_1 &k_2&k_3}$
at modular depth three either as coefficients of $\eee_{k_i}^{(j_i)}$ from the 
generating series $\Jeqv (\eee_k;\tau)$, as discussed in section \ref{sec:4}, or as
coefficients of $(X_i{-}\tau  Y_i)^{j_i} (X_i{-} \bar \tau  Y_i)^{k_i-j_i-2}$ from the 
equivariant integrals $M_{k_1,k_2,k_3}$, as investigated in section \ref{sec:8}.

\subsection{(Anti-)meromorphic building blocks}
\label{app:expeqv.1}

We start by making the structure of the
(anti-)meromorphic building blocks $ \hat \psi^{\rm sv}(\, \Imint(\eee_k))$
and $\Ip(\eee_k)$ of the generating series (\ref{eq:Iup}) more explicit
up to modular depth three. For this purpose, we parametrise the
change of alphabet $\hat \psi^{\rm sv}$ of section \ref{sec:holfrm}
as well as the letters $\eee_{\Delta^{\pm}}$ of the cuspidal kernels in
(\ref{MMVsec.06}) and (\ref{minusees}) via infinite families of real constants
$\chicst{j}{k}{j_1 &\ldots}{k_1 &\ldots}$, $\xicst{j}{\Delta}{j_1 &\ldots}{k_1 &\ldots}$ 
and $\etacst{j}{\Delta}{j_1 &\ldots}{k_1 &\ldots}$. With the conventions
(\ref{defword}) for words $\word [\begin{smallmatrix}  j_1 &j_2 &\ldots &j_\ell \\ k_1 &k_2 &\ldots &k_\ell \end{smallmatrix}]$
in $\eee_k^{(j)}$, we parametrise the change of alphabet $\hat \psi^{\rm sv}$ described
in more detail in section \ref{sec:4.2} via 
\beq
\hat \psi^{\rm sv}\big( \,\wword{j}{k}\big) = \wword{j}{k} + \! \!  \!  \sum_{k_1,k_2,j_1,j_2} \!  \! \! \chicst{j}{k}{j_1 &j_2}{k_1 &k_2}\wword{j_1 &j_2}{k_1 &k_2} + \! \!  \!  \! \!  \sum_{k_1,k_2,k_3,j_1,j_2,j_3}   \! \! \!   \! \! \chicst{j}{k}{j_1 &j_2 &j_3}{k_1 &k_2 &k_3}\wword{j_1 &j_2 &j_3}{k_1 &k_2 &k_3} + \ldots \, ,
\label{mtobeta.01}
\eeq
where the sums $ \sum_{k_1,k_2,j_1,j_2} $ and $ \sum_{k_1,k_2,k_3,j_1,j_2,j_3}$ here and below are understood
as covering the usual range of even $k_i \geq 4$ and $0\leq j_i \leq k_i{-}2$. The $ \chicst{j}{k}{j_1 &j_2}{k_1 &k_2} $
and $\chicst{j}{k}{j_1 &j_2 &j_3}{k_1 &k_2 &k_3}$ associated with words in up to three $\eee_k^{(j)}$ can be computed
from the series $\mathbb M^{\rm sv}(\hat z_i)$ in zeta generators in (\ref{defpsisv}), leading
to rational multiples of $\zeta_{2n_1+1}$ or $\zeta_{2n_1+1}\zeta_{2n_2+1}$. However, we do not need the explicit
form of $ \chicst{j}{k}{j_1 &j_2}{k_1 &k_2}$, $\chicst{j}{k}{j_1 &j_2 &j_3}{k_1 &k_2 &k_3}\in \mathbb R$ for the main 
goals of this appendix. We also do not use any symmetry properties of the $\chi^j_k$ 
under permutations of columns such as $\chicst{j}{k}{j_1 &j_2}{k_1 &k_2} = - \chicst{j}{k}{j_2 &j_1}{k_2 &k_1}$
which ensure that each order in (\ref{mtobeta.01}) is Lie-algebra valued.

A similar parametrisation is used for the letters $\eee_{\Delta^{\pm}}$ associated with cuspidal integration kernels,
\begin{align}
\wword{j}{\Delta^+}&=  \! \!  \sum_{k_1,k_2,j_1,j_2}  \! \! \xicst{j}{\Delta}{j_1 &j_2}{k_1 &k_2}\wword{j_1 &j_2}{k_1 &k_2} + \! \!   \! \!  \sum_{k_1,k_2,k_3,j_1,j_2,j_3}   \! \!   \! \! \xicst{j}{\Delta}{j_1 &j_2 &j_3}{k_1 &k_2 &k_3}\wword{j_1 &j_2 &j_3}{k_1 &k_2 &k_3} + \ldots \, ,
\label{mtobeta.02} \\
\wword{j}{\Delta^-}
&=  \! \!  \sum_{k_1,k_2,j_1,j_2}  \! \! \etacst{j}{\Delta}{j_1 &j_2}{k_1 &k_2}\wword{j_1 &j_2}{k_1 &k_2} + \! \!   \! \!  \sum_{k_1,k_2,k_3,j_1,j_2,j_3}   \! \!   \! \! \etacst{j}{\Delta}{j_1 &j_2 &j_3}{k_1 &k_2 &k_3}\wword{j_1 &j_2 &j_3}{k_1 &k_2 &k_3} + \ldots \, ,
\notag
\end{align}
where some of the $\xicst{j}{\Delta}{j_1 &j_2}{k_1 &k_2}$, $\xicst{j}{\Delta}{j_1 &j_2 &j_3}{k_1 &k_2 &k_3}$,
$ \etacst{j}{\Delta}{j_1 &j_2}{k_1 &k_2}$, $\etacst{j}{\Delta}{j_1 &j_2 &j_3}{k_1 &k_2 &k_3} \in \mathbb R$ can in
principle be inferred from the expressions (\ref{edeltaex}) for $\eee_{\Delta^{\pm}}$. By $\mf{sl}_2$-invariance of $\hat \psi^{\rm sv}$
and $\eee_{\Delta^{\pm}}^{(j)} = ({\rm ad}_{\eee_0})^j \eee_{\Delta^{\pm}}$, the $j=0$ instances of $\chicst{j}{k}{j_1 &\ldots}{k_1 &\ldots}$, $\xicst{j}{\Delta_k}{j_1 &\ldots}{k_1 &\ldots}$ 
and $\etacst{j}{\Delta_k}{j_1 &\ldots}{k_1 &\ldots}$ already determine those at non-zero $j$,
but this appendix will not make any use of this property.

By virtue of the expansions (\ref{mtobeta.01}) and (\ref{mtobeta.02}), we can bring the coefficients
of $\word [\begin{smallmatrix}  j_1 &j_2  \\ k_1 &k_2  \end{smallmatrix}]$ and
$\word [\begin{smallmatrix}  j_1 &j_2 &j_3 \\ k_1 &k_2 &k_3 \end{smallmatrix}]$ 
in $ \hat \psi^{\rm sv}(\, \Imint(\eee_k; \tau) )$ and $\Ip(\eee_k;\tau)$ into the following form:
\begin{align}
\Ip(\eee_k;\tau) &= 1 + \sum_{k_1,j_1} \word [\begin{smallmatrix}  j_1  \\ k_1 \end{smallmatrix}]
\eeno{j_1}{k_1}
+\! \!  \sum_{k_1,k_2,j_1,j_2}  \! \! \word [\begin{smallmatrix}  j_1 &j_2  \\ k_1 &k_2  \end{smallmatrix}] \bigg\{
\eeno{j_1 &j_2}{k_1 &k_2} + \sum_{k,j} \sum_{\Delta_k \in {\cal S}_k }  \xicst{j}{\Delta_k}{j_1 &j_2}{k_1 &k_2} \eeno{j}{\Delta_{k}}
\bigg\} \notag \\
&\quad + \! \!   \! \!  \sum_{k_1,k_2,k_3,j_1,j_2,j_3}   \! \!   \! \! \wword{j_1 &j_2 &j_3}{k_1 &k_2 &k_3}  \bigg\{
\eeno{j_1 &j_2 &j_3}{k_1 &k_2 &k_3} + \sum_{k,j} \sum_{\Delta_k \in {\cal S}_k }  \Big( 
\xicst{j}{\Delta_k}{j_1 &j_2 &j_3}{k_1 &k_2 &k_3} \eeno{j}{\Delta_{k}} \label{mtobeta.03}  \\
&\qquad \qquad \qquad \qquad \qquad 
+\xicst{j}{\Delta_k}{j_1 &j_2}{k_1 &k_2} \eeno{j &j_3}{\Delta_{k} &k_3}
+\xicst{j}{\Delta_k}{j_2 &j_3}{k_2 &k_3} \eeno{j_1 &j}{k_1 &\Delta_{k}}
\Big)\bigg\} + \ldots
\notag
\end{align}
and
\begin{align}
&\hat \psi^{\rm sv}(\, \Imint(\eee_k; \tau) ) = 1 + \sum_{k_1,j_1} \word [\begin{smallmatrix}  j_1  \\ k_1 \end{smallmatrix}]
\tce{j_1}{k_1}
+\! \!  \sum_{k_1,k_2,j_1,j_2}  \! \! \word [\begin{smallmatrix}  j_1 &j_2  \\ k_1 &k_2  \end{smallmatrix}]  
\tce{j_2 &j_1}{k_2 &k_1}  
\notag \\
&\quad +\! \!  \sum_{k_1,k_2,j_1,j_2}  \! \! \word [\begin{smallmatrix}  j_1 &j_2  \\ k_1 &k_2  \end{smallmatrix}] 
 \sum_{k,j} \bigg\{  \chicst{j}{k}{j_1 &j_2}{k_1 &k_2} \tce{j}{k}
+ \sum_{\Delta_k \in {\cal S}_k }  \etacst{j}{\Delta_k}{j_1 &j_2}{k_1 &k_2} \tce{j}{\Delta_{k}}
\bigg\} 
\notag \\
&\quad + \! \!   \! \!  \sum_{k_1,k_2,k_3,j_1,j_2,j_3}   \! \!   \! \! \wword{j_1 &j_2 &j_3}{k_1 &k_2 &k_3}  
\bigg\{ \tce{j_3 &j_2 &j_1}{k_3 &k_2 &k_1}  +  \sum_{k,j}  \chicst{j}{k}{j_1 &j_2 &j_3}{k_1 &k_2 &k_3} 
\tce{j}{k} \bigg\} \notag \\
&\quad + \! \!   \! \!  \sum_{k_1,k_2,k_3,j_1,j_2,j_3}   \! \!   \! \! \wword{j_1 &j_2 &j_3}{k_1 &k_2 &k_3}  
\sum_{k,j} 
\bigg\{ \chicst{j}{k}{j_1 &j_2 }{k_1 &k_2 } \tce{ j_3 &j}{ k_3 & k}
+ \chicst{j}{k}{j_2 &j_3 }{k_2 &k_3 } \tce{ j &j_1}{ k & k_1} \bigg\}
 \notag \\
&\quad + \! \!   \! \!  \sum_{k_1,k_2,k_3,j_1,j_2,j_3}   \! \!   \! \! \wword{j_1 &j_2 &j_3}{k_1 &k_2 &k_3} 
\sum_{k,j} \sum_{\Delta_k \in {\cal S}_k }  \Big( 
\etacst{j}{\Delta_k}{j_1 &j_2 &j_3}{k_1 &k_2 &k_3} \tce{j}{\Delta_{k}}  \label{mtobeta.04} \\
&\qquad \qquad \qquad 
+\etacst{j}{\Delta_k}{j_1 &j_2}{k_1 &k_2} \tce{j_3 &j}{ k_3 &\Delta_{k}}
+\etacst{j}{\Delta_k}{j_2 &j_3}{k_2 &k_3} \tce{j &j_1 }{ \Delta_{k} &k_1}
\Big) \notag \\
&\quad + \! \!   \! \!  \sum_{k_1,k_2,k_3,j_1,j_2,j_3}   \! \!   \! \! \wword{j_1 &j_2 &j_3}{k_1 &k_2 &k_3} 
\sum_{k,j} \sum_{\Delta_k \in {\cal S}_k } \tce{j  }{ \Delta_{k} } \notag \\
&\qquad \quad \times  \sum_{k',j'}  \Big(
\etacst{j}{\Delta_k}{j' &j_3}{k' &k_3} \chicst{j'}{k'}{j_1 &j_2}{ k_1 &k_2}
+ \etacst{j}{\Delta_k}{j_1 & j' }{k_1 &k'} \chicst{j'}{k'}{j_2 &j_3}{ k_2 &k_3}
\Big)
 + \ldots \, ,
\notag
\end{align}
where we employ the following shorthand to absorb factors of $(-1)^{j_i}$
\beq
\tce{j_1 &j_2 &\ldots &j_\ell}{k_1 &k_2 &\ldots &k_\ell}  = (-1)^{j_1+j_2+\ldots+j_\ell}
\overline{ \eeno{j_1 &j_2 &\ldots &j_\ell}{k_1 &k_2 &\ldots &k_\ell}  } 
\label{mtobeta.05}
\eeq
which applies in identical form to cases with $k_i \rightarrow \Delta_k$.

\subsection{\texorpdfstring{Assembling ${\cal E}^{\rm eqv}$}{Assembling Eeqv}}
\label{app:expeqv.2}

The next step is to combine the expansions (\ref{mtobeta.04}) and (\ref{mtobeta.05}) of the series in iterated integrals
with the constant series 

\begin{align}
\mathbb C^{\rm sv}(\eee_k) &= 1 + \sum_{k_1,j_1} \word [\begin{smallmatrix}  j_1  \\ k_1 \end{smallmatrix}]
\ccsv{j_1}{k_1}
+\! \! \! \sum_{k_1,k_2,j_1,j_2}  \! \!  \!\word [\begin{smallmatrix}  j_1 &j_2  \\ k_1 &k_2  \end{smallmatrix}] 
\ccsv{j_1 &j_2}{k_1 &k_2}  \notag \\
&\quad +   \! \!  \! \!   \! \!  \sum_{k_1,k_2,k_3,j_1,j_2,j_3}   \! \!   \! \!   \! \! \wword{j_1 &j_2 &j_3}{k_1 &k_2 &k_3} 
\ccsv{j_1 &j_2 &j_3}{k_1 &k_2 &k_3} +\ldots \, .
\label{mtobeta.06}
\end{align}
With these expansions at hand, one can obtain the following expressions for
the coefficients ${\cal E}^{\rm eqv}$ of $\Ieqv(\eee_k;\tau)$ by matching (\ref{eq:Iup}) with (\ref{eeqv.01})
\begin{align}
\eeqv{j_1}{k_1 } &=   \tce{j_1 }{k_1}  +\ccsv{j_1}{k_1} +  \eeno{j_1 }{k_1}\label{mtobeta.07} \\
\eeqv{j_1 &j_2 }{k_1 &k_2 } &=  \tce{ j_2 &j_1 }{k_2 &k_1}  +\ccsv{j_1 &j_2}{k_1 &k_2} +  \eeno{j_1 &j_2 }{k_1 &k_2} \notag \\
&\quad  +  \tce{j_1 }{k_1}  \ccsv{j_2}{k_2} +    \tce{j_1 }{k_1}   \eeno{j_2 }{k_2}
      +\ccsv{j_1}{k_1}  \eeno{j_2 }{k_2} \notag \\
&\quad + \sum_{k,j} \bigg\{ 
\chicst{j}{k}{j_1 &j_2}{k_1 &k_2} \tce{j}{k} +  \sum_{\Delta_k \in {\cal S}_k } 
\Big(
\xicst{j}{\Delta_k}{j_1 &j_2}{k_1 &k_2} \eeno{j}{\Delta_k} + \etacst{j}{\Delta_k}{j_1 &j_2}{k_1 &k_2} \tce{j}{\Delta_k}
\Big)
\bigg\}\, ,
\notag
\end{align}
where the case at modular depth two goes beyond the simpler classes of terms in (\ref{eeqv.02}).
The double sum over $j,k$ collapses to finitely many terms bounded by $k\leq k_1{+}k_2{-}2$ since all
of $\chicst{j}{k}{j_1 &j_2}{k_1 &k_2} $, $\xicst{j}{\Delta_k}{j_1 &j_2}{k_1 &k_2}$ and $\etacst{j}{\Delta_k}{j_1 &j_2}{k_1 &k_2} $
vanish otherwise. The same procedure gives rise to the following more lengthy expression at modular depth three:
\begin{align}
\eeqv{j_1 &j_2 &j_3}{k_1 &k_2 &k_3} &=  \tce{j_3 &j_2 &j_1 }{k_3 &k_2 &k_1}  +\ccsv{j_1 &j_2 &j_3}{k_1 &k_2 &k_3} +  \eeno{j_1 &j_2 &j_3}{k_1 &k_2 &k_3} +  \tce{j_1 }{k_1}  \ccsv{j_2}{k_2}   \eeno{j_3 }{k_3} \label{mtobeta.08} \\
&\quad  +  \tce{j_2 &j_1}{k_2 &k_1}  \ccsv{j_3}{k_3} +  \tce{j_1 }{k_1}  \ccsv{j_2 &j_3}{k_2 &k_3} 
+    \tce{j_2 &j_1 }{k_2 &k_1}   \eeno{j_3}{k_3} \notag \\
&\quad +    \tce{j_1 }{k_1}   \eeno{j_2 &j_3}{k_2 &k_3}
      +\ccsv{j_1 &j_2}{k_1 &k_2}  \eeno{j_3}{k_3} +\ccsv{j_1}{k_1}  \eeno{j_2 &j_3}{k_2 &k_3} \notag \\
&\quad + \sum_{k,j} \Big\{ 
 \chicst{j}{k}{j_1 &j_2 }{k_1 &k_2}  \ccsv{j_3}{k_3}  \tce{j}{k}  +  \chicst{j}{k}{j_1 &j_2 }{k_1 &k_2}  \eeno{j_3}{k_3}  \tce{j}{k} 
 \notag \\
&\quad \quad   + \chicst{j}{k}{j_1 &j_2 &j_3}{k_1 &k_2 &k_3} \tce{j}{k} 
+ \chicst{j}{k}{j_1 &j_2 }{k_1 &k_2 } \tce{j_3 &j}{k_3 &k} 
+ \chicst{j}{k}{j_2 &j_3}{k_2 &k_3} \tce{j &j_1}{k &k_1}   \Big\} \notag \\
&\quad + \sum_{k,j} \sum_{\Delta_k \in {\cal S}_{\Delta_k}} \Big\{  \etacst{j}{\Delta_k}{j_1 &j_2 }{k_1 &k_2}  \ccsv{j_3}{k_3}  \tce{j}{\Delta_k}  
+  \etacst{j}{\Delta_k}{j_1 &j_2 }{k_1 &k_2}  \eeno{j_3}{k_3}  \tce{j}{\Delta_k} 
 \notag \\
&\quad \quad   + 
\etacst{j}{\Delta_k}{j_1 &j_2 &j_3}{k_1 &k_2 &k_3} \tce{j}{\Delta_k} 
+ \etacst{j}{\Delta_k}{j_1 &j_2 }{k_1 &k_2 } \tce{j_3 &j}{k_3 &\Delta_k} 
+ \etacst{j}{\Delta_k}{j_2 &j_3}{k_2 &k_3} \tce{j &j_1}{\Delta_k &k_1} 
\Big\} \notag \\
&\quad + \sum_{k,j} \sum_{\Delta_k \in {\cal S}_{\Delta_k}} \Big\{  
\xicst{j}{\Delta_k}{j_2 &j_3 }{k_2 &k_3}  \ccsv{j_1}{k_1}  \eeno{j}{\Delta_k}  
+  \xicst{j}{\Delta_k}{j_2 &j_3 }{k_2 &k_3}  \tce{j_1}{k_1}  \eeno{j}{\Delta_k} 
 \notag \\
&\quad \quad   + 
\xicst{j}{\Delta_k}{j_1 &j_2 &j_3}{k_1 &k_2 &k_3} \eeno{j}{\Delta_k} 
+ \xicst{j}{\Delta_k}{j_1 &j_2 }{k_1 &k_2 } \eeno{j &j_3}{\Delta_k &k_3} 
+ \xicst{j}{\Delta_k}{j_2 &j_3}{k_2 &k_3} \eeno{j_1 &j}{k_1 &\Delta_k} 
\Big\} \notag \\
&\quad + \sum_{k,j} \sum_{\Delta_k \in {\cal S}_{\Delta_k}}  \tce{j }{\Delta_k } 
\sum_{k',j'}  \Big\{ \etacst{j}{\Delta_k}{j' &j_3}{k' &k_3} \chicst{j'}{k'}{j_1 &j_2}{ k_1 &k_2}
+ \etacst{j}{\Delta_k}{j_1 & j' }{k_1 &k'} \chicst{j'}{k'}{j_2 &j_3}{ k_2 &k_3} \Big\}\, .
\notag
\end{align}
Again, all the sums over $k,j$ and $k',j'$ are finite in view of the vanishing of 
$\chicst{j}{k}{j_1 &j_2}{k_1 &k_2} $, $\xicst{j}{\Delta_k}{j_1 &j_2}{k_1 &k_2}$, $\etacst{j}{\Delta_k}{j_1 &j_2}{k_1 &k_2} $
at $k> k_1{+}k_2{-}2$ as well as $\chicst{j}{k}{j_1 &j_2 &j_3}{k_1 &k_2 &k_3} $, $\xicst{j}{\Delta_k}{j_1 &j_2 &j_3}{k_1 &k_2 &k_3}$, $\etacst{j}{\Delta_k}{j_1 &j_2 &j_3}{k_1 &k_2 &k_3} $ at $k> k_1{+}k_2{+}k_3{-}4$.

\subsection{\texorpdfstring{Comparing $M_{k}$ and $M_{k_1,k_2}$ with ${\cal E}^{\rm eqv}$}{Comparing M(k) and M(k1,k2) with Eeqv}}
\label{app:expeqv.3}

We shall next apply the expressions (\ref{mtobeta.07}) for ${\cal E}^{\rm eqv}$ to
compare different representations of the equivariant integrals $M_k[X,Y;\tau]$
and $M_{k_1,k_2}[X_1,Y_1,X_2,Y_2;\tau]$ in section \ref{sec:mk.1}. The established
$\beta^{\rm eqv}$-representations in (\ref{mkdepth1}) and (\ref{eq:Mbeqv}) \cite{Dorigoni:2022npe}
are equivalent to
\begin{align}
M_{k}[X,Y;\tau] &=  -\frac{1}{4} (k{-}1)!
\sum_{j=0}^{k-2} (-1)^{j}\binom{ k{-}2}{j } 
\bigg( \frac{  Y }{2\pi i} \bigg)^{j}  
X^{k-2-j} \,  \eeqv{j }{ k} \, ,
\notag \\
M_{k_1,k_2}[X_1,Y_1,X_2,Y_2;\tau] &=  \frac{1}{16} (k_1{-}1)!(k_2{-}1)!
\sum_{j_1=0}^{k_1-2} \sum_{j_2=0}^{k_2-2}
(-1)^{j_1+j_2}\binom{ k_1{-}2}{ j_1 } \binom{ k_2{-}2}{ j_2 } \notag \\*
&\quad \times 
\bigg( \frac{  Y_1 }{2\pi i} \bigg)^{j_1}  
X_1^{k_1-2-j_1} 
 \bigg( \frac{  Y_2 }{2\pi i} \bigg)^{j_2}
  X_2^{k_2-2-j_2} \,  \eeqv{j_2  &j_1 }{ k_2 &k_1} \, .
\label{mtobeta.10}
\end{align}
By equating these expressions to the original construction of
$M_k$ and $M_{k_1,k_2}$ through the $(1,0)$-form $\underline{\rm G}_k[X,Y;\tau] $ 
in (\ref{defundG}), we generate identities that facilitate the proof of the
$\beta^{\rm eqv}$ representation (\ref{olidreams}) of $M_{k_1,k_2,k_3}$ at modular 
depth three.

At modular depth one, matching the original expression in (\ref{mkdepth1})
with the first line of (\ref{mtobeta.10}) (using the expression (\ref{mtobeta.07}) for $\eeqv{j }{ k}$)
yields
\begin{align}
M_{k}[X,Y;\tau] &=   -\frac{1}{2} \int_\tau^{i \infty} \underline{\rm G}_k[X,Y;\tau_1] 
- \frac{1}{2} \int_{\bar \tau}^{-i \infty} \overline{ \underline{\rm G}_k[X,Y;\tau_1]} 
- c_k[X,Y] \label{mtobeta.11}\\
&=   -\frac{1}{4} (k{-}1)!
\sum_{j=0}^{k-2} (-1)^{j}\binom{ k{-}2 }{ j } 
\bigg( \frac{  Y }{2\pi i} \bigg)^{j}  
X^{k-2-j} \big( 
 \eeno{j}{k} +
\tce{j }{k}  +\ccsv{j}{k} 
\big) \, .
\notag
\end{align}
This can in fact be refined to three separate identities comparing terms with holomorphic,
antiholomorphic or no iterated Eisenstein integrals:
\begin{align}
 \int_\tau^{i \infty} \underline{\rm G}_k[X,Y;\tau_1]  &= 
 \frac{1}{2} (k{-}1)!
\sum_{j=0}^{k-2} (-1)^{j}\binom{ k{-}2}{ j } 
\bigg( \frac{  Y }{2\pi i} \bigg)^{j}  
X^{k-2-j} \, \eeno{j}{k} \, ,
\label{mtobeta.12} \\
 \int_{\bar \tau}^{-i \infty} \overline{ \underline{\rm G}_k[X,Y;\tau_1]} &= 
 \frac{1}{2} (k{-}1)!
\sum_{j=0}^{k-2} (-1)^{j}\binom{ k{-}2}{ j } 
\bigg( \frac{  Y }{2\pi i} \bigg)^{j}  
X^{k-2-j} \, \tce{j }{k}   \, , 
 \notag \\
c_k[X,Y] &= \frac{1}{4} (k{-}1)!
\sum_{j=0}^{k-2} (-1)^{j}\binom{ k{-}2}{ j } 
\bigg( \frac{  Y }{2\pi i} \bigg)^{j}  
X^{k-2-j} \, \ccsv{j}{k} \, .
\notag
\end{align}
The analogous comparison at modular depth two reads (see (\ref{Kdepthtwo}) and (\ref{eq:Mbeqv}))
\begin{align}
&M_{k_1,k_2}[\tau] =  
\frac{1}{4}  \, \int_{\tau}^{i\infty}  \underline{\rm G}_{k_1}[\tau_1]  \!\int_{\tau_1}^{i\infty} \underline{\rm G}_{k_2}[\tau_2]  
   \, + \frac{1}{4} \, \int_{\tau}^{i\infty}   \underline{\rm G}_{k_1}[\tau_1]  
 \times \! \int_{\bar \tau}^{-i\infty} \overline{ \underline{\rm G}_{k_2}[\tau_2] } 
\label{mtobeta.13}
 \\
 &\quad + \frac{1}{4} \, \int_{\bar \tau}^{-i\infty} \overline{ \underline{\rm G}_{k_2}[\tau_2] }   \!
\int_{\bar \tau_2}^{-i\infty} \overline{ \underline{\rm G}_{k_1}[\tau_1] }
 +  \frac{1}{2}\, c_{k_2}  \int_{\tau}^{i\infty}  \underline{\rm G}_{k_1}[\tau_1]  
  \, +  \frac{1}{2}\, c_{k_1}  \int_{\bar \tau}^{-i\infty} \overline{ \underline{\rm G}_{k_2}[\tau_2] }
\notag \\
&\quad 
- c_{k_1,k_2} 
- \frac{1}{2} \int_{\bar \tau}^{-i\infty} \!   \overline{\underline{\rm g}_{k_1,k_2}[\tau_1]}
- \frac{1}{2} \int_{\tau}^{i\infty}  \underline{\rm f}_{k_1,k_2}[\tau_1]
\notag \\
&=  \frac{1}{16} (k_1{-}1)!(k_2{-}1)!
\sum_{j_1=0}^{k_1-2} \sum_{j_2=0}^{k_2-2}
(-1)^{j_1+j_2}\binom{ k_1{-}2}{ j_1 } \binom{ k_2{-}2}{ j_2 }  \bigg( \frac{  Y_1 }{2\pi i} \bigg)^{j_1}  
X_1^{k_1-2-j_1} 
 \bigg( \frac{  Y_2 }{2\pi i} \bigg)^{j_2}
  X_2^{k_2-2-j_2} 
  \notag \\*
 &\quad \times \bigg\{ \eeno{j_2 &j_1 }{k_2 &k_1} 
  +    \tce{j_2 }{k_2}   \eeno{j_1 }{k_1}
  + \tce{ j_1 &j_2 }{k_1 &k_2}  
   +\ccsv{j_2}{k_2}  \eeno{j_1 }{k_1}
   +  \tce{j_2 }{k_2}  \ccsv{j_1}{k_1}
    +\ccsv{j_2 &j_1}{k_2 &k_1}   \notag \\*
&\quad   \quad    \quad  + \sum_{k,j} \bigg[
\chicst{j}{k}{j_2 &j_1}{k_2 &k_1} \tce{j}{k} +  \sum_{\Delta_k \in {\cal S}_k } 
\Big(
 \etacst{j}{\Delta_k}{j_2 &j_1}{k_2 &k_1} \tce{j}{\Delta_k} + \xicst{j}{\Delta_k}{j_2 &j_1}{k_2 &k_1} \eeno{j}{\Delta_k} 
\Big)
\bigg] \bigg\} \, ,
  \notag
\end{align}
where we stopped displaying the dependence on $X_i,Y_i$ to avoid cluttering (see the comments
below (\ref{def:ck}) for our conventions). The first five terms on the two sides are readily seen to
match individually by (\ref{mtobeta.12}) or immediate consequences such as
\begin{align}
\int_{\tau}^{i\infty}  \underline{\rm G}_{k_1}[\tau_1]  \!\int_{\tau_1}^{i\infty} \underline{\rm G}_{k_2}[\tau_2] 
&= \frac{1}{4} (k_1{-}1)!(k_2{-}1)!
\sum_{j_1=0}^{k_1-2} \sum_{j_2=0}^{k_2-2}
(-1)^{j_1+j_2}\binom{ k_1{-}2}{j_1 } \binom{ k_2{-}2}{ j_2 } \notag \\
&\quad\times \bigg( \frac{  Y_1 }{2\pi i} \bigg)^{j_1}  
X_1^{k_1-2-j_1}  \bigg( \frac{  Y_2 }{2\pi i} \bigg)^{j_2}
 X_2^{k_2-2-j_2}  \,  \eeno{j_2 &j_1 }{k_2 &k_1} \, .
 \label{mtobeta.14}
\end{align}
The sixth terms $\sim c_{k_1,k_2}$ and $\ccsv{j_2 &j_1}{k_2 &k_1}$ on 
both sides of (\ref{mtobeta.13}) match by (\ref{de2csv}). Then, by separately
equating the leftover terms involving holomorphic and antiholomorphic integrals,
we deduce two new identities from (\ref{mtobeta.13}):
\begin{align}
 \int_{\bar \tau}^{-i\infty} \!   \overline{\underline{\rm g}_{k_1,k_2}[\tau_1]} &=
  - \frac{1}{8} (k_1{-}1)!(k_2{-}1)!
\sum_{j_1=0}^{k_1-2} \sum_{j_2=0}^{k_2-2}
(-1)^{j_1+j_2}\binom{ k_1{-}2}{ j_1 } \binom{ k_2{-}2}{ j_2 } 
 \bigg( \frac{  Y_1 }{2\pi i} \bigg)^{j_1}  
X_1^{k_1-2-j_1}  \notag \\
&\quad \times \bigg( \frac{  Y_2 }{2\pi i} \bigg)^{j_2}
 X_2^{k_2-2-j_2}  
 \sum_{k,j} \bigg\{
 \chicst{j}{k}{j_2 &j_1}{k_2 &k_1} \tce{j}{k}
+ \sum_{\Delta_k \in {\cal S}_k } 
 \etacst{j}{\Delta_k}{j_2 &j_1}{k_2 &k_1} \tce{j}{\Delta_k}   
\bigg\} \, , \notag \\
\int_{\tau}^{i\infty}  \underline{\rm f}_{k_1,k_2}[\tau_1] &= - \frac{1}{8} (k_1{-}1)!(k_2{-}1)!
\sum_{j_1=0}^{k_1-2} \sum_{j_2=0}^{k_2-2}
(-1)^{j_1+j_2}\binom{ k_1{-}2}{ j_1 } \binom{ k_2{-}2}{ j_2 }  \label{mtobeta.15}  \\
&\quad \times \bigg( \frac{  Y_1 }{2\pi i} \bigg)^{j_1}  
X_1^{k_1-2-j_1}  \bigg( \frac{  Y_2 }{2\pi i} \bigg)^{j_2}
 X_2^{k_2-2-j_2} 
 \sum_{k,j} \sum_{\Delta_k \in {\cal S}_k}  \xicst{j}{\Delta_k}{j_2 &j_1}{k_2 &k_1} \eeno{j}{\Delta_k} \, .
\notag 
\end{align}
The sums over $k,j$ in the second and fourth line are both bounded by $k\leq k_1{+}k_2{-}2$.
Note that the two types of contributions $ \chicst{j}{k}{j_2 &j_1}{k_2 &k_1} \tce{j}{k}$ and
$\etacst{j}{\Delta_k}{j_2 &j_1}{k_2 &k_1} \tce{j}{\Delta_k} $ to the integral over
$\overline{\underline{\rm g}_{k_1,k_2}[\tau_1]}$ correspond to the Eisenstein part 
and the cuspidal part of the underlying antiholomorphic modular form.

\subsection{\texorpdfstring{Comparing $M_{k_1,k_2,k_3}$ with ${\cal E}^{\rm eqv}$}{Comparing M(k1,k2,k3) with Eeqv}}
\label{app:expeqv.4}

At modular depth three, the $\beta^{\rm eqv}$-representation of $M_{k_1,k_2,k_3}$
in (\ref{olidreams}) that we wish to prove is equivalent to the following 
generalisation of (\ref{mtobeta.10}):
\begin{align}
M_{k_1,k_2,k_3}[\tau] &= - \frac{1}{64} (k_1{-}1)!(k_2{-}1)!(k_3{-}1)!
\sum_{j_1=0}^{k_1-2} \sum_{j_2=0}^{k_2-2} \sum_{j_3=0}^{k_3-2}
(-1)^{j_1+j_2+j_3}\binom{ k_1{-}2}{ j_1 } \!\binom{ k_2{-}2}{ j_2 } \! \binom{ k_3{-}2}{ j_3 } \notag \\
&\quad \times 
\bigg( \frac{  Y_1 }{2\pi i} \bigg)^{j_1}  
\! X_1^{k_1-2-j_1} 
 \bigg( \frac{  Y_2 }{2\pi i} \bigg)^{j_2}
 \! X_2^{k_2-2-j_2} 
   \bigg( \frac{  Y_3 }{2\pi i} \bigg)^{j_3}
 \! X_3^{k_3-2-j_3} 
  \,  \eeqv{j_3 &j_2  &j_1 }{ k_3 &k_2 &k_1} \, . 
\label{mtobeta.20}
\end{align}
The key idea of the proof is to insert the representation (\ref{mtobeta.08})
of $\eeqv{j_3 &j_2  &j_1 }{ k_3 &k_2 &k_1}$ and to compare term by term 
with the original definition of $M_{k_1,k_2,k_3}$ via (\ref{KtoM3}) with $K_{k_1,k_2,k_3}$
given by (\ref{k123expr}). Based on (\ref{mtobeta.12}), (\ref{mtobeta.15}) and their
corollaries for products and double or triple integrals, we can make 21 out of the 27 terms
in the expression (\ref{mtobeta.08}) for $\eeqv{j_3 &j_2  &j_1 }{ k_3 &k_2 &k_1}$
match with $K_{k_1,k_2,k_3}$ in (\ref{k123expr}). 
After setting these matching terms aside, the leftover task in proving (\ref{olidreams})
is to show that one can find solutions for $c_{k_1,k_2,k_3} $,
$  \overline{\underline{\rm g}_{k_1,k_2,k_3}[\tau_1]}$
and $ \underline{\rm f}_{k_1,k_2,k_3}[\tau_1]$ such that
\begin{align}
&M_{k_1,k_2,k_3}[\tau] - K_{k_1,k_2,k_3}[\tau] = 
- c_{k_1,k_2,k_3} 
- \frac{1}{2}  \int_{\bar \tau}^{-i\infty} \!   \overline{\underline{\rm g}_{k_1,k_2,k_3}[\tau_1]}
- \frac{1}{2} \int_{\tau}^{i\infty}  \underline{\rm f}_{k_1,k_2,k_3}[\tau_1]
\label{mtobeta.21}  \\
&\quad = - \frac{1}{64} (k_1{-}1)!(k_2{-}1)!(k_3{-}1)!
\sum_{j_1=0}^{k_1-2} \sum_{j_2=0}^{k_2-2} \sum_{j_3=0}^{k_3-2}
(-1)^{j_1+j_2+j_3}\binom{ k_1{-}2}{ j_1 } \binom{ k_2{-}2}{ j_2 } \binom{ k_3{-}2}{ j_3 } \notag \\
&\quad\quad \times 
\bigg( \frac{  Y_1 }{2\pi i} \bigg)^{j_1}  
X_1^{k_1-2-j_1} 
\bigg( \frac{  Y_2 }{2\pi i} \bigg)^{j_2}
 X_2^{k_2-2-j_2} 
 \bigg( \frac{  Y_3 }{2\pi i} \bigg)^{j_3}
  X_3^{k_3-2-j_3} \, \bigg\{  \ccsv{j_3 &j_2  &j_1 }{ k_3 &k_2 &k_1}  \notag \\
&\quad\quad \quad + \sum_{k,j} \bigg[   \chicst{j}{k}{ j_3 &j_2 &j_1}{ k_3 &k_2 &k_1} \tce{j}{k}
  + \sum_{ \Delta_k \in {\cal S}_k }   \Big(
   \xicst{j}{\Delta_k}{j_3 &j_2 &j_1}{k_3 &k_2 &k_1} \eeno{j}{\Delta_k}
   +   \etacst{j}{\Delta_k}{j_3 &j_2 &j_1}{k_3 &k_2 &k_1} \tce{j}{\Delta_k} 
\Big) \notag \\
&\quad\quad \quad \quad \quad \quad + \sum_{ \Delta_k \in {\cal S}_k } \tce{j}{\Delta_k} \sum_{k',j'}
\Big(
\etacst{j}{\Delta_k}{j' &j_1}{k' &k_1} \chicst{j'}{k'}{j_3 &j_2}{ k_3 &k_2}
+ \etacst{j}{\Delta_k}{j_3 & j' }{k_3 &k'} \chicst{j'}{k'}{j_2 &j_1}{ k_2 &k_1}
\Big) \bigg] \bigg\} \, .\notag
\end{align}
It is a non-trivial confirmation that all iterated integrals beyond depth one 
in the expression (\ref{mtobeta.08}) for $\eeqv{j_3 &j_2  &j_1 }{ k_3 &k_2 &k_1}$
have been successfully lined up with the expression (\ref{k123expr}) for $K_{k_1,k_2,k_3}$.
It is always possible to separately match the depth-one integrals ${\cal E}$, their complex conjugates $\pm \tilde {\cal E}$
and the $\tau$-independent terms in (\ref{mtobeta.21}) -- this process will in fact determine the
quantities $ \underline{\rm f}_{k_1,k_2,k_3}[\tau_1]$, $  \overline{\underline{\rm g}_{k_1,k_2,k_3}[\tau_1]}$
and $c_{k_1,k_2,k_3} $. The 
$\tau$-independent terms in (\ref{mtobeta.21})
imply the identity (\ref{de3csv}) between $c_{k_1,k_2,k_3}$ and $ \ccsv{j_3 &j_2  &j_1 }{ k_3 &k_2 &k_1}$
noted in the main text, and the equality of the (anti-)holomorphic integral amounts to
\begin{align}
 &\int\limits_{\bar \tau}^{-i\infty} \!   \overline{\underline{\rm g}_{k_1,k_2,k_3}[\tau_1]} =
 \frac{(k_1{-}1)!(k_2{-}1)!(k_3{-}1)!}{32} 
\sum_{j_1=0}^{k_1-2} \sum_{j_2=0}^{k_2-2} \sum_{j_3=0}^{k_3-2}
(-1)^{j_1+j_2+j_3}\binom{ k_1{-}2}{ j_1 }  \! \binom{ k_2{-}2}{ j_2 }  \! \binom{ k_3{-}2}{ j_3 } \notag \\
&\quad \times 
\bigg( \frac{  Y_1 }{2\pi i} \bigg)^{j_1}  
X_1^{k_1-2-j_1} 
\bigg( \frac{  Y_2 }{2\pi i} \bigg)^{j_2}
 X_2^{k_2-2-j_2} 
 \bigg( \frac{  Y_3 }{2\pi i} \bigg)^{j_3}
  X_3^{k_3-2-j_3} 
 \sum_{k,j} \bigg\{
   \chicst{j}{k}{ j_3 &j_2 &j_1}{ k_3 &k_2 &k_1} \tce{j}{k}    \label{mtobeta.23}  \\
   & \quad  \quad  
  + \sum_{ \Delta_k \in {\cal S}_k } \tce{j}{\Delta_k}  \bigg[  \etacst{j}{\Delta_k}{j_3 &j_2 &j_1}{k_3 &k_2 &k_1} 
+ \sum_{k',j'} \Big( \etacst{j}{\Delta_k}{j' &j_1}{k' &k_1} \chicst{j'}{k'}{j_3 &j_2}{ k_3 &k_2}
+ \etacst{j}{\Delta_k}{j_3 & j' }{k_3 &k'} \chicst{j'}{k'}{j_2 &j_1}{ k_2 &k_1}
\Big) \bigg] \bigg\} \,,
\notag
\end{align}  
as well as
\begin{align}
& \int\limits_{\tau}^{i\infty}  \underline{\rm f}_{k_1,k_2,k_3}[\tau_1] =
 \frac{(k_1{-}1)!(k_2{-}1)!(k_3{-}1)!}{32} 
\sum_{j_1=0}^{k_1-2} \sum_{j_2=0}^{k_2-2} \sum_{j_3=0}^{k_3-2}
(-1)^{j_1+j_2+j_3}\binom{ k_1{-}2}{ j_1 } \! \binom{ k_2{-}2}{ j_2 }  \!\binom{ k_3{-}2 }{ j_3 } \notag \\
&\quad   \times 
\bigg( \frac{  Y_1 }{2\pi i} \bigg)^{j_1}  
\! X_1^{k_1-2-j_1} 
\bigg( \frac{  Y_2 }{2\pi i} \bigg)^{j_2}
\! X_2^{k_2-2-j_2} 
 \bigg( \frac{  Y_3 }{2\pi i} \bigg)^{j_3}
\!  X_3^{k_3-2-j_3}   \sum_{k,j} \sum_{\Delta_k \in {\cal S}_k }  \xicst{j}{\Delta_k}{j_3 &j_2 &j_1}{k_3 &k_2 &k_1} \eeno{j}{\Delta_k}
  \label{mtobeta.24}    \, .
\end{align}  
The sums over $k,j$ and $k',j'$ are again finite by the vanishing conditions
stated below (\ref{mtobeta.08}). In conclusion, the expansion (\ref{olidreams}) of $M_{k_1,k_2,k_3}$
in terms of $\beqv{j_3&j_2&j_1}{k_3&k_2&k_1}$ obtained from the generating series $\mathbb J^{\rm eqv}$
has been shown to agree with the original construction in section \ref{sec:modcomp3} through the quantity $K_{k_1,k_2,k_3}$
in (\ref{k123expr}). As a side effect, we can explicitly determine the quantities 
$  \overline{\underline{\rm g}_{k_1,k_2,k_3}[\tau_1]}$ and $ \underline{\rm f}_{k_1,k_2,k_3}[\tau_1]$ via
 (\ref{mtobeta.23}) and (\ref{mtobeta.24}) in terms of information $\hat \psi^{\rm sv}$ and
 $\eee_{\Delta^{\pm}}$ from the generating series $\mathbb J^{\rm eqv}$.

\section{Proof of Lemma~\ref{lem2}}
\label{app:Tcocyc}

In this appendix, we shall prove that the $T$-cocycle at infinity defined by (\ref{eq:Tcocyle}) 
can be brought into the compact form (\ref{mt.13}). Given the definition of $\hat N$ and 
$\hat N_+$ in (\ref{mt.12}), we see that the expansion \eqref{mt.11} can be decomposed according to the double grading $r = \# \hat N_+$ and $J = \# \eee_0$. Let us then consider the particular contribution with fixed gradings $r,J$:
\begin{align}\label{eq:Omgrad}
  \Pexp \bigg( \int_{i\infty}^{i\infty-1} \mathbb{A}_+(\tau_1) \bigg) \Big\vert_{\substack{r = \# \hat N_+ \\  \!\!\!\!\!J = \# \eee_0}} &= \sum_{j_1,j_2,\ldots,j_r=0}^{\infty}\delta(j_1{+}\cdots{+} j_r {-}J)
(2\pi i )^{J+r} \\
&\quad \times \bigg( \prod_{i=1}^r \frac{ 1 }{j_i! \sum_{m=i}^r (j_m{+}1)} \bigg)    \hat N_+^{(j_1)} \hat N_+^{(j_2)} \ldots \hat N_+^{(j_r)}\,.
\notag
\end{align}
We can now use the simple fact that the repeated adjoint action $X^{(j)}= \ad_{\eee_0}^j X$ of $\eee_0$ on any algebra element $X$ can be rewritten as
\begin{equation}
X^{(j)} = \sum_{p=0}^j \binom{j}{p} (-1)^{j-p} \eee_0^p \,X\, \eee_0^{j-p} = \sum_{a_1,a_2=0}^\infty \delta(a_1{+}a_2 {-} j) \binom{a_1 {+}a_2}{a_2} (-1)^{a_2} \eee_0^{a_1} X \, \eee_0^{a_2}\,,
\end{equation}
leading to
\begin{align}\label{eq:NNexp}
 \hat N_+^{(j_1)} \hat N_+^{(j_2)} \ldots \hat N_+^{(j_r)} = &\sum_{a_1,\ldots ,a_{2r}=0}^\infty \prod_{i=1}^{r} \Big[  \delta(a_{2i-1}{+}a_{2i} {-} j_i) \binom{a_{2i-1} {+}a_{2i}}{a_{2i}} (-1)^{a_{2i}} \Big]\\ 
 &\nn \times \eee_0^{a_1} \hat N_+ \eee_0^{a_2+a_3} \hat N_+ \eee_0^{a_4+a_5}\cdots \hat N_+\eee_0^{a_{2r-2}+a_{2r-1}} \hat N_+ \eee_0^{a_{2r}}\,.
\end{align}
We now change summation variables and define
\begin{equation}
b_i \coloneqq  a_{2i-1}+a_{2i-2}\,,\qquad i=1,2,\ldots,r{+}1\,,\label{eq:changesum}
\end{equation} 
with the convention $a_{0}=a_{2r+1}=0$ such that $b_1 = a_1$ and  $b_{r+1}=a_{2r}$.
We can eliminate the $j_i$ variables using the Kronecker delta 
$\delta(j_1{+}\cdots{+} j_r {-}J)$  which, when combined with $\prod_i \delta(a_{2i-1}{+}a_{2i} {-} j_i)$, reduces to $\delta(b_1{+}\cdots {+}b_{r+1} {-} J)$.

Furthermore, we can also get rid of the variables $a_{2i-1}$ for $i=2,\ldots,r$ in favour of the new variables $b_i\geq0$ with $i=1,\ldots,r{+}1$. We still have to perform the sum over the remaining variables $a_{2i-2}$ with $i=2,\ldots,r$ over the range $0\leq a_{2i-2} \leq b_i$  as we can see from \eqref{eq:changesum}.

It is convenient to focus on the combinatorial factors appearing in \eqref{eq:Omgrad} and express them first in terms of the variables $a_{i}$:
\begin{align}
\bigg( \prod_{i=1}^r \frac{ 1 }{j_i! \sum_{m=i}^r (j_m{+}1)} \bigg)  = \prod_{i=1}^r \frac{1}{(a_{2i-1}{+}a_{2i})! (r{+}1{-}i{+}\sum_{m=2i-1}^{2r} a_m )} \,.
\end{align}
When we combine this expression with the additional factors coming from \eqref{eq:NNexp} we obtain
\begin{align}
&\nn\prod_{i=1}^r\bigg[  \frac{1}{(a_{2i-1}{+}a_{2i})! (r{+}1{-}i{+}\sum_{m=2i-1}^{2r} a_m )} \binom{a_{2i-1} {+}a_{2i}}{a_{2i}} (-1)^{a_{2i}} \bigg]\\
&=(-1)^{\sum_{i=1}^r a_{2i}}  \prod_{i=1}^{r} \frac{1}{(a_{2i-1})! \,(a_{2i})!} \prod_{i=1}^r  \frac{1}{(r{+}1{-}i{+}\sum_{m=2i-1}^{2r} a_m )}\,.
\end{align}
The next step is to substitute $a_{2i+1} = b_{i+1} - a_{2i}$ for $i=1,\ldots,r{-}1$, remembering the conditions $a_1 = b_1$ and $a_{2r}=b_{r+1}$. After a shift in the label $i$ and a rearrangement of the terms in the first product, we arrive at
\begin{equation}
\frac{1}{b_1!} \bigg[\prod_{i=1}^{r-1} \frac{(-1)^{a_{2i}}}{(b_{i+1} {-} a_{2i})! \,(a_{2i})!} \bigg]\frac{(-1)^{b_{r+1}}}{b_{r+1}!} \prod_{i=1}^r  \frac{1}{(r{+}1{-}i{-}a_{2i-2}{+}\sum_{m=i}^{r+1} b_m )}\,.\label{eq:inter1}
\end{equation}
Note that for $i=1$ the last fraction produces
\begin{equation}
\frac{1}{(r{-}a_{0}{+}\sum_{m=1}^{r+1} b_m )} = \frac{1}{(J{+}r)}\,,
\end{equation}
since $a_{0}=0$ and the $b$ variables are constrained by $\delta(b_1{+}\cdots {+}b_{r+1} {-} J)$.

We can now consider one by one the remaining summations over $0\leq a_{2i} \leq b_{i+1}$ with $i=1,\ldots,r{-}1$:
\begin{equation}
\sum_{a_{2i} = 0}^{b_{i+1}} \frac{ (-1)^{a_{2i}} }{(b_{i+1} {-} a_{2i})! \,(a_{2i})!} \frac{1}{ (r{-}i{-}a_{2i}{+}\sum_{m=i+1}^{r+1} b_m )} =(-1)^{b_{i+1}} \frac{\Gamma(r{-}i {+}\sum_{m=i+2}^{r+1} b_m)}{\Gamma(r{+}1{-}i {+}\sum_{m=i+1}^{r+1} b_m)}\,.
\end{equation}
When we combine all these factors together we see that summing \eqref{eq:inter1} over all $a_{2i}$ with $i=1,\ldots,r{-}1$ yields the telescopic product
\begin{align}
&\nn \frac{(-1)^{b_{r+1}}}{(b_1)! (b_{r+1})! } \frac{1}{(J{+}r)} \prod_{i=1}^{r-1} (-1)^{b_{i+1}} \frac{\Gamma(r{-}i {+}\sum_{m=i+2}^{r+1} b_m)}{\Gamma(r{+}1{-}i {+}\sum_{m=i+1}^{r+1} b_m)}=\frac{1}{(b_1)! (b_{r+1})! } \frac{1}{(J{+}r)} (-1)^{\sum_{i=2}^{r+1} b_i}   \\
& \ \ \ \ \times \frac{\Gamma(r{-}1 {+}\sum_{m=3}^{r+1}b_m)}{\Gamma(r {+}\sum_{m=2}^{r+1} b_m)}\frac{\Gamma(r{-}2 {+}\sum_{m=4}^{r+1} b_m)}{\Gamma(r{-}1 {+}\sum_{m=3}^{r+1} b_m)}\cdots \frac{\Gamma(r{-}(r{-}1) {+}\sum_{m=(r-1)+2}^{r+1} b_m)}{\Gamma(r{+}1{-}(r{-}1) {+}\sum_{m=(r-1)+1}^{r+1} b_m)}\\
&\nn =\frac{1}{(b_1)! (b_{r+1})! } \frac{1}{(J{+}r)} (-1)^{\sum_{i=2}^{r+1} b_i}  \frac{\Gamma(r{-}(r{-}1) {+}\sum_{m=(r-1)+2}^{r+1} b_m)}{\Gamma(r {+}\sum_{m=2}^{r+1} b_m)} = \frac{(-1)^{J-b_1}}{(b_1)! (J{+}r{-}b_1{-}1)! (J{+}r)}\,,
\end{align}
where again we used $\delta(b_1{+}\cdots {+}b_{r+1} {-} J)$.

We finally rewrite \eqref{eq:Omgrad} in these new variables 
\begin{align}\label{eq:Omgradother}
& \Pexp  \bigg( \int_{i\infty}^{i\infty-1} \mathbb{A}_+(\tau_1) \bigg) \Big\vert_{\substack{r = \# \hat N_+ \\  \!\!\!\!\!J = \# \eee_0}}  \\
&\nn = \sum_{b_1,\ldots,b_{r+1}=0}^\infty \delta(b_1{+}\cdots {+}b_{r+1} {-} J) \frac{ (2\pi i )^{J+r} (-1)^{J-b_1}}{(b_1)! (J{+}r{-}b_1{-}1)! (J{+}r)} \eee_0^{b_1}\hat N_+ \eee_0^{b_2}\hat N_+\cdots \hat N_+ \eee_0^{b_{r+1}}\,.
\end{align}
It is much easier to compute
\begin{align}
& e^{2\pi i \eee_0} 
 e^{2\pi i (\hat N_+-\eee_0)} \Big\vert_{\substack{r = \# \hat N_+ \\  \!\!\!\!\!J = \# \eee_0}} = \sum_{k_1,k_2=0}^\infty (2\pi i)^{k_1+k_2} \Big[\eee_0^{k_1}  (\hat N_+-\eee_0)^{k_2}\Big]\Big\vert_{\substack{r = \# \hat N_+ \\  \!\!\!\!\!J = \# \eee_0}} \label{eq:expexp} \\
 &\nn= \sum_{k_2=r}^{J+r} \frac{(2\pi i )^{J+r}}{\big(J{-}(k_2{-}r) \big)! k_2!} \eee_0^{J-(k_2-r)} \Big[ (\hat N_+-\eee_0)^{k_2}\Big\vert_{r = \# \hat N_+} \Big]\\
  &= \sum_{\ell_0,\ldots ,\ell_{r+1}=0}^{J}\delta(\ell_0{+}\ell_1{+} \ldots {+}\ell_{r+1}{-}J) \frac{(2\pi i )^{J+r}}{(J{+}r{-}\ell_0)! \ell_0!}  (-1)^{J-\ell_0} \eee_0^{\ell_0+\ell_1 }\hat N_+ \eee_0^{\ell_2} \hat N_+\cdots \hat N_+ \eee_0^{\ell_{r+1}}\,,\notag
\end{align}
where we used the fact that $k_2$ must be at least $r$ as to obtain enough powers of $(\hat N_+{-}\eee_0)$ such that $r = \# \hat N_+$ can be satisfied, and it must be at most $J{+}r$ otherwise we would get too many $\eee_0$ to satisfy $J = \# \eee_0$. In the last line we simply changed variables $k_2\to \ell_0 = J{-}(k_2{-}r)\in\{0,\ldots,J\}.$

The final step is to change summation variables $b_1 \coloneqq  \ell_0 {+}\ell_1$ while $b_i = \ell_i$ for $i=2,\ldots,r{+}1$.
If we substitute for $\ell_1= b_1 {-}\ell_0$ and perform the sum over $\ell_0 \in \{0,\ldots,b_1\}$ we simply have:
\begin{equation}
\sum_{\ell_0=0}^{b_1} \frac{(-1)^{J-\ell_0}}{(J{+}r{-}\ell_0)! \ell_0!}  =\frac{(-1)^{J-b_1}}{b_1! (J{+}r{-}b_1{-}1)!(J{+}r)  }\,,
\end{equation}
so that our expression \eqref{eq:expexp} identically reproduces \eqref{eq:Omgrad}.


\section{Zeta generators at modular depth three}
\label{app:F}

In this appendix, we add explicit examples at modular depth three
to the discussion of zeta generators in section \ref{sec:rev.up}.
The subsequent expressions apply to the uplift $\hat \sigma_w$
adapted to the free-algebra generators $\eee_k$ and can be
straightforwardly specialised to the zeta generators $\sigma_w$
associated with the derivations $\ep_k$ on ${\rm Lie}[a,b]$.


\subsection{\texorpdfstring{$[\zetaE_w,\eee_k]$ beyond modular depth two}{[zeta(w),e(k)] beyond modular depth two}}
\label{app:zep3}

For the arithmetic terms $\zetaE_w$ of $\sigmaE_w$, the
modular-depth-two contributions to their commutators with $\eee_k$
are known in a simple closed form (\ref{zetgen.07}) in
terms of the $t^d$ operation in (\ref{eq:tdpq}). 
We shall here give the terms of modular depth three in 
$[\zetaE_w,\eee_{k}] $ for all \degree s $2w{+}k \leq 18$
and refer to tentative extra contributions of modular depth four via ellipses. 
\begin{itemize}
\item $w=3$
\begin{align}
[\zetaE_3,\eee_{4}] &= \frac{\BF_6}{\BF_4} t^4(\eee_{4}, \eee_{6}) \, ,
 \label{zwep3.51}  \\
[\zetaE_3,\eee_{6}] &=   \frac{ \BF_8}{\BF_6}t^4(\eee_{4}, \eee_{8})
 - \frac{9 \BF_4^2}{10 \BF_6}  t^2(\eee_4,t^3(\eee_4,\eee_4))  \, ,
 \notag \\
[\zetaE_3,\eee_{8}]   &=  
 \frac{\BF_{10}}{\BF_8} t^4(\eee_{4}, \eee_{10})
+ \frac{ \BF_4 \BF_6 }{\BF_8}
\bigg\{ {-}\frac{15}{7} t^2(\eee_4, t^3(\eee_4,\eee_6))
+\frac{9}{4} t^3(\eee_4,t^2(\eee_4,\eee_6)) \bigg\} \, ,
 \notag \\
[\zetaE_3,\eee_{10}]  &=  \frac{\BF_{12}}{\BF_{10}} t^4(\eee_{4}, \eee_{12})
+ \frac{ \BF_4 \BF_8 }{\BF_{10}}
\bigg\{ {-}\frac{7}{3} t^2(\eee_4, t^3(\eee_4,\eee_8))
+\frac{12}{5} t^3(\eee_4,t^2(\eee_4,\eee_8)) \bigg\}  \notag \\
&\quad + \frac{ \BF_6^2 }{\BF_{10}}
\bigg\{
{-} \frac{25}{27} t^2 (\eee_6,t^3(\eee_4,\eee_6))
+\frac{2}{9}  t^3(\eee_6,t^2(\eee_4,\eee_6))
\bigg\} \, ,
\notag \\
[\zetaE_3,\eee_{12}]   &=  \frac{\BF_{14}}{\BF_{12}} t^4(\eee_{4}, \eee_{14})
+\frac{ \BF_4 \BF_{10} }{\BF_{12}}
\bigg\{
{-}\frac{27}{11} t^2(\eee_{4}, t^3(\eee_4,\eee_{10}))  + \frac{ 5}{2}  t^3(\eee_{4}, t^2(\eee_4,\eee_{10})) \bigg\} \notag \\
&\quad+  \frac{ \BF_6 \BF_8 }{\BF_{12}}
\bigg\{
{-}\frac{35}{44}   t^2(\eee_{6}, t^3(\eee_4,\eee_{8}))
+\frac{5}{33}  t^3(\eee_{6}, t^2(\eee_4,\eee_{8})) \notag \\
&\quad\quad\quad\quad\quad \quad
- \frac{35}{33}  t^2(\eee_{8}, t^3(\eee_4,\eee_{6}))
+\frac{7}{22}  t^3(\eee_{8}, t^2(\eee_4,\eee_{6}))
\bigg\} +\ldots \, .\notag
\end{align}
\item $w=5$
\begin{align}
[\zetaE_5,\eee_4]  &=  \frac{\BF_8}{\BF_4}t^6(\eee_{6}, \eee_{8})
+\frac{ \BF_6 \BF_2^3 }{2 \BF_4^2 } \, t^4(\eee_6,t^3(\eee_4,\eee_4))  \label{zwep3.52}\\
&\quad
- \BF_4 \bigg\{
\frac{9}{10} t^3(\eee_4,t^4(\eee_4,\eee_6))
+\frac{1}{5} t^4(\eee_4,t^3(\eee_4,\eee_6))
\bigg\} \, , \notag \\
[\zetaE_5,\eee_6]  &= \frac{\BF_{10}}{\BF_6} t^6(\eee_{6}, \eee_{10})
+ \frac{ \BF_8 \BF_2^3 }{2 \BF_4 \BF_6 } \, t^4(\eee_8,t^3(\eee_4,\eee_4))
 \notag \\
&\quad - \BF_4\bigg\{
2 t^3(\eee_6,t^4(\eee_4,\eee_6))
+\frac{20}{7} t^4(\eee_6,t^3(\eee_4,\eee_6))
+\frac{5}{2} t^5(\eee_6,t^2(\eee_4,\eee_6))
\bigg\} +\ldots \, ,\notag \\
[\zetaE_5,\eee_8] &= \frac{\BF_{12}}{\BF_8} t^6(\eee_{6}, \eee_{12})
+ \frac{ \BF_{10} \BF_2^3 }{2 \BF_4 \BF_8 } \, t^4(\eee_{10},t^3(\eee_4,\eee_4))
- \frac{125 \BF_6^2 }{84 \BF_8} \, t^2(\eee_6,t^5(\eee_6,\eee_6))  \notag \\
&\quad +\BF_4\bigg\{
{-}\frac{15}{14} t^2(\eee_4,t^5(\eee_6,\eee_8))
+\frac{3}{8}  t^3(\eee_4,t^4(\eee_6,\eee_8))
-\frac{1}{30}  t^4(\eee_4,t^3(\eee_6,\eee_8)) \notag \\
&\quad \quad \quad \quad
- \frac{27}{20}  t^3(\eee_8,t^4(\eee_4,\eee_6))
-\frac{5}{6} t^4(\eee_8,t^3(\eee_4,\eee_6))
-\frac{1}{6} t^5(\eee_8,t^2(\eee_4,\eee_6))
\bigg\}+\ldots \, .
\notag 
\end{align}
\item $w=7$
\begin{align}
[\zetaE_7,\eee_4]  &=  \frac{\BF_{10}}{\BF_4} t^8(\eee_{8}, \eee_{10})
+\frac{ \BF_8 \BF_2^2 }{\BF_6} \, t^6(\eee_8,t^3(\eee_4,\eee_6))
+ \frac{ \BF_6 \BF_2^2 }{2 \BF_4} \, t^4(\eee_6,t^5(\eee_6,\eee_6))
\notag\\*
&\quad - \BF_6 \bigg\{
\frac{15}{14} \, t^3(\eee_4, t^6(\eee_6,\eee_8)) 
+ \frac{5}{14} \, t^4(\eee_4, t^5(\eee_6,\eee_8)) \notag \\*
&\quad\quad\quad\quad
+\frac{5}{7} \, t^5(\eee_4, t^4(\eee_6,\eee_8))
+\frac{3}{28} \,t^6(\eee_4, t^3(\eee_6,\eee_8))
\bigg\}+\ldots\, .
  \label{zwep3.53}
\end{align}
\end{itemize}
These expressions are derived from the commutation relation (\ref{zetgen.06})
with $\hat N$ based on the terms of modular depth two in (\ref{zetgen.08}),
see \cite{Dorigoni:2023part1} for further details and generalisations.

\subsection{Expansion of zeta generators}
\label{app:exd3}

We shall next supplement the closed formula (\ref{zetgen.08}) for the expansion of
zeta generators up to modular depth two by contributions at
modular depth three. The following examples are
given up to and including \degree\ 14
\begin{align}
\sigmaE_3  &= -\frac{1}{2} \eee_4^{(2)} + \zetaE_3 +   \frac{ s^5_{4,4} }{1440} 
-\frac{[\eee_4,\eee _6^{(1)}]}{120960} +\frac{[\eee_4^{(1)},\eee _6] }{30240}
+\frac{[\eee_4,\eee _8^{(1)}]}{7257600} 
-\frac{[\eee_4^{(1)},\eee _8] }{1209600}
\label{sigexp.1} \\
&\quad  -\frac{[\eee_4,\eee _{10}^{(1)}]}{383\,201\,280}
+\frac{[\eee_4^{(1)},\eee _{10}]}{47\,900\,160} -\frac{[\eee_4,[\eee_4,\eee _6]]}{58\,060\,800}+\ldots \, ,
\notag
\\
\sigmaE_5 &=  -\frac{1}{4!} \eee_6^{(4)} +  \frac{5  s^3_{4,4} }{24} 
+ \zetaE_5 + \frac{ s^5_{4,6}}{720}  -   \frac{ s^7_{6,6} }{60480} 
+ \frac{1}{6 912} \big(
  [ \eee_4^{(1)} , [\eee_4^{(1)} , \eee_4^{(0)}]]
  + 2 [ \eee_4^{(0)}, [\eee_4^{(0)}, \eee_4^{(2)}]]
\big)  \notag \\
&\quad + \frac{ [ \eee_6, \eee_8^{(3)}] }{ 145152000 }
 -  \frac{ [\eee_6^{(1)}, \eee_{8}^{(2)}]}{36288000} 
 +  \frac{ [\eee_6^{(2)}, \eee_8^{(1)}] }{14515200} 
 -  \frac{ [\eee_6^{(3)}, \eee_8]}{7257600}
\notag \\
&\quad 
- \frac{ [ \eee_4 , [ \eee_4 , \eee_6^{(2)} ] ]}{2073600} 
- \frac{ 23 [\eee_4, [ \eee_4^{(1)} , \eee_6^{(1)} ]] }{24192000} 
+ \frac{ 289 [ \eee_4 , [ \eee_4^{(2)}, \eee_6 ]] }{48384000}  \notag \\
&\quad 
+ \frac{ 139 [\eee_4^{(1)}, [ \eee_4 , \eee_6^{(1)} ]]}{72576000} 
- \frac{ [\eee_4^{(1)}, [ \eee_4^{(1)} , \eee_6 ]]  }{4147200} 
- \frac{ 1007 [\eee_4^{(2)}, [ \eee_4 , \eee_6  ]] }{145152000}
 + \ldots \, ,
\notag \\
\sigmaE_7  &=  - \frac{1}{6!} \eee_8^{(6)} +  \frac{7s^3_{4,6}}{24}   
+    \frac{ s^5_{4,8} }{720}  - \frac{s^5_{6,6} }{288} 
  -      \frac{661 s^{3}(\eee_ 4, t^{3}(\eee_4, \eee_4)) }{14400}    \notag \\
&\quad+ \zetaE_7 -  \frac{s^7_{6,8} }{30240}  
+\frac{7s^{5}(\eee_ 4, t^{3}(\eee_4, \eee_6)) }{17280}
+\ldots \, ,
\notag
\end{align}
where $s^d_{p,q}=s^d(\eee_p, \eee_q)$ are defined by (\ref{eq:sdpq}).
The next examples are displayed up to and including \degree\ 16
(omitting tentative contributions of modular depth four and degree 16):
\begin{align}
\sigmaE_9  &=  -\frac{1}{ 8! } \eee_{10}^{(8)} + \frac{5 s^3_{4,8}}{18}  + \frac{7 s^3_{6,6} }{72} 
+ \frac{s^5_{4,10}}{720}   - \frac{7 s^5_{6,8}}{1440} 
\label{sigexp.0} \\*
&\quad 
+ \frac{34921 s^{2}(\eee_ 4, t^{4}(\eee_4, \eee_6))}{1134000}
+ \frac{ 2587 s^{3}(\eee_ 4, t^{3}(\eee_4, \eee_6)) }{37800} - 
\frac{ 529 s^{4}(\eee_ 4, t^{2}(\eee_4, \eee_6)) }{14400} 
\notag\\*
&\quad   -  \frac{s^7_{6,10}}{30240}  + \frac{s^7_{8,8}}{12096}
+\frac{ s^{5}(\eee_ 4, t^{3}(\eee_4, \eee_8)) }{2592}
+\frac{ 7s^{5}(\eee_ 4, t^{3}(\eee_6, \eee_6))  }{51840 } \notag \\*
&\quad - 
\frac{ 34921 s^{4}(\eee_ 6, t^{4}(\eee_6, \eee_4)) }{47628000} - 
\frac{ 2587 s^{5}(\eee_ 6, t^{3}(\eee_6, \eee_4)) }{1587600} + 
\frac{ 529 s^{6}(\eee_ 6, t^{2}(\eee_6, \eee_4)) }{604800 }  
+\ldots
  \notag \\
\sigmaE_{11} &= -\frac{1}{ 10! } \eee_{12}^{(10)} 
+  \frac{11 s^3_{4,10}}{40} + \frac{11 s^3_{6,8} }{60} 
 + \frac{ 242407}{14735232} s^{2}(\eee_ 4, t^{2}(\eee_4, \eee_6))
  +   \frac{s^5_{4,12} }{720} -\frac{s^5_{6,10}}{216}  - \frac{7 s^5_{8,8}}{4320}  \notag \\
&\quad 
  +\frac{11090423 s^{2}(\eee_ 4, t^{4}(\eee_4, \eee_8)) }{309439872} + 
\frac{ 3197 s^{3}(\eee_ 4, t^{3}(\eee_4, \eee_8))  }{57600} - 
\frac{ 2983 s^{4}(\eee_ 4, t^{2}(\eee_4, \eee_8))  }{86400} \notag \\
&\quad  + 
\frac{ 148753 s^{3}(\eee_ 4, t^{3}(\eee_6, \eee_6))  }{7367616} + 
\frac{ 490853 s^{3}(\eee_ 6, t^{3}(\eee_6, \eee_4))  }{17191104} +
\frac{ 156805 s^{4}(\eee_ 6, t^{2}(\eee_6, \eee_4))  }{14735232}
+\ldots \,.\notag 
\end{align}


\begin{thebibliography}{100}

\bibitem{Green:1999pv}
M.~B. Green and P.~Vanhove, ``{The Low-energy expansion of the one loop type II
  superstring amplitude},''
  \href{http://dx.doi.org/10.1103/PhysRevD.61.104011}{{\em Phys.Rev.} {\bf D61}
  (2000)  104011},
\href{http://arxiv.org/abs/hep-th/9910056}{{\tt arXiv:hep-th/9910056
  [hep-th]}}.

\bibitem{Green:2008uj}
M.~B. Green, J.~G. Russo, and P.~Vanhove, ``{Low energy expansion of the
  four-particle genus-one amplitude in type II superstring theory},''
  \href{http://dx.doi.org/10.1088/1126-6708/2008/02/020}{{\em JHEP} {\bf 02}
  (2008)  020},
\href{http://arxiv.org/abs/0801.0322}{{\tt arXiv:0801.0322 [hep-th]}}.

\bibitem{DHoker:2015gmr}
E.~D'Hoker, M.~B. Green, and P.~Vanhove, ``{On the modular structure of the
  genus-one Type II superstring low energy expansion},''
  \href{http://dx.doi.org/10.1007/JHEP08(2015)041}{{\em JHEP} {\bf 08} (2015)
  041},
\href{http://arxiv.org/abs/1502.06698}{{\tt arXiv:1502.06698 [hep-th]}}.

\bibitem{Gerken:2018jrq}
J.~E. Gerken, A.~Kleinschmidt, and O.~Schlotterer, ``{Heterotic-string
  amplitudes at one loop: modular graph forms and relations to open strings},''
  \href{http://dx.doi.org/10.1007/JHEP01(2019)052}{{\em JHEP} {\bf 01} (2019)
  052},
\href{http://arxiv.org/abs/1811.02548}{{\tt arXiv:1811.02548 [hep-th]}}.

\bibitem{DHoker:2015wxz}
E.~D'Hoker, M.~B. Green, {\"O}.~G{\"u}rdogan, and P.~Vanhove, ``Modular graph
  functions,'' \href{http://dx.doi.org/10.4310/CNTP.2017.v11.n1.a4}{{\em
  Commun. Num. Theor. Phys.} {\bf 11} (2017)  165--218},
\href{http://arxiv.org/abs/1512.06779}{{\tt arXiv:1512.06779 [hep-th]}}.

\bibitem{DHoker:2016mwo}
E.~D'Hoker and M.~B. Green, ``Identities between modular graph forms,''
  \href{http://dx.doi.org/10.1016/j.jnt.2017.11.015}{{\em J. Number Theory}
  {\bf 189} (2018)  25--80},
\href{http://arxiv.org/abs/1603.00839}{{\tt arXiv:1603.00839 [hep-th]}}.

\bibitem{DHoker:2019blr}
E.~D'Hoker and M.~B. Green, ``{Exploring transcendentality in superstring
  amplitudes},'' \href{http://dx.doi.org/10.1007/JHEP07(2019)149}{{\em JHEP}
  {\bf 07} (2019)  149},
\href{http://arxiv.org/abs/1906.01652}{{\tt arXiv:1906.01652 [hep-th]}}.

\bibitem{Dorigoni:2022iem}
D.~Dorigoni, M.~B. Green, and C.~Wen, ``{The SAGEX review on scattering
  amplitudes Chapter 10: Selected topics on modular covariance of type IIB
  string amplitudes and their~~supersymmetric Yang\textendash{}Mills duals},''
  \href{http://dx.doi.org/10.1088/1751-8121/ac9263}{{\em J. Phys. A} {\bf 55}
  (2022) no.~44, 443011}, \href{http://arxiv.org/abs/2203.13021}{{\tt
  arXiv:2203.13021 [hep-th]}}.

\bibitem{Broedel:2018izr}
J.~Broedel, O.~Schlotterer, and F.~Zerbini, ``{From elliptic multiple zeta
  values to modular graph functions: open and closed strings at one loop},''
  \href{http://dx.doi.org/10.1007/JHEP01(2019)155}{{\em JHEP} {\bf 01} (2019)
  155},
\href{http://arxiv.org/abs/1803.00527}{{\tt arXiv:1803.00527 [hep-th]}}.

\bibitem{Gerken:2020xfv}
J.~E. Gerken, A.~Kleinschmidt, C.~R. Mafra, O.~Schlotterer, and B.~Verbeek,
  ``{Towards closed strings as single-valued open strings at genus one},''
  \href{http://dx.doi.org/10.1088/1751-8121/abe58b}{{\em J. Phys. A} {\bf 55}
  (2022) no.~2, 025401}, \href{http://arxiv.org/abs/2010.10558}{{\tt
  arXiv:2010.10558 [hep-th]}}.

\bibitem{Brown:mmv}
F.~Brown, ``{Multiple modular values and the relative completion of the
  fundamental group of $\mathcal{M}_{1,1}$},''
  \href{http://arxiv.org/abs/1407.5167}{{\tt arXiv:1407.5167 [math.NT]}}.

\bibitem{Zerbini:2015rss}
F.~Zerbini, ``{Single-valued multiple zeta values in genus 1 superstring
  amplitudes},'' \href{http://dx.doi.org/10.4310/CNTP.2016.v10.n4.a2}{{\em
  Commun. Num. Theor. Phys.} {\bf 10} (2016)  703--737},
\href{http://arxiv.org/abs/1512.05689}{{\tt arXiv:1512.05689 [hep-th]}}.

\bibitem{Brown:2017qwo}
F.~Brown, ``{A class of non-holomorphic modular forms I},''
  \href{http://dx.doi.org/10.1007/s40687-018-0130-8}{{\em Res. Math. Sci.} {\bf
  5} (2018)  5:7}, \href{http://arxiv.org/abs/1707.01230}{{\tt arXiv:1707.01230
  [math.NT]}}.

\bibitem{Brown:2017qwo2}
F.~Brown, ``{A class of non-holomorphic modular forms II : equivariant iterated
  Eisenstein integrals},'' \href{http://dx.doi.org/10.1017/fms.2020.24}{{\em
  Forum~of~Mathematics,~Sigma} {\bf 8} (2020)  1},
\href{http://arxiv.org/abs/1708.03354}{{\tt arXiv:1708.03354 [math.NT]}}.

\bibitem{Zagier:2019eus}
D.~Zagier and F.~Zerbini, ``{Genus-zero and genus-one string amplitudes and
  special multiple zeta values},''
  \href{http://dx.doi.org/10.4310/CNTP.2020.v14.n2.a4}{{\em Commun. Num. Theor.
  Phys.} {\bf 14} (2020) no.~2, 413--452},
\href{http://arxiv.org/abs/1906.12339}{{\tt arXiv:1906.12339 [math.NT]}}.

\bibitem{Drewitt:2021}
J.~Drewitt, ``Laplace-eigenvalue equations for length three modular iterated
  integrals,''
  \href{http://dx.doi.org/https://doi.org/10.1016/j.jnt.2021.11.005}{{\em
  Journal of Number Theory} {\bf 239} (2022)  78--112},
  \href{http://arxiv.org/abs/2104.09916}{{\tt arXiv:2104.09916 [math.NT]}}.

\bibitem{DHoker:2017zhq}
E.~D'Hoker and W.~Duke, ``Fourier series of modular graph functions,''
  \href{http://dx.doi.org/10.1016/j.jnt.2018.04.012}{{\em J. Number Theory}
  {\bf 192} (2018)  1--36}, \href{http://arxiv.org/abs/1708.07998}{{\tt
  arXiv:1708.07998 [math.NT]}}.

\bibitem{DHoker:2019xef}
E.~D'Hoker and M.~B. Green, ``{Absence of irreducible multiple zeta-values in
  melon modular graph functions},''
  \href{http://dx.doi.org/10.4310/CNTP.2020.v14.n2.a2}{{\em Commun. Num. Theor.
  Phys.} {\bf 14} (2020) no.~2, 315--324},
\href{http://arxiv.org/abs/1904.06603}{{\tt arXiv:1904.06603 [hep-th]}}.

\bibitem{Vanhove:2020qtt}
P.~Vanhove and F.~Zerbini, ``{Building blocks of closed and open string
  amplitudes},'' \href{http://dx.doi.org/10.22323/1.383.0022}{{\em PoS} {\bf
  MA2019} (2022)  022}, \href{http://arxiv.org/abs/2007.08981}{{\tt
  arXiv:2007.08981 [hep-th]}}.

\bibitem{Schnetz:2013hqa}
O.~Schnetz, ``{Graphical functions and single-valued multiple
  polylogarithms},'' \href{http://dx.doi.org/10.4310/CNTP.2014.v8.n4.a1}{{\em
  Commun. Num. Theor. Phys.} {\bf 08} (2014)  589--675},
\href{http://arxiv.org/abs/1302.6445}{{\tt arXiv:1302.6445 [math.NT]}}.

\bibitem{Brown:2013gia}
F.~Brown, ``{Single-valued Motivic Periods and Multiple Zeta Values},''
  \href{http://dx.doi.org/10.1017/fms.2014.18}{{\em SIGMA} {\bf 2} (2014)
  e25},
\href{http://arxiv.org/abs/1309.5309}{{\tt arXiv:1309.5309 [math.NT]}}.

\bibitem{Schlotterer:2012ny}
O.~Schlotterer and S.~Stieberger, ``{Motivic Multiple Zeta Values and
  Superstring Amplitudes},''
  \href{http://dx.doi.org/10.1088/1751-8113/46/47/475401}{{\em J. Phys.} {\bf
  A46} (2013)  475401},
\href{http://arxiv.org/abs/1205.1516}{{\tt arXiv:1205.1516 [hep-th]}}.

\bibitem{Stieberger:2013wea}
S.~Stieberger, ``{Closed superstring amplitudes, single-valued multiple zeta
  values and the Deligne associator},''
  \href{http://dx.doi.org/10.1088/1751-8113/47/15/155401}{{\em J. Phys.} {\bf
  A47} (2014)  155401},
\href{http://arxiv.org/abs/1310.3259}{{\tt arXiv:1310.3259 [hep-th]}}.

\bibitem{Stieberger:2014hba}
S.~Stieberger and T.~R. Taylor, ``{Closed String Amplitudes as Single-Valued
  Open String Amplitudes},''
  \href{http://dx.doi.org/10.1016/j.nuclphysb.2014.02.005}{{\em Nucl. Phys.}
  {\bf B881} (2014)  269--287},
\href{http://arxiv.org/abs/1401.1218}{{\tt arXiv:1401.1218 [hep-th]}}.

\bibitem{Schlotterer:2018abc}
O.~Schlotterer and O.~Schnetz, ``{Closed strings as single-valued open strings:
  A genus-zero derivation},''
  \href{http://dx.doi.org/10.1088/1751-8121/aaea14}{{\em J. Phys.} {\bf A52}
  (2019) no.~4, 045401},
\href{http://arxiv.org/abs/1808.00713}{{\tt arXiv:1808.00713 [hep-th]}}.

\bibitem{Vanhove:2018mzv}
P.~Vanhove and F.~Zerbini, ``{Single-valued hyperlogarithms, correlation
  functions and closed string amplitudes},''
  \href{http://dx.doi.org/10.4310/ATMP.2022.v26.n2.a5}{{\em Adv. Theor. Math.
  Phys.} {\bf 26} (2022)  455--530},
  \href{http://arxiv.org/abs/1812.03018}{{\tt arXiv:1812.03018 [hep-th]}}.

\bibitem{Brown:2019wna}
F.~Brown and C.~Dupont, ``{Single-valued integration and superstring amplitudes
  in genus zero},'' \href{http://dx.doi.org/10.1007/s00220-021-03969-4}{{\em
  Commun. Math. Phys.} {\bf 382} (2021) no.~2, 815--874},
  \href{http://arxiv.org/abs/1910.01107}{{\tt arXiv:1910.01107 [math.NT]}}.

\bibitem{Dorigoni:2022npe}
D.~Dorigoni, M.~Doroudiani, J.~Drewitt, M.~Hidding, A.~Kleinschmidt,
  N.~Matthes, O.~Schlotterer, and B.~Verbeek, ``{Modular graph forms from
  equivariant iterated Eisenstein integrals},''
  \href{http://dx.doi.org/10.1007/JHEP12(2022)162}{{\em JHEP} {\bf 12} (2022)
  162}, \href{http://arxiv.org/abs/2209.06772}{{\tt arXiv:2209.06772
  [hep-th]}}.

\bibitem{Gerken:2019cxz}
J.~E. Gerken, A.~Kleinschmidt, and O.~Schlotterer, ``{All-order differential
  equations for one-loop closed-string integrals and modular graph forms},''
  \href{http://dx.doi.org/10.1007/JHEP01(2020)064}{{\em JHEP} {\bf 01} (2020)
  064},
\href{http://arxiv.org/abs/1911.03476}{{\tt arXiv:1911.03476 [hep-th]}}.

\bibitem{Gerken:2020yii}
J.~E. Gerken, A.~Kleinschmidt, and O.~Schlotterer, ``{Generating series of all
  modular graph forms from iterated Eisenstein integrals},''
  \href{http://dx.doi.org/10.1007/JHEP07(2020)190}{{\em JHEP} {\bf 07} (2020)
  no.~07, 190}, \href{http://arxiv.org/abs/2004.05156}{{\tt arXiv:2004.05156
  [hep-th]}}.

\bibitem{Goncharov:2005sla}
A.~Goncharov, ``{Galois symmetries of fundamental groupoids and noncommutative
  geometry},'' \href{http://dx.doi.org/10.1215/S0012-7094-04-12822-2}{{\em Duke
  Math.J.} {\bf 128} (2005)  209},
\href{http://arxiv.org/abs/math/0208144}{{\tt arXiv:math/0208144 [math.AG]}}.

\bibitem{Brown:2011mot}
F.~Brown, ``{Mixed Tate motives over $\mathbb Z$},''
  \href{http://dx.doi.org/10.4007/annals.2012.175.2.10}{{\em Ann.\ Math.} {\bf
  175} (2012) no.~2, 949--976}, \href{http://arxiv.org/abs/1102.1312}{{\tt
  arXiv:1102.1312 [math.AG]}}.

\bibitem{Brown:2011ik}
F.~Brown, \href{http://dx.doi.org/10.2969/aspm/06310031}{``On the decomposition
  of motivic multiple zeta values,''} in {\em Galois-{T}eichm\"{u}ller theory
  and arithmetic geometry}, vol.~63 of {\em Adv. Stud. Pure Math.}, pp.~31--58.
\newblock Math. Soc. Japan, Tokyo, 2012.
\newblock \href{http://arxiv.org/abs/1102.1310}{{\tt arXiv:1102.1310
  [math.NT]}}.

\bibitem{Dorigoni:2023part1}
D.~Dorigoni, M.~Doroudiani, J.~Drewitt, M.~Hidding, A.~Kleinschmidt,
  O.~Schlotterer, L.~Schneps, and B.~Verbeek, ``{Canonicalizing zeta
  generators: genus zero and genus one},''
  \href{http://arxiv.org/abs/2406.05099}{{\tt arXiv:2406.05099 [math.QA]}}.

\bibitem{Tsunogai}
H.~Tsunogai, ``On some derivations of {L}ie algebras related to {G}alois
  representations,'' \href{http://dx.doi.org/10.2977/prims/1195164794}{{\em
  Publ. Res. Inst. Math. Sci.} {\bf 31} (1995) no.~1, 113--134}.

\bibitem{Pollack}
A.~Pollack, ``{Relations between derivations arising from modular forms}.''
  \url{https://dukespace.lib.duke.edu/dspace/handle/10161/1281}, 2009.
\newblock Undergraduate thesis, Duke University.

\bibitem{Dorigoni:2021ngn}
D.~Dorigoni, A.~Kleinschmidt, and O.~Schlotterer, ``{Poincar\'e series for
  modular graph forms at depth two. Part II. Iterated integrals of cusp
  forms},'' \href{http://dx.doi.org/10.1007/JHEP01(2022)134}{{\em JHEP} {\bf
  01} (2022)  134}, \href{http://arxiv.org/abs/2109.05018}{{\tt
  arXiv:2109.05018 [hep-th]}}.

\bibitem{Brown2019}
F.~Brown, ``{From the Deligne-Ihara conjecture to multiple modular values},''
  \href{http://arxiv.org/abs/1904.00179}{{\tt arXiv:1904.00179 [math.AG]}}.

\bibitem{DeligneTBP}
P.~P. Deligne, ``Le groupe fondamental de la droite projective moins trois
  points,'' in {\em Galois Groups over $\mathbb Q$}, Y.~Ihara, K.~Ribet, and
  J.-P. Serre, eds., pp.~79--297.
\newblock Springer US, New York, NY, 1989.

\bibitem{Ihara:1990}
Y.~Ihara, ``Braids, {G}alois groups, and some arithmetic functions,'' in {\em
  Proceedings of the {I}nternational {C}ongress of {M}athematicians, {V}ol.
  {I}, {II} ({K}yoto, 1990)}, pp.~99--120.
\newblock Math. Soc. Japan, Tokyo, 1991.

\bibitem{IharaTakao:1993}
Y.~Ihara and N.~Takao. {Seminar talk}, 1993.

\bibitem{Gonchtalk}
A.~B. Goncharov, ``Multiple $\zeta$-values, {G}alois groups, and geometry of
  modular varieties,'' in {\em European Congress of Mathematics},
  C.~Casacuberta, R.~M. Mir{\'o}-Roig, J.~Verdera, and S.~Xamb{\'o}-Descamps,
  eds., pp.~361--392.
\newblock Birkh{\"a}user Basel, Basel, 2001.

\bibitem{GKZ:2006}
H.~Gangl, M.~Kaneko, and D.~Zagier,
  \href{http://dx.doi.org/10.1142/9789812774415\_0004}{``Double zeta values and
  modular forms,''} in {\em Automorphic forms and zeta functions}, pp.~71--106.
\newblock World Sci. Publ., Hackensack, NJ, 2006.

\bibitem{Schneps:2006}
L.~Schneps, ``On the {P}oisson bracket on the free {L}ie algebra in two
  generators,'' {\em J. Lie Theory} {\bf 16} (2006) no.~1, 19--37.

\bibitem{Brown:Anatomy}
F.~Brown, ``{Talk ``Anatomy of the motivic Lie algebra'' given at the program
  ``Grothendieck-Teichm\"uller Groups, Deformation and Operads'' (Newton
  Institute, Cambridge, UK)}.'' \url{https://sms.cam.ac.uk/media/1459610},
  2013.

\bibitem{brown_2017}
F.~Brown, ``Zeta elements in depth 3 and the fundamental lie algebra of the
  infinitesimal tate curve,'' \href{http://dx.doi.org/10.1017/fms.2016.29}{{\em
  Forum of Mathematics, Sigma} {\bf 5} (2017)  },
  \href{http://arxiv.org/abs/1504.04737}{{\tt arXiv:1504.04737 [math.NT]}}.

\bibitem{BaumardSchneps:2015}
S.~Baumard and L.~Schneps, ``On the derivation representation of the
  fundamental {L}ie algebra of mixed elliptic motives,''
  \href{http://dx.doi.org/10.1007/s40316-015-0040-8}{{\em Ann. Math. Qu\'{e}.}
  {\bf 41} (2017) no.~1, 43--62}, \href{http://arxiv.org/abs/1510.05549}{{\tt
  arXiv:1510.05549 [math.QA]}}.

\bibitem{hain_matsumoto_2020}
R.~Hain and M.~Matsumoto, ``Universal mixed elliptic motives,''
  \href{http://dx.doi.org/10.1017/S1474748018000130}{{\em Journal of the
  Institute of Mathematics of Jussieu} {\bf 19} (2020) no.~3, 663--766},
  \href{http://arxiv.org/abs/1512.03975}{{\tt arXiv:1512.03975 [math.AG]}}.

\bibitem{Mafra:2022wml}
C.~R. Mafra and O.~Schlotterer, ``{Tree-level amplitudes from the pure spinor
  superstring},'' \href{http://dx.doi.org/10.1016/j.physrep.2023.04.001}{{\em
  Phys. Rept.} {\bf 1020} (2023)  1--162},
  \href{http://arxiv.org/abs/2210.14241}{{\tt arXiv:2210.14241 [hep-th]}}.

\bibitem{Frost:2023stm}
H.~Frost, M.~Hidding, D.~Kamlesh, C.~Rodriguez, O.~Schlotterer, and B.~Verbeek,
  ``{Motivic coaction and single-valued map of polylogarithms from zeta
  generators},'' \href{http://dx.doi.org/10.1088/1751-8121/ad5edf}{{\em J. Phys.} {\bf A57}
  (2024) no.~31, 31LT01},
\href{http://arxiv.org/abs/2312.00697}{{\tt arXiv:2312.00697 [hep-th]}}.  
  

\bibitem{svpolylog}
F.~Brown, ``{Polylogarithmes multiples uniformes en une variable},'' {\em C. R.
  Acad. Sci. Paris} {\bf Ser. I 338} (2004)  527--532.

\bibitem{Ihara:stable}
Y.~Ihara, ``On the stable derivation algebra associated with some braid
  groups,'' {\em Israel Journal of Mathematics} {\bf 80} (1992)  35--153.

\bibitem{Furusho2000TheMZ}
H.~Furusho, ``The multiple zeta value algebra and the stable derivation
  algebra,'' {\em Publ. Res. Inst. Math. Sci.} {\bf 39} (2003) no.~4, 695--720.
  \url{http://projecteuclid.org/euclid.prims/1145476044}.

\bibitem{Manin:1973}
J.~I. Manin, ``Periods of cusp forms, and {$p$}-adic {H}ecke series,'' {\em
  Mat. Sb. (N.S.)} {\bf 92(134)} (1973)  378--401, 503.

\bibitem{Bourjaily:2022bwx}
J.~L. Bourjaily {\em et al.}, ``{Functions Beyond Multiple Polylogarithms for
  Precision Collider Physics},'' in {\em {Snowmass 2021}}.
\newblock 3, 2022.
\newblock \href{http://arxiv.org/abs/2203.07088}{{\tt arXiv:2203.07088
  [hep-ph]}}.

\bibitem{Abreu:2022mfk}
S.~Abreu, R.~Britto, and C.~Duhr, ``{The SAGEX review on scattering amplitudes
  Chapter 3: Mathematical structures in Feynman integrals},''
  \href{http://dx.doi.org/10.1088/1751-8121/ac87de}{{\em J. Phys. A} {\bf 55}
  (2022) no.~44, 443004}, \href{http://arxiv.org/abs/2203.13014}{{\tt
  arXiv:2203.13014 [hep-th]}}.

\bibitem{Gerken:review}
J.~E. Gerken, ``{Modular Graph Forms and Scattering Amplitudes in String
  Theory},'' \href{http://arxiv.org/abs/2011.08647}{{\tt arXiv:2011.08647
  [hep-th]}}.

\bibitem{Berkovits:2022ivl}
N.~Berkovits, E.~D'Hoker, M.~B. Green, H.~Johansson, and O.~Schlotterer,
  ``{Snowmass White Paper: String Perturbation Theory},'' in {\em {2022
  Snowmass Summer Study}}.
\newblock 3, 2022.
\newblock \href{http://arxiv.org/abs/2203.09099}{{\tt arXiv:2203.09099
  [hep-th]}}.

\bibitem{DHoker:2022dxx}
E.~D'Hoker and J.~Kaidi, ``{Lectures on modular forms and strings},''
  \href{http://arxiv.org/abs/2208.07242}{{\tt arXiv:2208.07242 [hep-th]}}.

\bibitem{ZagierF}
D.~Zagier, ``Periods of modular forms and {J}acobi theta functions,''
  \href{http://dx.doi.org/10.1007/BF01245085}{{\em Invent. Math.} {\bf 104}
  (1991) no.~3, 449--465}.

\bibitem{Maass}
H.~Maass, {\em Lectures on modular functions of one complex variable}, vol.~29
  of {\em Tata Institute of Fundamental Research Lectures on Mathematics and
  Physics}.
\newblock Tata Institute of Fundamental Research, Bombay, second~ed., 1983.
\newblock With notes by Sunder Lal.

\bibitem{Gerken:2018zcy}
J.~E. Gerken and J.~Kaidi, ``{Holomorphic subgraph reduction of higher-point
  modular graph forms},'' \href{http://dx.doi.org/10.1007/JHEP01(2019)131}{{\em
  JHEP} {\bf 01} (2019)  131},
\href{http://arxiv.org/abs/1809.05122}{{\tt arXiv:1809.05122 [hep-th]}}.

\bibitem{Basu:2015ayg}
A.~Basu, ``{Poisson equation for the Mercedes diagram in string theory at genus
  one},'' \href{http://dx.doi.org/10.1088/0264-9381/33/5/055005}{{\em Class.
  Quant. Grav.} {\bf 33} (2016) no.~5, 055005},
\href{http://arxiv.org/abs/1511.07455}{{\tt arXiv:1511.07455 [hep-th]}}.

\bibitem{Kleinschmidt:2017ege}
A.~Kleinschmidt and V.~Verschinin, ``{Tetrahedral modular graph functions},''
  \href{http://dx.doi.org/10.1007/JHEP09(2017)155}{{\em JHEP} {\bf 09} (2017)
  155},
\href{http://arxiv.org/abs/1706.01889}{{\tt arXiv:1706.01889 [hep-th]}}.

\bibitem{DHoker:2015sve}
E.~D'Hoker, M.~B. Green, and P.~Vanhove, ``{Proof of a modular relation between
  1-, 2- and 3-loop Feynman diagrams on a torus},''
  \href{http://dx.doi.org/10.1016/j.jnt.2017.07.022}{{\em J.\ Number Theory}
  (2018)  381},
\href{http://arxiv.org/abs/1509.00363}{{\tt arXiv:1509.00363 [hep-th]}}.

\bibitem{Basu:2016kli}
A.~Basu, ``{Proving relations between modular graph functions},''
  \href{http://dx.doi.org/10.1088/0264-9381/33/23/235011}{{\em Class. Quant.
  Grav.} {\bf 33} (2016) no.~23, 235011},
\href{http://arxiv.org/abs/1606.07084}{{\tt arXiv:1606.07084 [hep-th]}}.

\bibitem{DHoker:2016quv}
E.~D'Hoker and J.~Kaidi, ``{Hierarchy of Modular Graph Identities},''
  \href{http://dx.doi.org/10.1007/JHEP11(2016)051}{{\em JHEP} {\bf 11} (2016)
  051},
\href{http://arxiv.org/abs/1608.04393}{{\tt arXiv:1608.04393 [hep-th]}}.

\bibitem{Richards:2008jg}
D.~M. Richards, ``{The One-Loop Five-Graviton Amplitude and the Effective
  Action},'' \href{http://dx.doi.org/10.1088/1126-6708/2008/10/042}{{\em JHEP}
  {\bf 10} (2008)  042},
\href{http://arxiv.org/abs/0807.2421}{{\tt arXiv:0807.2421 [hep-th]}}.

\bibitem{Green:2013bza}
M.~B. Green, C.~R. Mafra, and O.~Schlotterer, ``{Multiparticle one-loop
  amplitudes and S-duality in closed superstring theory},''
  \href{http://dx.doi.org/10.1007/JHEP10(2013)188}{{\em JHEP} {\bf 10} (2013)
  188},
\href{http://arxiv.org/abs/1307.3534}{{\tt arXiv:1307.3534 [hep-th]}}.

\bibitem{Basu:2016mmk}
A.~Basu, ``{Simplifying the one loop five graviton amplitude in type IIB string
  theory},'' \href{http://dx.doi.org/10.1142/S0217751X17500749}{{\em Int. J.
  Mod. Phys.} {\bf A32} (2017) no.~14, 1750074},
\href{http://arxiv.org/abs/1608.02056}{{\tt arXiv:1608.02056 [hep-th]}}.

\bibitem{Gerken:2020aju}
J.~E. Gerken, ``{Basis Decompositions and a Mathematica Package for Modular
  Graph Forms},'' \href{http://dx.doi.org/10.1088/1751-8121/abbdf2}{{\em J.
  Phys. A} {\bf 54} (2021) no.~19, 195401},
  \href{http://arxiv.org/abs/2007.05476}{{\tt arXiv:2007.05476 [hep-th]}}.

\bibitem{Ganglzagier}
D.~Zagier and H.~Gangl, ``Classical and elliptic polylogarithms and special
  values of {$L$}-series,'' in {\em The arithmetic and geometry of algebraic
  cycles ({B}anff, {AB}, 1998)}, vol.~548 of {\em NATO Sci. Ser. C Math. Phys.
  Sci.}, pp.~561--615.
\newblock Kluwer Acad. Publ., Dordrecht, 2000.

\bibitem{Nilsnewarticle}
N.~Matthes, ``On the algebraic structure of iterated integrals of quasimodular
  forms,'' \href{http://dx.doi.org/10.2140/ant.2017.11.2113}{{\em Algebra \&
  Number Theory} {\bf 11-9} (2017)  2113--2130},
  \href{http://arxiv.org/abs/1708.04561}{{\tt arXiv:1708.04561}}.

\bibitem{Dorigoni:2021jfr}
D.~Dorigoni, A.~Kleinschmidt, and O.~Schlotterer, ``{Poincar\'e series for
  modular graph forms at depth two. Part I. Seeds and Laplace systems},''
  \href{http://dx.doi.org/10.1007/JHEP01(2022)133}{{\em JHEP} {\bf 01} (2022)
  133}, \href{http://arxiv.org/abs/2109.05017}{{\tt arXiv:2109.05017
  [hep-th]}}.

\bibitem{Broedel:2015hia}
J.~Broedel, N.~Matthes, and O.~Schlotterer, ``{Relations between elliptic
  multiple zeta values and a special derivation algebra},''
  \href{http://dx.doi.org/10.1088/1751-8113/49/15/155203}{{\em J. Phys.} {\bf
  A49} (2016) no.~15, 155203},
\href{http://arxiv.org/abs/1507.02254}{{\tt arXiv:1507.02254 [hep-th]}}.

\bibitem{LNT}
J.-G. Luque, J.-C. Novelli, and J.-Y. Thibon, ``{Period polynomials and Ihara
  brackets},'' {\em J. Lie Theory} {\bf 17} (2007)  229--239,
  \href{http://arxiv.org/abs/math/0606301}{{\tt arXiv:math/0606301
  [math.CO,math.NT]}}.

\bibitem{Hain}
R.~Hain, ``Notes on the universal elliptic {KZB} connection,''
  \href{http://dx.doi.org/10.4310/PAMQ.2020.v16.n2.a2}{{\em Pure Appl. Math.
  Q.} {\bf 16} (2020) no.~2, 229--312},
  \href{http://arxiv.org/abs/1309.0580}{{\tt arXiv:1309.0580 [math.AG]}}.

\bibitem{WWWe}
J.~Broedel, N.~Matthes, and O.~Schlotterer.
  \url{https://tools.aei.mpg.de/emzv}.

\bibitem{KZB}
D.~Calaque, B.~Enriquez, and P.~Etingof,
  \href{http://dx.doi.org/10.1007/978-0-8176-4745-2\_5}{``Universal {KZB}
  equations: the elliptic case,''} in {\em Algebra, arithmetic, and geometry:
  in honor of {Y}u. {I}. {M}anin. {V}ol. {I}}, vol.~269 of {\em Progr. Math.},
  pp.~165--266.
\newblock Birkh\"{a}user Boston, Boston, MA, 2009.
\newblock \href{http://arxiv.org/abs/math/0702670}{{\tt arXiv:math/0702670}}.

\bibitem{EnriquezEllAss}
B.~Enriquez, ``Elliptic associators,''
  \href{http://dx.doi.org/10.1007/s00029-013-0137-3}{{\em Selecta Math. (N.S.)}
  {\bf 20} (2014) no.~2, 491--584}, \href{http://arxiv.org/abs/1003.1012}{{\tt
  arXiv:1003.1012 [math.QA]}}.

\bibitem{Schneps:2015mzv}
L.~Schneps, ``{Elliptic double shuffle, Grothendieck-Teichm{\"u}ller and mould
  theory},'' \href{http://dx.doi.org/10.1007/s40316-020-00141-7}{{\em Ann.
  Math. Qu\'ebec} {\bf 44(2)} (2020)  261--289},
  \href{http://arxiv.org/abs/1506.09050}{{\tt arXiv:1506.09050 [math.NT]}}.

\bibitem{brown2017a}
F.~Brown, ``A class of non-holomorphic modular forms {{III}}: Real analytic
  cusp forms for {$\mathrm{SL}_2(\mathbb{Z})$},''
  \href{http://dx.doi.org/10.1007/s40687-018-0151-3}{{\em Research in the
  Mathematical Sciences} {\bf 5} (2018) no.~3, Paper No. 34, 36},
  \href{http://arxiv.org/abs/1710.07912}{{\tt arXiv:1710.07912}}.

\bibitem{Abreu:2017enx}
S.~Abreu, R.~Britto, C.~Duhr, and E.~Gardi, ``{Algebraic Structure of Cut
  Feynman Integrals and the Diagrammatic Coaction},''
  \href{http://dx.doi.org/10.1103/PhysRevLett.119.051601}{{\em Phys. Rev.
  Lett.} {\bf 119} (2017) no.~5, 051601},
  \href{http://arxiv.org/abs/1703.05064}{{\tt arXiv:1703.05064 [hep-th]}}.

\bibitem{Abreu:2017mtm}
S.~Abreu, R.~Britto, C.~Duhr, and E.~Gardi, ``{Diagrammatic Hopf algebra of cut
  Feynman integrals: the one-loop case},''
  \href{http://dx.doi.org/10.1007/JHEP12(2017)090}{{\em JHEP} {\bf 12} (2017)
  090}, \href{http://arxiv.org/abs/1704.07931}{{\tt arXiv:1704.07931
  [hep-th]}}.

\bibitem{Britto:2021prf}
R.~Britto, S.~Mizera, C.~Rodriguez, and O.~Schlotterer, ``{Coaction and
  double-copy properties of configuration-space integrals at genus zero},''
  \href{http://dx.doi.org/10.1007/JHEP05(2021)053}{{\em JHEP} {\bf 05} (2021)
  053}, \href{http://arxiv.org/abs/2102.06206}{{\tt arXiv:2102.06206
  [hep-th]}}.

\bibitem{Saad:2020mzv}
A.~Saad, ``{Multiple zeta values and iterated Eisenstein integrals},''
  \href{http://arxiv.org/abs/2009.09885}{{\tt arXiv:2009.09885 [math.NT]}}.

\bibitem{LMSonEMZV}
P.~Lochak, N.~Matthes, and L.~Schneps, ``Elliptic multizetas and the elliptic
  double shuffle relations,''
  \href{http://dx.doi.org/10.1093/imrn/rnaa060}{{\em International Mathematics
  Research Notices} {\bf 2021} (2021)  695--753},
  \href{http://arxiv.org/abs/1703.09410}{{\tt arXiv:1703.09410 [math.NT]}}.

\bibitem{Fleig:2015vky}
P.~Fleig, H.~P.~A. Gustafsson, A.~Kleinschmidt, and D.~Persson, {\em Eisenstein
  series and automorphic representations}, vol.~176 of {\em Cambridge Studies
  in Advanced Mathematics}.
\newblock Cambridge University Press, Cambridge, 2018.
\newblock \href{http://arxiv.org/abs/1511.04265}{{\tt arXiv:1511.04265
  [math.NT]}}.
\newblock With applications in string theory.

\bibitem{Klinger-Logan:2018sjt}
K.~Klinger-Logan, ``{Differential equations in automorphic forms},''
  \href{http://dx.doi.org/10.4310/CNTP.2018.v12.n4.a4}{{\em Commun. Num. Theor.
  Phys.} {\bf 12} (2018)  767--827}.

\bibitem{Green:2014yxa}
M.~B. Green, S.~D. Miller, and P.~Vanhove, ``{$SL(2, \mathbb{Z})$-invariance
  and D-instanton contributions to the $D^6 R^4$ interaction},''
  \href{http://dx.doi.org/10.4310/CNTP.2015.v9.n2.a3}{{\em Commun. Num. Theor.
  Phys.} {\bf 09} (2015)  307--344},
\href{http://arxiv.org/abs/1404.2192}{{\tt arXiv:1404.2192 [hep-th]}}.

\bibitem{Fedosova:2023cab}
K.~Fedosova, K.~Klinger-Logan, and D.~Radchenko, ``{Convolution identities for
  divisor sums and modular forms},''
  \href{http://arxiv.org/abs/2312.00722}{{\tt arXiv:2312.00722 [math.NT]}}.

\bibitem{Alday:2023pet}
L.~F. Alday, S.~M. Chester, D.~Dorigoni, M.~B. Green, and C.~Wen, ``{Relations
  between integrated correlators in $ \mathcal{N} $ = 4 supersymmetric
  Yang-Mills theory},'' \href{http://dx.doi.org/10.1007/JHEP05(2024)044}{{\em
  JHEP} {\bf 05} (2024)  044}, \href{http://arxiv.org/abs/2310.12322}{{\tt
  arXiv:2310.12322 [hep-th]}}.

\bibitem{Dorigoni:2022bcx}
D.~Dorigoni, A.~Kleinschmidt, and R.~Treilis, ``{To the cusp and back:
  resurgent analysis for modular graph functions},''
  \href{http://dx.doi.org/10.1007/JHEP11(2022)048}{{\em JHEP} {\bf 11} (2022)
  048}, \href{http://arxiv.org/abs/2208.14087}{{\tt arXiv:2208.14087
  [hep-th]}}.

\bibitem{Dorigoni:2023nhc}
D.~Dorigoni and R.~Treilis, ``{Two string theory flavours of generalised
  Eisenstein series},'' \href{http://dx.doi.org/10.1007/JHEP11(2023)102}{{\em
  JHEP} {\bf 11} (2023)  102}, \href{http://arxiv.org/abs/2307.07491}{{\tt
  arXiv:2307.07491 [hep-th]}}.

\bibitem{Eichler:1957}
M.~Eichler, ``Eine {V}erallgemeinerung der {A}belschen {I}ntegrale,''
  \href{http://dx.doi.org/10.1007/BF01258863}{{\em Math. Z.} {\bf 67} (1957)
  267--298}.

\bibitem{Shimura:1959}
G.~Shimura, ``Sur les int\'{e}grales attach\'{e}es aux formes automorphes,''
  \href{http://dx.doi.org/10.2969/jmsj/01140291}{{\em J. Math. Soc. Japan} {\bf
  11} (1959)  291--311}.

\bibitem{DHoker:2018mys}
E.~D'Hoker, M.~B. Green, and B.~Pioline, ``Asymptotics of the {$D^8{\cal R}^4$}
  genus-two string invariant,''
  \href{http://dx.doi.org/10.4310/CNTP.2019.v13.n2.a3}{{\em Commun. Num. Theor.
  Phys.} {\bf 13} (2019) no.~2, 351--462},
\href{http://arxiv.org/abs/1806.02691}{{\tt arXiv:1806.02691 [hep-th]}}.

\bibitem{Dhoker:2020gdz}
E.~D'Hoker, A.~Kleinschmidt, and O.~Schlotterer, ``{Elliptic modular graph
  forms. Part I. Identities and generating series},''
  \href{http://dx.doi.org/10.1007/JHEP03(2021)151}{{\em JHEP} {\bf 03} (2021)
  151}, \href{http://arxiv.org/abs/2012.09198}{{\tt arXiv:2012.09198
  [hep-th]}}.

\bibitem{new:eMGF}
M.~Hidding, O.~Schlotterer, and B.~Verbeek, ``{Elliptic modular graph forms II:
  Iterated integrals},'' \href{http://arxiv.org/abs/2208.11116}{{\tt
  arXiv:2208.11116 [hep-th]}}.

\bibitem{DHoker:2017pvk}
E.~D'Hoker, M.~B. Green, and B.~Pioline, ``{Higher genus modular graph
  functions, string invariants, and their exact asymptotics},''
  \href{http://dx.doi.org/10.1007/s00220-018-3244-3}{{\em Commun. Math. Phys.}
  {\bf 366} (2019) no.~3, 927--979},
\href{http://arxiv.org/abs/1712.06135}{{\tt arXiv:1712.06135 [hep-th]}}.

\bibitem{DHoker:2020uid}
E.~D'Hoker and O.~Schlotterer, ``{Identities among higher genus modular graph
  tensors},'' \href{http://dx.doi.org/10.4310/CNTP.2022.v16.n1.a2}{{\em Commun.
  Num. Theor. Phys.} {\bf 16} (2022) no.~1, 35--74},
  \href{http://arxiv.org/abs/2010.00924}{{\tt arXiv:2010.00924 [hep-th]}}.

\bibitem{Pioline:2015qha}
B.~Pioline, ``{A Theta lift representation for the Kawazumi-Zhang and Faltings
  invariants of genus-two Riemann surfaces},''
  \href{http://dx.doi.org/10.1016/j.jnt.2015.12.021}{{\em J. Number Theor.}
  {\bf 163} (2016)  520--541},
\href{http://arxiv.org/abs/1504.04182}{{\tt arXiv:1504.04182 [hep-th]}}.

\bibitem{Enriquez:Emzv}
B.~Enriquez, ``Analogues elliptiques des nombres multiz\'etas,''
  \href{http://dx.doi.org/10.24033/bsmf.2718}{{\em Bull. Soc. Math. France}
  {\bf 144} (2016) no.~3, 395--427}, \href{http://arxiv.org/abs/1301.3042}{{\tt
  arXiv:1301.3042 [math.NT]}}.

\bibitem{Mafra:2019ddf}
C.~R. Mafra and O.~Schlotterer, ``{All-order alpha'-expansion of one-loop
  open-string integrals},''
  \href{http://dx.doi.org/10.1103/PhysRevLett.124.101603}{{\em Phys. Rev.
  Lett.} {\bf 124} (2020) no.~10, 101603},
\href{http://arxiv.org/abs/1908.09848}{{\tt arXiv:1908.09848 [hep-th]}}.

\bibitem{Mafra:2019xms}
C.~R. Mafra and O.~Schlotterer, ``{One-loop open-string integrals from
  differential equations: all-order $\alpha$'-expansions at $n$ points},''
  \href{http://dx.doi.org/10.1007/JHEP03(2020)007}{{\em JHEP} {\bf 03} (2020)
  007},
\href{http://arxiv.org/abs/1908.10830}{{\tt arXiv:1908.10830 [hep-th]}}.

\bibitem{Stieberger:2022lss}
S.~Stieberger, ``{A Relation between One-Loop Amplitudes of Closed and Open
  Strings (One-Loop KLT Relation)},''
  \href{http://arxiv.org/abs/2212.06816}{{\tt arXiv:2212.06816 [hep-th]}}.

\bibitem{Stieberger:2023nol}
S.~Stieberger, ``{One-Loop Double Copy Relation in String Theory},''
  \href{http://dx.doi.org/10.1103/PhysRevLett.132.191602}{{\em Phys. Rev.
  Lett.} {\bf 132} (2024) no.~19, 191602},
  \href{http://arxiv.org/abs/2310.07755}{{\tt arXiv:2310.07755 [hep-th]}}.

\bibitem{Bhardwaj:2023vvm}
R.~Bhardwaj, A.~Pokraka, L.~Ren, and C.~Rodriguez, ``{A double copy from
  twisted (co)homology at genus one},''
  \href{http://dx.doi.org/10.1007/JHEP07(2024)040}{{\em JHEP} {\bf 07} (2024)
  040}, \href{http://arxiv.org/abs/2312.02148}{{\tt arXiv:2312.02148 [hep-th]}}.  

  
\bibitem{Mazloumi:2024wys}
P.~Mazloumi and S.~Stieberger, ``{One-loop Double Copy Relation from Twisted
  (Co)homology},'' \href{http://arxiv.org/abs/2403.05208}{{\tt arXiv:2403.05208
  [hep-th]}}.

\bibitem{DHoker:2019mib}
E.~D'Hoker, ``{Integral of two-loop modular graph functions},''
  \href{http://dx.doi.org/10.1007/JHEP06(2019)092}{{\em JHEP} {\bf 06} (2019)
  092},
\href{http://arxiv.org/abs/1905.06217}{{\tt arXiv:1905.06217 [hep-th]}}.

\bibitem{DHoker:2021ous}
E.~D'Hoker and N.~Geiser, ``{Integrating three-loop modular graph functions and
  transcendentality of string amplitudes},''
  \href{http://dx.doi.org/10.1007/JHEP02(2022)019}{{\em JHEP} {\bf 02} (2022)
  019}, \href{http://arxiv.org/abs/2110.06237}{{\tt arXiv:2110.06237
  [hep-th]}}.

\bibitem{Doroudiani:2023bfw}
M.~Doroudiani, ``{Integral of depth zero to three basis of Modular Graph
  Functions},'' \href{http://dx.doi.org/10.1007/JHEP07(2024)029}{{\em JHEP} {\bf
  07} (2024)  029}, \href{http://arxiv.org/abs/2311.07287}{{\tt
  arXiv:2311.07287 [hep-th]}}.

\bibitem{zagier1981rankin}
D.~Zagier, ``The rankin-selberg method for automorphic functions which are not
  of rapid decay,'' {\em Journal of the Faculty of Science, the University of
  Tokyo. Sect. 1 A, Mathematics} {\bf 28} (1982)  415--437.
  \url{https://api.semanticscholar.org/CorpusID:122093412}.

\bibitem{Angelantonj:2011br}
C.~Angelantonj, I.~Florakis, and B.~Pioline, ``{A new look at one-loop
  integrals in string theory},''
  \href{http://dx.doi.org/10.4310/CNTP.2012.v6.n1.a4}{{\em Commun. Num. Theor.
  Phys.} {\bf 6} (2012)  159--201},
\href{http://arxiv.org/abs/1110.5318}{{\tt arXiv:1110.5318 [hep-th]}}.

\bibitem{Angelantonj:2012gw}
C.~Angelantonj, I.~Florakis, and B.~Pioline, ``{One-Loop BPS amplitudes as
  BPS-state sums},'' \href{http://dx.doi.org/10.1007/JHEP06(2012)070}{{\em
  JHEP} {\bf 06} (2012)  070}, \href{http://arxiv.org/abs/1203.0566}{{\tt
  arXiv:1203.0566 [hep-th]}}.

\bibitem{Basu:2016xrt}
A.~Basu, ``{Poisson equation for the three loop ladder diagram in string theory
  at genus one},'' \href{http://dx.doi.org/10.1142/S0217751X16501694}{{\em Int.
  J. Mod. Phys.} {\bf A31} (2016) no.~32, 1650169},
\href{http://arxiv.org/abs/1606.02203}{{\tt arXiv:1606.02203 [hep-th]}}.

\bibitem{Diamantis:2020}
N.~Diamantis, ``Modular iterated integrals associated with cusp forms,''
  \href{http://dx.doi.org/10.1515/forum-2021-0224}{{\em Forum Mathematicum}
  {\bf 34} (2022)  157--174}, \href{http://arxiv.org/abs/2009.07128}{{\tt
  arXiv:2009.07128 [math.NT]}}.

\bibitem{Blumlein:2009cf}
J.~Bl{\"u}mlein, D.~J. Broadhurst, and J.~A.~M. Vermaseren, ``{The Multiple
  Zeta Value Data Mine},''
  \href{http://dx.doi.org/10.1016/j.cpc.2009.11.007}{{\em Comput. Phys.
  Commun.} {\bf 181} (2010)  582--625},
\href{http://arxiv.org/abs/0907.2557}{{\tt arXiv:0907.2557 [math-ph]}}.

\bibitem{Brown2014MotivicPA}
F.~Brown, ``Motivic periods and the projective line minus three points,'' in
  {\em {Proceedings of the ICM 2014}}.
\newblock 2014.
\newblock \href{http://arxiv.org/abs/1407.5165}{{\tt arXiv:1407.5165
  [math.NT]}}.
\newblock \url{https://api.semanticscholar.org/CorpusID:118359180}.

\bibitem{brown2015notes}
F.~Brown, ``Notes on motivic periods,''
  \href{http://dx.doi.org/10.4310/CNTP.2017.v11.n3.a2}{{\em Commun. Number
  Theory Phys.} {\bf 11} (2017) no.~3, 557--655},
  \href{http://arxiv.org/abs/1512.06410}{{\tt arXiv:1512.06410 [math.NT]}}.

\bibitem{goncharov2001multiple}
A.~B. Goncharov, ``Multiple polylogarithms and mixed tate motives,''
  \href{http://arxiv.org/abs/math/0103059}{{\tt arXiv:math/0103059 [math.AG]}}.

\bibitem{Broadhurst:1996kc}
D.~J. Broadhurst and D.~Kreimer, ``{Association of multiple zeta values with
  positive knots via Feynman diagrams up to 9 loops},''
  \href{http://dx.doi.org/10.1016/S0370-2693(96)01623-1}{{\em Phys. Lett. B}
  {\bf 393} (1997)  403--412}, \href{http://arxiv.org/abs/hep-th/9609128}{{\tt
  arXiv:hep-th/9609128}}.

\bibitem{Jianqiang}
J.~Zhao, {\em {Multiple zeta functions, multiple polylogarithms, and their
  special values}}.
\newblock World Scientific, New Jersey, 2016.

\bibitem{GilFresan}
J.~I.~B. Gil and J.~Fresan, {\em {Multiple zeta values: from numbers to
  motives}}.
\newblock Clay Mathematics Proceedings, to appear.
\newblock \url{http://javier.fresan.perso.math.cnrs.fr/mzv.pdf}.

\bibitem{Broedel:2016kls}
J.~Broedel, M.~Sprenger, and A.~Torres~Orjuela, ``{Towards single-valued
  polylogarithms in two variables for the seven-point remainder function in
  multi-Regge-kinematics},''
  \href{http://dx.doi.org/10.1016/j.nuclphysb.2016.12.016}{{\em Nucl. Phys. B}
  {\bf 915} (2017)  394--413}, \href{http://arxiv.org/abs/1606.08411}{{\tt
  arXiv:1606.08411 [hep-th]}}.

\bibitem{DelDuca:2016lad}
V.~Del~Duca, S.~Druc, J.~Drummond, C.~Duhr, F.~Dulat, R.~Marzucca,
  G.~Papathanasiou, and B.~Verbeek, ``{Multi-Regge kinematics and the moduli
  space of Riemann spheres with marked points},''
  \href{http://dx.doi.org/10.1007/JHEP08(2016)152}{{\em JHEP} {\bf 08} (2016)
  152}, \href{http://arxiv.org/abs/1606.08807}{{\tt arXiv:1606.08807
  [hep-th]}}.

\end{thebibliography}

\providecommand{\href}[2]{#2}\begingroup\raggedright\endgroup

\end{document}